\RequirePackage[l2tabu]{nag}		
\documentclass[a4paper,14pt,leqno,openbib,oldfontcommands]{memoir} 
\usepackage{dsfont}
\usepackage{datetime}
\usepackage{ifpdf}
\ifpdf
\fi
\ifdraftdoc 
	\usepackage{draftwatermark}				
	\SetWatermarkScale{0.3}
	\SetWatermarkText{\bf Draft: \today}
\fi
\newsubfloat{figure}
\newsubfloat{table}
\settrimmedsize{297mm}{210mm}{*}
\settypeblocksize{634pt}{448.13pt}{*} 
\setulmargins{4cm}{*}{*} 
\setlrmargins{*}{*}{1.5} 
\setmarginnotes{17pt}{51pt}{\onelineskip} 
\setheadfoot{\onelineskip}{2\onelineskip} 
\setheaderspaces{*}{2\onelineskip}{*} 
\checkandfixthelayout
\usepackage{fouriernc}
\usepackage[T1]{fontenc}
\OnehalfSpacing 
\setsecnumdepth{subsection} 
\maxsecnumdepth{subsubsection}
\usepackage{calc,soul,fourier}
\makeatletter 
\newlength\dlf@normtxtw 
\newsavebox{\feline@chapter} 
\newcommand\feline@chapter@marker[1][4cm]{%
	\sbox\feline@chapter{%
		\resizebox{!}{#1}{\fboxsep=1pt%
			\colorbox{gray}{\color{white}\thechapter}%
		}}%
		\rotatebox{90}{%
			\resizebox{%
				\heightof{\usebox{\feline@chapter}}+\depthof{\usebox{\feline@chapter}}}%
			{!}{\scshape\so\@chapapp}}\quad%
		\raisebox{\depthof{\usebox{\feline@chapter}}}{\usebox{\feline@chapter}}%
} 
\newcommand\feline@chm[1][4cm]{%
	\sbox\feline@chapter{\feline@chapter@marker[#1]}%
	\makebox[0pt][c]{
		\makebox[1cm][r]{\usebox\feline@chapter}%
	}}
\makechapterstyle{daleifmodif}{

	\renewcommand\printchapternum{\null\hfill\feline@chm[2.5cm]\par}

} 
\makeatother 
\chapterstyle{daleifmodif}
\makepagestyle{myvf} 
\makeoddfoot{myvf}{}{\thepage}{} 
\makeevenfoot{myvf}{}{\thepage}{} 
\makeheadrule{myvf}{\textwidth}{\normalrulethickness} 
\makeevenhead{myvf}{\small\textsc{\leftmark}}{}{} 
\makeoddhead{myvf}{}{}{\small\textsc{\rightmark}}
\newcommand{\clearemptydoublepage}{\newpage{\thispagestyle{empty}\cleardoublepage}}
\makeindex
\usepackage{import}

\usepackage{lipsum}					
\usepackage{amsfonts} 					
\usepackage[centertags]{amsmath}			
\usepackage{stmaryrd}					
\usepackage{amssymb}					
\usepackage{amsthm}					
\usepackage{newlfont}					
\usepackage{layouts}					
\usepackage{graphicx}					
\usepackage{longtable,rotating}			
\usepackage[utf8]{inputenc}			
\usepackage{colortbl}					
\usepackage{wasysym}					
\usepackage{mathrsfs}					
\usepackage{float}						
\usepackage{verbatim}					
\usepackage{upgreek }					
\usepackage{latexsym}					
\usepackage[square,numbers,
		     sort&compress]{natbib}		
\usepackage{url}						
\usepackage[spanish,english]{babel}		
\usepackage{color}                    				
\usepackage[colorlinks=true,
		     allcolors=black]{hyperref}              
\usepackage{memhfixc}					
\usepackage{enumerate}					
\usepackage{footnote}					
\usepackage{microtype}					
\usepackage{rotfloat}					
\usepackage{alltt}						
\usepackage[version=0.96]{pgf}			
\usepackage{tikz}						
\usetikzlibrary{arrows,shapes,snakes,
		       automata,backgrounds,
		       petri,topaths}				
\newcommand{\pgftextcircled}[1]{                                                                    
    \setbox0=\hbox{#1}%
    \dimen0\wd0%
    \divide\dimen0 by 2%
    \begin{tikzpicture}[baseline=(a.base)]%
        \useasboundingbox (-\the\dimen0,0pt) rectangle (\the\dimen0,1pt);
        \node[circle,draw,outer sep=0pt,inner sep=0.1ex] (a) {#1};
    \end{tikzpicture}
}
\newcommand{\blackged}{\hfill$\blacksquare$}
\newcommand{\whiteged}{\hfill$\square$}
\newcounter{proofcount}

\let\oldsqrt\sqrt
\def\sqrt{\mathpalette\DHLhksqrt}
\def\DHLhksqrt#1#2{%
\setbox0=\hbox{$#1\oldsqrt{#2\,}$}\dimen0=\ht0
\advance\dimen0-0.2\ht0
\setbox2=\hbox{\vrule height\ht0 depth -\dimen0}%
{\box0\lower0.4pt\box2}}
\newcommand{\mycaption}[2][\@empty]{
	\captionnamefont{\scshape} 
	\changecaptionwidth
	\captionwidth{0.9\linewidth}
	\captiondelim{.\:} 
	\indentcaption{0.75cm}
	\captionstyle[\centering]{}
	\setlength{\belowcaptionskip}{10pt}
	\ifx \@empty#1 \caption{#2}\else \caption[#1]{#2}
}
\newcommand{\mysubcaption}[2][\@empty]{
	\subcaptionsize{\small}
	\hangsubcaption
	\subcaptionlabelfont{\rmfamily}
	\sidecapstyle{\raggedright}
	\setlength{\belowcaptionskip}{10pt}
	\ifx \@empty#1 \subcaption{#2}\else \subcaption[#1]{#2}
}
\usepackage{lettrine}
\newcommand{\initial}[1]{%
	\lettrine[lines=3,lhang=0.33,nindent=0em]{
		\color{gray}
     		{\textsc{#1}}}{}}
\theoremstyle{plain}

\theoremstyle{plain}

\theoremstyle{plain}
\theoremstyle{definition}

\theoremstyle{plain}

\theoremstyle{plain}

\theoremstyle{plain}

\usepackage{lipsum}
\begin{document}
%
%
%
%
%
\frontmatter
\pagenumbering{roman}
%
%
%
%
%
%
\begin{titlingpage}
\begin{SingleSpace}
\calccentering{\unitlength} 
\begin{adjustwidth*}{\unitlength}{-\unitlength}
\vspace*{13mm}
\begin{center}
\rule[0.5ex]{\linewidth}{2pt}\vspace*{-\baselineskip}\vspace*{3.2pt}
\rule[0.5ex]{\linewidth}{1pt}\\[\baselineskip]
{\HUGE PERTURBATIVE ASPECTS OF}\\ 
\vglue4mm
{\HUGE THE DECONFINEMENT TRANSITION}\\[4mm]
\vglue2mm
{\Large \textit{-- Physics beyond the Faddeev-Popov model --}}\\
\rule[0.5ex]{\linewidth}{1pt}\vspace*{-\baselineskip}\vspace{3.2pt}
\rule[0.5ex]{\linewidth}{2pt}\\
\vspace{6.5mm}
{\large By}\\
\vspace{6.5mm}
{\large\textsc{Urko Reinosa}}\\
\vspace{11mm}
\includegraphics[scale=0.5]{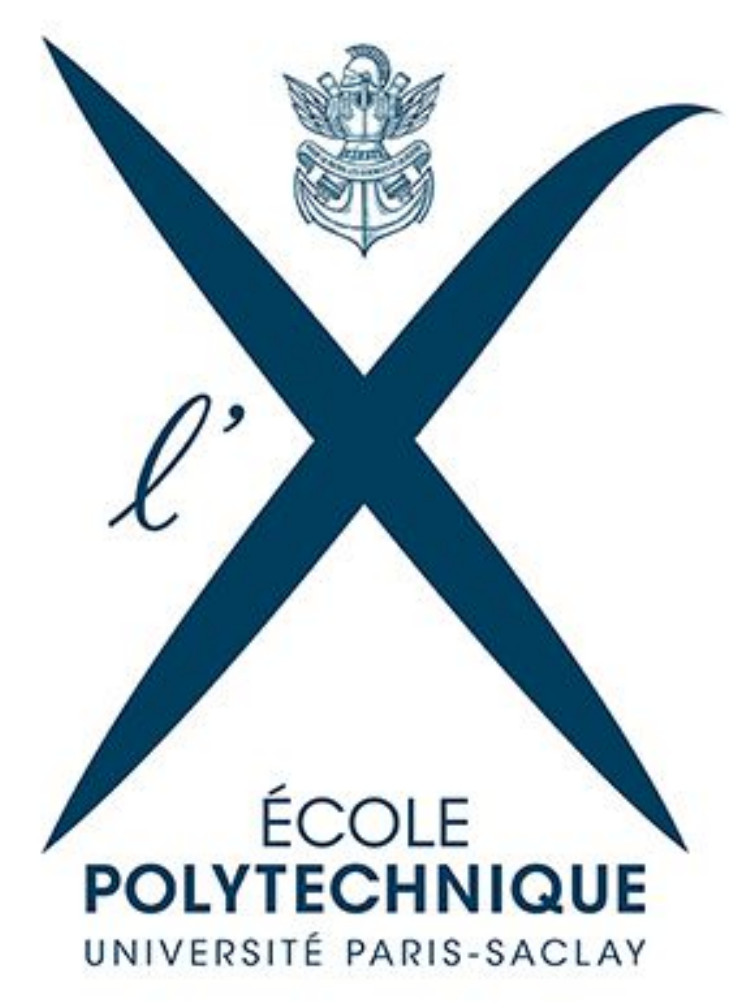}\\
\vspace{6mm}
{\large Centre de Physique Th\'eorique\\
\textsc{CNRS, Ecole Polytechnique,}\\
\textsc{Institut Polytechnique de Paris}}\\
\vspace{11mm}
\begin{minipage}{17cm}
Habilitation thesis defended at Universit\'e Pierre et Marie Curie, Paris.
\end{minipage}\\
\vspace{9mm}
{\large\textsc{4th of June 2019}}
\vspace{12mm}
\end{center}
\end{adjustwidth*}
\end{SingleSpace}
\end{titlingpage}
\clearemptydoublepage
%
%
%

\chapter*{Abstract}

\initial{I}n the case of non-abelian gauge theories, the standard Faddeev-Popov gauge-fixing procedure in the Landau gauge is known to be incomplete due to the presence of gauge-equivalent gluon field configurations that fulfil the gauge condition, also known as Gribov copies. A widespread belief is that the proper analysis of the low energy properties of non-abelian theories in this gauge requires, therefore, the extension of the gauge-fixing procedure, beyond the Faddeev-Popov recipe.

This manuscript reviews various applications of the Curci-Ferrari model, a phenomenological proposal for such an extension, based on the decoupling properties of Landau gauge correlators as computed on the lattice. In particular, we investigate the predictions of the model concerning the confinement/deconfinement transition of strongly interacting matter at finite temperature, first in the case of pure Yang-Mills theory for various gauge groups, and then in a formal regime of Quantum Chromodynamics where all quarks are considered heavy. We show that most qualitative aspects and also many quantitative features of the deconfinement transition in these theories can be accounted for within the Curci-Ferrari model, with only one additional parameter, adjusted from comparison to lattice simulations. Moreover, these features emerge in a systematic and controlled perturbative expansion, as opposed to the ill-defined perturbative expansion within the Faddeev-Popov model in the infrared.

The applications of the Curci-Ferrari model at finite temperature and/or density require one to consider a background extension of the Landau gauge, the so-called Landau-deWitt gauge. Therefore, besides the above mentioned applications, the manuscript is intended as a thorough but pedagogical introduction to these techniques at finite temperature and/or density, including the rationale for introducing the so-called background field effective action, the role of the Weyl chambers in discussing the various physical symmetries of the problem and the complications that emerge due to the sign problem in the case of a real quark chemical potential (in the QCD case). It also investigates the fate of the decoupling correlation functions in the presence of a background as computed in the Curci-Ferrari model and conjectures a specific behavior for the corresponding functions evaluated on the lattice, in the case of the SU($2$) gauge group.

\clearpage
\clearemptydoublepage
%
%
%

\chapter*{Dedication and acknowledgements}

\initial{T}he time that I have spent writing this manuscript strangely reminds me that of the imaginary time formalism with periodic boundary conditions used throughout this work: it does not flow in the usual direction, but rather orthogonal to it, and it is characterized by perpetual, almost periodic, repetitions until the final result is achieved. Now that I am freed from the constraints of this unusual time, I would like to express my most sincere gratitude to all the people that supported me on many different levels, prior, during and after the writing process. This manuscript would clearly not be the same without them.

First of all, my warmest thanks to the members of the jury, Maxim Chernodub, Fran\c{c}ois Gelis, Antal Jakovac, Jean-Lo\"\i c Kneur, Dominique Mouhanna and Samuel Wallon, who accepted the invitation to read and review the manuscript and to participate to the oral defence. Thanks also to Fr\'ed\'eric Fleuret, R\'egine Perzynski, Marco Picco and Nathalie Suirco, for their many advises and for accepting to move the deadline by a few months, in order to allow me to cope with some real time, real life imperatives. Thanks finally to Juana Isabel Mallmann for making the defence in Alan Turing building's possible and incredibly smooth.

I want to express my deepest gratitude to my direct collaborators on the topics covered in this thesis: Julien Serreau, Matthieu Tissier and Nicol\'as Wschebor. I consider myself very lucky to have crossed paths with these top-notch researchers and I am very proud of the work that we have accomplished together. Thanks also to their respective laboratories (and by this I do not mean the walls but the people within them) where a substantial part of the present work was  developed: Laboratoire de Physique Th\'eorique de la Mati\`ere Condens\'ee (Universit\'e Paris Sorbonne), Laboratoire AstroParticule et Cosmologie (Universit\'e Paris-Diderot), Instituto de F\'isica (Universidad de la R\'epublica, Montevideo). Of course, I have also benefited from interactions with many other collaborators, which have contributed to shaping my knowledge in one way or another and whose insight certainly contributes to the flavor (or I should say the color) of this thesis. In particular, a special thanks goes to Jean-Paul Blaizot, David Dudal, Jan M. Pawlowski, Marcela Pel\'aez and Zsolt Sz\'ep.

This thesis was written partly at the ``Centre de Physique Th\'eorique'' (CPHT) on Ecole Polytechnique campus (even though many hours were also spent at the central library, amidst the noisy workers). I would like to thank the whole CPHT members for the outstanding working atmosphere that they contribute to create, quite above average I would say. In particular, I want to thank Florence Auger, Fadila Debbou and Malika Lang for their promptness in dealing with any kind of administrative issue. Thanks also to our Director, Jean-Ren\'e Chazottes, and to Ecole Polytechnique, for accepting to pay my registration at the University.

A special thanks goes of course to past and present members of the Particle Physics group at CPHT, Tran Truong, Georges Grunberg, Tri Nang Pham, Bernard Pire, Claude Roiesnel, St\'ephane Munier, Cyrille Marquet and C\'edric Lorc\'e (plus postdoctoral fellows as well as master and PhD students) for the enjoyable working atmosphere and for accepting me as a member of the group despite the crazy taste for gluonic mass operators (hopefully, this thesis will succeed in conveying the message that this is not that crazy after all). An even more special thanks goes to Bernard Pire and St\'ephane Munier who have invested a considerable amount of energy for many years now into strengthening the visibility of our group/laboratory. I owe them much since my hiring twelve years ago. They have always been available to listen to my queries and to provide good pieces of advise.

Thank you to all the friends and colleagues that, every now and then, inquired about the progress of the thesis, in particular those, Laetitia and Mathieu, Xoana and Daniel, that lend their houses so that I could find a few hours of focus on the writing of the manuscript .

Finally, my most special thank you to my dear wife and kids: Rosa, Aitor, Guillem, Rosa (did I mention periodic boundary conditions?). Your patience has been infinite. Mine, certainly the inverse.

\vglue60mm

\hglue90mm {\it To my parents,}

\clearpage
\clearemptydoublepage
%
%
\renewcommand{\contentsname}{Table of Contents}
\maxtocdepth{subsection}
\tableofcontents*
\addtocontents{toc}{\par\nobreak \mbox{}\hfill{\bf Page}\par\nobreak}
\clearemptydoublepage
%
%
%
%
\mainmatter

%
%
%
\let\textcircled=\pgftextcircled
\chapter{Introduction: The many paths to QCD}
\label{chap:intro}

\initial{Q}uantum Chromodynamics, or QCD for short, is by now well accepted as the fundamental theory governing the strong force. According to this theory, the elementary particles sensible to the strong interaction, known as {\it quarks} and {\it anti-quarks}, carry a generalized notion of charge, the {\it color}, that allows them to exchange quanta, known as {\it gluons}, in a way similar to the exchange of photons by electrons and positrons in Quantum Electrodynamics (QED).  A crucial difference with this latter theory is, however, that the gluons themselves are carriers of color, allowing them to self-interact. Correspondingly, the SU(3) symmetry group associated to the color charge is non-abelian, in contradistinction with the abelian U(1) group at the basis of QED. The fundamental theory of the strong interaction appears, therefore, as the non-abelian generalization of Quantum Electrodynamics.

Although conceptually quite appealing, this generalization hides, in fact, the long process that lead to the construction of QCD as the fundamental theory of the strong interaction, from the thorough study of the many observed particles that reacted to the strong force and the proposal of the quark model as a way to bring order to complexity \cite{GellMann:1962xb,GellMann:1964nj,Zweig:1964jf}, to the experimental evidence for the existence of quarks \cite{Bloom:1969kc,Breidenbach:1969kd}, the proposal of a new type of charge with an associated non-abelian symmetry group \cite{Fritzsch:1973pi} and the final formulation of QCD in the form of a non-abelian gauge theory. 

The main reason explaining this long process is that, unlike the other theories describing the fundamental forces of Nature, the elementary bricks of QCD are not directly observable. Instead, quarks and anti-quarks appear to us in the form of a large fauna of bound states or resonances, the {\it hadrons,} of which the protons, the neutrons and the pions are just a few representatives. Moreover, these compound particles come with the added mystery to always appear in a color-neutral form. This property, known as (color) {\it confinement} \cite{Greensite:2011zz}, as evaded a fully satisfactory theoretical grasp since the advent of QCD, and, even though it is now pretty much accepted to be a mathematical property of the theory \cite{Creutz:1980zw}, its rigorous first principle derivation is one of the open challenges in theoretical particle physics. The challenge is rooted in the fact that the coupling of the strong interaction is much larger than the corresponding coupling of the electromagnetic interaction and, perturbation theory, so useful in this latter case, is admittedly of no use here.

Confinement characterizes, however, the low energy regime of the strong interaction. In the opposite, high energy limit, QCD displays a totally different behavior. Indeed, as any other relativistic field theory, the coupling of the interaction varies, or ``runs'', with the energy scale relevant to the particular process under scrutiny. In the case of QCD, the running coupling decreases and approaches zero logarithmically for asymptotically large values of the energy. This special property is known as {\it asymptotic freedom} \cite{Politzer:1973fx,Gross:1973id} and turns QCD at high energies into a weakly interacting system of quarks and gluons, thus providing access to some of its properties from first principle perturbative calculations.\footnote{In practice, a given process involves both hard and soft scales. If so-called {\it factorization} applies, one can rewrite the process as a convolution between hard and soft components. The former, because they involve large momentum transfers, can be treated within perturbative QCD. The latter, although non-perturbative, are universal and evolve with the running scale in a way that can again be determined within perturbative QCD.}

Another exciting property of the high energy regime can be revealed by imagining coupling the system to a thermostat. Indeed, owing to asymptotic freedom, one expects the interaction to decrease as the temperature is increased, up to the point where quarks and anti-quarks cannot remain bound anymore inside of hadrons. The low temperature {\it confined phase} is then expected to evolve into a {\it deconfined phase}, sometimes dubbed as {\it quark-gluon plasma}, in which quarks and gluons are liberated and color neutrality constraints do not apply anymore. Similarly, one expects to find a deconfined phase at large matter densities, the so-called {\it color-flavor locked phase} \cite{Alford:1998mk}, although it is superconducting in nature and therefore rather different from the deconfined phase at high temperature. 

Deconfined phases of matter are believed to be relevant in various physical situations of interest, for instance during the thermal history of the early Universe \cite{Kolb:1990vq}, or in the core of certain ultra dense stellar objets \cite{Glendenning:1997wn}. The quest for the quark-gluon plasma and its decay into a confining phase (as the system cools down) is also the central motivation for heavy-ion collision experiments, at RHIC (BNL, Brookhaven, USA), LHC (Cern, Geneva, Switzerland) or FAIR (GSI, Darmstadt, Germany). Beyond providing the experimental evidence for the existence of deconfined phases of matter, the ultimate goal of these experiments is to acquire valuable insight into the QCD phase diagram, not only as a function of the temperature, but also as a function of additional external parameters. For instance, the possible presence of a critical end-point \cite{Karsch:2001nf,Gavai:2004sd} terminating a line of first order phase transitions in the plane defined by temperature and density has been and remains nowadays a hotly debated issue. The classification and analysis of exotic phases along the density axis or as a function of a possible magnetic field is also of topical relevance \cite{McLerran:2008ua,Kojo:2009ha,Kharzeev:2013jha,Mizher:2010zb,Chernodub:2010qx}.

Parallel to these large experimental programs, an intense theoretical activity has been devoted to extract the properties of the QCD phase diagram from first principle calculations.\footnote{We shall not review here the many interesting approaches that are based on low energy models of QCD. See for instance \cite{Buballa:2003qv,Ratti:2006wg,Rossner:2007ik,Schaefer:2007pw,Herbst:2010rf} and references therein.} As already mentioned, perturbation theory is certainly a valuable tool in inferring the behavior of the system at asymptotically large temperatures or densities. However, for lower values of these parameters, it has to face the counter-effect of asymptotic freedom, namely that the coupling becomes larger and larger, eventually invalidating the use of any perturbative expansion. As a matter of fact, if one insists in decreasing the energy further, the perturbative running coupling diverges at a finite scale, known as $\Lambda_{QCD}\simeq 200$\,MeV that sets an apparently impassable barrier for perturbative methods. Moreover, because this energy scale is not much different from the deconfinement temperature, see below, one concludes that neither the deconfinement transition nor the confined phase are amenable to perturbative methods. In fact, even when combined with Hard-Thermal-Loop re-summation techniques at high temperature \cite{Braaten:1989mz}, perturbation theory is believed to give a good insight on the deconfined phase only down to temperatures of the order of two to three times the transition temperature \cite{Andersen:2002ey,Andersen:2011sf}. We will have more to add below on the relevance of perturbative approaches but the above considerations are usually taken as the starting point for the development of non-perturbative tools. 

The most famous of them is certainly lattice QCD for it yields a full numerical solution to the theory within a given space-time discretisation \cite{Wilson:1974sk,Montvay:1994cy}. It is based on the Euclidean functional integral formulation of QCD and the probabilistic interpretation of the latter, that allows for the use of importance sampling Monte-Carlo techniques. After various decades of improvements in order to solve many practical implementation issues, lattice QCD has now reached the era of precision, providing much physical insight as well as a wealth of valuable data that other approaches can use for benchmarking. In the vacuum, this robust approach gives a very good account of the hadronic spectrum of the theory \cite{Bernard:2001av} as well as compelling evidence that confinement is a property of QCD \cite{Creutz:1980zw}. At finite temperature, it provides a clear evidence for the existence of a transition between confined and deconfined phases at a temperature of around $T_d\simeq 154\,$MeV \cite{Karsch:2000kv}. The transition is not a sharp transition, however, rather a crossover \cite{Bazavov:2014pvz,Borsanyi:2013bia} characterized by a rapid but smooth variation of the thermodynamical properties of the system, which can also be accurately evaluated within lattice simulations \cite{Borsanyi:2012ve}.

One of the main drawbacks of lattice simulations is that they rely crucially on the probabilistic interpretation of the functional integral. Away from this comfort zone, that is whenever the functional to be integrated is not positive definite, they suffer a tremendous loss of accuracy, known as {\it sign problem} \cite{deForcrand:2010ys,Philipsen:2010gj}. The latter prevents the investigation of many interesting quantities such as the QCD phase diagram for moderate to large densities (and small temperatures) or the evaluation of dynamical quantities such as transport coefficients. Many approaches have been devised to circumvent or at least tame the lattice sign problem, such as Taylor expansions around small densities, re-weightings of the functional integral, numerical continuations from imaginary chemical potentials, and more recently the use of Lefschetz thimbles \cite{Tanizaki:2015rda} or of complex Langevin dynamics \cite{Aarts:2013uxa}. Although valuable progress could be achieved within each of these approaches, no complete solution to the QCD sign problem is available so far.

The second possible class of methods beyond perturbation theory go under the name of continuum (or functional) non-perturbative methods. To avoid the sign problem, these approaches do not aim at a direct numerical evaluation of the QCD functional integral but rather at the (approximate) resolution of sets of exact equations characterizing the dynamics of the system. These equations can be for instance the set of quantum equations of motion, known as Dyson-Schwinger equations \cite{Schwinger:1951ex,Schwinger:1951hq,Dyson:1949ha,vonSmekal97,Alkofer00}, or the hierarchy of renormalization group equations that one derives from the Wetterich equation \cite{Wetterich:1992yh,Ellwanger96,Freire:2000bq}. Related approaches include the use of $n$-particle-irreducible effective actions \cite{Fukushima:2012qa,Fister:2013bh}, the Hamiltonian formalism and its variational principle \cite{Feuchter:2004mk,Reinhardt:2013iia}, or Dyson-Schwinger equations modified through the pinch technique \cite{Binosi:2009qm}.

One common denominator to most continuum approaches is that the elementary quantities they give directly access to are the correlation functions of the system. The latter obey infinite hierarchies of coupled equations that cannot, in general, be solved exactly. As such, there exists in general no simple systematics for improvement, as opposed to the perturbative or lattice approaches. Instead, one usually resorts to truncations of the infinite hierarchy of equations, dictated either by physical intuition or computability criteria. The quality of these truncations needs in any case to be tested {\it a posteriori.} An added difficulty in the case of QCD is that the correlation functions are not uniquely defined. Indeed, within any gauge theory such as QCD, the definition of correlation functions makes sense only within a specified gauge, turning the correlation functions into subtly gauge-dependent quantities, as opposed to the physical, gauge-invariant observables that one can access from lattice simulations. In principle, observables can be reconstructed in terms of the gauge-variant correlation functions but this requires an accurate determination of the latter through appropriate truncations, the quality of which can again be tested only {\it a posteriori.} Comparison between various continuum approaches or confrontation to lattice results (when available) is therefore a crucial element in finding the appropriate truncations.

Here, lattice simulations come specially in handy. Indeed, the latter can not only be formulated in a gauge invariant setting but also within a specified gauge. In particular, over the past twenty years, an intense activity has been devoted to the evaluation of Landau gauge correlation functions, both from lattice simulations and from continuum non-perturbative approaches. The reason for choosing the Landau gauge is two-fold. First, this gauge can be formulated as an extremization problem which is perfectly suited for a lattice implementation \cite{Boucaud08,Boucaud:2011ug,Cucchieri_08b,Cucchieri_08,Cucchieri09,Bornyakov2008,Bornyakov09,Bogolubsky09,Dudal10,Maas:2011se}. Second, and even more importantly, the Landau gauge correlation functions have been thought to provide direct access to the physics of confinement. Indeed, in their seminal work \cite{Kugo:1979gm}, by combining the symmetry properties of the Landau gauge together with the hypothesis of confinement, Kugo and Ojima could predict a very characteristic behavior of the two-point correlation functions for the ghost fields (certain additional degrees of freedom that are introduced when specifying to the Landau gauge). According to them, the ghost two-point function should be strongly enhanced at low momenta, with respect to the two-point function in the absence of interactions. Correspondingly, the gluon propagator should vanish at low momenta. This characteristic low momentum behavior, known nowadays as {\it scaling}, could be observed in the various continuum non-perturbative approaches mentioned above \cite{vonSmekal97,Alkofer00,Zwanziger01,Fischer03,Fischer04,Fischer:2006vf}. 

This solution, however, does not seem to be the one seen in lattice simulations where the gluon two-point function and the ghost dressing function (the ratio of the ghost two-point function to its non-interacting version) both saturate to finite non-zero values at low momentum, defining what is nowadays referred to as a {\it decoupling} behavior \cite{Aguilar04,Boucaud06,Aguilar07,Aguilar08,Boucaud08}. It has eventually been shown that non-perturbative continuum approaches can accommodate both scaling and decoupling type solutions \cite{Fischer:2008uz}, with however the price of modifying the boundary conditions of the corresponding infinite hierarchy of equations. Since then, the agreement between lattice and non-perturbative approaches has considerably increased, providing better control on the truncations that are considered and opening the way to a myriad of applications of functional methods to QCD. In particular, many aspects of the QCD phase diagram can now be addressed with these methods \cite{Braun:2007bx,Marhauser:2008fz,Braun:2009gm,Braun:2010cy,Pawlowski:2010ht,Fischer:2010fx,Fischer:2018sdj,Reinhardt:2017pyr,Fu:2019hdw}.

This manuscript reviews the results obtained within yet a third route, as originally proposed in \cite{Tissier:2010ts,Tissier:2011ey}. To understand it better, we need to take a few steps back. We argued above that perturbation theory, although quite relevant at high energies, seems to contain the seeds of its own breakdown as the energy is lowered, since the perturbative running coupling increases and eventually diverges at a finite scale. It is to be noted, however, that, in a gauge theory, the very definition of perturbation theory requires a gauge to be specified. Put it differently, there is not a unique perturbative expansion in QCD, but infinitely many, in fact as many as there are ways to fix the gauge. It is true that the way the coupling runs is universal at high energies.\footnote{The $\beta$ function that controls this running is shown to be two-loop universal.} However, both the actual value of the coupling (at all energies) and its running at low energies are not universal. In general, they depend both on the renormalization scheme and on the gauge that one works with. Therefore, the properties of the perturbative expansion, including its range of validity, depend on the precise procedure used to fix the gauge.

Gauge-fixing is usually performed by following a standard approach, know as the {\it Faddeev-Popov procedure.} The outcome of this procedure is that, in practice, one should not work with the original QCD action but, rather, with a gauge-fixed version of it, known as the {\it Faddeev-Popov action.} In principle, these two formulations are identical. In practice, however, the Faddeev-Popov construction relies on certain mathematical assumptions which are known not to be realized due to the so-called {\it Gribov copy problem} or {\it Gribov ambiguity} \cite{Gribov:1977wm}. It is generally accepted that this mathematical subtlety can be neglected in the high energy regime of the theory. However, in the opposite limit, no one really knows how it could impact the gauge-fixed implementation of QCD. Even the breakdown of perturbation theory at low energies could be questioned and some quantities could become amenable to perturbative methods. The thesis to be defended in this manuscript is that certain aspects, in particular the physics of the deconfinement transition, could become akin to perturbative methods once the standard Faddeev-Popov gauge-fixing procedure is appropriately extended.

In fact, we know already that this perturbative scenario is too na\"ive for the strict QCD case \cite{Pelaez:2017bhh}. There are compelling evidences, however, that the scenario could apply to the gluonic or {\it pure gauge} sector of QCD. This manuscript aims at reviewing some of these evidences. We mention of course that the study of the pure gauge sector is not a purely academic question, disconnected from QCD. The corresponding Yang-Mills theory remains non-trivial due to the self-interaction of the gluons and, to some extent, it is believed to capture some of the non-trivial features of QCD, in a simplified setting. In particular, unlike the physical QCD case, the deconfinement transition appears here as a genuine phase transition, associated with the breaking of a symmetry, the so-called {\it center symmetry} of Yang-Mills theory at finite temperature, that can be probed with order parameters such as the {\it Polyakov loop}. Moreover, as discussed in \cite{Pelaez:2017bhh,Maelger:2019cbk}, a perturbative grasp on the gluon dynamics could open the way to the study of some of the properties of QCD, if not with perturbative methods, at least by means of a systematic expansion scheme, controlled by small parameters. 

Roughly speaking, the approaches beyond the standard gauge-fixing procedure can be classified into two categories: semi-constructive approaches on the one hand, that aim at resolving the Gribov problem, at least in some approximate form, and more phenomenologically inspired approaches on the other hand, that aim at constraining or even falsifying the operators that could appear beyond the Faddeev-Popov prescription. In this second type of approaches, constraints could come from experimental measurements but also from lattice simulations. In particular, gauge-fixed lattice simulations are a method of choice in constraining whatever model beyond the Faddeev-Popov prescription, precisely because they themselves do not rely on the Faddeev-Popov construction. 

Among the possible phenomenological models beyond the Faddeev-Popov action, the so-called Curci-Ferrari model \cite{Curci:1976bt} (also known as Fradkin-Tuytin model \cite{Fradkin:1970ib}) is a particularly interesting one. This model was originally introduced as a renormalizable infrared regularization of the Faddeev-Popov action, in the form of a mass term for the gluon field that eventually needed to be taken to zero. More recently, it has been proposed as a model beyond the Faddeev-Popov procedure, with the important difference that the Curci-Ferrari mass remains here a free parameter.\footnote{What is really free is the value of the Curci-Ferrari mass at a given scale. Interestingly enough, the running Curci-Ferrari mass runs to $0$ both at high and at low energies.} This proposal is grounded on how (surprisingly) well the lattice Landau gauge decoupling type correlators in the vacuum can be accommodated by the one-loop correlation functions of the model \cite{Tissier:2010ts,Tissier:2011ey}. This applies not only to the two-point correlator functions, but also to the three-point correlators \cite{Pelaez:2013cpa}. Moreover, the model being renormalizable, there is only one additional parameter as compared to the Faddeev-Popov action, that can be adjusted from the comparisons to lattice results in the vacuum. The success of the model in reproducing the vacuum Landau gauge correlators from high to low momentum scales relies on the existence of {\it infrared safe} renormalization group trajectories along which the coupling of the interaction remains moderate.\footnote{In particular, no Landau pole is found, as opposed to renormalization group trajectories in the Faddeev-Popov model.}

As a further stringent test of the model, the goal of this manuscript is to review how consistent are its perturbative predictions away from the vacuum, in particular with regard to the deconfinement transition and the corresponding phase structure. The plan is as follows:

$\bullet$ In the next chapter, after reviewing the basic properties and limitations of the standard gauge-fixing procedure, we discuss some of the possible approaches beyond it. We also introduce the approach to be examined in this manuscript, based on the Curci-Ferrari model, and review some of the results obtained with this approach in the vacuum, regarding the Landau gauge correlation functions. A first attempt at studying finite temperature effects is also reviewed, based on the hypothesis that the finite temperature Landau gauge correlators could carry some imprint of the deconfinement transition. It turns out that this analysis is inconclusive essentially because the Landau gauge, so useful in the vacuum, does not properly capture the order parameter associated to the deconfinement phase transition. In order to understand the limitations of the Landau gauge, we recall some basic considerations related to center symmetry and the Polyakov loop in Chapter 3. We also explain why these basic properties are difficult to capture not only within the Landau gauge but in fact within any standard gauge-fixed setting.

$\bullet$ The previous difficulties can be solved by generalizing the gauge-fixing in the presence of a background field. Even though part of this is known material, we dedicate Chapters 4 and 5 to a self-contained investigation of the use of background field methods at finite temperature. In particular, we stress the importance of a description of the states of the system that is free of the redundancy associated to gauge invariance, as well as the role of {\it self-consistent backgrounds} as alternative order parameters for center symmetry. We also discuss other symmetry constraints such as charge conjugation, homogeneity and isotropy, which are rarely discussed in the literature. As far as possible, we try to critically discuss the various implicit assumptions that are usually made when applying background field methods at finite temperature, in particular regarding the properties of the gauge-fixed measure.

$\bullet$ The perturbative study of the Yang-Mills deconfinement transition within the Curci-Ferrari model in the presence of a background is given in Chapter 6, at leading order, together with a comparison to other approaches. The convergence properties of the approach are investigated in Chapter 7 where we evaluate the next-to-leading order corrections. Chapter 8 investigates further the relation between center symmetry and the deconfinement transition by discussing the case of the SU(4) gauge group.

$\bullet$ We also discuss the perturbative predictions of the model for a theory cousin to Yang-Mills theory, namely QCD in the regime where all quarks masses are considered heavy. Although this does not correspond to the physical QCD case, this formal regime of QCD has received a lot of attention lately since it possesses a rich phase structure that can be probed with the same order parameters as in the pure Yang-Mills case and which can be used as a benchmarking of any method that one plans to extend to the real QCD case. We analyze to which extent the phase structure in this regime can be described using perturbation theory within the Curci-Ferrari model. This is done in Chapter 10 after some additional material is provided in Chapter 9 on the use of background field methods at finite density.

$\bullet$ Finally, in Chapter 11, we revisit the question of a possible imprint of the deconfinement transition on the two-point correlation functions but this time from the point of view of the background extension of the Landau gauge. In particular, under certain natural assumptions a {\it bona fide} gauge fixing should satisfy, we postulate a specific behavior of the gluon propagator at the deconfinement transition, in the case of the SU(2) gauge group, that realizes what was originally searched for in the Landau gauge.

$\bullet$ Conclusions are presented in Chapter 12 where the main results are summarized and an outlook is proposed.\\

This manuscript reviews results covered in Refs. \cite{Reinosa:2013twa,Reinosa:2014ooa,Reinosa:2014zta,Reinosa:2015gxn,Reinosa:2015oua,Reinosa:2016iml,Reinosa:2017qtf}. Beyond the mere review, we have tried as much as possible to provide a self-contained document, in particular with regard to the used methodology. We have also included some unpublished work, such as for instance the material in Sections 6.3.4, 8.3.4, 10.2 or Appendices A and D. The manuscript was written during the first half of 2019 and defended the 4th of June of that same year. Since this defended version, a new chapter has been added (chapter 8) together with some recent relevant references.

\clearemptydoublepage
%
%
%
%
\let\textcircled=\pgftextcircled
\chapter{Faddeev-Popov gauge-fixing\\ and the Curci-Ferrari model}
\label{chap:intro}

\initial{T}he point of view taken in this manuscript is, one, that tackling the low energy properties of non-abelian gauge theories in the continuum requires extending the standard, but incomplete, Faddeev-Popov gauge-fixing procedure, and, second, that once such an extension is found, a new perturbative scheme could become available in the infrared. In this first chapter, we introduce the Curci-Ferrari action, as a phenomenological model for an extension of the Faddeev-Popov action in the Landau gauge.

For the sake of completeness, we first review the standard gauge-fixing procedure together with its main properties and limitations. We then discuss some of the approaches that have been devised in order to go beyond it and, finally, particularize to the Curci-Ferrari model. We recall that the latter does a pretty reasonable job in reproducing some of the known low energy properties of Yang-Mills theory in the vacuum, already at one-loop order. We also discuss the difficulties that appear when using the model at finite temperature, serving as the main motivation for the developments in the rest of the manuscript.

\section{Standard gauge-fixing}
For simplicity, let us review the standard gauge-fixing procedure in the case of Yang-Mills theory (YM), the theory obtained from QCD after neglecting the dynamics of quarks. Of course, a similar discussion could be carried out in the presence of matter fields.

\subsection{Gauge-invariance}
Yang-Mills theory describes the dynamics of a non-abelian gauge field $A_\mu^a(x)$. The index $a$ corresponds to an internal degree of freedom (color) and labels the generators $it^a$ of a non-abelian gauge group, SU(N) in what follows. The non-abelian structure is encoded in the structure constants $f^{abc}$ such that $[t^a,t^b]=if^{abc}t^c$, and the dynamics is specified by the action
\begin{equation}\label{eq:SYM}
S_{YM}[A]=\frac{1}{4g_0^2}\int d^dx\,\,F_{\mu\nu}^a(x)\,F_{\mu\nu}^a(x)\,,
\end{equation}
with $F_{\mu\nu}^a\equiv\partial_\mu A_\nu^a-\partial_\nu A_\mu^a+f^{abc}A_\mu^b\,A_\nu^c$ the non-abelian generalization of the QED field-strength tensor. Since the applications to be discussed in this manuscript concern the equilibrium properties of the QCD/YM system, we have here chosen the Euclidean version of the action and, therefore, there is no distinction between covariant and contravariant indices \cite{Laine:2016hma}. Moreover, the gauge field should be taken periodic along the Euclidean time direction, with a period equal to the inverse temperature $\beta\equiv 1/T$. Correspondingly, the integration symbol $\int d^dx$ needs to be understood as $\int_0^\beta d\tau \int d^{d-1}x$ and we have $A_\mu^a(\tau+\beta,\vec{x})=A_\mu^a(\tau,\vec{x})$.

The main feature of the YM action is of course that it is gauge-invariant. This is most easily seen by rewriting  (\ref{eq:SYM}) in an intrinsic form that does not depend on the particular coordinate system used to describe the color degrees of freedom.\footnote{Later, this will also facilitate the change from the standard, Cartesian bases to the so-called Cartan-Weyl bases. We shall introduce and use of these bases in subsequent chapters.} One interprets any colored object $X^a$ as an element of the SU(N) Lie algebra $\smash{X\equiv iX^at^a}$. To any two such elements $X$ and $Y$, one then associates the Killing form\footnote{The minus sign is chosen such that $(it^a;it^b)=\delta^{ab}$.}
\begin{equation}
(X;Y)\equiv -2\,\mbox{tr}\,XY\,,
\end{equation}
which allows to rewrite the Yang-Mills action as
\begin{equation}
S_{YM}[A]=\frac{1}{4g_0^2}\int d^dx\,\,\Big(F_{\mu\nu}(x);F_{\mu\nu}(x)\Big)\,,
\end{equation}
with $F_{\mu\nu}=\partial_\mu A_\nu-\partial_\nu A_\mu-[A_\mu,A_\nu]$. In this intrinsic representation, a gauge transformation of the gauge-field is defined to be
\begin{equation}\label{eq:gauge_transfo}
A^U_\mu(x)\equiv U(x)\,A_\mu(x)\,U^\dagger(x)-U(x)\,\partial_\mu U^\dagger(x)\,,
\end{equation}
with $U(x)\in$ SU(N), $\forall x$. It is then easily verified that the field strength tensor transforms correspondingly as
\begin{equation}
F_{\mu\nu}^U(x)=U(x)\,F_{\mu\nu}(x)\,U^\dagger(x)\,.
\end{equation}
From its definition, the Killing form is trivially invariant under color rotations in the sense $(UXU^\dagger;UYU^\dagger)=(X;Y)$. It follows, as announced, that the action (\ref{eq:SYM}) is gauge invariant. In more pompous terms, it is constant along a given {\it orbit,} defined as the collection of configurations $A^U$ as $U$ spans the possible gauge transformations, for a given $A$.

\subsection{Observables}
The gauge-invariance of the YM action is just the expression of a certain arbitrariness in the choice of the gauge-field configuration that describes a given physical situation. Observables cannot depend on this arbitrariness and, therefore, are to be represented by gauge-invariant functionals.\footnote{We are deliberately being vague here concerning the type of gauge transformations that should be considered. At finite temperature, where the gluon field is periodic along the Euclidean time direction, the true, unphysical gauge transformations are also periodic. There are more general gauge transformations that preserve the periodicity of the fields but those are associated to physical transformations in a sense to be clarified in the next chapter.} To any such observable ${\cal O}[A]$, one associates an expectation value as
\begin{equation}\label{eq:Ovev}
\langle{\cal O}\rangle\equiv\frac{\int {\cal D}A\,{\cal O}[A]\,e^{-S_{YM}[A]}}{\int {\cal D}A\,e^{-S_{YM}[A]}}\,.
\end{equation}
where $S_{{YM}}[A]$ and ${\cal D}A$ are also gauge-invariant.\\

The usual difficulty with the above definition is that it involves two indefinite integrals. Indeed, since ${\cal O}[A]$, $S_{{YM}}[A]$ and ${\cal D}A$ are gauge-invariant, the integrals sum redundantly over the orbits of the gauge group and are thus proportional to the (infinite) volume of the group. Even though these two infinities should formally factorize and cancel between the numerator and the denominator of (\ref{eq:Ovev}), their presence prevents any expansion of $\langle{\cal O}\rangle$ in terms of expectation values of gauge-variant functionals, typically products of gauge fields at the same spacetime point, as needed by most continuum approaches. 

One possibility to tackle this problem is to find a way to rewrite the definition (\ref{eq:Ovev}) identically as
\begin{equation}\label{eq:Ogf}
\langle{\cal O}\rangle=\frac{\int {\cal D}_{\mbox{\scriptsize gf}}[A]\,{\cal O}[A]\,e^{-S_{{YM}}[A]}}{\int {\cal D}_{\mbox{\scriptsize gf}}[A]\,e^{-S_{{YM}}[A]}}\,,
\end{equation}
where the measure ${\cal D}_{\mbox{\scriptsize gf}}[A]$ is restricted to gauge field configurations obeying a certain, {\it gauge-fixing} condition $F[A]=0$. Good gauge-fixing conditions should be such that $F[A^U]\neq F[A]$, and, therefore, ${\cal D}_{\mbox{\scriptsize gf}}[A^U]\neq {\cal D}_{\mbox{\scriptsize gf}}[A]$. In this case, the expression (\ref{eq:Ogf}) for the expectation value extends to gauge-variant functionals and can then be used as a starting point for developing continuum methods.\\

Of course, the crux of the problem lies in the construction of the gauge-fixed measure ${\cal D}_{\mbox{\scriptsize gf}}[A]$. The usual strategy is the so-called Faddeev-Popov procedure \cite{Faddeev:1967fc,Pokorski} which we now recall.

\subsection{Faddeev-Popov procedure}
In the Faddeev-Popov approach, under the assumption that the constraint $F[A]=0$ admits a unique solution on each orbit, one writes
\begin{equation}\label{eq:unity}
1=\int{\cal D}U\,\Delta[A^U]\,\delta\big(F[A^U]\big)\,,
\end{equation}
where $\Delta[A^U]$ is the determinant of $\delta F[A^U]/\delta U$, the so-called Faddeev-Popov operator.\footnote{Strictly speaking, it is the absolute value of the determinant that should appear in Eq.~(\ref{eq:unity}). However, since it is assumed that there is only one solution to $F[A]=0$ on each orbit, and if one further assumes that this solution changes continuously as one changes the orbit, the sign of the determinant is constant and, therefore, irrelevant in the Faddeev-Popov approach. Moreover, that $\Delta[A^U]$ does not depend independently on $A$ and $U$ follows from the identity $A^{VU}=(A^U)^V$ which implies $F[A^{\delta VU}]=F[(A^U)^{\delta V}]$. }

Plugging the identity (\ref{eq:unity}) into the definition (\ref{eq:Ovev}) and using the gauge invariance of ${\cal O}[A]$, $S_{{YM}}[A]$ and ${\cal D}A$, one obtains
\begin{equation}\label{eq:OFPm1}
\langle{\cal O}\rangle_{\mbox{\tiny FP}}=\frac{\int{\cal D}U\int {\cal D}A^U\,{\cal O}[A^U]\,\Delta[A^U]\,\delta(F[A^U])\,e^{-S_{{YM}}[A^U]}}{\int{\cal D}U\int {\cal D}A^U\,\Delta[A^U]\,\delta(F[A^U])\,e^{-S_{{YM}}[A^U]}}\,.
\end{equation}
Changing variables from $A^U$ to $A$, the volume of the gauge group factorizes and cancels between the numerator and the denominator. One then arrives at the following {\it gauge-fixed} expression for the expectation value:
\begin{equation}\label{eq:OFP0}
\langle{\cal O}\rangle_{\mbox{\tiny FP}}=\frac{\int {\cal D}_{\mbox{\tiny FP}}[A]\,{\cal O}[A]\,e^{-S_{{YM}}[A]}}{\int {\cal D}_{\mbox{\tiny FP}}[A] \,e^{-S_{{YM}}[A]}}\,,
\end{equation}
where the Faddeev-Popov gauge-fixed measure is defined as
\begin{equation}\label{eq:FP_measure}
{\cal D}_{\mbox{\tiny FP}}[A]\equiv {\cal D}A\,\Delta[A]\,\delta(F[A])\,.
\end{equation}
A similar analysis applies to the partition function except for an overall volume factor which, however, does not affect the temperature-dependent part of the free-energy density. One finds $Z={\cal V}\times Z_{\mbox{\tiny FP}}$, with
\begin{equation}\label{eq:FPZ}
Z_{\mbox{\tiny FP}}=\int {\cal D}_{\mbox{\tiny FP}}[A]\,e^{-S_{{YM}}[A]}\,,
\end{equation}
and ${\cal V}$ the volume of the group of gauge transformations, such that $-\ln\,{\cal V}\propto \beta\Omega$ with $\beta\Omega$ the Euclidean spacetime volume.\\

We mention that the previous derivation relies on a rather strong assumption, namely that the solution to the gauge-fixing condition $F[A]=0$ is unique along a given orbit. As pointed out by Gribov \cite{Gribov:1977wm}, this assumption is generally wrong due to the existence, instead, of multiple solutions, the so-called Gribov copies. For this reason, we have denoted by $\langle{\cal O}\rangle_{\mbox{\tiny FP}}$ the gauge-fixed expression for the expectation value of an observable as obtained from the Faddeev-Popov approach, which may differ from $\langle {\cal O}\rangle$ as originally defined in Eq.~(\ref{eq:Ovev}). We shall come back to this important point in the next section. For the time being, we continue reviewing the properties of the Faddeev-Popov approach.

\subsection{Faddeev-Popov action}
The previous formulation is not very practical due to the presence of both the determinant $\Delta[A]$ and the functional Dirac distribution $\delta(F[A])$ in the gauge-fixed measure (\ref{eq:FP_measure}). As it is well known, one can transform the latter into a standard field theory with the price of introducing additional fields.

First, the factor $\delta(F[A])$ can be treated using a Nakanishi-Lautrup field $h^a$ as
\begin{equation}
\delta(F[A])=\int {\cal D}h\,\exp\left\{-\int d^dx\,ih^a(x)\,F^a[A](x)\right\}.
\end{equation}
Second, the determinant $\Delta[A]$ can be evaluated by assuming that $F[A]=0$ due to the presence of the factor $\delta(F[A])$. One finds
\begin{equation}
\Delta[A]=\mbox{det}\,\left.\frac{\delta F^a[A^U](x)}{\delta\theta^b(y)}\right|_{\theta=0}=\mbox{det}\,\left.\int d^dz\,\frac{\delta F^a[A](x)}{\delta A_\mu^c(z)}\frac{\delta(A^U)_\mu^c(z)}{\delta\theta^b(y)}\right|_{\theta=0}\!\!,
\end{equation}
with $U=e^{i\theta^a t^a}$. Since $(A^U)^c_\mu(z)$ is nothing but the covariant derivative $D_\mu\theta^c(z)\equiv \partial_\mu\theta^c(z)+f^{cde}A_\mu^d(z)\theta^e(z)$ at leading order in $\theta$, one arrives eventually at
\begin{equation}
\Delta[A]=\mbox{det}\,\left(\int d^dz\,\frac{\delta F^a[A](x)}{\delta A_\mu^c(z)}D_\mu^{cb}\delta(z-y)\right).
\end{equation}
Finally, by introducing (Grassmanian) ghost and anti-ghost fields $c^a$ and $\bar c^a$, this rewrites
\begin{equation}
\Delta[A]=\int {\cal D}[c,\bar c]\,\exp\left\{\int d^dx\int d^dy\,\,\bar c^a(x)\,\frac{\delta F^a[A](x)}{\delta A_\mu^b(y)}\,D_\mu c^b(y)\right\}.
\end{equation}
All together, the gauge-fixed expression for the expectation value of an observable in the Faddeev-Popov approach reads
\begin{equation}\label{eq:OFP}
\langle{\cal O}\rangle_{\mbox{\tiny FP}}=\frac{\int {\cal D}[A,c,\bar c,h]\,{\cal O}[A]\,e^{-S_{\mbox{\tiny FP}}[A,c,\bar c,h]}}{\int {\cal D}[A,h,c,\bar c]\,e^{-S_{\mbox{\tiny FP}}[A,c,\bar c,h]}}\,,
\end{equation}
where $S_{\mbox{\tiny FP}}[A,c,\bar c,h]\equiv S_{{YM}}[A]+\delta S_{\mbox{\tiny FP}}[A,c,\bar c,h]$ is the so-called Faddeev-Popov action, with
\begin{equation}
\delta S_{\mbox{\tiny FP}}\equiv-\int d^dx\int d^dy\,\bar c^a(x)\,\frac{\delta F^a[A](x)}{\delta A_\mu^b(y)}\,D_\mu c^b(y)+\int d^dx\,ih^a(x)\,F^a[A](x)\,.\label{eq:FP}
\end{equation}
Similarly, one rewrites the Faddeev-Popov partition function (\ref{eq:FPZ}) as $Z_{\mbox{\tiny FP}}=\int {\cal D}[A,c,\bar c,h]\,e^{-S_{\mbox{\tiny FP}}[A,c,\bar c,h]}$.\\

The Landau gauge to be considered in this manuscript corresponds to the choice $F^a[A](x)=\partial_\mu A^a_\mu(x)$. The corresponding Faddeev-Popov operator is $\partial_\mu D^{ab}_\mu\delta(x-y)$, and therefore
\begin{equation}\label{eq:FP_Landau}
\delta S_{\mbox{\tiny FP}}=\int d^dx\,\Big\{\partial_\mu\bar c^a(x)\,D_\mu c^b(x)+ih^a(x)\,\partial_\mu A_\mu^a(x)\Big\}\,,
\end{equation}
where an integration by parts has been used in the ghost term.

\subsection{BRST symmetry}
As we now recall, the Faddeev-Popov action possesses a very important symmetry, the so-called BRST symmetry \cite{Becchi:1975nq}, at the origin of many properties, including its renormalizability \cite{Becchi:1975nq,ZinnJustin:1974mc,Weinberg}.

Suppose that we perform an infinitesimal gauge transformation of the form $\delta A_\mu^a=\bar\eta\,D_\mu c^a\equiv \bar\eta\,sA_\mu^a$, with $\bar\eta$ a constant Grassmanian parameter. The Yang-Mills part of the action is of course invariant by construction, so let us focus on the gauge-fixing part $\delta S_{\mbox{\tiny FP}}$. We find
\begin{eqnarray}\label{eq:svar_FP}
s\delta S_{\mbox{\tiny FP}} & \!\!\!\!=\!\!\!\! & \int d^dx\int d^dy\,(ih^a(x)-s\bar c^a(x))\,\frac{\delta F^a[A](x)}{\delta A_\mu^b(y)}\,D_\mu c^b(y)\nonumber\\
& \!\!\!\!+\!\!\!\! & \int d^dx\int d^dy\,\bar c^a(x)\,\frac{\delta F^a[A](x)}{\delta A_\mu^b(y)}\,s^2A^b_\mu(y)\nonumber\\
 & \!\!\!\!+\!\!\!\! & \int d^dx\,sih^a(x)\,F^a[A](x)\,,
\end{eqnarray}
where we have used that 
\begin{eqnarray}
& & \int d^dy\int d^dz\,\frac{\delta F^{a}[A](x)}{\delta A_\mu^b(y)\delta A_\nu^c(z)}\,D_\nu c^c(z)\,D_\mu c^b(y)\nonumber\\
& & \hspace{0.5cm}=\,-\int d^dz\int d^dy\,\frac{\delta F^{a}[A](x)}{\delta A_\nu^c(z)\delta A_\mu^b(y)}\,D_\mu c^b(y)\,D_\nu c^c(z)\nonumber\\
& & \hspace{0.5cm} =\,-\int d^dy\int d^dz\,\frac{\delta F^{a}[A](x)}{\delta A_\mu^b(y)\delta A_\nu^c(z)}\,D_\nu c^c(z)\,D_\mu c^b(y)=0\,.
\end{eqnarray}
We see that the gauge-fixing contribution is invariant if we set $s\bar c^a=ih^a$, $sih^a=0$ and choose $sc^a$ such that $s^2A_\mu^a=0$. A simple calculation reveals that this is ensured if $sc^a=\frac{1}{2}\,f^{abc}c^bc^c$. \\

In summary, the Faddeev-Popov action is invariant under
\begin{equation}
sA_\mu^a=D_\mu c^a, \quad sc^a=\frac{1}{2}\,f^{abc}c^bc^c, \quad s\bar c^a=ih^a, \quad sih^a=0\,,
\end{equation}
which is known as BRST symmetry. It is easily verified that $s^2$ vanishes not only over $A_\mu^a$ but over the whole functional space and the BRST symmetry $s$ is then said to be nilpotent. In fact, the BRST invariance of the Faddeev-Popov action can be understood in terms of the nilpotency of $s$ and the gauge-invariance of $S_{{YM}}$ since the action rewrites
\begin{equation}\label{eq:sS}
S_{\mbox{\tiny FP}}=S_{{YM}}+s\int d^dx\,\bar c^a(x)\,F^a[A](x)\,,
\end{equation}
and therefore $sS_{\mbox{\tiny FP}}=sS_{{YM}}=0$.

\subsection{Gauge-fixing independence}
For the gauge-fixing procedure to make sense at all, it should be of course such that the expectation value of a gauge-invariant functional does not depend on the choice of gauge-fixing condition. In the Faddeev-Popov approach, this basic property is not guaranteed a priori since, as mentioned above, the procedure relies on an incorrect assumption. It is one of the merits of the BRST symmetry to ensure nonetheless the gauge-fixing independence of the observables in the Faddeev-Popov framework.\\

\noindent{To see how this works, let us first derive the following lemma:}\\

\vglue-8mm

\noindent{$\_\_\_\_\_\_\_\_\_\_\_\_\_\_\_\_\_\_\_\_\_\_\_\_\_\_\_\_\_\_\_\_\_\_\_\_\_\_\_\_\_\_\_\_\_\_\_\_\_\_\_\_\_\_\_\_\_\_\_\_\_\_\_\_\_\_\_$}\\
Given an infinitesimal symmetry $\varphi\to\varphi+\delta\varphi$ of a theory $S[\varphi]$, the expectation value of the infinitesimal variation $\delta{\cal O}[\varphi]$ needs to vanish.\\
\vglue-11mm
\noindent{$\_\_\_\_\_\_\_\_\_\_\_\_\_\_\_\_\_\_\_\_\_\_\_\_\_\_\_\_\_\_\_\_\_\_\_\_\_\_\_\_\_\_\_\_\_\_\_\_\_\_\_\_\_\_\_\_\_\_\_\_\_\_\_\_\_\_\_$}\\

\vglue-4mm

\noindent{This lemma is easily shown by writing}
\begin{equation}
\int {\cal D}\varphi\,{\cal O}[\varphi]\,e^{S[\varphi]}=\int {\cal D}\varphi\,{\cal O}[\varphi+\delta\varphi]\,e^{S[\varphi]}\,,
\end{equation}
where one performs a change of variables in the form of an infinitesimal transformation and uses the assumed invariance of $S[\varphi]$ under this transformation.\footnote{It is to be mentioned that this standard derivation does not pay too much attention to how the measure and the integration domain are transformed. This is of course because, in general, they are both invariant. However, in the case of a BRST transformation, the initial integration domain for the gauge-field $A$ is made of purely numerical functions, whereas the transformed domain is a more general space, including still commuting but non-numerical contributions of the form $\bar\eta\,Dc$. Similarly the measure transforms non-trivially. We argue in Appendix~\ref{chap:BRST} that the above lemma is unaffected by these subtleties.}

Let us now apply the lemma to prove the gauge-fixing independence of the observables in the Faddeev-Popov approach. Consider first the Faddeev-Popov partition function $Z_{\mbox{\tiny FP}}$ and let $\lambda$ denote any parameter that may enter the definition of the gauge-fixing. From Eq.~(\ref{eq:sS}), we find
\begin{equation}
\frac{d}{d\lambda}\ln Z_{\mbox{\tiny FP}}=\left\langle s\int d^dx\,\left(\bar c^a(x)\,\frac{dF^a[A](x)}{d\lambda}\right)\right\rangle.
\end{equation}
Being the expectation value of an infinitesimal symmetry transformation, the right-hand side needs to vanish according to the lemma, from which we deduce that $d Z_{\mbox{\tiny FP}}/d\lambda=0$ and thus that the partition function does not depend on the choice of gauge-fixing functional $F[A]$. This result extends in fact to any observable. Using the previous result together with $s{\cal O}[A]=0$, we find indeed
\begin{equation}
\frac{d}{d\lambda}\langle{\cal O}\rangle_{\mbox{\tiny FP}}=\left\langle s\left({\cal O}[A]\int d^dx\,\bar c^a(x)\frac{dF^a[A](x)}{d\lambda}\right)\right\rangle,
\end{equation}
which vanishes for the same reason.

\section{Infrared completion of the gauge-fixing}
As we have already mentioned, the derivation of the Faddeev-Popov action (\ref{eq:FP}), together with its main properties, relies on a strong assumption, which is known to not always hold true \cite{Gribov:1977wm}. In particular, it is incorrect in the case of the Landau gauge \cite{Niemi:1981yh} to which we restrict from now on. 

\subsection{Gribov copies}
The point is that, contrary to what is assumed in the Faddeev-Popov construction, the gauge-fixing condition $F[A]=0$ admits multiple solutions along a given orbit, the so-called Gribov copies. The existence of the latter invalidates the use of Eq.~(\ref{eq:unity}) and, therefore, the identification of $\langle{\cal O}\rangle$ and $\langle {\cal O}\rangle_{\mbox{\tiny FP}}$, as given respectively by Eqs.~(\ref{eq:Ovev}) and (\ref{eq:OFP}). 

The identification $\langle {\cal O}\rangle_{\mbox{\tiny FP}}\approx \langle {\cal O}\rangle$ is nonetheless believed to be legitimate at high energies since only a perturbative region of the space of gauge-field configurations contributes to the functional integral and the Gribov copies can be neglected. In this case, one can evaluate $\langle{\cal O}\rangle$ using the perturbative expansion of $\langle{\cal O}\rangle_{\mbox{\tiny FP}}$ which is controlled at high energies thanks to the asymptotic freedom property \cite{Gross:1973id}. 

In contrast, in the low energy regime, the situation is much less clear. The perturbative expansion of $\langle{\cal O}\rangle_{\mbox{\tiny FP}}$ is useless due to the presence of an infrared Landau pole. At the same time, there is no argument anymore in favour of the identification $\langle{\cal O}\rangle_{\mbox{\tiny FP}}\approx\langle{\cal O}\rangle$. As a matter of fact, undoing the step from Eq.~(\ref{eq:OFP0}) to Eq.~(\ref{eq:OFPm1}) and integrating over the gauge group by taking into account possible copies, one finds
\begin{equation}\label{eq:undo_FP}
\langle{\cal O}\rangle_{\mbox{\tiny FP}}=\frac{\int {\cal D}A\,{\cal O}[A]\,e^{-S_{{YM}}[A]}\,\sum_{i(A)} s_{i(A)}}{\int {\cal D}A\,e^{-S_{{YM}}[A]}\,\sum_{i(A)} s_{i(A)}}\,,
\end{equation}
where $i(A)$ labels the Gribov copies along the orbit of $A$ and where $s_{i(A)}$ is the sign of the Faddeev-Popov determinant on copy $i(A)$. If the quantity $\sum_{i(A)} s_{i(A)}$ were non-zero and of constant sign along the various orbits, the identification of $\langle{\cal O}\rangle_{\mbox{\tiny FP}}$ and $\langle{\cal O}\rangle$ would indeed be correct. Unfortunately, for compact gauge groups, the sum vanishes instead, leading to a $0/0$ indetermination, the so-called Neuberger zero problem \cite{Neuberger:1986vv,Neuberger:1986xz}, which prevents the formal identification of $\langle{\cal O}\rangle_{\mbox{\tiny FP}}$ and $\langle{\cal O}\rangle$ beyond the one discussed above in the high energy, perturbative domain. Similarly, the Faddeev-Popov partition function vanishes.

This calls for constructing alternative gauge-fixing procedures that take into account the effect of the Gribov copies, at least in some approximate form.\footnote{Another interesting approach relies in decomposing the gauge group into a subgroup where the gauge-fixing problem is trivial and a quotient group where the Neuberger zero is absent \cite{vonSmekal:2007ns,vonSmekal:2008es}.} The quest for such an infrared completion of the Faddeev-Popov gauge-fixing is not only formal. In fact, according to certain scenarios, once such a gauge-fixing is found, a new perturbative expansion could become available in the low energy regime \cite{Tissier:2010ts,Tissier:2011ey,Pelaez:2013cpa}.\\

Various strategies have been devised to include the Gribov copies in a more rigorous way. Let us briefly review some of them.

\subsection{Gribov-Zwanziger approach}
One possibility is to restrict the functional integration over gauge-field configurations to a subdomain such as some (but in general not all) copies are excluded. This is for instance achieved in the Gribov-Zwanziger approach where so-called infinitesimal copies are excluded by restricting to a region such that the Faddeev-Popov operator $-\partial_\mu D_\mu$ is positive definite \cite{Zwanziger:1989mf,Zwanziger:1992qr}. 

Just as in the Faddeev-Popov approach, the Gribov-Zwanziger procedure can be formulated as a local and renormalizable theory. A nilpotent BRST symmetry can be identified but it is non-local \cite{Capri:2015ixa}. More recently, a proposal has been made to rewrite the Gribov-Zwanziger action into an alternative local form that displays a local and nilpotent BRST symmetry \cite{Capri:2016aqq}. This rewriting requires however neglecting once more the presence of copies.

One very appealing feature of the Gribov-Zwanziger approach is that a mass scale is dynamically generated and determined only in terms of the Yang-Mills coupling, a feature that any {\it bona fide} gauge-fixing should possess. However, in its original formulation, the Gribov-Zwanziger action predicts correlation functions at odds with the ones obtained on the lattice. It has since then been refined by the inclusion of condensates in order to improve the comparison with lattice results \cite{Dudal:2008sp,Dudal:2008rm}.

\subsection{Serreau-Tissier approach}
Another possible strategy, followed for instance by Serreau and Tissier \cite{Serreau:2012cg}, see also \cite{vonSmekal:2008en} for a similar idea, is to sum formally over all copies, just as in Eq.~(\ref{eq:undo_FP}) but with a weighting factor $w[A^{U_{i(A)}}]$ that avoids the Neuberger zero problem. To ensure that the result is an identical rewriting of Eq.~(\ref{eq:Ovev}), one considers an average rather than a sum. 

To this purpose, one inserts in Eq.~(\ref{eq:Ovev}) the identity
\begin{equation}\label{eq:unity2}
1=\frac{\sum_{i(A)} s_{i(A)} w[A^{U_{i(A)}}]}{\sum_{i(A)} s_{i(A)}w[A^{U_{i(A)}}]}=\frac{\int {\cal D}U \Delta [A^U]\,\delta\big(
F[A^U]\big)\,w[A^U]}{\int {\cal D}U' \Delta [A^{U'}]\,\delta\big(
F[A^{U'}]\big)\,w[A^{U'}]}\,.
\end{equation}
Under the integral over $A$, the volume of the gauge group can be factored out in the integral over $U$, while leaving the integral over $U'$ unaffected thanks to the change of variables $U'\to U'U^\dagger$. One eventually arrives at
\begin{equation}
\langle{\cal O}\rangle=\frac{\int {\cal D}_{\mbox{\tiny ST}}[A]\,{\cal O}[A]\,e^{-S_{{YM}}[A]}}{\int {\cal D}_{\mbox{\tiny ST}}[A] \,e^{-S_{{YM}}[A]}}\,,
\end{equation}
with the Serreau-Tissier measure defined as
\begin{equation}
{\cal D}_{\mbox{\tiny ST}}[A]\equiv {\cal D}A\,\Delta[A]\,\delta(F[A])\,e^{-\ln\int {\cal D}U' \Delta [A^{U'}]\,\delta(F[A^{U'}])\,w[A^{U'}]}\,.
\end{equation}
Note that we have not included any labelling of $\langle{\cal O}\rangle$ since the rewriting is exact at this point. 

In principle, the logarithm of the $U'$-integral in the above expression provides a non-trivial correction to the Faddeev-Popov action. In practice, however, it is not possible to evaluate this correction exactly and the following strategy has been adopted, based on the well known replica trick of Statistical Physics \cite{replica}. One writes
\begin{equation}
{\cal D}_{\mbox{\tiny ST}}[A]=\lim_{z\to 0}{\cal D}^z_{\mbox{\tiny ST}}[A]\,,
\end{equation}
with
\begin{equation}
{\cal D}^z_{\mbox{\tiny ST}}[A]\equiv{\cal D}A\,\Delta[A]\,\delta(F[A])\left(\int {\cal D}U' \Delta [A^{U'}]\,\delta\big(F[A^{U'}]\big)\,w[A^{U'}]\right)^{z-1}\,.
\end{equation}
For any strictly positive integer value of $z$, the right-hand-side can be rewritten as a standard field theory involving $z$ replicated versions of the ghost, anti-ghosts and Nakanishi-Lautrup fields, in addition to $z-1$ replicas for the field $U'$.  

This formulation in terms of replicas can be used for practical calculations. The subtle question is, however, how the results obtained for an integer number of replicas $z\in\mathds{N}^*$ can be extrapolated to the zero replica limit $z\to 0$, which requires $z\in\mathds{R}$. To date, it is not clear how this limit should be taken.\footnote{An analytic function $f(z)$ is not uniquely determined from its values for $z\in\mathds{N}^*$.} For this reason we shall write the last step of the Serreau-Tissier approach as
\begin{equation}
\langle{\cal O}\rangle_{\mbox{\tiny ST}}=\lim_{z\to 0}\frac{\int {\cal D}^z_{\mbox{\tiny ST}}[A]\,{\cal O}[A]\,e^{-S_{{YM}}[A]}}{\int {\cal D}^z_{\mbox{\tiny ST}}[A] \,e^{-S_{{YM}}[A]}}\,,
\end{equation}
with a labelled expectation value $\langle{\cal O}\rangle_{\mbox{\tiny ST}}$ to emphasize that, just as in the case of the Faddeev-Popov approach, the identification between $\langle{\cal O}\rangle_{\mbox{\tiny ST}}$ and $\langle{\cal O}\rangle$ is still open to debate.

\subsection{Curci-Ferrari approach}
In addition to these semi-constructive strategies, one can envisage a more phenomenological approach. Indeed, since the standard Faddeev-Popov action is meant to be modified from a proper account of the Gribov copies, one can try to propose possible corrections in the form of operators added to the Faddeev-Popov action\footnote{While fulfilling of course the basic symmetries compatible with the gauge-fixing at hand.} and constrain the corresponding couplings from comparison to experiment and/or numerical simulations. All things considered, the idea is pretty similar to the search of  theories beyond the Standard Model, of course at a more pedestrian level and with the important difference that, in the present case, the action for the physical theory is known, be it the Yang-Mills action or the QCD action. What is searched after is a gauge-fixed version of this physical action, beyond the standard Faddeev-Popov recipe. 

For the Landau gauge, one proposal which has been explored in recent years is the one based on the Curci-Ferrari model \cite{Curci:1976bt,Fradkin:1970ib}. It consists in supplementing the Faddeev-Popov action (\ref{eq:FP_Landau}) with a mass term for the gluon field, $S_{\mbox{\tiny CF}}[A,c,\bar c,h]\equiv S_{{YM}}[A]+\delta S_{\mbox{\tiny CF}}[A,c,\bar c,h]$, with
\begin{equation}\label{eq:CF}
\delta S_{\mbox{\tiny CF}}=\int d^dx\,\left\{\partial_\mu\bar c^a(x)\,D_\mu c^a(x)+ih^a(x)\,\partial_\mu A_\mu^a(x)+\frac{1}{2}m^2 A_\mu^a(x)A_\mu^a(x)\right\}.
\end{equation}
We stress that this is just a model and the question to be asked here is not whether this action represents a {\it bona fide} gauge-fixing but, rather, whether the Curci-Ferrari mass term could represent a dominant contribution to the unknown gauge-fixed action beyond the Faddeev-Popov terms. Of course, other operators could be dominant and the only way to test this is to confront the predictions of the Curci-Ferrari model with those of alternative approaches. The rest of the manuscript will be essentially concerned with this question.

Yet, it is interesting to note that the Curci-Ferrari model possesses quite a number of convenient features. In particular, it is stable under renormalization, which means that there is only one additional parameter to be dealt with. That this is so can be seen as a consequence of the fact that the model possesses a BRST-like symmetry $s_m$. Indeed, if we take $s_mA_\mu^a=sA_\mu^a$ (we want of course the Yang-Mills part to be invariant on its own), the Curci-Ferrari mass term transforms as
\begin{equation}
s_m\int d^dx\,\frac{1}{2}m^2 A_\mu^a(x)A_\mu^a(x)=-\int d^dx\,m^2 c^a(x)\,\partial_\mu A_\mu^a(x)\,, 
\end{equation}
where we have used an integration by parts. The obtained variation is very similar to the third term of Eq.~(\ref{eq:svar_FP}). It follows that the Curci-Ferrari action is invariant under the modified BRST transformation
\begin{equation}
s_mA_\mu^a=D_\mu c^a, \quad s_mc^a=\frac{1}{2}f^{abc}c^bc^c, \quad s_m\bar c^a=ih^a, \quad s_mih^a=m^2 c^a\,,
\end{equation}
where $s_m$ differs from $s$ only on its action over the Nakanishi-Lautrup field. Related to this symmetry are two non-renormalization theorems, that greatly facilitate the renormalization of the Curci-Ferrari model \cite{Wschebor:2007vh,Tissier:2011ey}.

Other properties of the Faddeev-Popov action do not survive the presence of the Curci-Ferrari mass term. In particular, the modified BRST symmetry is not nilpotent anymore since $s^2_m\bar c^a=m^2 c^a$. Moreover, the ``gauge-fixed'' part of the action does not rewrite as a variation under this modified BRST transformation. It should be stressed, however, that it is not clear whether the standard BRST symmetry should be manifest in the infrared, precisely because it appears as a result of neglecting the Gribov copies.\footnote{Interestingly enough, the complicated landscape of Gribov copies is very similar to the landscape of extrema of the energy functional in disordered systems. The latter are usually treated using the Parisi-Sourlas procedure \cite{Parisi:1979ka}, which takes the same steps as the Faddeev-Popov procedure. In this context, it is known that the associated BRST-like symmetry is spontaneously broken below a critical dimension \cite{Tissier:2011zz}. One could very well imagine a similar breaking as a function of the energy in the context of Yang-Mills theories.}\\

Let us close this section by mentioning that the Curci-Ferrari model bears interesting relations to the other approaches described above. For instance, as we have mentioned, the standard Gribov-Zwanziger action needs to be refined with the use of condensates that modify the usual Gribov-Zwanziger propagator as
\begin{equation}
G_{\mbox{\tiny GZ}}(Q)=\frac{Q^2}{Q^4+\gamma^4} \,\,\, \rightarrow \,\,\, G_{\mbox{\tiny RGZ}}(Q)=\frac{Q^2+M^2}{Q^4+(M^2+m^2)Q^2+M^2m^2+\gamma^4}\,.
\end{equation}
A recent investigation on the dynamical generation of these condensates finds that $M^2$ is larger than both the Gribov parameter $\gamma^2$ and the gluon condensate $m^2$ \cite{Dudal:2019ing}. Although the differences are not dramatic, it is interesting to note that for large enough $M^2$, the refined Gribov Zwanziger propagator approaches the Curci-Ferrari one, so not only the corresponding fields decouple, but the Gribov parameter disappears from the tree-level propagator.

Another connection exists with the Serreau-Tissier approach. Indeed, by choosing the weighting functional as
\begin{equation}
w[A]\equiv \exp\left\{-\int d^dx\,\,\frac{1}{2}\,\rho\,A_\mu^a(x)A_\mu^a(x)\right\},
\end{equation}
it could be shown that the perturbative evaluation of any correlation function involving only the original, non-replicated fields, is equivalent to the evaluation of the same correlation function within the Curci-Ferrari model with $\smash{m\equiv z\,\rho}$ \cite{Serreau:2012cg}. The na\"\i ve zero replica limit $z\to 0$ brings us back to the Faddeev-Popov action. However, one could consider the limit in a different way, by first interpreting $z\rho$ as the bare mass of the Curci-Ferrari model and by using it to absorb the corresponding divergences, after which the limit $z\to 0$ would again be trivial but would lead to the Curci-Ferrari model instead.\footnote{The fact that the zero replica limit does not commute with other limits is a well known fact. In applications to disordered systems, it is generally true that one needs to take the infinite volume limit before the zero replica limit.} Of course, the above argument does not fix the value of the renormalized Curci-Ferrari mass in terms of the Yang-Mills coupling, a necessary condition for the procedure to correspond to a {\it bona fide} gauge-fixing, but it gives some indication that the Curci-Ferrari model is maybe not that far from this goal. As a matter of fact, a mechanism for the dynamical generation of such a mass could be identified in a gauge cousin to the Landau gauge \cite{Tissier:2017fqf}.

\subsection{Connection to other approaches}
In addition to the above connections, the Curci-Ferrari model could also be relevant for other continuum approaches including the functional renormalization group \cite{Ellwanger96,Fischer:2008uz,Cyrol:2016tym}, Dyson-Schwinger equations \cite{Alkofer00,vonSmekal:1997ohs,vonSmekal:1997ern,Boucaud:2008ky,Boucaud:2011ug}, the pinch technique \cite{Aguilar:2008xm,Binosi:2009qm,RodriguezQuintero:2010wy}, or the variational approach of \cite{Quandt:2013wna}.

All these approaches take the Faddeev-Popov action as a starting point and aim ideally at circumventing the difficulties of the latter in the infrared by going beyond perturbation theory. However, because it requires introducing a cut-off, the practical implementation of these approaches necessarily leads to an explicit breaking of the BRST symmetry and implies de facto that the starting action is more general than the Faddeev-Popov one. In particular, in all present continuum approaches to Yang-Mills theory, a bare mass term is introduced in one way or an other to deal with the quadratic divergences that the use of a cut-off entails. This mass term can take the form of a subtraction in the Dyson-Schwinger and pinch technique approaches \cite{Huber:2014tva,Aguilar:2008xm}, an additional parameter $m_\Lambda$ at the UV scale in the functional renormalization group approach \cite{Cyrol:2016tym}, or even an explicit mass counterterm in the variational approach \cite{Quandt:2013wna}. 

We stress here that this is not a problem {\it per se}. One is always allowed (and sometimes forced) to use a regulator that breaks certain symmetries of the problem. This just means that the theory space that needs to be considered is larger. The relevant question is rather how the additional couplings are fixed in terms of the Yang-Mills coupling. The answer to this question depends on whether or not BRST symmetry survives in the infrared. In the case it does, one should fix all the additional parameters such that the consequences of BRST symmetry are fulfilled. For instance, in the functional renormalization group approach of \cite{Cyrol:2016tym}, one proposal is to adjust the $m_\Lambda$ parameter so as to trigger the scaling solution, if one deems the latter a consequence of BRST symmetry (from the Kugo-Ojima scenario \cite{Kugo:1979gm}). 

However, in the case where BRST symmetry does not survive in the infrared, the appropriate adjustment of the additional couplings remains an open question and, for all practical purposes so far, the bare mass remains a free parameter. For instance, in the functional renormalization group approach, if one wants to reproduce the lattice correlation functions, the parameter $m_\Lambda$ needs to be adjusted to fit the lattice data. The same is true in the case of the pinch technique with the parameter $\Delta_{\mbox{\scriptsize reg}}^{-1}(0)$ introduced in \cite{Aguilar:2008xm}. It could still be that the lattice decoupling solution can be generated within a BRST invariant framework but this requires subtle mechanisms such as the irregularities proposed in \cite{Cyrol:2016tym} or the Schwinger mechanism discussed in \cite{Binosi:2012sj,Aguilar:2015bud,Aguilar:2016ock}. To date, and to our knowledge, none of these mechanisms was seen to be realized.  In this type of scenarios, studying the Curci-Ferrari model perturbatively can bring an interesting perspective to the rest of continuum approaches, as a way to disentangle what is genuinely non-perturbative from what could become perturbative once a mechanism for the generation of a Curci-Ferrari mass has been identified. We shall present some of these comparisons in subsequent chapters.

As an added note to the original manuscript, let us mention some recent and interesting development that appearead one year after the writing of this thesis. In the context of Dyson-Schwinger equations, new truncations seem to indicate that the earlier ambiguities related to the removal of quadratic divergences do not seem to impact much the gluon propagator or its corresponding dressing function \cite{Huber:2020keu}. Whether the full arbitrariness that the subtraction
of quadratic divergences entails has been tested in \cite{Huber:2020keu} as well as how the observed insensivity to this subtraction
depends on the specifics of the truncation and how it can be implemented in other non-perturbative continuum approaches remain open
questions.

\section{Review of results}
Let us conclude this first chapter by reviewing some of the tests that the Curci-Ferrari model has already passed. We shall also discuss some cases where it is not fully conclusive, as a motivation for the developments to be presented in the rest of the manuscript.

The more direct way to test the Curci-Ferrari model is to compare its predictions for Landau gauge-fixed correlation functions with the corresponding predictions obtained from the lattice: on the one hand, gauge-fixed correlation functions are the best quantities to test the specificities of a given gauge; on the other hand, gauge-fixed lattice simulations are less sensitive to the Gribov ambiguity since gauge-fixing is done by selecting one copy per orbit.\footnote{It is statistically rare that two configurations are chosen on the same gauge orbit.} 

\subsection{Zero temperature}
\begin{figure}[t!]
\begin{center}
\includegraphics[height=0.33\textheight]{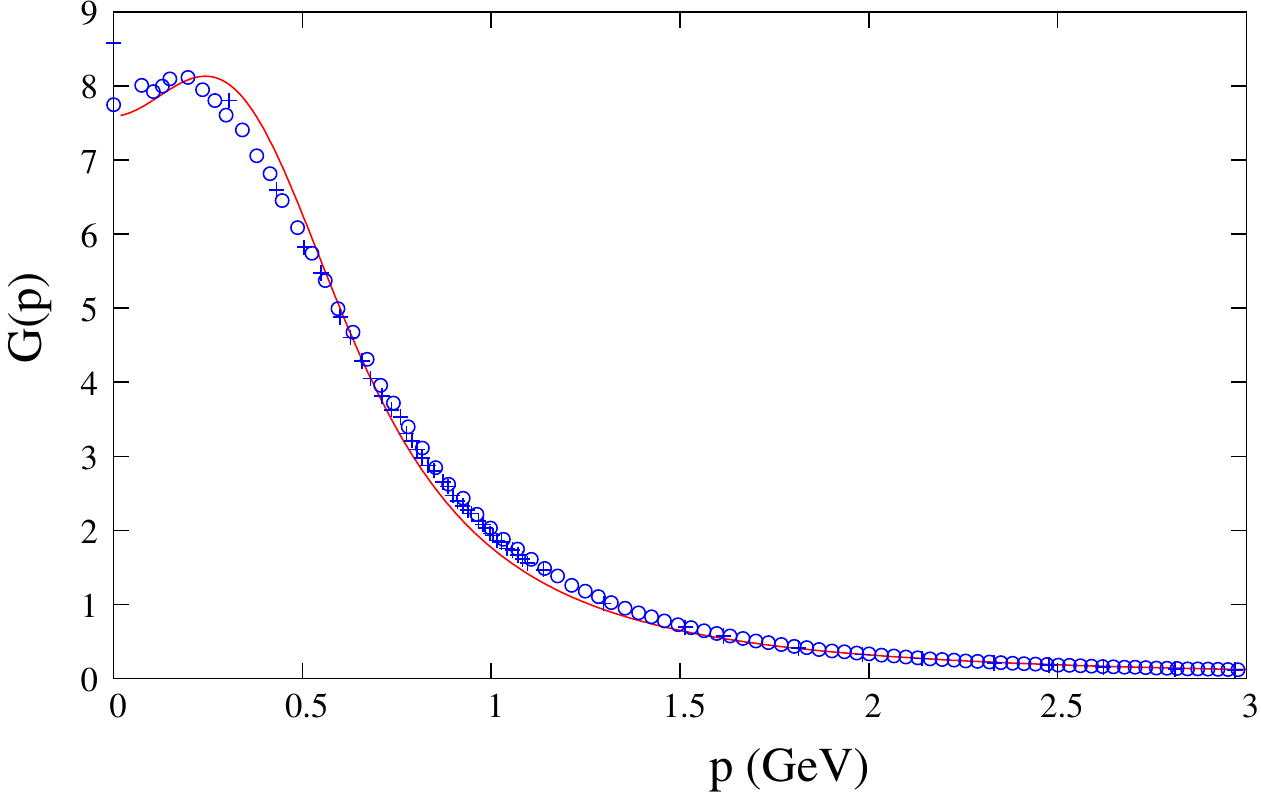}\\
\vglue4mm
\hglue10mm\includegraphics[height=0.33\textheight]{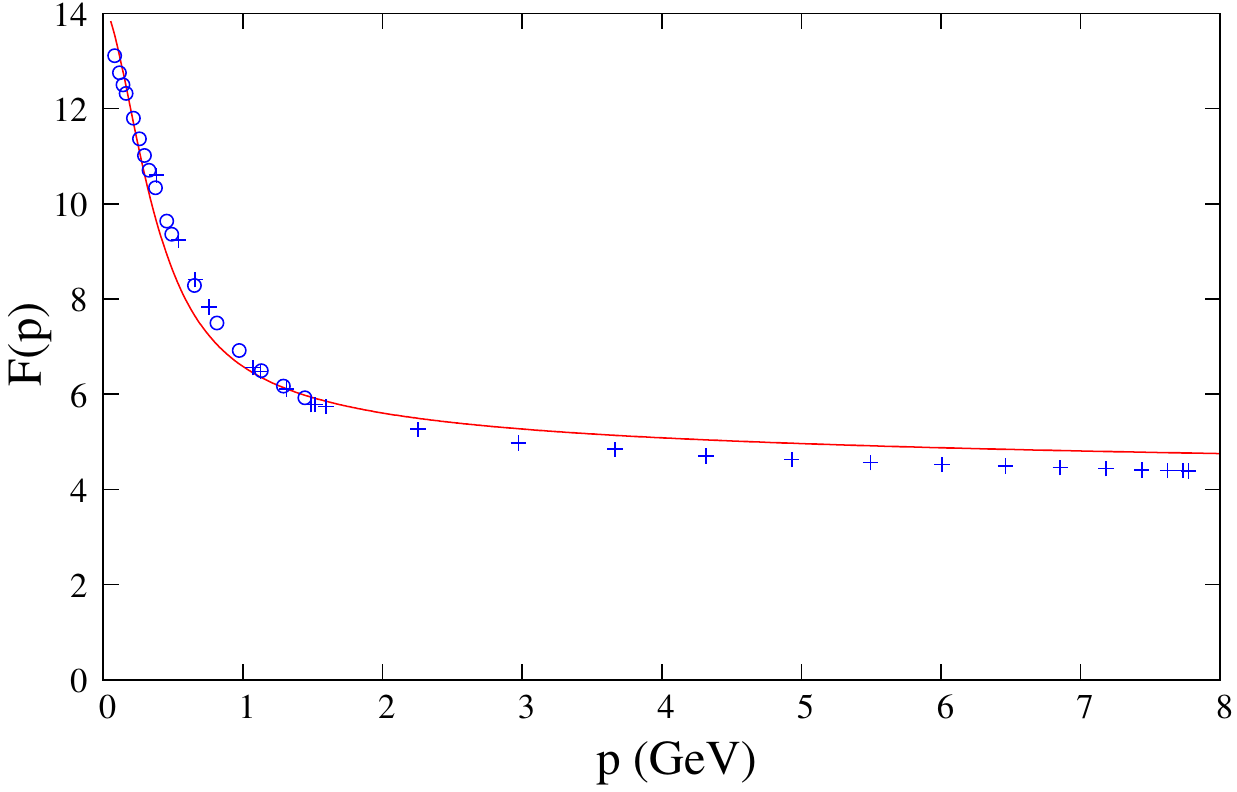}
\mycaption[Vacuum Curci-Ferrari two-point correlators]{SU(3) vacuum gluon propagator and ghost dressing function as computed in the Curci-Ferrari model at one-loop order, compared to Landau gauge-fixed lattice simulations of the same quantities. The parameters are defined in the infrared safe scheme of \cite{Tissier:2011ey} and take the values $g\simeq 4.9$ and $m\simeq 540\,$MeV for a renormalization scale $\mu\simeq 1\,$GeV.}
\label{fig:CF_two}
\end{center}
\end{figure}

\begin{figure}[t!]
\begin{center}
\includegraphics[height=0.33\textheight]{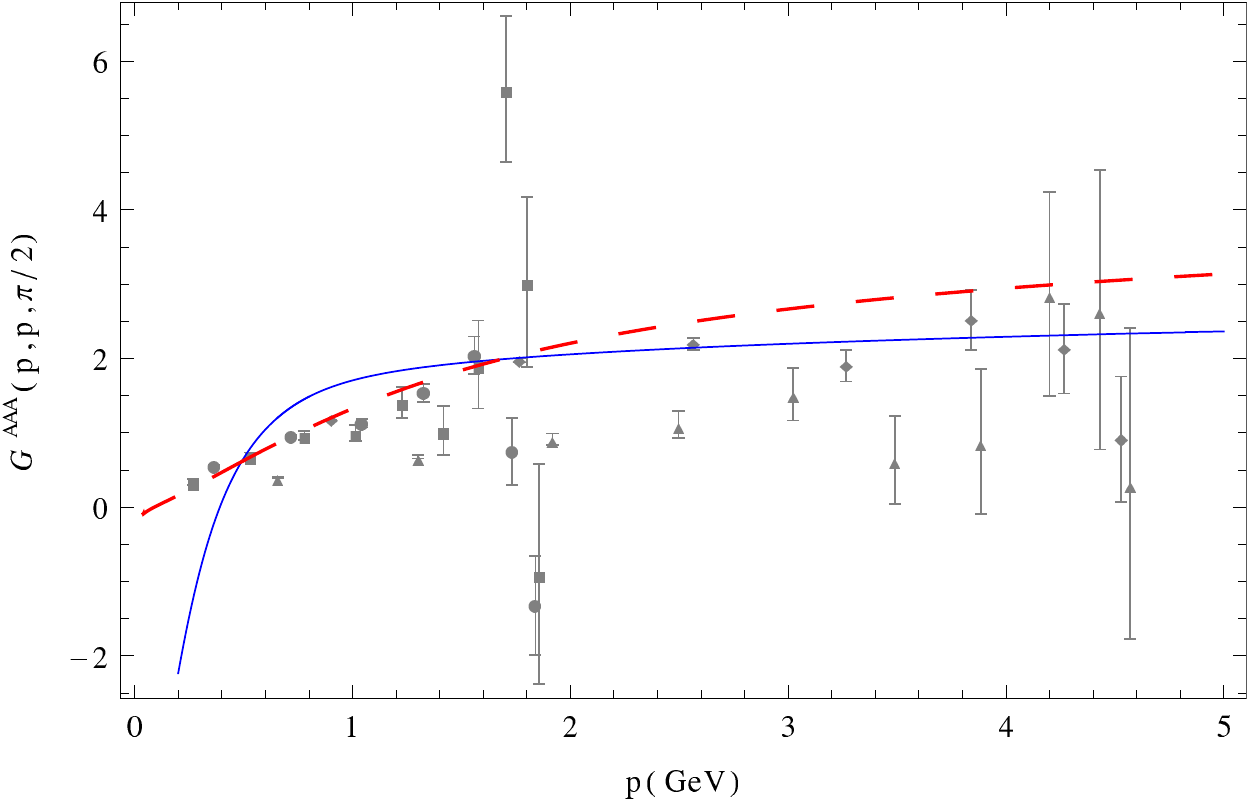}\\
\vglue4mm
\hglue10mm\includegraphics[height=0.33\textheight]{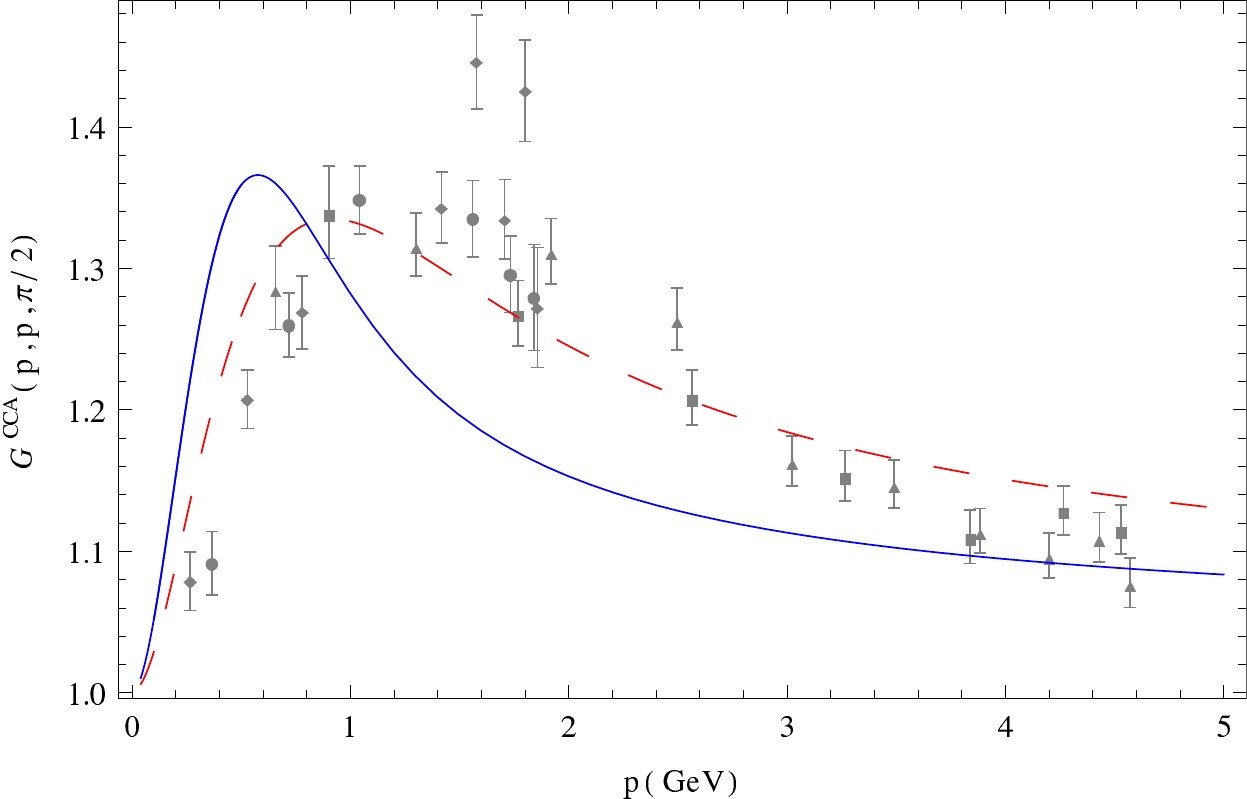}
\mycaption[Vacuum Curci-Ferrari three-point vertices]{SU(3) vacuum three-gluon and ghost-antighost-gluon vertices for certain configurations of the external momenta as computed in the Curci-Ferrari model at one-loop order, compared to Landau gauge-fixed lattice simulations of the same quantities.}
\label{fig:CF_three}
\end{center}
\end{figure}

In a series of works \cite{Tissier:2010ts,Tissier:2011ey,Pelaez:2013cpa}, the lattice results for the vacuum two-point correlation functions in the Landau gauge \cite{Cucchieri_08b,Cucchieri_08,Bornyakov2008,Cucchieri09,Bogolubsky09,Bornyakov09,Dudal10,Maas:2011se}, as well as for the three-point vertices, have been systematically compared to the one-loop perturbative predictions of the Curci-Ferrari model. Some of these comparisons are shown in Figs.~\ref{fig:CF_two} and \ref{fig:CF_three}. 

The agreement for the two-point functions is rather impressive. A recent two-loop calculation confirms (and even improves) these results showing that the very good agreement at one-loop was not accidental \cite{Gracey2l}. The agreement is less impressive for the three-point vertices but still qualitatively very good, given the simplicity of the one-loop approach and the uncertainties of the lattice simulations. In particular, the Curci-Ferrari model predicts a zero crossing of the three-gluon vertex, as also seen in other approaches \cite{Athenodorou:2016oyh}. The comparison to other continuum approaches has been more systematically discussed in \cite{Reinosa:2017qtf}. For instance, the qualitative dependence of the results on the value of the Curci-Ferrari mass has been found to be similar to the dependence of the functional renormalization group results with respect to the value of the initial parameter $m_\Lambda$ alluded to above.

\begin{figure}[t!]
\begin{center}
\includegraphics[height=0.42\textheight]{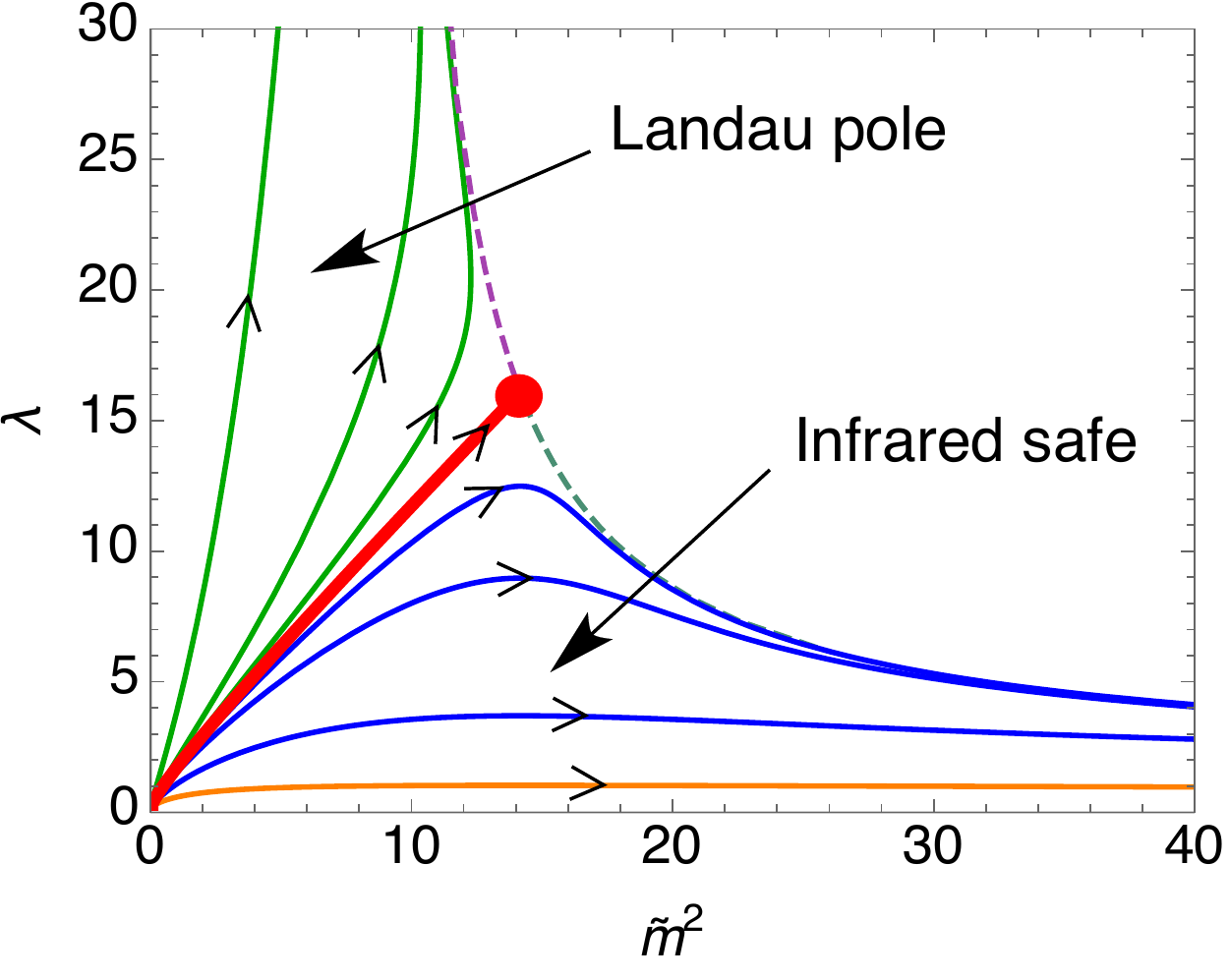}\\
\mycaption[One-loop flow structure of the Curci-Ferrari model]{The one-loop flow structure of the Curci-Ferrari model in the plane $(\tilde m^2,\lambda)$, where $\tilde m^2(\mu)\equiv m^2/\mu^2$ and $\lambda(\mu)\equiv g^2(\mu)N/(16\pi^2)$.}
\label{fig:CF_flow}
\end{center}
\end{figure}

\begin{figure}[t!]
\begin{center}
\includegraphics[height=0.42\textheight]{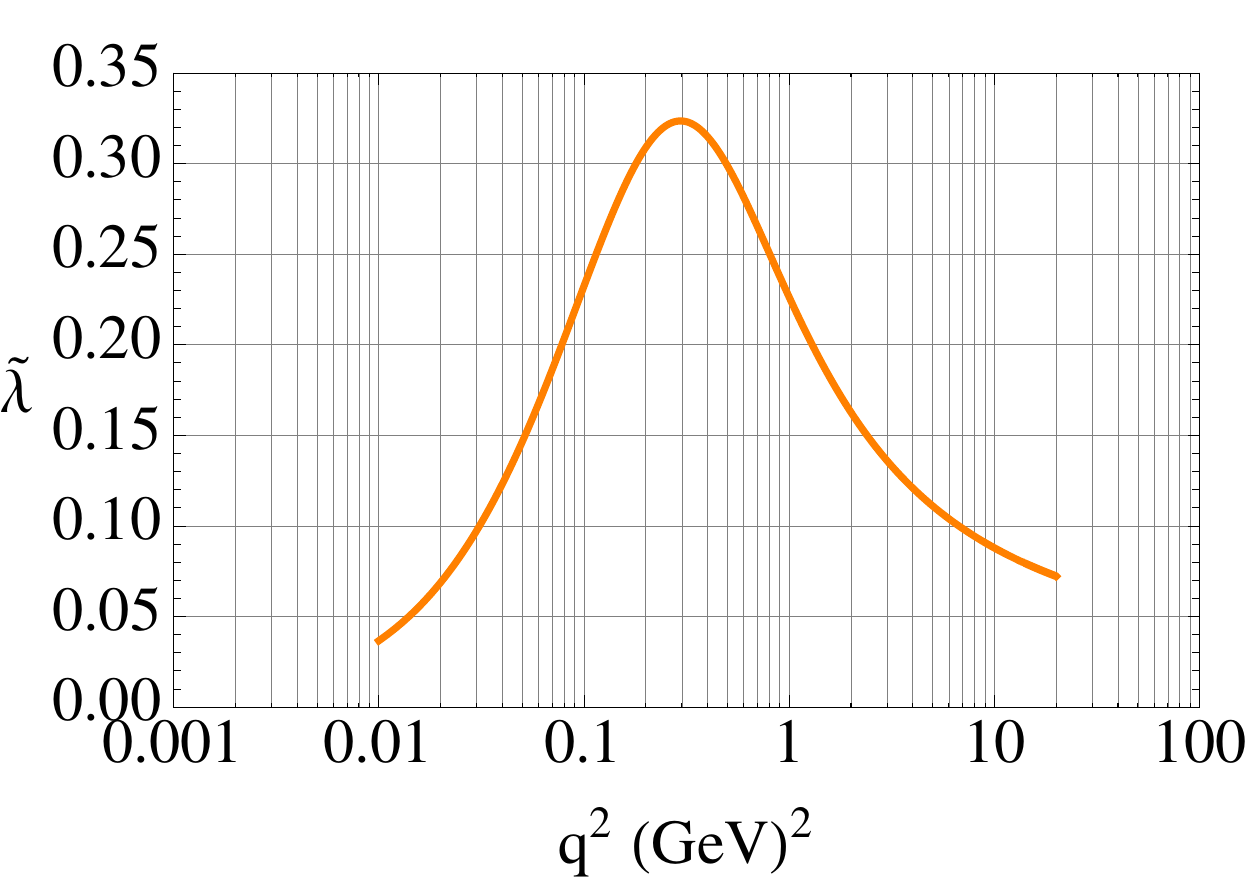}\\
\mycaption[One-loop running coupling from the Curci-Ferrari model]{Running of the relevant expansion parameter of the Curci-Ferrari model $\tilde\lambda(\mu)\equiv \lambda(\mu)/(1+\tilde m^2(\mu))$ along the trajectory that best fits the lattice data. The quantities $\tilde m^2(\mu)$ and $\lambda(\mu)$ are defined as in Fig.~\ref{fig:CF_flow}.}
\label{fig:CF_running}
\end{center}
\end{figure}

It is also to be mentioned that the flow structure of the Curci-Ferrari model has been studied at one-loop order \cite{Tissier:2011ey,Serreau:2012cg,Reinosa:2017qtf} (and also, more recently, at two-loop order \cite{Gracey2l}), see Fig.~\ref{fig:CF_flow}. Interestingly enough, it is found that, in addition to renormalization group trajectories displaying an infrared Landau pole, reminiscent from the one in the Fadeev-Popov approach, there exists a family of trajectories which can be defined over all energy scales. In particular, the one trajectory that best fits the lattice results remains always in the perturbative regime, see Fig.~\ref{fig:CF_running}, supporting the idea that the perturbative expansion can be used at all scales in the Curci-Ferrari model.

\subsection{Finite temperature}
Gauge-fixed lattice simulations also provide results for the Yang-Mills two-point correlation functions at finite temperature \cite{Heller:1995qc,Heller:1997nqa,Cucchieri:2000cy,Cucchieri:2001tw,Cucchieri:2007ta,Fischer:2010fx,Cucchieri:2011di,Cucchieri:2012nx,Aouane:2011fv,Maas:2011ez,Silva:2013maa}, as a further testing ground for the various continuum approaches. 

One particularly scrutinized quantity has been the longitudinal susceptibility given by the zero momentum value of the longitudinal gluon propagator,\footnote{In the Landau gauge, the gluon propagator is ($4$d) transverse. At finite temperature, however, this transverse component decomposes into so-called ($3$d) longitudinal and transverse components.} $\chi={\cal G}_L(p^2=0)$. Early lattice results showed a rather sharp variation of this quantity around the Yang-Mills deconfinement transition \cite{Cucchieri:2011di,Maas:2011ez} and opened the way to speculations about the possibility of accessing the physical transition from the study of gauge-variant quantities \cite{Fister:2011uw,Huber:2013yqa,Quandt:2015aaa}.

\begin{figure}[t!]
\begin{center}  
\includegraphics[height=0.33\textheight]{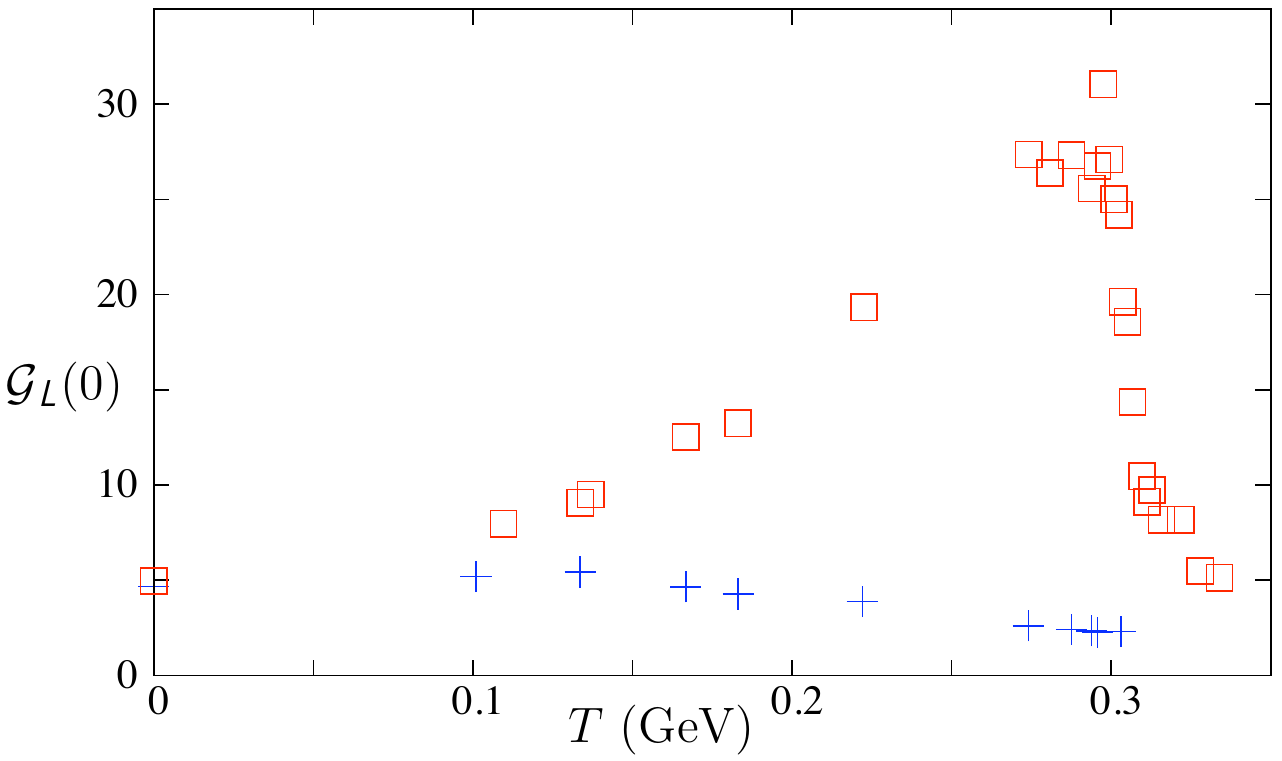}\\
\mycaption[Zero-momentum value of the longitudinal gluon propagator]{Zero momentum value of the longitudinal gluon propagator (susceptibility), as obtained from the Curci-Ferrari model at one-loop and compared to the results of Landau gauge-fixed lattice simulations.}
\label{fig:CF_Debye}
\end{center}
\end{figure}

The fit of the lattice data to the one-loop Curci-Ferrari two-point functions at finite temperature is globally satisfactory, see \cite{Reinosa:2013twa} for more details. However, the temperature dependence of the susceptibility differs substantially from the one obtained in lattice simulations, see Fig.~\ref{fig:CF_Debye}. This, however, does not necessarily point out to a limitation of the Curci-Ferrari approach since the other continuum approaches also fail to reproduce this particular feature of the lattice results.\footnote{It should be mentioned that the longitudinal gluon propagator is very sensitive to the details of the lattice simulation. Improved lattice results show a much less pronounced sensitivity around the transition. It would be interesting to update the comparisons between the continuum and the lattice approaches in light of these more recent results.}

Rather, it has been suggested in \cite{Fister:2011uw} that the discrepancy between continuum and lattice results may be due to the fact that the order parameter associated to the deconfinement transition is not properly accounted for in the Landau gauge (and in fact in most standard gauge fixings). As put forward in \cite{Braun:2007bx}, one way to tackle this problem is to extend the standard Landau gauge into the so-called Landau-deWitt gauge that allows for the possibility of a background gauge field configuration. Following this idea, we shall extend the Curci-Ferrari model in the presence of a background and investigate the perturbative predictions of the model at finite temperature, in particular with regard to the QCD/YM theory phase diagrams. Before embarking on this adventure, in the next three chapters, we shall make a technical digression on the application of background field methods at finite temperature, explaining in detail why and how they allow to circumvent the limitations of standard gauge fixings.
\clearemptydoublepage
%
%
%
%
\let\textcircled=\pgftextcircled
\chapter{Deconfinement transition\\ and center symmetry}
\label{chap:generalities}

\initial{T}he goal of this chapter is to motivate why standard gauge fixings do not provide the appropriate framework in order to study the confinement/deconfinement transition in the continuum. To this purpose, we shall start reviewing some basic knowledge of finite temperature QCD and the associated phase transition.

In fact, the physical QCD transition does not correspond to a sharp transition between two distinct phases of matter but, rather, to a smooth crossover of the various thermodynamical observables characterizing the system \cite{Borsanyi:2013bia}. To some extent, this obscures the theoretical understanding of the transition itself. It proves convenient, therefore, to consider particular limits for which the transition becomes a sharp one, testable with robust order parameters. One may then hope to access the physical case more simply from these limiting cases.

For the most part, this manuscript will be concerned with the case of pure Yang-Mills theory which can be seen as the limit of QCD with infinitely heavy quarks: $S_{QCD}\to S_{YM}$. In this simplified setting, the transition can be probed with the help of an observable known as the Polyakov loop. Moreover, it can be interpreted as the spontaneous breaking of a special symmetry known as center symmetry. This chapter is mainly devoted to a review of these concepts for they play a major role in the rest of the manuscript, even when quarks are included back in the analysis. 

Although the applications to be covered in the present manuscript will concern mainly the SU($2$) and SU($3$) gauge groups, we shall introduce the various relevant concepts and later derive the various relevant formulas within the general case of SU(N). Some general discussion and an application to the SU($4$) gauge group can be found in Chapter~\ref{chap:more}.

\section{The Polyakov loop}
\label{sec:polyakov}
Let us start recalling the definition of the Polyakov loop and how it relates to the confinement/deconfinement transition \cite{Polyakov:1978vu}. 

\subsection{Definition}
For a given gauge field configuration $A_\mu(x)=iA_\mu^a(x) t^a$ on the Lie algebra, we define the Wilson line wrapped along the compact Euclidean time direction as
\begin{equation}
 L_A(\vec{x})\equiv {\cal P}\exp\left\{\int_0^\beta d\tau\, A_0(\tau,\vec{x})\right\}.
\end{equation}
Here, ${\cal P}$ denotes the path-ordering along the Euclidean time direction that arranges Lie algebra elements $A_0(\tau,\vec{x})$ from left to right according to the decreasing value of the argument $\tau$. We also define the normalized trace of $L_A(\vec{x})$ as $\Phi_A(\vec{x})\equiv \mbox{tr}\,L_A(\vec{x})\,/N$. Finally, the {\it Polyakov loop} $\ell$ is defined as the expectation value of $\Phi_A(\vec{x})$:
\begin{equation}\label{eq:l_def}
 \ell\equiv \big\langle \Phi_A(\vec{x})\big\rangle_{YM}\equiv\frac{\int {\cal D}A\,{\cal \Phi}_A(\vec{x})\,e^{-S_{YM}[A]}}{\int {\cal D}A\,e^{-S_{YM}[A]}}\,.
\end{equation}
We mention that the Wilson line $L_A(\vec{x})$ is a unitary matrix and, then, the modulus of its normalized trace $\Phi_A(\vec{x})$ is not larger than unity. Moreover, since the averaging weight in Eq.~(\ref{eq:l_def}) is positive, it follows that $|\ell|\leq 1$.

\subsection{Order parameter interpretation}
It can be shown that $\ell$ is related to the free-energy cost $\Delta F$ for bringing a static test quark into the thermal bath of gluons as \cite{Polyakov:1978vu,Svetitsky:1985ye}
\begin{equation}
 \ell= e^{-\beta\Delta F}.
\end{equation}
The interpretation of the Polyakov loop as an order parameter for the confinement/deconfinement transition follows immediately. Indeed, if we assume that the system is in the confined phase, no free quarks are allowed to roam around in the thermal bath ($\Delta F=\infty$) and then $\ell= 0$. In contrast, in the deconfined phase, quarks can be introduced in the system provided some finite amount of energy is paid ($\Delta F<\infty$) and then $\ell\neq 0$.

The behavior of the Polyakov loop as a function of the temperature has been studied on the lattice \cite{Svetitsky:1985ye,Kaczmarek:2002mc,Lucini:2005vg,Greensite:2012dy,Smith:2013msa} where it yields a second order type transition in the case of the SU($2$) gauge group and a first order type transition for SU(N) with $N\geq 3$. It is one of the purposes of this manuscript to review how these seemingly non-perturbative results can be accessed from perturbative methods, in the framework of the Curci-Ferrari model.

\subsection{Gauge transformations}
It will be useful in the following to recall how $L_A(\vec{x})$ transforms under a gauge transformation (\ref{eq:gauge_transfo}). To that aim, we introduce a slightly more general Wilson line
\begin{equation}
 L_A(\sigma_f,\sigma_i,\vec{x})\equiv {\cal P}\exp\left\{\int_{\sigma_i}^{\sigma_f} d\tau\, A_0(\tau,\vec{x})\right\},
\end{equation}
with $0<\sigma_i<\sigma_f<\beta$. 

Owing to the definition of the path-ordering operation ${\cal P}$, we find the two conditions
\begin{eqnarray}
 \frac{\partial}{\partial\sigma_f}L_A(\sigma_f,\sigma_i,\vec{x}) & \!\!\!\!\!=\!\!\!\!\! & A_0(\sigma_f,\vec{x})\,L_A(\sigma_f,\sigma_i,\vec{x})\,,\\
 L_A(\sigma_i,\sigma_i,\vec{x}) & \!\!\!\!\!=\!\!\!\!\! & \mathds{1}\,,
\end{eqnarray}
which characterize $L_A(\sigma_f,\sigma_i,\vec{x})$ entirely. In particular, it is easily checked that $U(\sigma_f,\vec{x})\,L_A(\sigma_f,\sigma_i,\vec{x})\,U^\dagger(\sigma_i,\vec{x})$ obeys the same two conditions as the Wilson line $L_{A^U}(\sigma_f,\sigma_i,\vec{x})$ and, therefore, that
\begin{equation}\label{eq:id}
 L_{A^U}(\sigma_f,\sigma_i,\vec{x})=U(\sigma_f,\vec{x})\,L_A(\sigma_f,\sigma_i,\vec{x})\,U^\dagger(\sigma_i,\vec{x})\,.
\end{equation}
The corresponding transformation rule for $L_A(\vec{x})$ is obtained by setting $\sigma_i=0$ and $\sigma_f=\beta$.

\section{Center symmetry}
\label{sec:center}
The previous considerations relate the confinement/deconfinement transition to the behavior of the Polyakov loop but do not explain why the latter could vanish in some instances. As we now recall, this is related to a very special symmetry of the Euclidean YM action (\ref{eq:SYM}) at finite temperature known as center symmetry \cite{Svetitsky:1985ye,Greensite:2011zz,Pisarski:2002ji}.

\subsection{The role of boundary conditions}
The Yang-Mills action is invariant under gauge transformations (\ref{eq:gauge_transfo}). At finite temperature, the only, however important, constraint on the field $U(x)$ is that it should preserve the periodic boundary conditions of the gauge fields along the Euclidean time direction, that is, it should be such that $A^U_\mu(\tau+\beta,\vec{x})=A^U_\mu(\tau,\vec{x})\,,\forall A_\mu(x)$. After some simple algebra, this condition rewrites
\begin{equation}
 \partial_\mu Z(\tau,\vec{x}) -[A_\mu(x),Z(\tau,\vec{x})]=0\,,\forall A_\mu(x)\,,
\end{equation}
where $Z(\tau,\vec{x})\equiv U^\dagger(\tau,\vec{x})\,U(\tau+\beta,\vec{x})$. Since this has to be true for any possible gauge field configuration, it follows that $Z(\tau,\vec{x})$ should be a constant SU(N) matrix $Z$ commuting with all the generators of the su(N) Lie algebra and, therefore, with all the elements of the SU(N) group.\\

In conclusion, Euclidean YM theory at finite temperature is invariant under gauge transformations that are $\beta$-periodic in time up to an element $Z$ of the center of the gauge group:
\begin{equation}\label{eq:periodic}
 U(\tau+\beta,\vec{x})=U(\tau,\vec{x})\,Z,\,\forall\tau,\,\forall\vec{x}\,.
\end{equation}
The center elements are easily seen to be of the form $Z=e^{i\frac{2\pi}{N}k}\mathds{1}$, with $k=0,\dots,N-1$ such that $\det\,Z=1$. These transformations form a group, referred to as the group of {\it twisted gauge transformations,} denoted ${\cal G}$ in what follows.

\subsection{Relation to the deconfinement transition}
To unveil the relevance of the twisted gauge transformations for the deconfinement transition, let us see how the Polyakov loop transforms under the action of ${\cal G}$. First, owing to Eq.~(\ref{eq:id}), we find the transformation rule
\begin{eqnarray}\label{eq:transfo_PhiA}
 \Phi_{A^U}(\vec{x}) & \!\!\!\!\!=\!\!\!\!\! & \frac{1}{N}\,\mbox{tr}\,U(\beta,\vec{x})\,L_A(\vec{x})\,U^\dagger (0,\vec{x})\\
& \!\!\!\!\!=\!\!\!\!\! & \frac{1}{N}\,\mbox{tr}\,U^\dagger (0,\vec{x})\,U(\beta,\vec{x})\,L_A(\vec{x})=e^{i\frac{2\pi}{N}k}\,\Phi_A(\vec{x})\,,\nonumber
\end{eqnarray}
where we have used the cyclic property of the trace as well as Eq.~(\ref{eq:periodic}).

In order to derive the corresponding transformation rule for $\ell$, we note that, in general, the definition (\ref{eq:l_def}) makes sense only in the presence of an infinitesimal, symmetry-breaking source term added to the action: $S_{YM}\to S_{YM}-0^+e^{i\alpha}\int d\vec{x}\,\Phi_A(\vec{x})$. To account for the presence of this external source of modulus $0^+$ and direction $e^{i\alpha}$, we denote the Polyakov loop more rigorously as $\ell_\alpha$. 

With these words of caution in mind, we can now perform a change of variables in Eq.~(\ref{eq:l_def}) in the form of a twisted gauge transformation:
\begin{eqnarray}
\ell_\alpha & \!\!\!\!\!=\!\!\!\!\! & \frac{\int {\cal D}A^U\,\Phi_{A^U}(\vec{x})\,e^{-S_{YM}[A^U]+0^+e^{i\alpha}\int d\vec{x}\, \Phi_{A^U}(\vec{x})}}{\int {\cal D}A^U\,e^{-S_{YM}[A^U]+0^+e^{i\alpha}\int d\vec{x}\, \Phi_{A^U}(\vec{x})}}\\
& \!\!\!\!\!=\!\!\!\!\! & \frac{\int {\cal D}A\,\Phi_{A^U}(\vec{x})\,e^{-S_{YM}[A]+0^+e^{i\alpha}\int d\vec{x}\, \Phi_{A^U}(\vec{x})}}{\int {\cal D}A\,e^{-S_{YM}[A]+0^+e^{i\alpha}\int d\vec{x}\, \Phi_{A^U}(\vec{x})}}=e^{i\frac{2\pi}{N}k}\ell_{\alpha+\frac{2\pi}{N}k}\,,\forall k\,.\nonumber
\end{eqnarray}
In the last steps, we have used the transformation rule (\ref{eq:transfo_PhiA}), the invariance of both ${\cal D}A$ and $S_{YM}[A]$ under twisted gauge transformations and the fact that the periodic boundary conditions of the fields are preserved. We have thus found that, under a phase multiplication of the external source by $e^{-i\frac{2\pi}{N}k}$, the Polyakov loop transforms as 
\begin{equation}\label{eq:l_transfo}
 \ell_{\alpha-\frac{2\pi}{N}k}=e^{i\frac{2\pi}{N}k}\ell_\alpha\,.
\end{equation}

\vglue4mm

\noindent{From here, two scenarios are possible:}
\begin{itemize}
\item If the symmetry ${\cal G}$ is realized in the Wigner-Weyl sense, then the definition (\ref{eq:l_def}) makes sense without the need of the symmetry-breaking source. The transformation rule (\ref{eq:l_transfo}) becomes then a constraint equation for the Polyakov loop implying that the latter has to vanish.

\item In contrast, in the case where the symmetry is spontaneously broken in the Nambu-Goldstone sense, one expects degenerate ground states with different values of $\ell$ connected to each other by (\ref{eq:l_transfo}). The Polyakov loop does not need to be zero anymore and the system is in the deconfined phase.\footnote{The statement is certainly true for the SU($2$) and SU($3$) groups. The general discussion in the case of SU(N) is slightly more involved, as we discuss in Chapter~\ref{chap:more}.}
\end{itemize}

\pagebreak

We have thus related the existence of a confined/deconfined phase to the explicit/broken realization of the symmetry group ${\cal G}$. This connection offers a promising way for studying the confinement/deconfinement transition which we shall exploit in subsequent chapters.

\subsection{The center symmetry group}
For later considerations, it will be important to realize that not all the transformations of ${\cal G}$ are physical. To see this, consider the subgroup ${\cal G}_0$ of {\it periodic gauge transformations,} corresponding to twisted gauge transformations with $k=0$. The operator $\Phi_A(\vec{x})$ is an observable (in the sense that its expectation value $\ell$ measures a physical quantity) which is left invariant by the elements of ${\cal G}_0$. It is therefore legitimate to assume that the elements of ${\cal G}_0$ correspond to true, and therefore unphysical, gauge transformations.

In contrast, even though they also look like gauge transformations, the elements of ${\cal G}$ that do not belong to ${\cal G}_0$ correspond to physical transformations because they lead to explicit modifications of the observable $\Phi_A(\vec{x})$. In fact, two transformations $U_1,U_2\in {\cal G}$ that are related to each other by $U_1=U_2\,U_3$, with $U_3\in {\cal G}_0$, lead to the same modification of $\Phi_A(\vec{x})$, and, therefore, have the same physical meaning. It follows that the true physical content of ${\cal G}$ lies within the quotient set ${\cal G}/{\cal G}_0$.  Moreover, since the subgroup ${\cal G}_0$ is normal within ${\cal G}$ (that is $\forall U\in{\cal G}$, $\forall U_0\in{\cal G}_0$, $UU_0U^\dagger\in {\cal G}_0$), this quotient set ${\cal G}/{\cal G}_0$ inherits a group structure from ${\cal G}$, isomorphic to $\mathds{Z}_N$. This is known as the {\it center symmetry group.}\\

In summary, two transformations of ${\cal G}$ that belong to the same class in ${\cal G}/{\cal G}_0$ should be seen as redundant versions of the same, physical center transformation. In particular, the elements of ${\cal G}_0$ do not correspond to physical transformations since they define the neutral element of ${\cal G}/{\cal G}_0$. These simple remarks will play a major role in the next two chapters.

\section{Center symmetry and gauge fixing}
\label{sec:gauge}
The results that we have recalled so far rely on the gauge-invariant formulation of the theory. In practice, however, continuum approaches to the deconfinement transition require working in a given gauge. It is therefore important to assess whether and how the previous results extend to a gauge-fixed setting. 

\subsection{Gauge-fixed measure}
Gauge fixing can be interpreted as the replacement of the gauge-invariant measure ${\cal D}A$ that enters the evaluation of any observable by a non gauge-invariant measure ${\cal D}_{\mbox{\scriptsize gf}}\,[A]$, without affecting the value of the observable. For instance, in the case of the Polyakov loop, one writes
\begin{equation}\label{eq:obs}
 \ell=\frac{\int {\cal D}A\,\Phi_A(\vec{x})\,e^{-S_{YM}[A]}}{\int {\cal D}A\,e^{-S_{YM}[A]}}=\frac{\int {\cal D}_{\mbox{\scriptsize gf}}\,[A]\,\Phi_A(\vec{x})\,e^{-S_{YM}[A]}}{\int {\cal D}_{\mbox{\scriptsize gf}}\,[A]\,e^{-S_{YM}[A]}}\,,
\end{equation}
where the appropriate infinitesimal symmetry-breaking source, see the discussion above, has been left implicit. 

The subtleties related to the construction of bona-fide and practical gauge-fixed measures that realize the above type of identities have been discussed in the previous chapter and we shall not repeat them here.\footnote{We recall, however, that the gauge-fixed measure ${\cal D}_{\mbox{\tiny gf}}\,[A]$ is usually not local. Its localization requires one to introduce integrations over auxiliary fields, such as the Faddeev-Popov ghost and anti-ghost fields, or the Nakanishi-Lautrup field. Conveniently, however, the formal discussion to be presented in this and the next two chapters can be done without introducing these fields.} The only point that should be stressed here is that, even though both integral representations of the Polyakov loop lead to the transformation rule (\ref{eq:l_transfo}), at the heart of its order parameter interpretation, the origin of this transformation rule is slightly different in each case. The crux of the problem is whether the transformation rule (\ref{eq:l_transfo}) can be maintained in the presence of approximations.
 
\subsection{Gauge fixing and approximations}
As we have seen above, in the non gauge-fixed setting, the transformation rule (\ref{eq:l_transfo}) is already essentially manifest at the level of the corresponding integrand in Eq.~(\ref{eq:obs}) because both ${\cal D}A$ and $S_{YM}[A]$ are invariant under twisted gauge transformations while $\Phi_A(\vec{x})$ transforms according to Eq.~(\ref{eq:transfo_PhiA}). In this case, were we to approximate the evaluation of the functional integral by selecting only some of the gauge field configurations (as one would do on the lattice), the transformation rule (\ref{eq:l_transfo}) would remain true if, for any gauge-field configuration $A_\mu(x)$ considered in the evaluation of the integral, we would also include a representative of each of the possible twisted gauge transformations of $A_\mu(x)$, for $k=1,\dots,N-1$. This can always be achieved in practice without summing over all configurations under the functional integral.

The situation is drastically different within a gauge-fixed setting. Indeed, because twisted gauge transformations look so much like true gauge transformations, it is generally true that standard gauge-fixing conditions, and in turn the corresponding gauge-fixed measures ${\cal D}_{\mbox{\scriptsize gf}}\,[A]$, are not invariant under twisted gauge transformations. In this case, the integrand in the gauge-fixed integral representation of $\ell$ does not have the expected transformation rule.\footnote{Due to the existence of Gribov copies, it could be that, for certain configurations $A_\mu(x)$, there exist transformations $U\in {\cal G}$, such that ${\cal D}_{\mbox{\tiny gf}}\,[A^U]={\cal D}_{\mbox{\tiny gf}}\,[A]$. This does not solve the problem, however, because it is not guaranteed that this property applies to an arbitrary gauge field configuration and to a collection of transformations representing each of the physical center transformations.} The latter emerges only after integrating over \underline{all} gauge field configurations, and, because the functional integral is never computed exactly, the transformation rule of $\ell$ is usually not manifest.\\

In the next chapter, we show how these shortcomings can be cured by upgrading standard gauge fixings into background-extended gauge fixings.

\clearemptydoublepage
%
%
%
%
\let\textcircled=\pgftextcircled
\chapter{Background field gauges:\\ States and Symmetries}
\label{chap:bg}

\initial{I}n this chapter, we recall how the shortcomings of standard gauge fixings concerning the description of the confinement/deconfinement transition in the continuum can be cured by means of a background field. We put special emphasis on the justification of some of the steps that are usually taken when implementing background field methods at finite temperature. In particular, we discuss the underlying hypotheses concerning the gauge-fixed measure and we stress the importance of finding a good, redundancy-free description of the states of the system that then allows for the analysis of center symmetry breaking.

We try to follow a semi-deductive approach: we start recalling the basic properties of background field gauges, argue why these properties are a priori not sufficient to solve the problem identified at the end of the previous chapter and, finally, show how the problem is solved thanks to the notion of self-consistent backgrounds. The main point of this chapter is that such backgrounds play the role of order parameters for center symmetry. We give a formal but relatively compact proof of this known result and postpone a more explicit illustration to the next chapter. 

Self-consistent backgrounds offer an alternative route to the study of the confinement/deconfinement transition \cite{Braun:2007bx} that we shall pursue in subsequent chapters within the framework of the Curci-Ferrari model. The present chapter is also the opportunity to introduce various related constructs, such as the background field effective action or the background-dependent Polyakov loop that shall ubiquitously appear in the rest of the manuscript. For later use, we also extend the considerations on center symmetry to other symmetries, in particular to charge conjugation.

\section{The role of the background field}\label{sec:bg}
In the previous chapter, we claimed that the limitations of standard gauge fixings regarding the continuum description of the deconfinement transition could be cured with the help of background field gauges. Let us now review some of the properties of this type of gauges that shall eventually help reaching this conclusion. We stress however that these properties, although helpful, will not be sufficient. The ultimate solution to the problem will be presented in Sec.~\ref{sec:states}.

\subsection{Center-invariant gauge-fixed measure}
As we saw in the previous chapter, the problem with standard gauge fixings is that the gauge-fixing conditions and, in turn, the associated gauge-fixed measures ${\cal D}_{\mbox{\scriptsize gf}}\,[A]$, are not invariant under twisted gauge transformations. This makes it difficult to ensure the appropriate transformation rule of the Polyakov loop under center transformations (and, therefore, its order parameter interpretation), unless the Polyakov loop is computed exactly. The idea behind introducing a background is  to upgrade the gauge fixing in such a way that the corresponding gauge-fixed measure becomes invariant (in a sense to be specified below), with the hope that the transformation rule of the Polyakov loop applies even in the presence of approximations.

For the sake of illustrating the procedure, let us consider the case of the Landau gauge. Given a gauge field configuration $A_\mu(x)$ such that $\partial_\mu A_\mu(x)=0$ and a twisted gauge transformation $U\in {\cal G}$, one has typically $\partial_\mu A_\mu^U(x)\neq 0$ and, therefore, ${\cal D}_{\mbox{\scriptsize Landau}}\,[A^U]\neq {\cal D}_{\mbox{\scriptsize Landau}}\,[A]$. To cure this obviously inconvenient feature, one introduces an arbitrary (periodic) background field configuration $\bar A_\mu(x)$ and replaces the Landau gauge-fixing operator $\partial_\mu A_\mu(x)$ by a covariant version of it, namely ${\bar D}_\mu(A_\mu(x)-{\bar A}_\mu(x))$, where $\bar D_\mu\equiv\partial_\mu\,\_\,-[\bar A_\mu(x),\,\_\,]$ denotes the adjoint covariant derivative in the presence of the background $\bar A_\mu(x)$. The condition
\begin{equation}\label{eq:LdW}
 {\bar D}_\mu(A_\mu(x)-{\bar A}_\mu(x))=0\,,
\end{equation}
defines the so-called Landau-deWitt (LdW) gauge and is invariant under simultaneous twisted gauge transformations of $A_\mu(x)$ and $\bar A_\mu(x)$:
\begin{eqnarray}
 A^U_\mu(x) & \!\!\!\!\!=\!\!\!\!\! & U(x)A_\mu(x)\,U^\dagger(x)-U(x)\,\partial_\mu U^\dagger(x)\,,\label{eq:transfo_sim1}\\
 \bar A^U_\mu(x) & \!\!\!\!\!=\!\!\!\!\! & U(x)\bar A_\mu(x)\,U^\dagger(x)-U(x)\,\partial_\mu U^\dagger(x)\,.\label{eq:transfo_sim2}
\end{eqnarray}
The corresponding gauge-fixed measure is then such that
\begin{equation}
 {\cal D}_{\mbox{\scriptsize LdW}}\,[A^U;\bar A^U]={\cal D}_{\mbox{\scriptsize LdW}}\,[A;\bar A]\,,\,\,\forall U\in {\cal G}\,.
\end{equation}
It is in this sense that the gauge-fixed measure associated to the background extension of the Landau gauge is invariant under twisted gauge transformations. 

The procedure can be easily extended to any other gauge fixing and one generally arrives at a gauge-fixed measure such that
\begin{equation}\label{eq:inv}
 {\cal D}_{\mbox{\scriptsize gf}}\,[A^U;\bar A^U]={\cal D}_{\mbox{\scriptsize gf}}\,[A;\bar A]\,,\,\,\forall U\in {\cal G}\,.
\end{equation}
For later use we stress that the background should be interpreted as an infinite collection of gauge-fixing parameters. As such, any {\it bona fide} background gauge fixing should be such that the observables do not dependent on the chosen background.

\subsection{Center-invariant effective action}\label{sec:EA}
One interesting feature of the invariance property (\ref{eq:inv}) is that it is preserved by fluctuations. This is expected because the symmetry is realized linearly. To check this is in more detail, let us add a source term to the Yang-Mills action and define the generating functional $\hat W[J;\bar A]$ as
\begin{equation}\label{eq:def_W}
 \hat W[J;\bar A]\equiv\ln\int {\cal D}_{\mbox{\scriptsize gf}}\,[A;\bar A]\,\exp\left\{-S_{YM}[A]+\int d^dx\,\,(J_\mu;a_\mu)\right\},
\end{equation}
where we have introduced the field $a_\mu(x)\equiv A_\mu(x)-\bar A_\mu(x)$. Being the difference of two gauge fields, it transforms under (\ref{eq:transfo_sim1})-(\ref{eq:transfo_sim2}) as $a^U_\mu(x)=U(x)\,a_\mu(x)\,U^\dagger(x)$.

Let us now evaluate $\hat W[J^U;\bar A^U]$ with $J^U_\mu(x)\equiv U(x)\,J_\mu(x)\,U^\dagger(x)$. Using a change of variables of the form (\ref{eq:transfo_sim1}) under the functional integral (\ref{eq:def_W}), together with the identity (\ref{eq:inv}) and the invariance of both the YM action and the source term, we find\footnote{Needless to say, it is also important that the considered transformation $U$ preserves the periodic boundary conditions of the gauge field along the temporal direction.}
\begin{equation}\label{eq:id_W}
 \hat W[J^U;\bar A^U]=\hat W[J;\bar A]\,,\,\,\forall U\in {\cal G}\,.
\end{equation}
Next, we define the effective (or quantum) action $\hat\Gamma[a;\bar A]$ as the Legendre transform of $\hat W[J;\bar A]$ with respect to the source:
\begin{equation}\label{eq:def_Gamma}
 \hat \Gamma[a;\bar A]\equiv-\hat W[J[a;\bar A];\bar A]+\int d^dx\,\,(J_\mu[a;\bar A];a_\mu)\,,
\end{equation}
where $J_\mu[a;\bar A]$ is obtained by inverting the relation
\begin{equation}
a_\mu[J;\bar A](x)=\frac{\delta \hat W}{\delta J_\mu(x)}\,,
\end{equation}
for a fixed background. From the identity (\ref{eq:id_W}), it follows that
\begin{equation}
 a[J^U;\bar A^U]=a^U[J;\bar A]\,,\,\,\forall U\in{\cal G}\,,
\end{equation}
which, upon inversion, reads
\begin{equation}\label{eq:J_transfo}
 J[a^U;\bar A^U]=J^U[a;\bar A]\,,\,\,\forall U\in {\cal G}\,.
\end{equation}
Finally, combining Eqs.~(\ref{eq:id_W}), (\ref{eq:def_Gamma}) and (\ref{eq:J_transfo}), we arrive at
\begin{equation}
 \hat \Gamma[a^U;\bar A^U]=\hat \Gamma[a;\bar A]\,,\,\,\forall U\in {\cal G}\,,
\end{equation}
which shows that the symmetry (\ref{eq:transfo_sim1})-(\ref{eq:transfo_sim2}) is indeed preserved by fluctuations, as already anticipated.

In what follows, we shall use a source term of the form $(J_\mu;A_\mu)$, which means that we work instead with the generating functional $W[J;\bar A]\equiv\hat W[J;\bar A]+\int d^dx\,\,(J_\mu;\bar A_\mu)$. It is easily seen that the corresponding effective action is related to the previous one by $\Gamma[A;\bar A]=\hat\Gamma[A-\bar A;\bar A]$ and obeys, therefore, the symmetry identity \cite{Weinberg:1996kr}
\begin{equation}\label{eq:id_Gamma}
 \Gamma[A^U;\bar A^U]=\Gamma[A;\bar A]\,,\,\,\forall U\in {\cal G}\,.
\end{equation}

\subsection{Background gauge fixing and approximations}
Let us now turn back to our original problem of maintaining the transformation rule of the Polyakov loop in the presence of approximations and see if the just derived properties help in that matter.

If we denote by $\ell[\bar A]$ the Polyakov loop computed in some approximation that preserves Eq.~(\ref{eq:inv}),\footnote{This is easily achieved in practice because the symmetry is realized linearly.} then, following similar steps as in the previous chapter, we can derive the transformation rule
\begin{equation}
 \ell_{\alpha-\frac{2\pi}{N}k}[\bar A^U]=e^{i\frac{2\pi}{N}k}\ell_\alpha[\bar A]\,,\,\,\forall U\in {\cal G}\,,
\end{equation}
where $e^{i\frac{2\pi}{N}k}\mathds{1}=U^\dagger(0,\vec{x})U(\beta,\vec{x})$ and $e^{i\alpha}$ is the direction of the necessary, infinitesimal, symmetry-breaking source. In the Wigner-Weyl realization of the symmetry, the Polyakov loop does not depend on $\alpha$, and then
\begin{equation}\label{eq:no_constraint}
 \ell[\bar A^U]=e^{i\frac{2\pi}{N}k}\ell[\bar A]\,,\,\,\forall U\in{\cal G}\,.
\end{equation}
As interesting as these relations might be, they do not allow us, however, to solve our original problem. The reason is that Eq.~(\ref{eq:no_constraint}) is not really a constraint but rather a relation between two different approximated evaluations of the Polyakov loop. Only when the latter is computed exactly, does $\ell[\bar A]$ become independent of $\bar A$ (seen as an infinite collection of gauge-fixing parameters) and Eq.~(\ref{eq:no_constraint}) implies that the Polyakov loop has to vanish in the center-symmetric phase. We are thus back to our original problem: even within a background-extended gauge fixing, the transformation properties of the Polyakov loop seem difficult to maintain in the presence of approximations.

This negative result may look discouraging. However, we can try to attack the problem in a different manner which exploits beneficially the previous results. Since the confinement/deconfinement transition corresponds to a change of the state of the system from a center-symmetric to a center-breaking state, we only need to find a procedure that allows us to discriminate, at each temperature, between these two types of states, or, equivalently, to identify the center-symmetric states among all possible states. In the case of standard gauge fixings, the gauge-fixed measure is not invariant under twisted gauge transformations and neither is the corresponding effective action. Since the states of the system are usually extracted from the effective action, it is clear that such a framework does not facilitate the identification of center-symmetric states. In contrast, the quantum realization of center symmetry in background gauge fixings, as encoded in Eq.~(\ref{eq:id_Gamma}), is certainly a favourable condition for such an identification to be possible. 

\subsection{Center-symmetric states: first try}\label{sec:lim}
Let us make a first try in identifying the center-symmetric states. It will lead to a loose end, but it will help understanding the correct procedure to be presented in the next section.

A priori, the actual state of the system is obtained at the minimum of the effective action. More precisely, in the background framework considered here, one needs to minimize the effective action $\Gamma[A,\bar A]$ with respect to the field $A$, for a given background $\bar A$. The actual state of the system appears then as a certain background-dependent configuration $A_{\mbox{\scriptsize min}}[\bar A]$ such that
\begin{equation}\label{eq:def_min}
 \Gamma[A_{\mbox{\scriptsize min}}[\bar A];\bar A]\leq \Gamma[A;\bar A]\,,\,\,\forall A\,.
\end{equation}
We now would like to asses whether such a state $A_{\mbox{\scriptsize min}}[\bar A]$ is center-symmetric or center-breaking.

To this purpose, we apply a twisted gauge transformation $U$ to $A_{\mbox{\scriptsize min}}[\bar A]$. The invariance property (\ref{eq:id_Gamma}) means that the transformed configuration $A_{\mbox{\scriptsize min}}^U[\bar A]$ is also a minimum, but of the functional $\Gamma[\,\_\,;\bar A^U]$. Indeed, we have
\begin{equation}
\forall A\,,\,\,\Gamma[A,\bar A^U]=\Gamma[A^{U^\dagger};\bar A]\geq \Gamma[A_{\mbox{\scriptsize min}}[\bar A];\bar A]=\Gamma[A^U_{\mbox{\scriptsize min}}[\bar A];\bar A^U]\,,
\end{equation}
where we have successively used Eqs.~(\ref{eq:id_Gamma}), (\ref{eq:def_min}) and again (\ref{eq:id_Gamma}). It follows that a state of the system, as described in the presence of the background $\bar A$, is transformed into another state, but described in the presence of a different background $\bar A^U$. This change of description as we transform the state of the system makes it difficult to identify the states that are invariant and, therefore, to draw any conclusion on the possible breaking of center symmetry at some temperature.\\
 
The situation is similar to that in Eq.~(\ref{eq:no_constraint}) where the background is changed as a center transformation is applied. This is of course reminiscent of the fact that the invariance property (\ref{eq:id_Gamma}) requires the background to be transformed as well. In the next section, we show nonetheless that the problem can be circumvented by restricting the discussion to so-called self-consistent backgrounds.

\section{Self-consistent backgrounds}\label{sec:states}
{\it Self-consistent backgrounds} are specific background configurations $\bar A_s$ defined by the property
\begin{equation}\label{eq:self}
 \bar A_s=A_{\mbox{\scriptsize min}}[\bar A_s]\,.
\end{equation}
As it is well known, the minimum of the effective action coincides with the one-point function and, therefore, the self-consistency condition also rewrites $\smash{\bar A_s=\langle A\rangle_{\bar A_s}}$, where $\langle\dots\rangle_{\bar A_s}$ denotes the expectation value in the presence of the gauge-fixed measure ${\cal D}_{\mbox{\scriptsize gf}}\,[A;\bar A_s]$. So, in a sense, the strategy of restricting to self-consistent backgrounds corresponds to constantly adapting the gauge fixing (through the choice of the background configuration) in such a way that the one-point function is always known exactly.\footnote{In particular, it follows that self-consistent backgrounds depend on the temperature, as opposed to the fixed backgrounds considered in the previous section.} This has certainly an obvious practical interest. 

However, the main benefit of self-consistent backgrounds lies in that they allow for the identification of center-symmetric states. In fact, as we now explain, self-consistent backgrounds are order parameters for center symmetry and can be used, therefore, in place of the Polyakov loop. In order to reach this conclusion, we first introduce the background field effective action that leads both to a simple characterization of the self-consistent backgrounds and to their interpretation as the actual states of the system.

\subsection{Background field effective action}
The {\it background field effective action} is nothing but the effective action $\Gamma[A,\bar A]$ evaluated along the subspace $A=\bar A$:
\begin{equation}\label{eq:def_Gtilde}
 \tilde\Gamma[\bar A]\equiv\Gamma[A=\bar A;\bar A]\,.
\end{equation}  
It obeys the invariance property
\begin{equation}\label{eq:id_Gtilde}
 \tilde \Gamma[\bar A^U]=\tilde\Gamma[\bar A]\,,\,\,\forall U\in {\cal G}\,,
\end{equation}
as it can easily be shown using Eqs.~(\ref{eq:id_Gamma}) and (\ref{eq:def_Gtilde}). 

Let us now see how it allows for a characterization of the self-consistent backgrounds. One important ingredient to find such characterization will be that the value of the effective action at its minimum does not depend on the chosen background:
\begin{equation}\label{eq:inv_bg}
 \Gamma[A_{\mbox{\scriptsize min}}[\bar A];\bar A]=\Gamma[A_{\mbox{\scriptsize min}}[\bar A'];\bar A']\,,\,\,\forall \bar A'\,.
\end{equation}
This is because minimizing the effective action with respect to $A$ corresponds to taking the zero-source limit in Eq.~(\ref{eq:def_Gamma}), that is to evaluating the free-energy of the system up to a factor $\beta$.\footnote{It is assumed here that the gauge-fixed measure ${\cal D}_{\mbox{\tiny gf}}\,[A]$ includes appropriate normalization factors, such that the zero-source limit of Eq.~(\ref{eq:def_W}) is indeed the free-energy up to a factor $-\beta$ and a trivial shift related to the volume of the gauge group.} As any other physical observable, the free-energy cannot depend on the background $\bar A$. This is exactly what is encoded in Eq.~(\ref{eq:inv_bg}).\\

With this property in mind, let us derive the promised characterization of self-consistent backgrounds. We first show that a self-consistent background $\bar A_s$ is necessarily an absolute minimum of $\tilde\Gamma[\bar A]$. To this purpose, we write the following chain of relations
\begin{eqnarray}
  \tilde\Gamma[\bar A_s] & \!\!\!\!\!=\!\!\!\!\! & \Gamma[\bar A_s,\bar A_s]\\
& \!\!\!\!\!=\!\!\!\!\! & \Gamma[A_{\mbox{\tiny min}}[\bar A_s];\bar A_s]\nonumber\\
& \!\!\!\!\!=\!\!\!\!\! & \Gamma[A_{\mbox{\tiny min}}[\bar A];\bar A]\,,\,\,\forall \bar A\nonumber\\
& \!\!\!\!\!\leq\!\!\!\!\! & \Gamma[\bar A;\bar A]\,,\,\,\forall \bar A\nonumber\\
& \!\!\!\!\!\leq\!\!\!\!\! & \tilde\Gamma[\bar A]\,,\,\,\forall \bar A\,,\nonumber
\end{eqnarray}
where we have successively used Eqs.~(\ref{eq:def_Gtilde}), (\ref{eq:self}), (\ref{eq:inv_bg}), (\ref{eq:def_min}) and again (\ref{eq:def_Gtilde}). Thus, as announced, a self-consistent background $\bar A_s$ is an absolute minimum of the functional $\tilde\Gamma[\bar A]$.

Next, we prove the following version of the reciprocal property: assuming that there exists at least one self-consistent background $\bar A_s$, then any absolute minimum $\bar A_m$ of $\tilde\Gamma[\bar A]$ is also a self-consistent background. To this purpose, we note that
\begin{equation}\label{eq:def_Abarm}
  \tilde\Gamma[\bar A_m]\leq\tilde\Gamma[\bar A]\,,\forall\bar{A}\,,
\end{equation}
and write
\begin{eqnarray}
  \tilde\Gamma[\bar A_m] & \!\!\!\!\!\leq\!\!\!\!\! & \tilde\Gamma[\bar A_s]\\
& \!\!\!\!\!\leq\!\!\!\!\! & \Gamma[\bar A_s,\bar A_s]\nonumber\\
& \!\!\!\!\!\leq\!\!\!\!\! & \Gamma[A_{\mbox{\tiny min}}[\bar A_s],\bar A_s]\nonumber\\
& \!\!\!\!\!\leq\!\!\!\!\! & \Gamma[A_{\mbox{\tiny min}}[\bar A_m],\bar A_m]\,.\nonumber
\end{eqnarray}
This time, we have successively used Eqs.~(\ref{eq:def_Abarm}), (\ref{eq:def_Gtilde}), (\ref{eq:self}) and (\ref{eq:inv_bg}). This implies that $\bar A_m=A_{\mbox{\tiny min}}[\bar A_m]$, and, therefore, that $\bar A_m$ is self-consistent, as announced.\\

In conclusion, we have shown that there are either no self-consistent backgrounds at all or that they are exactly given by the absolute minima of the functional $\tilde\Gamma[\bar A]$. We shall assume from here on that we are in the second scenario.\footnote{The present derivation assumes the unicity of the field configuration $A_{\mbox{\tiny min}}[\bar A]$ for each background configuration $\bar A$. Similar results can be obtained under the assumption of degenerate minima.}

\subsection{Background description of the states}
Now that we have characterized the self-consistent backgrounds as the absolute minima of the background field effective action, let us see how they provide a new description of the states of the system that does not suffer from the limitations identified in Sec.~\ref{sec:lim}. 

The point is that, when working with self-consistent backgrounds, all the relevant information can be obtained from the functional $\tilde\Gamma[\bar A]$ and the (self-consistent) background configurations $\bar A_s$ that minimize it, without any reference to the functional $\Gamma[A,\bar A]$ or to the field configuration $A_{\mbox{\scriptsize min}}[\bar A]$. In particular, up to a trivial factor, the free-energy of the system is given by
\begin{equation}
  \Gamma[A_{\mbox{\scriptsize min}}[\bar A_s];\bar A_s]=\Gamma[\bar A_s,\bar A_s]=\tilde\Gamma[\bar A_s]\,,
\end{equation}
that is, it is given by the background field effective action $\tilde\Gamma[\bar A]$ evaluated at any self-consistent background $\bar A_s$. Therefore, in a certain sense, one can interpret the space of background field configurations $\bar A$ over which $\tilde\Gamma[\bar A]$ is varied as the the space of all potentially available states, the actual state of the system corresponding to those particular (self-consistent) backgrounds that minimize $\tilde\Gamma[\bar A]$.

Now, the invariance property (\ref{eq:id_Gtilde}) means that, after applying a twisted gauge transformation $U\in {\cal G}$ to a given state $\bar A_s$, the newly obtained configuration $\bar A^U_s$ is also a minimum of the functional $\tilde\Gamma[\bar A]$ and, therefore, another acceptable state of the system. The important difference with respect to the discussion in Sec.~\ref{sec:lim} is that there is no change of description as one transforms the state since the functional that needs to be minimized remains the same. As a consequence, there is no obstacle anymore to the identification of center-symmetric states.

\subsection{Orbit description of the states}
Before we proceed to this identification, however, it is important to realize that the previous description of the states carries some redundancy. This relates to the fact that the symmetry identity (\ref{eq:id_Gtilde}) applies in particular to the periodic gauge transformations $U\in {\cal G}_0$. As we have seen in the previous chapter, these transformations need to be considered as true, and therefore unphysical, gauge transformations. Their presence only reflects the redundancy of the description in terms of gauge fields and, consequently, two background configurations that belong to the same orbit under the action of ${\cal G}_0$ should be considered as describing the same physical state. In other words, the physical states of the system correspond to the various possible ${\cal G}_0$-orbits in the space of background configurations and a given ${\cal G}_0$-orbit (a given physical state) admits various equivalent representations in terms of background configurations.

The redundancy in the description of the physical states by means of background configurations is similar to the redundancy in the description of physical center transformations by means of twisted gauge transformations that we discussed in the previous chapter. And just as it was possible to remove the redundancy inherent to ${\cal G}$ by considering the group ${\cal G}/{\cal G}_0$, it is possible to remove the redundancy inherent to the use of backgrounds by working instead with the corresponding ${\cal G}_0$-orbits. 

To see how this is achieved in practice, we first notice that the background field effective action can be defined directly on the ${\cal G}_0$-orbits since $\tilde\Gamma[\bar A]$ depends only on the orbit $\bar {\cal A}$ the background $\bar A$ belongs to. We can then define $\tilde\Gamma[\bar {\cal A}]\equiv\tilde\Gamma[\bar A]$, where $\bar A\in \bar {\cal A}$ is any background in the orbit $\bar {\cal A}$, and the physical state is obtained by minimizing $\tilde\Gamma[\bar {\cal A}]$ over the space of ${\cal G}_0$-orbits. Second, it is possible to define the action of a center transformation  ${\cal U}\in{\cal G}/{\cal G}_0$ directly on the ${\cal G}_0$-orbits. Indeed, owing to the property $\forall U\in{\cal G}$, $\forall U_0\in{\cal G}_0$, $UU_0U^\dagger\in {\cal G}_0$, all the backgrounds in a given ${\cal G}_0$-orbit are transformed under a twisted gauge transformation into backgrounds belonging to one and the same orbit, which means that one can directly define the action of a twisted gauge transformation on ${\cal G}_0$-orbits. Moreover, this action depends only on the class of ${\cal G}/{\cal G}_0$ the twisted gauge transformation belongs to, thereby defining the action of ${\cal G}/{\cal G}_0$ directly on ${\cal G}_0$-orbits. Finally, in this redundancy-free formulation, the symmetry identity (\ref{eq:id_Gtilde}) reads 
\begin{equation}\label{eq:id_Gtilde_2}
\tilde\Gamma[\bar {\cal A}^{\cal U}]=\tilde\Gamma[\bar {\cal A}]\,,\,\,\forall {\cal U}\in {\cal G}/{\cal G}_0\,.
\end{equation}

From all these considerations, it follows, as announced, that one can work exclusively in terms of ${\cal G}_0$-orbits and center transformations.

\subsection{Center-symmetric states: second try}
It should be clear by now that the center-symmetric states of the system correspond to the center-invariant ${\cal G}_0$-orbits, that is orbits $\bar {\cal A}$ such that
\begin{equation}\label{eq:sym_state_1}
  \forall {\cal U}\in {\cal G}/{\cal G}_0\,,\,\, \bar{\cal A}^{\cal U}=\bar {\cal A}\,.
\end{equation}
In terms of background configurations, this reads
\begin{equation}\label{eq:sym_state_2}
  \forall U\in {\cal G}\,,\,\, \exists U_0\in{\cal G}_0\,,\,\,\bar A_\mu^U(x)=\bar A_\mu^{U_0}(x)\,,
\end{equation}
that is center-symmetric states corresponds to backgrounds that are invariant under twisted gauge transformations, modulo periodic gauge transformations.\\

In the next chapter, we will see how to practically access the center-invariant ${\cal G}_0$-orbits using either Eq.~(\ref{eq:sym_state_1}) or Eq.~(\ref{eq:sym_state_2}). For the time being, what we can say is that we have classified the physical states of the system into center-invariant and center non-invariant ${\cal G}_0$-orbits. Therefore, the ${\cal G}_0$-orbits and, by extension, the background configurations, play the role of order parameters for the confinement/deconfinement transition, as announced earlier. This is the central result of this chapter, at the basis of the applications to be presented in subsequent chapters. For other proofs of this result in the case of the SU($2$) gauge group, we refer to \cite{Braun:2007bx,Marhauser:2008fz}.

\section{Other symmetries}
The previous considerations apply not only to center symmetry but, as we now explain, to any physical symmetry of the quantum action. After introducing some generalities, we discuss in particular the case of charge conjugation symmetry which plays an important role in subsequent applications.

\subsection{Generalities}
Consider a physical transformation ${\cal T}$ of the state of the system. In the space of background configurations, it writes $\bar A\to\bar A^{\cal T},$\footnote{We are implicitly assuming here that the background field effective action $\tilde\Gamma[\bar A]$ is invariant under ${\cal T}$, which happens in particular if the gauge-fixing condition, and in turn the gauge-fixed measure ${\cal D}_{\mbox{\scriptsize gf}}\,[A;\bar A]$, are invariant under $(A,\bar A)\to (A^{\cal T},\bar A^{\cal T})$, and if the symmetry is realized linearly.} where $\bar A^{\cal T}$ should not be mistaken with the notation $\bar A^U$. In fact, ${\cal T}$ is a formal group transformation, not necessarily an element of SU($N$). 

The transformation ${\cal T}$ being physical, the application of ${\cal T}$ followed by ${\cal T}^{-1}$ to a background representing a given ${\cal G}_0$-orbit (a given physical state) should lead to a background of the same ${\cal G}_0$-orbit, even when some additional true gauge transformations are inserted between ${\cal T}$ and ${\cal T}^{-1}$. Mathematically, this writes as
\begin{equation}\label{eq:toto}
  \forall U\in {\cal G_0}\,,\,\,\exists U'\in {\cal G}_0\,,\,\, ((\bar A^{\cal T}_\mu)^U)^{{\cal T}^{-1}}(x)=\bar A^{U'}_\mu(x)\,.
\end{equation}
From this, it follows immediately that the physical transformation ${\cal T}$ can be defined directly on the ${\cal G}_0$-orbits, as it should, of course, for a physical transformation.

Moreover, just as with physical center transformations, the group of physical transformations ${\cal T}$ can be extended into a group of transformations $T$ defined as the transformations ${\cal T}$ modulo elements of ${\cal G}_0$. Owing to Eq.~(\ref{eq:toto}), ${\cal G}_0$ is a normal subgroup of this extended group and the corresponding quotient is isomorphic to the group of physical transformations. The reason for introducing such an extended group is that, in the space of background configurations, any transformation of the extended group is a valid representation of the corresponding physical transformation.

The states of the system that are invariant under the physical transformation ${\cal T}$ correspond to the ${\cal T}$-invariant ${\cal G}_0$-orbits. Mathematically, this writes as
\begin{equation}\label{eq:1}
  \forall {\cal T}\!\!\!,\,\, \bar{\cal A}^{\cal T}=\bar{\cal A}\,,
\end{equation}
or, in terms of background configurations, as
\begin{equation}\label{eq:2}
  \forall T\!,\,\, \exists U_0\in{\cal G}_0\,,\,\, \bar A^T_\mu(x)=\bar A^{U_0}_\mu(x)\,.
\end{equation}
In particular, in terms of background configurations, physical invariance needs to be understood as invariance modulo true gauge transformations. The characterization (\ref{eq:1}) of invariant states is useful when one has a simple description of the orbits, such as the description in terms of Weyl chambers that we review in the next chapter. The characterization (\ref{eq:2}) is used when such a description is not available, for which we also give an example in the next chapter.

\subsection{Charge conjugation}
Let us now illustrate these general considerations with the important case of charge conjugation. On the gauge field, this transformation reads
\begin{equation}\label{eq:C1}
 A_\mu^{\cal C}(x)\equiv -A^{\mbox{\scriptsize t}}_\mu(x)\,,
\end{equation}
and changes a given representation into the corresponding contragredient representation. It is easily checked that this indeed corresponds to a physical symmetry in the sense of the condition (\ref{eq:toto}). 

In order to see the effect of this symmetry on the Polyakov loop, it is convenient to introduce the normalised trace of the Wilson line associated to the anti-quark $\bar\Phi_A(\vec{x})\equiv\Phi_{-A^{\mbox{\scriptsize t}}}(\vec{x})$ which transforms according to the contragredient representation.\footnote{In terms of the anti-path-ordering, this rewrites $\bar\Phi_A(\vec{x})=\mbox{\scriptsize tr}\,\bar {\cal P}\,\exp\left\{-\int_0^\beta d\tau\, A_0(\tau,\vec{x})\right\}$.} The corresponding Polyakov loop
\begin{equation}
 \bar\ell\equiv\big\langle\bar\Phi_A\big\rangle_{YM}
\end{equation}
is referred to as the {\it anti-Polyakov loop.} From Eq.~(\ref{eq:C1}) and the charge conjugation invariance of the YM action, it follows immediately that\footnote{Strictly speaking, this identity holds as long as there is no spontaneous breaking of center symmetry. When the symmetry is broken, the identity applies to one of the possible states. The other states are invariant under a combination of charge conjugation and a center transformation.}
\begin{equation}
 \bar\ell=\ell\,.\label{eq:bat}
\end{equation}
This relation is to be expected since charge conjugation symmetry is an unbroken symmetry of the YM system and, therefore, it should cost the same energy to bring a quark or an anti-quark into the thermal bath of gluons.  

On the other hand, because $A_\mu=iA_\mu^at^a$ is anti-hermitian, we have $-A_\mu^{\mbox{\scriptsize t}}(x)=A^*_\mu(x)$ and thus $\bar\Phi_A(\vec{x})=\Phi_A^*(\vec{x})$. This implies
\begin{equation}
 \bar\ell=\ell^*\,.\label{eq:bi}
\end{equation}
We mention that this result relies crucially  on the fact that the averaging measure under the functional integral is real. Combining it with the previous result, we deduce that the Polyakov and anti-Polyakov loops are real, which is a necessary condition for their interpretation as $e^{-\beta\Delta F}$. In Chapter \ref{chap:bfg_quarks}, we shall investigate how these properties change in the presence of quarks with a finite chemical potential.\\

As for the background gauge-fixed theory, assuming that the gauge-fixed action remains invariant under charge conjugation and because the symmetry is realized linearly, it is not difficult to argue that
\begin{equation}
 \tilde\Gamma[\bar A^C]=\tilde\Gamma[\bar A]\,,
\end{equation}
for any transformation $C$ equal to ${\cal C}$ modulo elements of ${\cal G}_0$. In terms of ${\cal G}_0$-orbits, this reads
\begin{equation}\label{eq:inv_C}
 \tilde\Gamma[\bar {\cal A}^{\cal C}]=\tilde\Gamma[\bar {\cal A}]\,,
\end{equation}
which expresses charge conjugation invariance at the quantum level. Charge conjugation invariant states are such that $\bar {\cal A}^{\cal C}=\bar {\cal A}$. We shall characterize them more precisely in the next chapter.

\section{Additional remarks}\label{sec:rem2}
Let us conclude this chapter with some critical remarks on the rationale behind the previous background field gauge construction.

\subsection{Back to the Polyakov loop}
The most important feature of the previous construction is that the interpretation of the ${\cal G}_0$-orbits as order parameters for the deconfinement transition remains valid in the presence of approximations, provided the latter maintain the symmetry identity (\ref{eq:id_Gtilde_2}). This is easily achieved in practice because the symmetry is realized linearly.\footnote{For instance, the perturbative expansion to be used in the following chapters will fulfil this property.} Moreover, since ${\cal G}_0$-orbits represent an alternative order parameter, we do not need to be concerned anymore with the Polyakov loop and the question of whether it is possible to maintain its transformation rule in the presence of approximations. It is interesting to note, however, that the same framework allows to answer this question positively.

To see this, let us allow for a non-zero source $J$ coupled to $a$ which can be seen as a functional $J[a;\bar A]$ of $a$ and $\bar A$, see the discussion of the effective action $\hat\Gamma[a;\bar A]$ in Sec.~\ref{sec:EA}. Suppose then that the source is chosen such that $a\equiv A-\bar A=0$, that is the background $\bar A$ is forced to be self-consistent with the help of the source $J[\bar A]\equiv J[0;\bar A]$. We can compute the Polyakov loop in the presence of this source. Because it depends on the background $\bar A$, in what follows, we refer to it as the {\it background-dependent Polyakov loop,} and denote it by $\ell[\bar A]$. 

Now, from Eq.~(\ref{eq:J_transfo}), it follows that
\begin{equation}\label{eq:id_J}
 J[\bar A^U]=J^U[\bar A]\,,\,\,\forall U\in {\cal G}\,,
\end{equation} 
from which it is easily deduced that
\begin{equation}\label{eq:rule}
 \ell[\bar A^U]=e^{i\frac{2\pi}{N}k}\ell[\bar A]\,,\,\,\forall U\in {\cal G}\,.
\end{equation}
In particular, since $\ell[\bar A^U]=\ell[\bar A]$ for $U\in {\cal G}_0$, $\ell[\bar A]$ can be defined directly on ${\cal G}_0$-orbits, $\ell[\bar {\cal A}]\equiv\ell[\bar A]$, and the transformation rule (\ref{eq:rule}) rewrites
\begin{equation}
 \ell[\bar {\cal A}^{\cal U}]=e^{i\frac{2\pi}{N}k}\ell[\bar {\cal A}]\,,\,\,\forall {\cal U}\in {\cal G}/{\cal G}_0\,.
\end{equation}
Now, if the orbit is center-invariant, it follows that
\begin{equation}
 \ell[\bar {\cal A}]=e^{i\frac{2\pi}{N}k}\ell[\bar {\cal A}]\,,\,\,\forall k\in\{0,1,\dots,N-1\}\,,
\end{equation}
and, therefore, $\ell[\bar {\cal A}]=0$. This property relies only on the identity (\ref{eq:id_J}) which is again easily satisfied in the presence of approximations. 

Similarly, under charge conjugation 
\begin{equation}\label{eq:cov_C}
 J[\bar A^C]=J^C[\bar A]\,.
\end{equation}
From this, it is easily deduced that
\begin{equation}
 \ell[\bar A^C]=\bar\ell[\bar A]=\ell^*[\bar A]\,.
\end{equation}
In terms of ${\cal G}_0$-orbits, this reads
\begin{equation}
 \ell[\bar {\cal A}^{\cal C}]=\bar\ell[\bar {\cal A}]=\ell^*[\bar {\cal A}]\,,
\end{equation}
and for a charge-conjugation-invariant orbit (which should represent the actual state of the YM system) we recover the above results for the physical Polyakov and anti-Polyakov loops.\\

It should be mentioned that the functional $\ell[\bar A]$ defined here is different from the one that we introduced in Sec.~3.1.3. The latter was just an approximated version of the physical Polyakov loop whose $\bar A$-dependence stem precisely from the use of approximations. Here, instead, the functional $\ell[\bar A]$ does depend on $\bar A$, even in the absence of approximations, because it is not the physical Polyakov loop but, rather, the value of the Polyakov loop as the system is forced into a state corresponding to the background $\bar A$ by means of a non-zero source $J[\bar A]$. The physical Polyakov loop is retrieved when evaluating $\ell[\bar A]$ for a self-consistent background $\bar A_s$ which is precisely such that $J[\bar A_s]=0$.\footnote{One may wonder why we do not need an infinitesimal, symmetry-breaking source here, similar to the one that we introduced in the previous chapter. In the case where the physical state of the system is represented by a background $\bar A_s$ belonging to a center-invariant orbit, the system is in the symmetric phase and the symmetry-breaking source is indeed not needed. For any other physical state, there will typically exist various orbits connected to each other by center transformations. The values of $\ell[\bar A]$ on each of these orbits represent the various vacua in the Nambu-Goldstone realization of the symmetry, as they would equivalently be reached from an infinitesimal, symmetry-breaking source.} If it so happens that this self-consistent background belongs to a center-invariant ${\cal G}_0$-orbit, it follows from the above that $\ell[\bar A_s]=0$, and so, even in the presence of approximations.

\subsection{Hypothesis on the gauge-fixed measure}

The justification of all the previous results relies on two natural but strong assumptions on the generating functional $W[J;\bar A]$ which in turn imply constraints on the gauge-fixed measure ${\cal D}_{\mbox{\scriptsize gf}}\,[A;\bar A]$. These assumptions are, first, that $W[J;\bar A]$ should be convex with respect to $J$ and, second, that $W[0;\bar A]$ should not depend on the background $\bar A$. It is important to keep these basic assumptions in mind since they are not necessarily implemented exactly in the practical realizations of the gauge fixing and/or in the presence of approximations and could therefore lead to some artefacts. We shall discuss some of these artefacts in Chap.~\ref{chap:correlators}. Here, we analyse the basic assumptions in more detail.

\paragraph{Convexity of $W[J;\bar A]$:}

The convexity of $W[J;\bar A]$ enters crucially in the identification of the self-consistent backgrounds as the absolute minima of the functional $\tilde\Gamma[\bar A]$. The reason is that the proof of the standard fact that the limit of zero sources corresponds to the minimization of the functional $\Gamma[A;\bar A]$ with respect to $A$ requires, in its simpler form, the convexity of $W[J;\bar A]$. Indeed, $W[J;\bar A]$ lies always above any of its tangential planes:
\begin{equation}
 W[J;\bar A]\geq W[J_0;\bar A]+\int d^dx\,\left(\left.\frac{\delta W}{\delta J(x)}\right|_{J_0};J(x)-J_0(x)\right).
\end{equation}
In terms of the effective action, it follows that
\begin{eqnarray}
 \Gamma\left[\frac{\delta W}{\delta J};\bar A\right] & \!\!\!\!\!\equiv\!\!\!\!\! & -W[J;\bar A]+\int d^dx\,\left(\frac{\delta W}{\delta J(x)};J(x)\right)\\
& \!\!\!\!\!\leq\!\!\!\! & -W[J_0;\bar A]+\int d^dx\,\left(\left.\frac{\delta W}{\delta J(x)}\right|_{J_0};J_0(x)\right)\nonumber\\
& & +\,\int d^dx\, \left(\frac{\delta W}{\delta J(x)}-\left.\frac{\delta W}{\delta J(x)}\right|_{J_0};J(x)\right),\nonumber
\end{eqnarray}
for any sources $J$ and $J_0$. Taking the limit $J\to 0$, we finally arrive at
\begin{equation}
 \Gamma\left[\left.\frac{\delta W}{\delta J}\right|_{J\to 0};\bar A\right]\leq  \Gamma\left[\left.\frac{\delta W}{\delta J}\right|_{J_0};\bar A\right],\,\,\forall J_0\,,
\end{equation}
which justifies the minimization principle referred to above.\\

The convexity of $W[J;\bar A]$ is a natural assumption to be made since it is a property of an ``ideal gauge fixing'' where one configuration per orbit is selected in the space of gauge field configurations. The gauge-fixed measure associated to this ideal gauge fixing is positive (since the original measure ${\cal D}A$ is positive). As we show in App.~\ref{app:convexity} using H\"older inequality, this implies the convexity of $W[J,\bar A]$.

In many practical implementations of the gauge fixing, however, the positivity of the gauge-fixed measure is not satisfied. For instance, in the Faddeev-Popov approach, positivity is violated due to the presence of the associated determinant which is known not to have a definite sign over the space of gauge-field configurations. The same is true a priori for the Serreau-Tissier approach due to the averaging over Gribov copies with an alternating sign, although it could be that the positivity violation remains moderate if certain copies dominate with respect to others. The situation seems more under control in the Gribov-Zwanziger approach where the functional integral is, in principle at least, restricted to the first Gribov region defined by the positivity of the Faddeev-Popov operator. However, the positivity of the measure is not necessarily guaranteed in practical implementations of the Gribov-Zwanziger approach, although the violation could again remain small. 

In fact, the positivity of the gauge-fixed measure is just a sufficient condition for identifying an order parameter. The order parameter interpretation remains correct provided the positivity violation remains moderate to ensure the convexity of $W[J;\bar A]$. Since a dynamical mass scale is expected to be generated (and is actually generated in some cases) that damps large enough values of the gauge field, we expect the positivity violation to remain indeed moderate. In the Curci-Ferrari approach, this damping factor is there from the beginning, so we expect the positivity requirement to be fulfilled if the Curci-Ferrari mass is large enough.\\

We mention finally that the non-positivity of the integration measure can also have a physical origin, as in the presence of dynamical quarks at finite baryonic density. In this case as well, the very foundation of the background field method needs to be critically revisited. We shall deal with this matter in Chapter \ref{chap:bfg_quarks}.

\paragraph{Background independence:}
The other crucial property that enters the justification of the background field approach is that of the background-independence of the free-energy. In the Faddeev-Popov approach this is ensured from the general properties derived in Chap.~\ref{chap:intro}. In other approaches, including any continuum approach using a free mass parameter, this question needs to be investigated. For the Curci-Ferrari approach, we do so in Chap.~\ref{chap:YM_1loop}.

\subsection{Landau gauge paradox}
Although compelling, the above rationale for using the background field method at finite temperature is not completely free of paradoxes. A particularly intriguing one is the case of the vanishing background configuration $\bar A=0$. Indeed, in the absence of background, the invariance under color rotations is manifest and since we do not expect this symmetry to be spontaneously broken, we should have $\langle A\rangle_{\bar A=0}=0$ which implies that $\bar A=0$ is a self-consistent background. As such, it should always appear as a minimum of $\tilde\Gamma[\bar A]$. This is {\it a priori} a welcome result since $\tilde\Gamma[\bar A=0]=\Gamma[A=0;\bar A=0]$ is nothing but the free-energy density of the system computed in the Landau gauge and should, therefore, coincide with the free-energy density computed in the Landau-deWitt gauge from whatever relevant non-trivial self-consistent background $\bar A_s$.

The problem is however that $\bar A=0$ cannot correspond to a center-symmetric state as we show in the next chapter. This poses therefore the question how, in a would be confining phase for which $\bar A_s\neq 0$, one could have both an absolute minimum of $\tilde\Gamma[\bar A]$ at a center-symmetric point and another absolute minimum, with the same value of $\tilde\Gamma[\bar A]$, at the center-breaking point $\bar A=0$. One possible scenario could be that $\bar A=0$ needs to be interpreted as an unstable state (in a sense that still needs to be defined) which only reaches the same value of $\tilde\Gamma[\bar A]$ as enough fluctuations are included in the evaluation of $\tilde\Gamma[\bar A]$. To some extent, this is similar to the flattening that occurs to the {\it mexican hat} potential in the O(N) model with spontaneously broken symmetry \cite{Tetradis:1992qt}. As matter of fact, we will see in the next chapters that $\bar A=0$ appears indeed as a maximum of $\tilde\Gamma[\bar A]$ in a fixed order approximation. How the effective action could flatten to bring $\bar A=0$ to the same potential depth as $\bar A_s$ and how this would affect the use of background field methods at finite temperature needs still to be clarified.

\clearemptydoublepage
%
%
%
%
\let\textcircled=\pgftextcircled
\chapter{Background field gauges:\\ Weyl chambers}
\label{chap:symmetries}

\initial{A}s we have seen previously, within any background-field gauge, the physical states of the Yang-Mills system can be interpreted in terms of ${\cal G}_0$-orbits in the space of background configurations, where ${\cal G}_0$ is the subgroup of periodic gauge transformations within the group ${\cal G}$ of twisted gauge transformations. These considerations are crucial to the discussion of the confinement/deconfinement transition since the physical states that are center-symmetric, and thus confining, correspond to the invariant ${\cal G}_0$-orbits under center transformations. Preparing the ground for subsequent applications, it is the purpose of the present chapter to recall how these invariant orbits are identified in practice. 

We shall do so using a simplifying assumption on the properties of the background based on the homogeneity and isotropy of the Yang-Mills system at finite temperature: we shall restrict to constant temporal backgrounds in the diagonal part of the algebra. As a consequence, our analysis of center symmetry will require only a subgroup $\tilde {\cal G}$ of the twisted gauge transformations, those that preserve this specific form of the background. We shall start by deriving the general form of these transformations in terms of the so-called Weyl transformations and what we refer to as the winding transformations. The analysis of these two types of transformations is considerably simplified if one operates a change of color basis from the conventional Cartesian bases to canonical or Cartan-Weyl bases. Since these bases play a major role in subsequent calculations, we spend some time recalling their definition and general properties. 

Of particular interest for the analysis of the confinement/deconfinement transition is the subgroup $\tilde {\cal G}_0$ of periodic transformations within $\tilde {\cal G}$. It will allow us to interpret the physical states of the system as the $\tilde {\cal G}_0$-orbits in the restricted space of constant, temporal and diagonal backgrounds, and the center-symmetric states as the center-invariant $\tilde {\cal G}_0$-orbits. The identification of the latter will then be achieved using the notion of Weyl chambers. In view of its importance when quarks are included back in the analysis, we shall also discuss charge conjugation symmetry using again the Weyl chambers. 

Finally, we shall return to the original assumption on the form of the background and critically discuss the homogeneity and isotropy constraints on the possible background configurations.

\section{Constant temporal backgrounds}
As we have seen, the actual state of the system is obtained by minimizing the background field effective action $\tilde\Gamma[\bar A]$ in the space of background field configurations, or, in a redundancy-free description, by minimizing the background field effective action $\tilde\Gamma[\bar {\cal A}]$ in the space of ${\cal G}_0$-orbits. As usual, the search for minima can be simplified using the symmetries of the system: any symmetry that is known not to be spontaneously broken, should be manifest at the level of the state (Wigner-Weyl realization) and therefore at the level of the minimizing background/orbit.

\subsection{Homogeneity and Isotropy}
In particular, the Yang-Mills system at finite temperature is homogenous and isotropic, that is, it is invariant under both temporal and spatial translations and also under spatial rotations. Because these symmetries are expected not to be broken spontaneously, it makes sense to restrict the search for minima of $\tilde\Gamma[\bar A]$ to constant temporal backgrounds of the form $\bar A_\mu(x)=\bar A_0\,\delta_{\mu0}$.

The discussion is in fact a little bit more subtle due to the general symmetry considerations of the previous chapter. In particular, the homogeneity and isotropy of the state translates into the invariance of the corresponding ${\cal G}_0$-orbit but not necessarily into the invariance of the various background configurations that represent the orbit. Thus, even though the orbit associated to a constant temporal background is both homogenous and isotropic, there could be other invariant orbits such that none of the representing backgrounds are invariant, but only invariant modulo a non-trivial element of ${\cal G}_0$. We shall return to this interesting possibility at the end of this chapter and assume for the moment (as usually done in the literature) that we can indeed restrict to ${\cal G}_0$-orbits that contain constant temporal backgrounds, or, more simply, that we can work exclusively with constant temporal backgrounds of the form $\bar A_\mu(x)=\bar A_0\,\delta_{\mu0}$.

Moreover, since color rotations are elements of ${\cal G}_0$, we can always assume that the constant Lie algebra element $\bar A_0$ has been color rotated to the diagonal part of the su(N) algebra.\footnote{This is because the elements of su(N) are anti-hermitian matrices and are thus diagonalizable by a unitary change of basis.} Denoting by $\{iH_j\}$ a basis of the latter, with $[H_j,H_k]=0$, see App.~\ref{app:sun}, the most general background that we shall consider is then of the form
\begin{equation}\label{eq:bg_res}
 \beta\bar A_\mu(x)=i r_jH_j\,\delta_{\mu0}\,,
\end{equation}
where the factor $\beta\equiv 1/T$ has been introduced to make the components $r_j$ dimensionless. In fact, the discussion from here on applies more generally to any semi-simple Lie algebra. Such algebras admit maximally commuting subalgebras, the so-called Cartan subalgebras, which any element of the algebra can be rotated into. In what follows, unless specifically stated, we assume that we work in this general context. Denoting by $d_C$ the dimension of the Cartan subalgebra, the possible background configurations are then represented by a vector $r\in\mathds{R}^{d_C}$, with $d_C=N-1$ in the case of SU(N). We refer to this copy of $\mathds{R}^{d_C}$ as the {\it restricted background space} and to the vector $r$ as the {\it restricted background.} 

Correspondingly, the analysis of the background effective action $\tilde\Gamma[\bar A]$ is reduced to that of a {\it background field effective potential} $V(r)$ and the background dependent Polyakov loop $\ell[\bar A]$ introduced in the previous chapter becomes a function $\ell(r)$ of the restricted background. Similarly, the symmetry transformations should also be restricted to those that preserve the form (\ref{eq:bg_res}) of the background and which we now characterize.

\subsection{Restricted twisted gauge transformations}
The twisted gauge transformations $U(x)$ that preserve the form (\ref{eq:bg_res}) of the background are such that $U(x)\,i r_jH_j\delta_{\mu0}\,U^\dagger(x)-U(x)\,\beta\,\partial_\mu U^\dagger(x)$ is constant, temporal and diagonal, for any restricted background $r$. We refer to these transformations as {\it restricted twisted gauge transformations.} They form a subgroup of ${\cal G}$, denoted $\tilde {\cal G}$ in the following.

If we take the index $\mu$ to be spatial in the condition above, we find that the transformations of $\tilde {\cal G}$ can only depend on $\tau$. Moreover, choosing $r=0$, we find that $\beta\,dU/d\tau=is_jH_j\,U$ for some $s\in\mathds{R}^{d_C}$. Integrating this equation yields
\begin{equation}\label{eq:twisted_res}
  U(\tau)=e^{i\frac{\tau}{\beta} s_jH_j}W\,,
\end{equation}
for some color rotation $W\in$ SU(N). Finally, for the above constraint to be valid for a given $r\neq 0$, we should have $W r_jH_jW^\dagger=s_jH_j$ for some $s$ depending on $r$. 

In summary, the elements of $\tilde {\cal G}$ take the form (\ref{eq:twisted_res}), with $W$ a color rotation that leaves the Cartan subalgebra globally invariant. Such type of color rotations are known as {\it Weyl transformations.} The transformations of the form $e^{i\frac{\tau}{\beta} s_jH_j}$ will be referred to as {\it winding transformations.}

\subsection{Charge conjugation}
Even though it will play a secondary role for the moment, let us also discuss charge conjugation, which we recall is defined as $A_\mu^{\cal C}(x)\equiv -A_\mu^{\mbox{\tiny t}}(x)$. In the case of SU($N$), since $H_j^{\mbox{\tiny t}}=H_j$, we have $A_\mu^j(x)\to -A_\mu^j(x)$ and charge conjugation leaves the Cartan subalgebra globally invariant.

We can therefore consider its action directly on the restricted background space, where it corresponds to the geometrical transformation $r\to -r$, that is the reflection about the origin. The invariance of the potential under charge conjugation reads
\begin{equation}\label{eq:V_CC}
 V(-r)=V(r)\,,
\end{equation}
and the constraint on the background-dependent Polyakov loops is
\begin{equation}
 \ell(-r)=\bar\ell(r)\,.
\end{equation}
We shall now study to which geometrical transformations in the restricted background space the Weyl and the winding transformations correspond to.

\section{Winding and Weyl transformations}
The winding transformation $e^{i\frac{\tau}{\beta} s_jH_j}$ acts on a given gauge field configuration $A_\mu(x)$ as
\begin{eqnarray}\label{eq:winding}
 \beta A_\mu(x) & \!\!\!\!\!\to\!\!\!\!\! & e^{i\frac{\tau}{\beta} s_jH_j}\beta A_\mu(x)\,e^{-i\frac{\tau}{\beta} s_jH_j}+is_jH_j\delta_{\mu0}\\
& & =\,e^{i\frac{\tau}{\beta} s_j\,\mbox{\scriptsize ad}_{H_j}}\beta A_\mu(x)+is_jH_j\delta_{\mu0}\,.\nonumber
\end{eqnarray}
Here, we have introduced $\mbox{ad}_{\theta}\equiv [\theta,\,\_\,]$ and we have used that $e^\theta Xe^{-\theta}=e^{\mbox{\scriptsize ad}_\theta}X$. The operators $\mbox{ad}_{H_j}$ commute with each other since $[\mbox{ad}_{H_j},\mbox{ad}_{H_k}]=\mbox{ad}_{[H_j,H_k]}=0$. To obtain a more explicit representation of the winding transformations, we can try, therefore, to diagonalize simultaneously the action of the operators $\mbox{ad}_{H_j}$ on the Lie algebra. This is precisely the role of the Cartan-Weyl bases. These bases will also be used below for the characterization of the Weyl transformations.\footnote{They will also lead to considerable simplifications in the calculations to be performed in subsequent chapters.}

\subsection{Cartan-Weyl bases}
A Cartan-Weyl basis is a basis $\{iH_j,iE_\alpha\}$ of the Lie algebra which extends the basis $\{iH_j\}$ of the  Cartan subalgebra and which diagonalizes simultaneously the adjoint action of all the elements of the Cartan subalgebra \cite{Zuber}:
\begin{eqnarray}
 \mbox{ad}_{H_j}H_k & \!\!\!\!\!\equiv\!\!\!\!\! & [H_j,H_k]=0\,,\\
 \mbox{ad}_{H_j}E_\alpha & \!\!\!\!\!\equiv\!\!\!\!\! & [H_j,E_\alpha]=\alpha_j\,E_\alpha\,.\label{eq:CW}
\end{eqnarray}
The labels $\alpha$ are vectors of $\mathds{R}^{d_C}$ whose components are the $\alpha_j$. This copy of $\mathds{R}^{d_C}$ is however not the same as the restricted background space defined above. For reasons that will become clear shortly, we refer to it as the {\it dual (restricted) background space.} It can be shown that roots always appear by pairs $(\alpha,-\alpha)$. They form what is known as the root diagram of the corresponding algebra.\\

\begin{figure}[t!]
\begin{center}
\includegraphics[height=0.42\textheight]{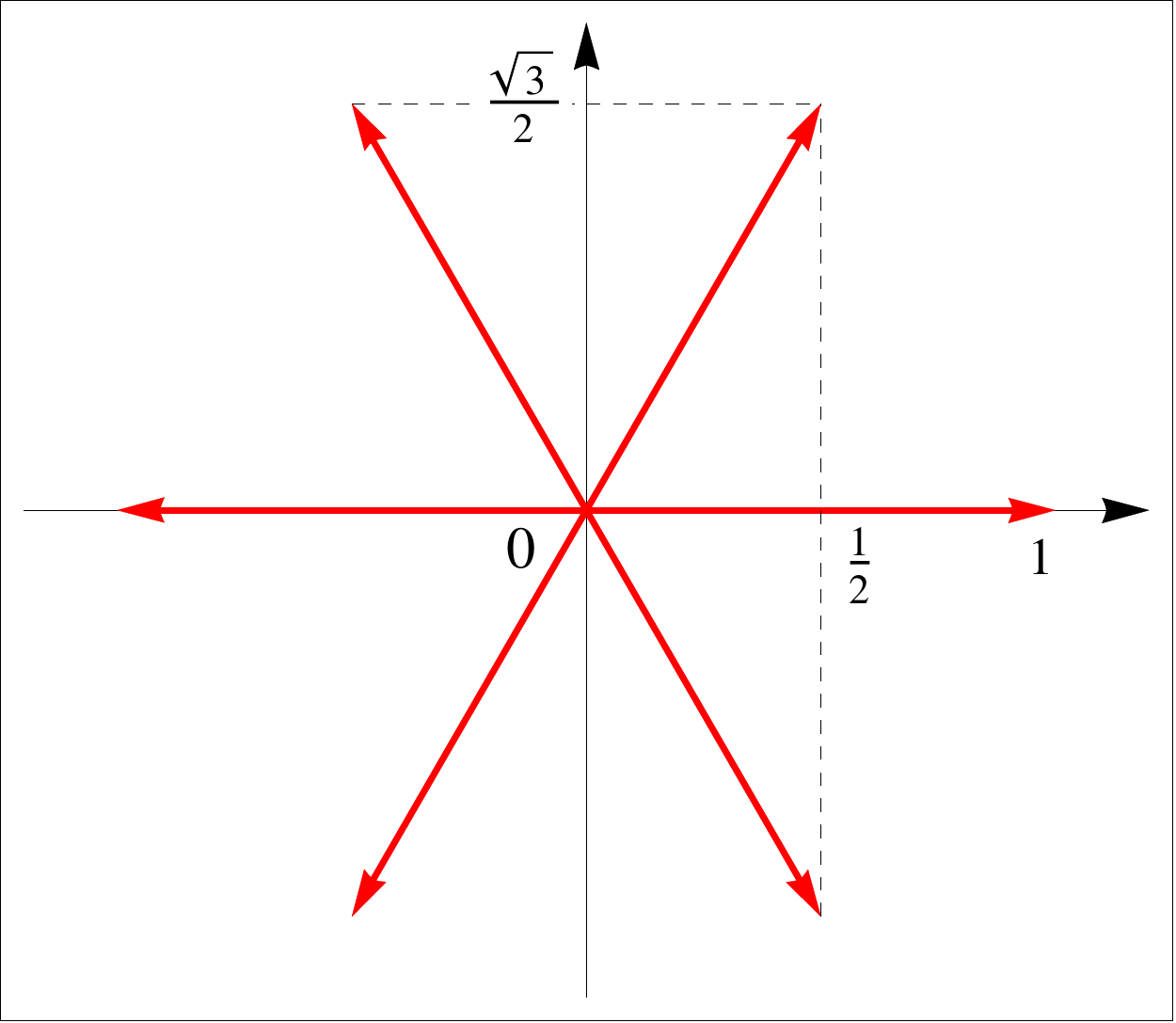}
\mycaption[Root diagram of SU($3$) and of its SU($2$) subgroups]{The SU($3$) root diagram. The root diagram of SU($2$) is also visible if one restricts to the horizontal axis. The other pairs of roots can be interpreted as corresponding to the other SU($2$) subgroups of SU($3$).}
\label{fig:roots}
\end{center}
\end{figure}

To take a few examples, consider the SU($2$) and SU($3$) cases. For SU($2$), the Cartan subalgebra is generated by the diagonal Pauli matrix $H_3=\sigma_3/2$. From the commutation relations $[\sigma_i/2,\sigma_j/2]=i\epsilon_{ijk}\sigma_k/2$, it is easily seen that one possible Cartan-Weyl basis is given by $\{i\sigma_3/2,i\sigma_{+}/2,i\sigma_{-}/2\}$ with $\sigma_{\pm}=(\sigma_1\pm i\sigma_2)/\sqrt{2}$, and roots $\pm 1$. In the case of $SU(3)$, the Cartan subalgebra is generated by the diagonal Gell-Mann matrices $H_3=\lambda_3/2$ and $H_8=\lambda_8/2$. Using the values for the SU($3$) structure constants, it is easily checked that the roots are $\pm (1,0)$, $\pm(1/2,\sqrt{3}/2)$ and $\pm(1/2,-\sqrt{3}/2)$. The corresponding root diagrams are represented in Fig.~\ref{fig:roots}. The root diagram of SU(N) is constructed in App.~\ref{app:sun} where we show in particular that all roots are of norm unity.\\

We mention that Cartan-Weyl bases are not bases of the original, real-valued Lie algebra but, rather, of the complexified Lie algebra. Therefore, in such bases, the components of an element $X$ of the original Lie algebra do not need to be real. However, one can always choose the basis elements $E_\alpha$ such that $X=i(X_jH_j+X_\alpha E_\alpha)$, with $X_j\in\mathds{R}$ and $X^*_\alpha=X_{-\alpha}$, see App.~\ref{app:sun} for a more detailed discussion.

\subsection{Winding transformations}
Decomposing the gauge field $A_\mu(x)$ into a Cartan-Weyl basis, it is easily checked that the winding transformation (\ref{eq:winding}) writes
\begin{eqnarray}
 \beta A_\mu^j(x) & \!\!\!\!\!\to\!\!\!\!\! & \beta A_\mu^j(x)+s_j\,,\label{eq:t1}\\
 \beta A_\mu^\alpha(x) & \!\!\!\!\!\to\!\!\!\!\! & \beta A_\mu^\alpha(x)\,e^{i\frac{\tau}{\beta}s\cdot\alpha}\,,\label{eq:t2}
\end{eqnarray}
where $s\cdot\alpha\equiv s_j\alpha_j$. Then, the components of $\beta A_\mu(x)$ along the Cartan subalgebra are just shifted by the corresponding components of $s$, while the orthogonal components\footnote{The reason why we dub these components `orthogonal' is given in App.~\ref{app:sun}} are multiplied by the phase $e^{i\frac{\tau}{\beta}s\cdot\alpha}$. 

So far, however, we did not implement the fact that winding transformations, like any other twisted gauge transformation, should preserve the periodicity of the gauge field along the temporal direction. This implies some constraints on the vector $s$ defining the winding transformation. Since the components of the gauge field along the Cartan subalgebra are just shifted by a constant, the constraints originate only from the orthogonal components. For these components to remain periodic, we need to require the following set of conditions:
\begin{equation}\label{eq:cond}
 s\cdot\alpha\in 2\pi\,\mathds{Z}\,,\,\,\forall\alpha\,.
\end{equation}
In fact, it is sufficient that these conditions be satisfied for a set of roots $\{\alpha^{(j)}\}$ that generate any other root as a linear combination with integer coefficients. In this case, the conditions (\ref{eq:cond}) define a lattice, dual to the one generated by the roots. If we introduce a basis $\{\bar\alpha^{(k)}\}$ of this dual lattice, such that
\begin{equation}
 \alpha^{(j)}\cdot\bar\alpha^{(k)}=2\pi\delta_{jk}\,,
\end{equation}
the general solution to (\ref{eq:cond}) takes then the form
\begin{equation}
 s=n_k\bar\alpha^{(k)}\,,
\end{equation}
where $n_k\in\mathds{Z}$ and a summation over $k$ is implied. 

As far as the restricted background $r$ is concerned, because it is diagonal by assumption, it transforms as (\ref{eq:t1}), that is as
\begin{equation}
r\to r+n_k\bar\alpha^{(k)}\,.
\end{equation}
We conclude that, in the restricted background space, the winding transformations are generated by translations along the vectors $\bar\alpha^{(k)}$. We can then assume that these vectors belong to the restricted background space. This explains why we dubbed the copy of $\mathds{R}^{d_C}$ containing the roots (and in particular the $\alpha^{(j)}$) as the dual (restricted) background space.\\

In the SU($2$) case, the Cartan subalgebra is one dimensional and the restricted background is given by a single real number $r$. We can take for instance $\alpha^{(1)}=1$ and, therefore, $\bar\alpha^{(1)}=2\pi$. It follows that winding transformations are generated by the shift $r\to r+2\pi$. In the SU(3) case, the Cartan subalgebra is two dimensional and the restricted background is given by a vector $r=(r_3,r_8)$. We can choose $\alpha^{(1)}=(1/2,-\sqrt{3}/2)$ and $\alpha^{(2)}=(1/2,\sqrt{3}/2)$. Then, winding transformations are generated by translations along the vectors $\bar\alpha^{(1)}=(2\pi,-2\pi/\sqrt{3})$ and $\bar\alpha^{(2)}=(2\pi,2\pi/\sqrt{3})$.

\subsection{Weyl transformations}
Let us now consider the Weyl transformations defined as the color rotations that leave the diagonal part of the algebra globally invariant. It is easily seen that these transformations cannot be infinitesimal and therefore, we need to consider finite transformations of the form $W=e^\theta$. The action of such a color transformation on any element $X$ of the algebra is 
\begin{equation}
 e^{\theta}Xe^{-\theta}=e^{\mbox{\scriptsize ad}_\theta}X=\sum_{n=0}^\infty \frac{1}{n!}\mbox{ad}_\theta^nX\,.
\end{equation}
Suppose now that we take $\theta$ of the form $\theta=i(\theta_\alpha E_\alpha+\theta_{-\alpha}E_{-\alpha})\equiv w_\alpha$, with $\theta_{-\alpha}=\theta_\alpha^*$ and where no summation over $\alpha$ is implied, hence the notation $w_\alpha$. A simple recursion using Eq.~(\ref{eq:CW}) shows that
\begin{equation}
\mbox{ad}^{2p}_{w_\alpha}(H_j)=i^{2p}\big(2|\theta_\alpha|^2\alpha^2\big)^p\frac{\alpha_k H_k}{\alpha^2}\alpha_j\,,
\end{equation}
for $p>0$, and
\begin{equation}
\mbox{ad}^{2p+1}_{w_\alpha}(H_j)=-i^{2p+1}\big(2|\theta_\alpha|^2\alpha^2\big)^p(\theta_\alpha E_\alpha-\theta_{-\alpha} E_{-\alpha})\alpha_j\,,
\end{equation}
for $p\geq 0$. It follows that
\begin{align}
 e^{w_\alpha}H_je^{-w_\alpha} & =  H_j+\Big(\cos\big(\sqrt{2\alpha^2}|\theta_\alpha|\big)-1\Big) \frac{\alpha_k H_k}{\alpha^2}\alpha_j\\
& - \frac{i\alpha_j}{\sqrt{2\alpha^2}|\theta_\alpha|}\sin\big(\sqrt{2\alpha^2}|\theta_\alpha|\big) (\theta_\alpha E_\alpha-\theta_{-\alpha}E_{-\alpha})\,.\nonumber
\end{align}
Now, choosing $|\theta_\alpha|=\pi/\sqrt{2\alpha^2}$, we can enforce the transformation to leave the diagonal part of the algebra globally invariant:
\begin{equation}
 H_j \to H_j-2\frac{\alpha_kH_k}{\alpha^2}\alpha_j\,.
\end{equation}
Considering this as a passive transformation, the restricted background $r$ transforms as
\begin{equation}\label{eq:reflect}
 r \to r-2\frac{\alpha\cdot r}{\alpha^2}\alpha\,,
\end{equation}
which is indeed a Weyl transformation as defined in \cite{Zuber}. In the restricted background space $\mathds{R}^{d_C}$, this corresponds to a reflection with respect to an hyperplane orthogonal to the root $\alpha$ and containing the origin.\footnote{In Eq.~(\ref{eq:reflect}), it seems that we are shifting a vector of the restricted space ($r$) by a vector of the dual space ($\alpha$). However, the root $\alpha$ enters this formula only through its direction $\alpha/||\alpha||$, the magnitude of the shift being proportional to the norm of $r$.}\\

Finally, we mention that there exist other color rotations that leave the Cartan subalgebra invariant, in a trivial way. Those correspond to rotations $e^{i\theta_jH_j}$ generated by the elements of the Cartan subalgebra itself. Those are of no interest for the present discussion since they do not lead to geometrical transformations of the restricted background. They will however play an important role in Chapter \ref{chap:YM_2loop} as we derive the Feynman rules using Cartan-Weyl bases.

\subsection{Summary}
From the above considerations, we derive the following invariance properties of the background field effective potential:
\begin{equation}
 V(r+\bar\alpha^{(k)})=V(r)\,,\,\,\forall\bar\alpha^{(k)}\,,
\end{equation}
as well as
\begin{equation}
 V\left(r-2\frac{\alpha\cdot r}{\alpha^2}\alpha\right)=V(r)\,,\,\,\forall\alpha\,.
\end{equation}
Correspondingly,
\begin{equation}\label{eq:phiofk}
 \ell(r+\bar\alpha^{(k)})=e^{i\varphi(k)}\ell(r)\,,\,\,\forall\bar\alpha^{(k)}\,,
\end{equation}
for some $\varphi(k)$ to be fixed in the next section, and
\begin{equation}
 \ell\left(r-2\frac{\alpha\cdot r}{\alpha^2}\alpha\right)=\ell(r)\,,\,\,\forall\alpha\,.
\end{equation}

\section{Weyl chambers and symmetries}
In the previous section, we could interpret the various elements of $\tilde {\cal G}$ as geometrical transformations in the restricted background space $\mathds{R}^{d_C}$. Of particular importance is the subgroup $\tilde {\cal G}_0$ of periodic transformations within $\tilde {\cal G}$ since the physical states are interpreted as the corresponding $\tilde {\cal G}_0$-orbits, and the center-symmetric states correspond to the center-invariant $\tilde {\cal G}_0$-orbits. We shall now characterize the elements of $\tilde {\cal G}_0$ and then identify the center-invariant $\tilde {\cal G}_0$-orbits using the concept of Weyl chambers.\\

In fact, we know already part of the elements of $\tilde {\cal G}_0$ since the Weyl transformations are trivially periodic. The other generators of $\tilde {\cal G}_0$ are the periodic winding transformations, which we now characterize.

\subsection{Periodic winding transformations}
To identify the periodic winding transformations, we assume that the $H_j$ can all be diagonalized.\footnote{In the case of SU(N), we have chosen the $H_j$ already in an explicitly diagonalized form.} Since they commute with each other, they can in fact be diagonalized simultaneously,
\begin{equation}
  H_j|\rho\rangle=\rho_j|\rho\rangle,
\end{equation}
where the labels $\rho$ are called the weights of the defining representation or {\it defining weights} for short, and are just a convenient way to gather all eigenvalues $\rho_j$ of a given eigenvector $|\rho\rangle$. The defining weights are again vectors of $\mathds{R}^{d_C}$ and can, therefore, be represented together with the root diagram. The reason for representing them on the same copy of $\mathds{R}^{d_C}$ will appear more clearly once we introduce the quarks back in our analysis.
\begin{figure}[t!]
\begin{center}
\includegraphics[height=0.42\textheight]{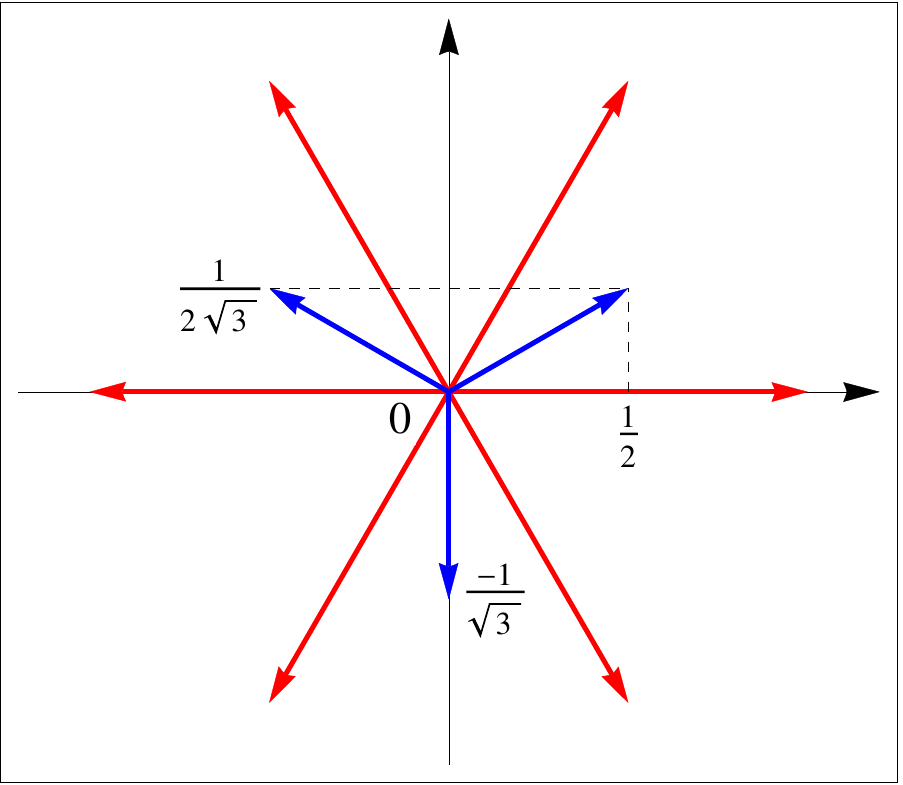}
\mycaption[Weigths of the defining representation of SU($3$)]{Weight diagram for the defining representation of SU($3$), together with the corresponding root diagram (which is nothing but the non-zero weight diagram of the adjoint representation).}
\label{fig:weights}
\end{center}
\end{figure}
In the SU($2$) case, the weights of the defining representation are $\pm 1/2$. In the SU($3$) case, we find $(1/2,1/(2\sqrt{3}))$, $(-1/2,1/(2\sqrt{3}))$ and $(0,-1/\sqrt{3})$, see Fig.~\ref{fig:weights} for a representation of the corresponding weight diagram and App.~\ref{app:sun} for the construction of the defining weights for any value of $N$.\\

In terms of the weights, a general winding transformation writes
\begin{equation}\label{eq:winding_decomp}
 e^{i\frac{\tau}{\beta}s_jH_j}=\sum_{\rho}e^{i\frac{\tau}{\beta}s\cdot\rho}|\rho\rangle\langle\rho|\,,
\end{equation}
where $s\cdot\rho\equiv s_j\rho_j$. Periodic winding transformations are obtained by requiring that
\begin{equation}
\label{eq:cond4}
 s\cdot\rho\in2\pi\,\mathds{Z}\,,\forall\rho.
\end{equation}
The solution to (\ref{eq:cond4}) can then be found by repeating the previous arguments used to solve (\ref{eq:cond}). We first introduce a set of weights or their opposites $\{\rho^{(j)}\}$ that generates any other weight through linear combination with integer coefficients. The general solution to (\ref{eq:cond4}) is then
\begin{equation}
 s=n_k\bar\rho^{(k)}\,,
\end{equation} 
with $n_k\in\mathds{Z}$ and where $\{\bar\rho^{(k)}\}$ is a basis of the lattice dual to the one generated by the weights:
\begin{equation}
 \rho^{(j)}\cdot\bar\rho^{(k)}=2\pi\delta_{jk}\,.
\end{equation}
It follows that, in the restricted background space, periodic winding transformations are generated by translations along the vectors $\bar\rho^{(k)}$.

We mention that, because periodic winding transformations are just particular winding transformations, the lattice generated by $\{\bar\rho^{(j)}\}$ needs to be a sub-lattice of the one generated by $\{\bar\alpha^{(j)}\}$. This can be understood from the fact that the adjoint representation is generally obtained by decomposing the tensor product of the defining representation and the associated contragredient representation (whose weights are opposite to those of the defining representation). We can thus construct the eigenstates of $\mbox{ad}_{H_j}$ in terms of the eigenstates of $H_j$ and we find that the roots can always be written as differences of weights of the defining representation (an explicit proof is given in App.~\ref{app:sun} for the SU(N) case). The lattice generated by $\{\alpha^{(j)}\}$ is then a sub-lattice of the one generated by $\{\rho^{(j)}\}$, which implies the opposite relation for the corresponding dual lattices.\\

In the SU($2$) case, we can choose $\rho^{(1)}=1/2$ and, therefore $\bar\rho^{(1)}=4\pi$. Thus, the periodic winding transformations are generated by the shift $r\to r+4\pi$. In the SU(3) case, we can choose $\rho^{(1)}=(1/2,-1/(2\sqrt{3}))$ and $\rho^{(2)}=(1/2,1/(2\sqrt{3}))$. Then, the periodic winding transformations are generated by translations along the vectors $\bar\rho^{(1)}=(2\pi,-2\pi\sqrt{3})$ and $\bar\rho^{(2)}=(2\pi,2\pi\sqrt{3})$. We note that, in both cases, we could choose the various bases such that $\bar\alpha^{(k)}=4\pi\rho^{(k)}$ and $\bar\rho^{(k)}=4\pi\alpha^{(k)}$, see the previous section for the values of $\alpha^{(k)}$ and $\bar\alpha^{(k)}$. This property generalizes to higher values of $N$, see App.~\ref{app:sun}. Then, if we define the {\it reduced background} as $\bar r\equiv r/4\pi$, we can identify the reduced and the dual background spaces and represent all the above geometrical transformations together with the root and weight diagrams: Weyl transformations correspond to reflections with respect to hyperplanes orthogonal to roots and containing the origin, periodic winding transformations are generated by translations along the roots, whereas generic winding transformations are generated by translations along the weights of the defining and contragredient representations. In the next section, for simplicity, we define the Weyl chambers in the reduced background space. However, when representing the latter graphically, we do so in the restricted background space by applying a factor $4\pi$.

We mention finally that, according to Eq.~(\ref{eq:winding_decomp}), the center element associated to a winding transformation of parameter $s$ is given by $e^{is\cdot\rho}$.\footnote{This center element does not depend on the choice of $\rho$ since $s=n_k\bar\alpha^{(k)}$ and the difference of two weights is always a root, which implies that $s\cdot\rho-s\cdot\rho'$ is a multiple of $2\pi$.} In particular, this fixes the phase in Eq.~(\ref{eq:phiofk}) to $\varphi(k)=\bar\alpha^{(k)}\cdot\rho=4\pi\rho^{(k)}\cdot\rho$ and we show in App.~\ref{app:sun} that, in the case of SU(N), this is always $\mp 2\pi/N$, with the sign depending on whether $\rho^{(k)}$ is a weight or the opposite of a weight.

\subsection{Weyl chambers and invariant states}
\begin{figure}[t]
\begin{center}
\hspace{1.0cm}\includegraphics[height=0.5\textheight]{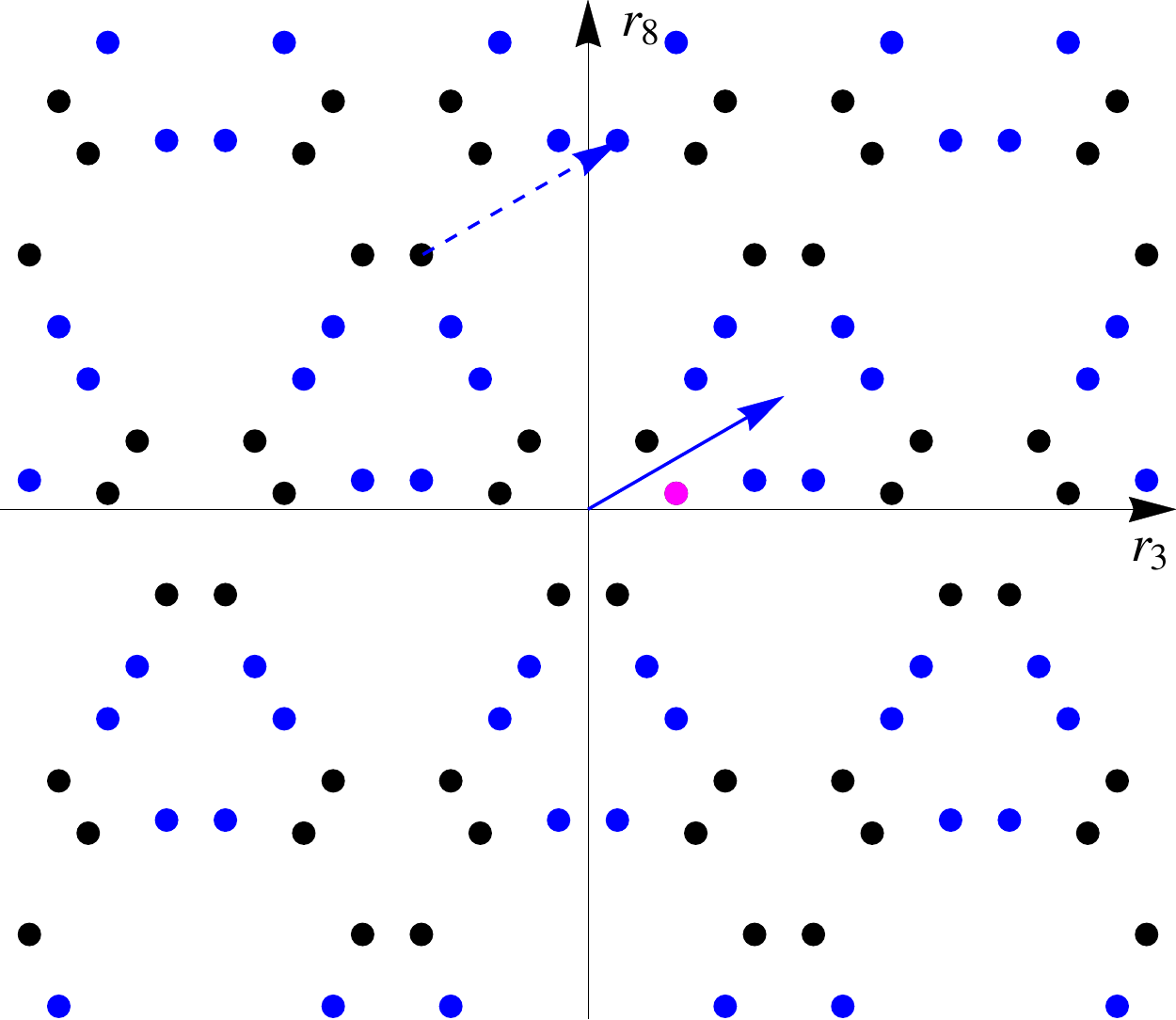}\\
\mycaption[$\tilde {\cal G}_0$-orbits and their transformation under $\tilde {\cal G}$]{A given $\tilde {\cal G}_0$-orbit (black dots) and its transformation (blue dots) under a non-periodic winding transformation (translation along the blue vector). To vary the original orbit, one varies locally one of its representing backgrounds (pink dot) with little control on how the orbit changes globally, making it difficult to identify the center-invariant orbits.}
\label{fig:pain}
\end{center}
\end{figure}

We are now ready to characterize the center-invariant $\tilde {\cal G}_0$-orbits, that is the center-symmetric states, in the reduced background space $\mathds{R}^{d_C}$. One possible strategy would be to work directly with the orbits as collections of background configurations and, by trial and error, find which orbits are globally invariant under center transformations. In Fig.~\ref{fig:pain}, we have tried to illustrate how painful such a procedure could be, in the SU($3$) case. The difficulty here is that, in order to explore the various orbits, one varies a given orbit locally (by changing any of the backgrounds that represent the orbit), with little control on how the orbit changes globally, and therefore little chance of finding the center-invariant orbits. We now introduce a simpler strategy based on the notion of Weyl chambers which allows us to treat the orbits as a whole.

The idea is to use the geometrical interpretation of the transformations of $\tilde {\cal G}_0$ in the reduced background space to subdivide the latter into cells which are connected to each other by the transformations of $\tilde {\cal G}_0$ and which are, therefore, physically equivalent. These are the so-called {\it Weyl chambers}\footnote{Here the definition of the Weyl chambers differs substantially from the one in Mathematics, where it corresponds to subdivisions associated to Weyl transformations only \cite{Zuber}.} whose points can be seen as representing the various possible $\tilde {\cal G}_0$-orbits and thus the various possible states of the system.

Now, as we have seen in the previous chapter, the action of a center transformation ${\cal U}\in{\cal G}/{\cal G}_0\simeq\tilde{\cal G}/\tilde {\cal G}_0$ can be defined directly on the ${\cal G}_0$-orbits and, similarly on the $\tilde {\cal G}_0$-orbits. This action should then appear as a geometrical transformation of a given Weyl chamber into itself. To see how this is achieved in practice, one first chooses any representative $U\in {\cal G}$ of the center transformation ${\cal U}$ and applies the transformation to the chosen Weyl chamber. This typically moves the Weyl chamber away from its original location. However, without altering the physical interpretation, we can bring the Weyl chamber back to its original location by using transformations of $\tilde {\cal G}_0$. In doing so, one defines the action of $\tilde {\cal G}$ directly on the chosen Weyl chamber. Moreover, two transformations of $\tilde {\cal G}$ that correspond to the same center transformation in $\tilde {\cal G}/\tilde {\cal G}_0$ lead to the same geometrical transformation of the Weyl chamber, thereby defining the action of the center symmetry group  $\tilde {\cal G}/\tilde {\cal G}_0$ on the chosen Weyl chamber.\\

In summary, center transformations correspond to certain geometrical transformations of the Weyl chambers into themselves, and the symmetric states appear as the fixed points of the Weyl chambers under these geometrical transformations. Similar considerations apply to other physical transformation, in particular to charge conjugation.

\subsection{Explicit construction for SU(N)}
In the case of $SU(N)$, we have seen that, in the reduced background space, the generators of $\tilde {\cal G}_0$  are the translations by a root and the reflections with respect to hyperplanes orthogonal to a root and containing the origin. In order to identify the Weyl chambers, it is convenient to find a generating set of $\tilde {\cal G}_0$ made only of reflections. Such a set is given by all the reflections with respect to hyperplanes orthogonal to a root and translated by any multiple of half that root. Indeed, it is easily checked that these reflections are elements of $\tilde {\cal G}_0$ and that the combination of two adjacent such reflections generates the translation by the corresponding root. The benefit of this generating set is that it allows to identify the Weyl chambers as the regions delimitated by the corresponding network of hyperplanes. Moreover, the reflections with respect to the facets of the so-obtained Weyl chambers allow to transform them into one another without altering the physical interpretation.

To further understand the structure of the Weyl chambers, consider first the network of hyperplanes generated by the roots of the basis $\{\alpha^{(j)}\}$. The intersections of these hyperplanes define a lattice generated by a certain basis $\{x^{(j)}\}$ and whose elements obey the constraints 
\begin{equation}
  x^{(j)}\cdot \alpha^{(k)}=0\,,\,\, \mbox{for } k\neq j \quad \mbox{and} \quad (x^{(j)}-\alpha^{(j)}/2)\cdot\alpha^{(j)}=0\,.
\end{equation} 
Using that the roots are all of norm unity, we find that the solution to these constraints is $x^{(j)}=\bar\alpha^{(j)}/4\pi=\rho^{(j)}$. This implies that the $SU(N)$ Weyl chambers are sub-pavings of the parallelepipeds generated by the weights $\rho^{(j)}$. To finally reach the Weyl chambers, we need to see how the remaining hyperplanes (those associated to roots other than the $\alpha^{(j)}$), further divide these parallelepipeds. We shall do this here in the cases $N=2$ and $N=3$, and leave the discussion of $N=4$ for a future version of the manuscript.\\

\begin{figure}[t]
\begin{center}
\hspace{0.4cm}\includegraphics[height=0.25\textheight]{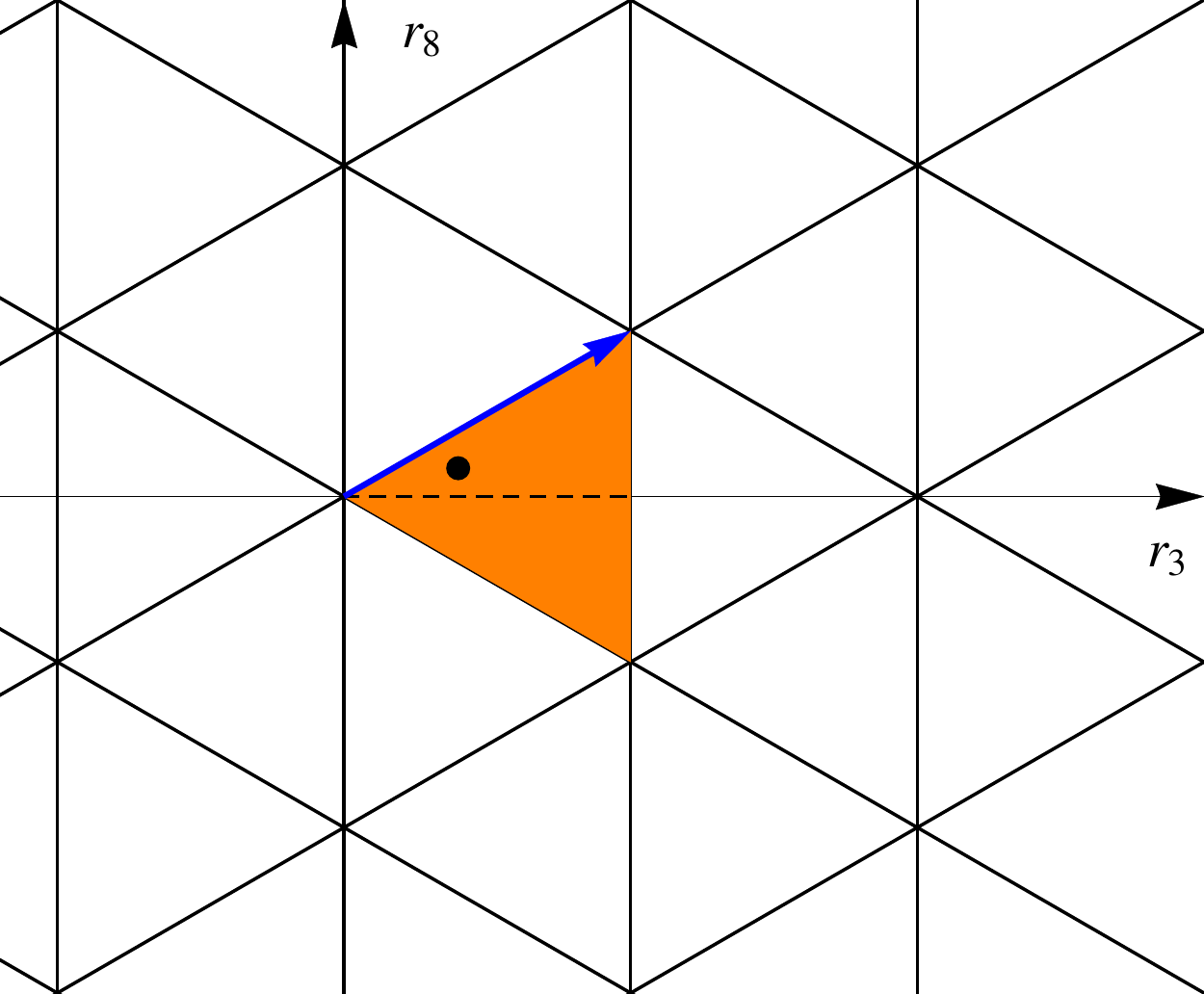}\hspace{0.9cm}\includegraphics[height=0.25\textheight]{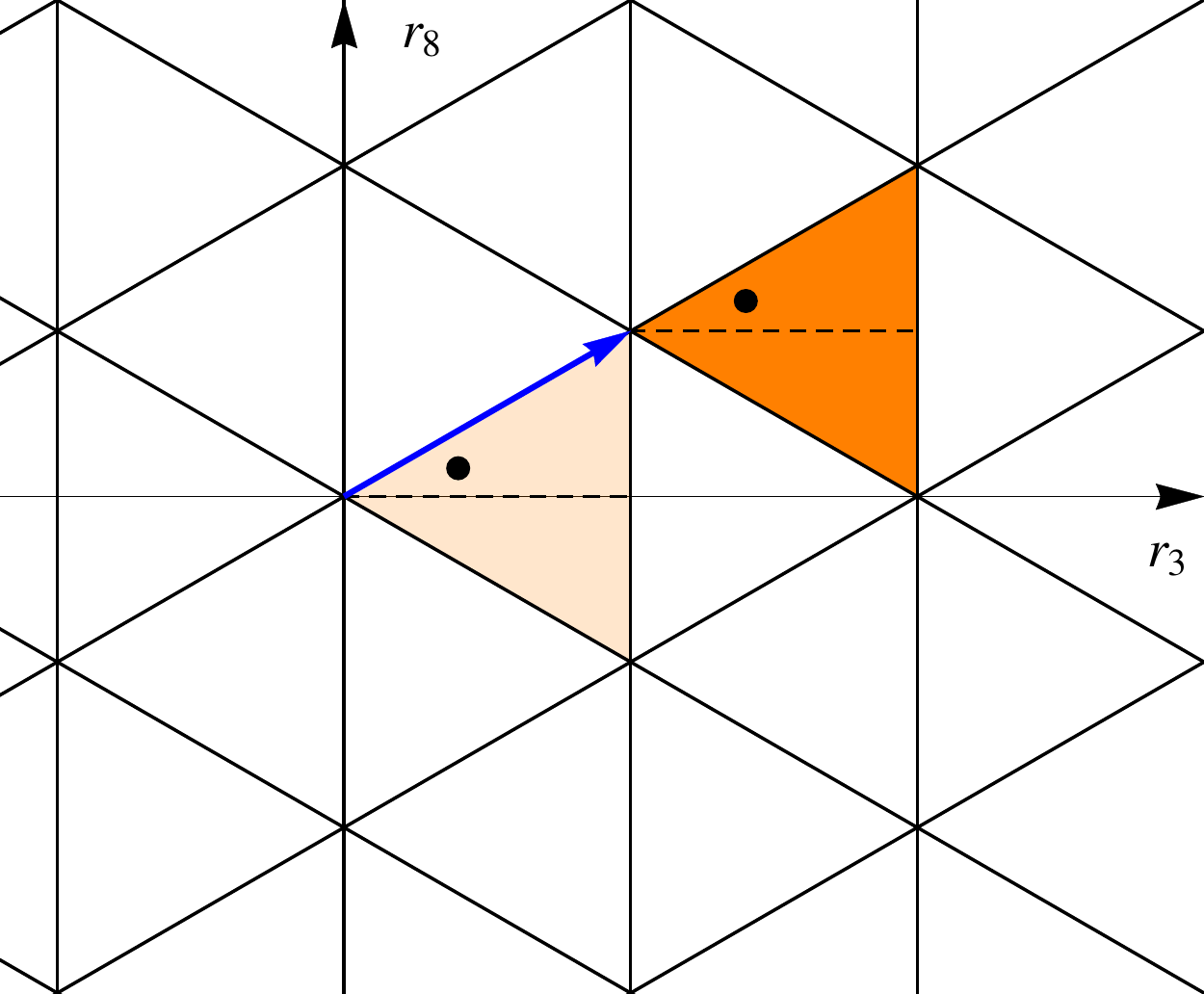}\\
\vspace{1.0cm}
\hspace{0.4cm}\includegraphics[height=0.25\textheight]{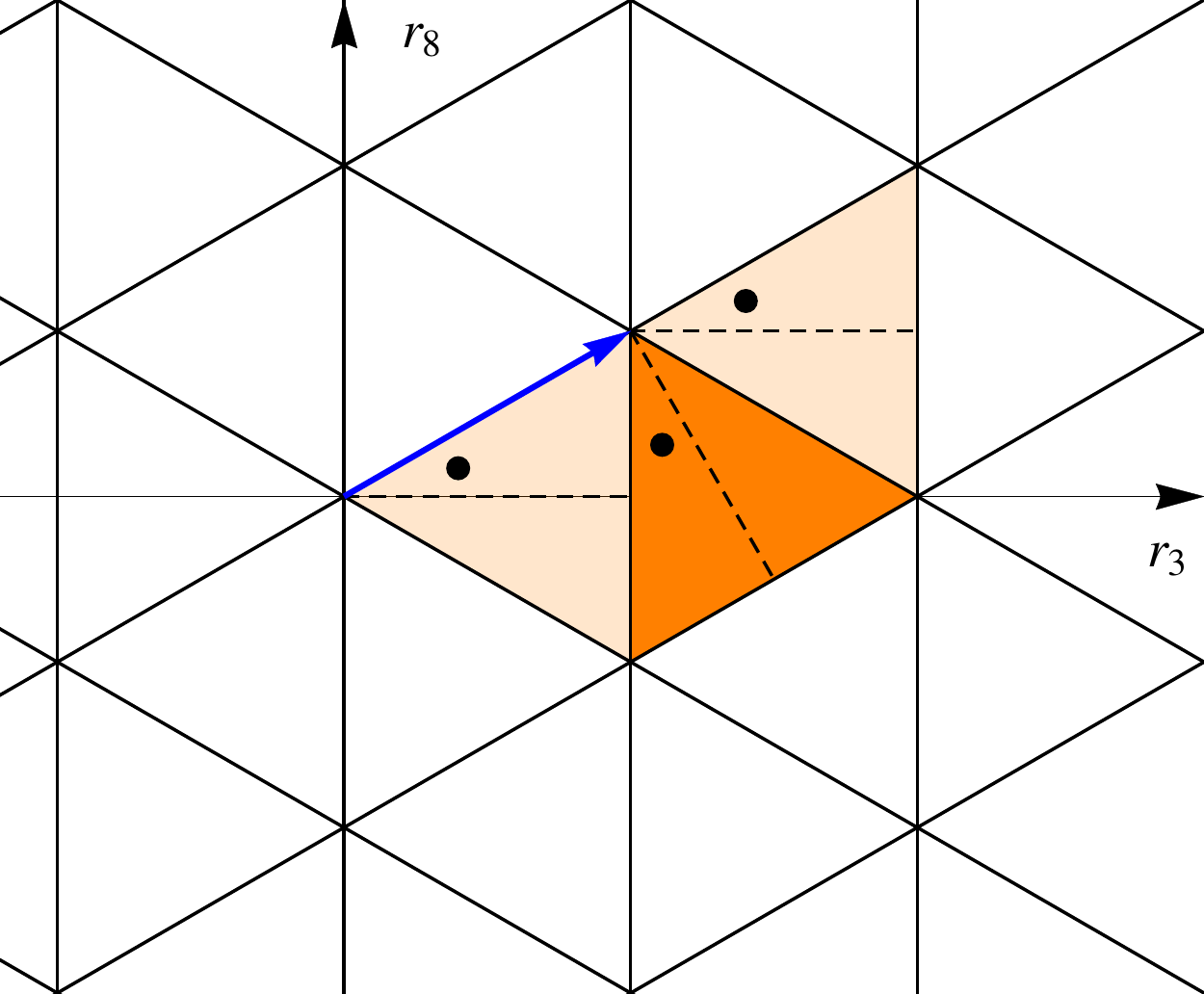}\hspace{0.9cm}\includegraphics[height=0.25\textheight]{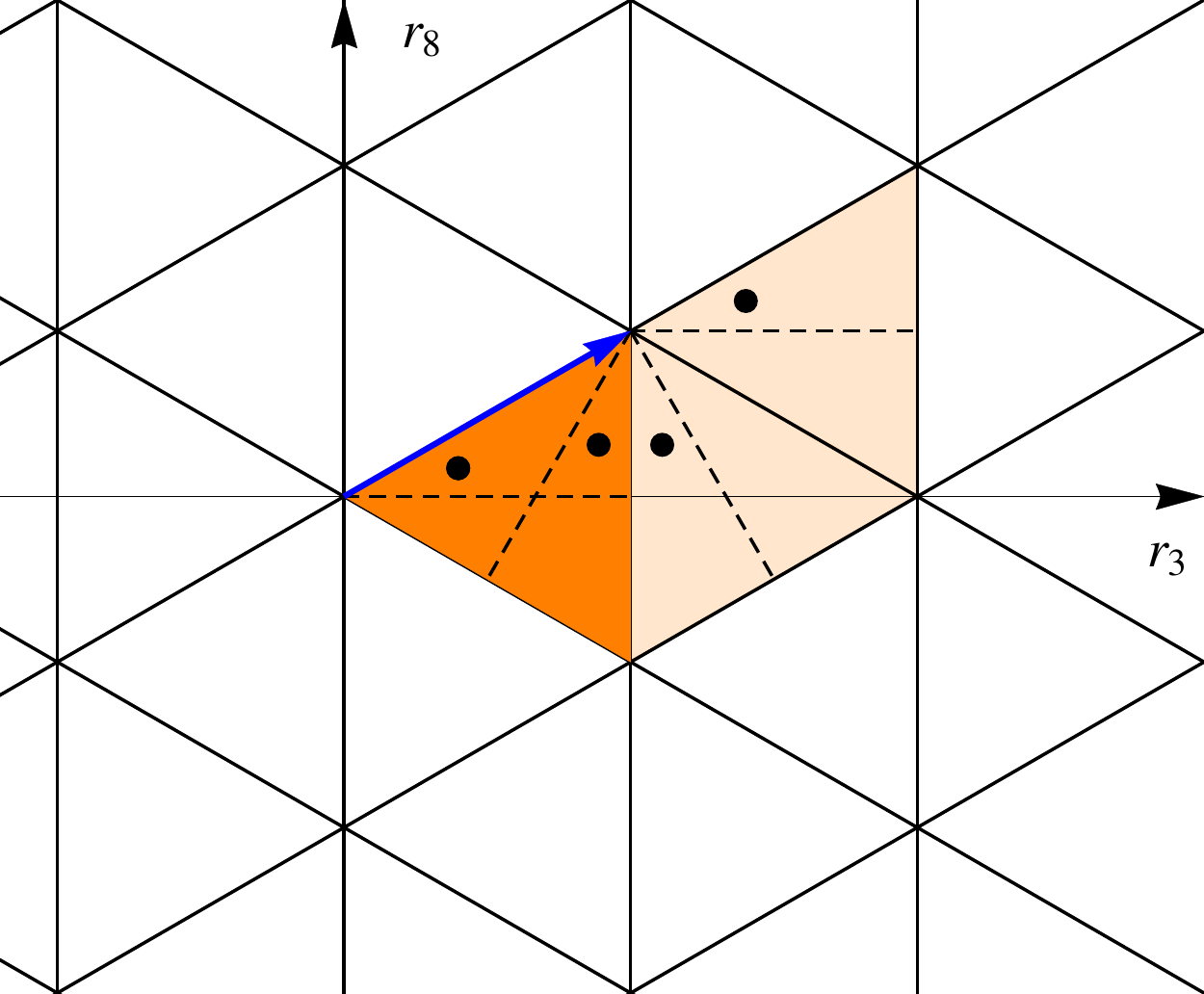}
\mycaption[SU($3$) Weyl chambers and their transformation under $\tilde {\cal G}/\tilde {\cal G}_0$]{SU($3$) Weyl chambers and their transformations under the action of $\tilde {\cal G}$ and $\tilde {\cal G}/\tilde {\cal G}_0$. The first figure shows the fundamental Weyl chamber together with one of its axes and an arbitrary point (representing an arbitrary state) to ease orientation as we transform the Weyl chamber. In each figure, the darker color indicates the actual position of the Weyl chamber, while the light colors indicate previous positions of the Weyl chamber.}
\label{fig:SU3_center}
\end{center}
\end{figure}

For SU($2$), there is only one family of hyperplanes, associated to $\alpha^{(1)}=1$. In the reduced background space, the Weyl chambers are then directly given by the parallelepipeds generated by $\rho^{(1)}=1/2$, that is the intervals $[k/2,(k+1)/2]$, with $k\in\mathds{Z}$. It will be enough to work in the interval $[0,1/2]$ which we refer to as the {\it fundamental Weyl chamber}. For SU($3$), there are three families of hyperplanes associated to $\alpha^{(1)}=(1/2,-\sqrt{3}/2)$, $\alpha^{(2)}=(1/2,\sqrt{3}/2)$ and $\alpha^{(3)}=-\alpha^{(1)}-\alpha^{(2)}$. The fundamental parallelepiped associated to the first two is given in terms of $\rho^{(1)}=(1/2,-1/(2\sqrt{3}))$ and $\rho^{(2)}=(1/2,1/(2\sqrt{3}))$. It is further divided by the family of hyperplanes associated to $\alpha^{(3)}$. One finds eventually that the Weyl chambers are equilateral triangles, see Fig.~\ref{fig:SU3_center}, the fundamental Weyl chamber being defined as the one with vertices $(0,0)$, $(1/2,-1/(2\sqrt{3}))$ and $(1/2,1/(2\sqrt{3}))$.\\

We can now locate the center-symmetric states. We have seen that translations along the $\rho^{(k)}$ (the edges of the fundamental parallelepiped) correspond to elementary center transformations with a phase $\varphi(k)=\mp 2\pi/N$ with the sign depending on whether $\rho^{(k)}$ is a weight or the opposite of a weight. Moreover, reflections with respect to the facets of the Weyl chamber are elements of $\tilde {\cal G}_0$, by construction, and allow one to move the Weyl chamber around without altering the physical interpretation. In the SU($2$) case, under the translation by $\rho^{(1)}$, the fundamental Weyl chamber $[0,1/2]$ is transformed into $[1/2,1]$. The latter can be brought back to its original location by means of a reflection with respect to $1/2$. As a result, we find that center transformations act on the fundamental Weyl chamber as $\bar r\to 1/2-\bar r$. There is only one fixed-point in the fundamental Weyl chamber, one confining state, $\bar r=1/4$ (or $r=\pi$). In the SU($3$) case, the fundamental equilateral triangle is such that the two edges connected to the origin correspond respectively to a weight and the opposite of a weight. Therefore, translations along these edges represent the two non-trivial elements of $\mathds{Z}_3$. After any of these translations, see Fig.~\ref{fig:SU3_center}, we need two reflections with respect to the edges to bring the fundamental triangle back to its original location, resulting in a rotation by an angle $\pm 2\pi/3$. Again, there is only one fixed-point under any of these transformations, the center of the triangle $\bar r=(1/3,0)$ (or $r=(4\pi/3,0)$). In order to analyse the deconfinement transition, we need therefore to evaluate the background field effective potential on the fundamental Weyl chamber and monitor the position of its minimum with respect to the center-symmetric point, as the temperature is varied. We shall do so in various situations starting from next chapter.\\

\begin{figure}[t]
\begin{center}
\hspace{0.4cm}\includegraphics[height=0.25\textheight]{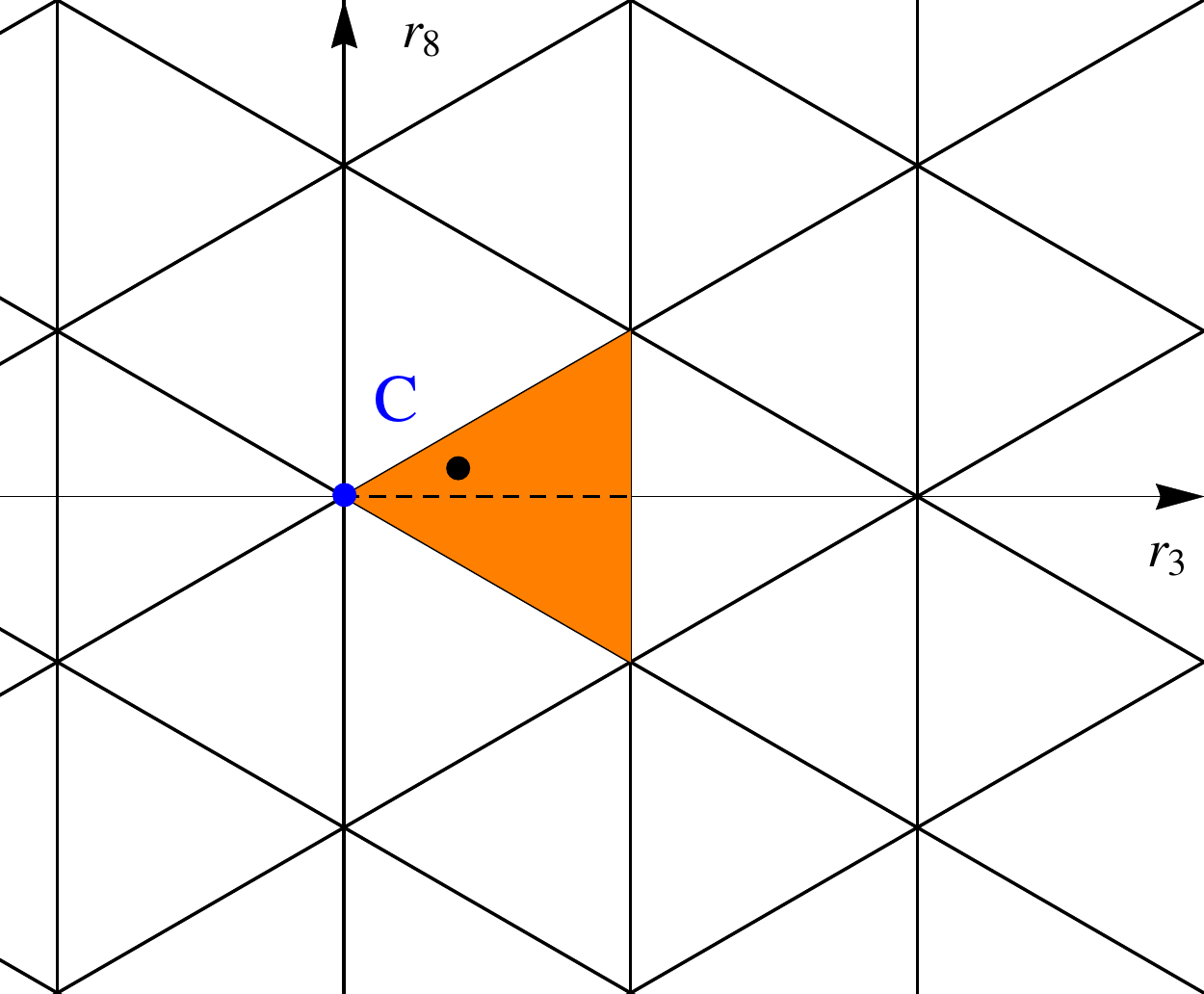}\hspace{0.9cm}\includegraphics[height=0.25\textheight]{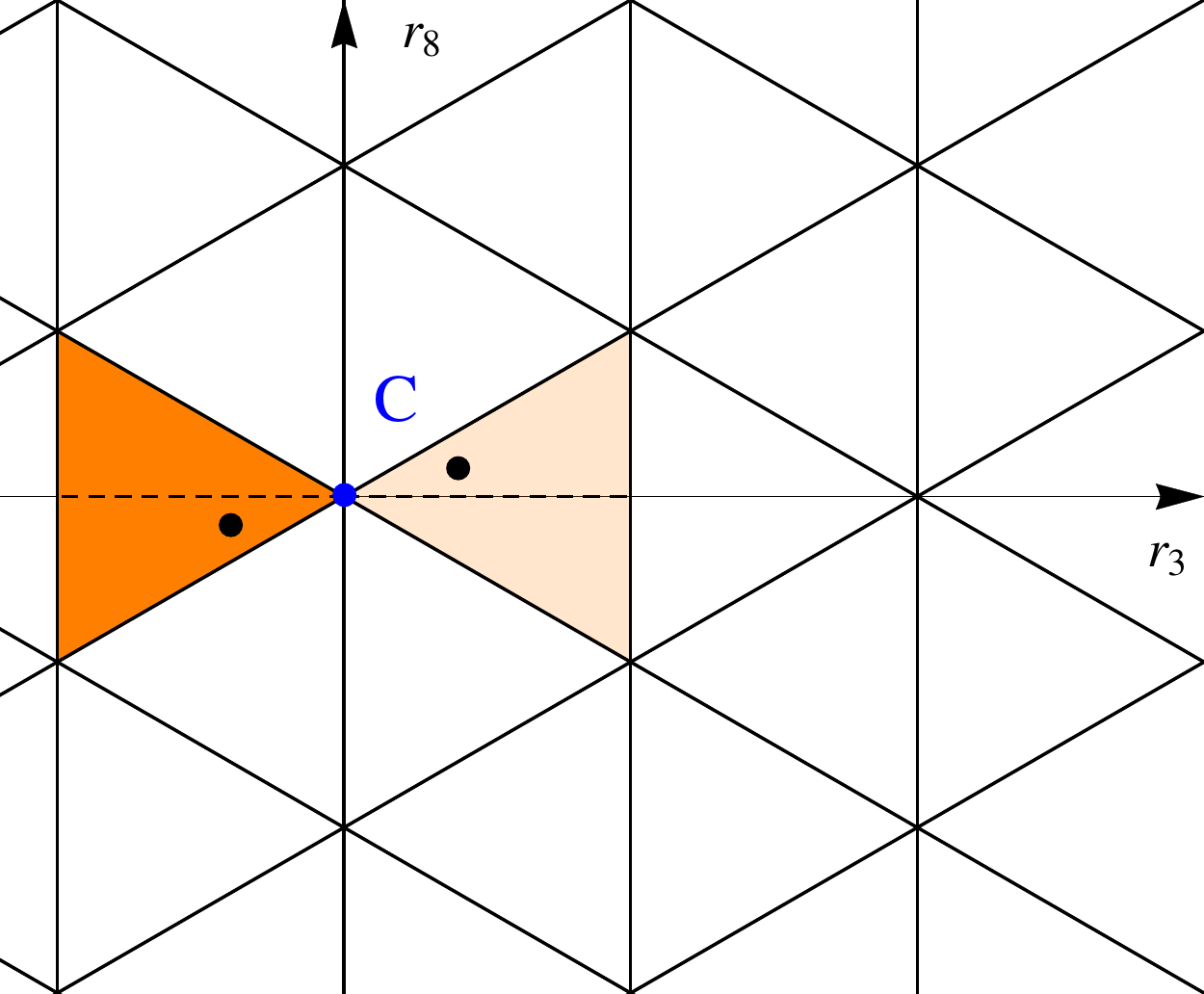}\\
\vspace{1.0cm}
\hspace{1.35cm}\includegraphics[height=0.25\textheight]{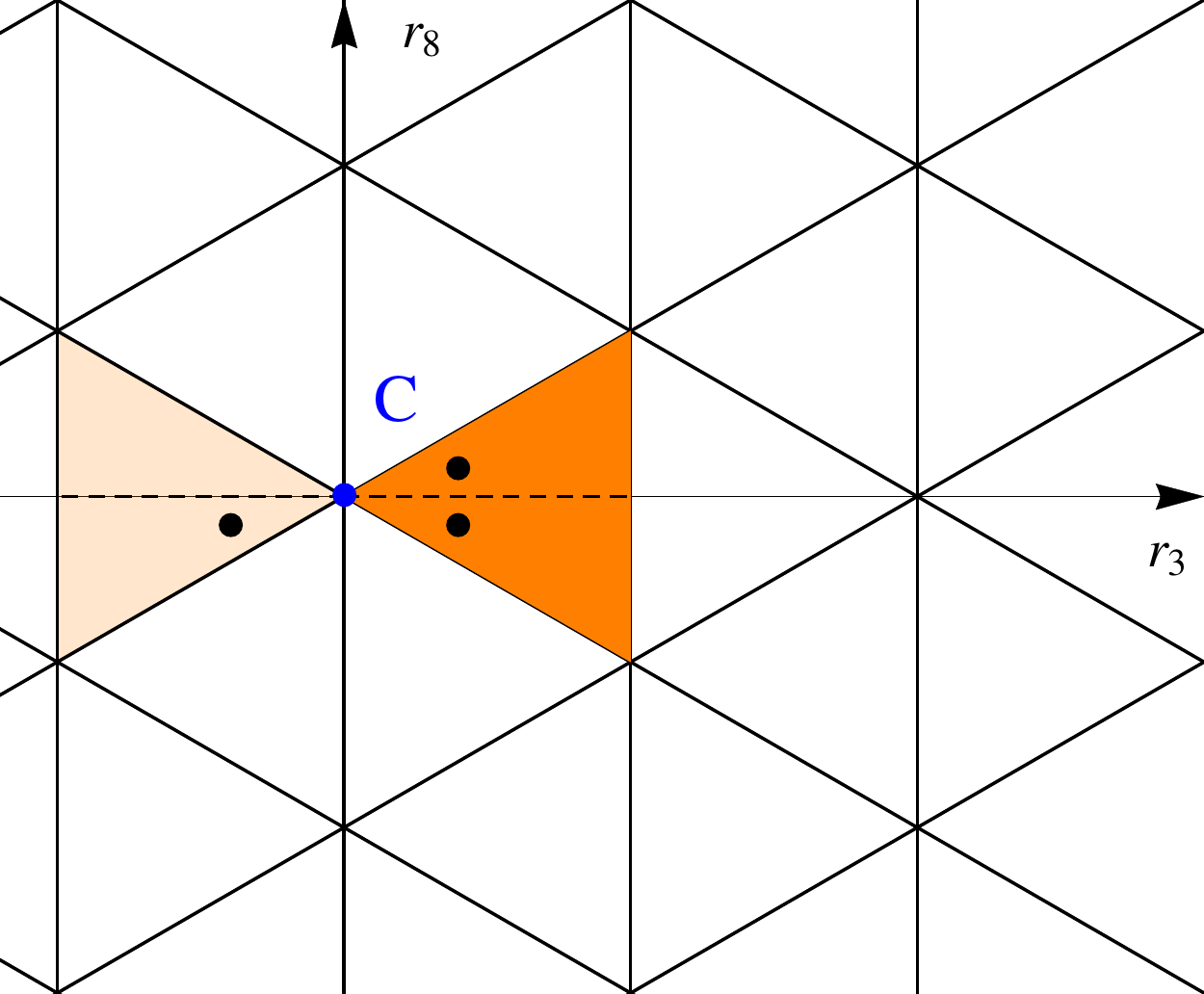}
\mycaption[SU($3$) Weyl chambers and their transformation under $C$]{SU($3$) Weyl chambers and their transformations under charge conjugation. The first figure shows the fundamental Weyl chamber together with one of its axes and an arbitrary point (representing an arbitrary state) to ease orientation as we transform the Weyl chamber. In each figure, the darker color indicates the actual position of the Weyl chamber, while the light colors indicate previous positions of the Weyl chamber.}
\label{fig:SU3_C}
\end{center}
\end{figure}

Similarly, we can locate the charge conjugation invariant states. In the SU($2$) case, charge conjugation is nothing but a Weyl reflection. There is therefore no constraint from charge conjugation invariance in that case. In the SU($3$) case, applying the transformation $r\to -r$ to the fundamental Weyl chamber and bringing it back to its original location by means of an element of $\tilde {\cal G}_0$, one finds that the physical charge conjugation is realized as a reflection of the fundamental Weyl chamber about the $r_8=0$ axis, see Fig.~\ref{fig:SU3_C}. In other words, charge conjugation invariance of the YM system implies that one can restrict the analysis to restricted backgrounds of the form $r=(r_3,0)$ in the fundamental Weyl chamber. We shall find in Chapter \ref{chap:YM_1loop} that this is indeed what is found from the minimization of the background field effective potential. This will remain true in the presence of quarks with a vanishing chemical potential. However, for a non-vanishing chemical potential, due to the explicit breaking of charge conjugation invariance, we expect the restricted background to develop a non-zero $r_8$ component. We shall see precisely how this happens in Chapters \ref{chap:bfg_quarks} and \ref{chap:quarks}.

\section{Euclidean spacetime symmetries}
To conclude this chapter, let us critically revisit the original assumption of a constant and temporal background. Recall that the reason behind this choice was that such backgrounds, and in turn, the corresponding ${\cal G}_0$-orbits, are invariant both under temporal and spatial translations and under spatial rotations, in line with the homogeneity and isotropy of the Yang-Mills system at finite temperature. 

However, as we also pointed out, there could be other states compatible with homogeneity and isotropy, such that none of the representing backgrounds are invariant under translations or rotations, but only invariant modulo non-trivial true gauge transformations. Using Eq.~(\ref{eq:2}), such backgrounds would be such that
\begin{equation}
\mbox{(P1)} \qquad \forall x,u\in [0,\beta]\times\mathds{R}^3,\,\exists U_u(x)\in {\cal G}_0, \bar A_\mu(x+u)=\bar A^{U_u}_\mu(x)\nonumber
\end{equation}
and
\begin{equation}
\mbox{(P2)} \qquad \forall x\in [0,\beta]\times\mathds{R}^3,\forall R\in SO(3),\,\exists U_R(x)\in {\cal G}_0, \, \left\{\begin{array}{l}
\bar A_0(R^{-1}x)=A^{U_R}_0(x)\\
R_{ij}\bar A_j(R^{-1}x)=\bar A^{U_R}_i(x)
\end{array}\right.\!\!,\nonumber
\end{equation}
where we have introduced the notation $R\,x\equiv (\tau,R\,\vec{x})$. It is trivially verified that any background in the same orbit than an explicitly translation and rotation invariant background obeys these properties. Here, however, we wonder about the existence of backgrounds that obey (P1) and (P2) but whose orbit does not contain any constant temporal representative. 

In App.~\ref{app:hom_iso}, we investigate the existence of such backgrounds, in the SU($2$) case. In addition to the orbits containing constant temporal backgrounds, we find orbits containing {\it space-color locked} backgrounds of the form
\begin{equation}
\bar A_0(x)=0\,, \quad \bar A_i^a(x)=\bar A\delta_i^a\,,
\end{equation}
with $\bar A\neq 0$. These backgrounds obey condition (P1) in a trivial way. They also obey condition (P2) because any spatial rotation can be absorbed back in the form of a color rotation. Moreover, their orbits do not contain any constant and temporal background representative because their field-strength tensor is non-zero, $F_{ij}^a=\bar A^2\varepsilon_{aij}$. 

Even though one should in principle evaluate the background field effective action on these configurations and compare it to the result obtained with constant temporal backgrounds, we expect the space-color locked configurations to play no role in the confining phase since they are not center-invariant. It would be interesting to extend the analysis to the SU(3) case and see if other types of configurations could be relevant in the confining phase.

\clearemptydoublepage
%
%
%
%
\let\textcircled=\pgftextcircled
\chapter{Yang-Mills deconfinement transition at leading order}
\label{chap:YM_1loop}

\initial{A}fter these various introductory chapters on the use of background field techniques at finite temperature,\footnote{Some additional features will be discussed in Chap.~\ref{chap:bfg_quarks} for the case of finite density.} we can at last put all the developed machinery into work and start investigating the confinement/deconfinement transition from a perturbative perspective. We shall do so using the Landau-deWitt gauge (the background extension of the Landau gauge) and its infrared completion in the form of a background extended Curci-Ferrari model. Indeed, given that the original Curci-Ferrari model opens a perturbative window at low energy in the Landau gauge at vanishing temperature, it is legitimate to expect that the corresponding background extension shares the same properties at finite temperature.

In fact, it is the purpose of the rest of the manuscript to review how many features of the QCD phase diagram, usually obtained from non-perturbative methods, can be captured from a simple perturbative expansion within the background extension of the Curci-Ferrari model. In this chapter, we do so at one-loop order, in the case of the pure Yang-Mills theory.  

After introducing some generalities about the Landau-deWitt gauge, we use the properties of the Cartan-Weyl bases introduced earlier to derive a general expression for the one-loop background-field effective potential, valid for any gauge group, and which obeys all the symmetries identified in Chapter \ref{chap:symmetries}. We then use this expression to investigate the confinement-deconfinement transition in the SU($2$) and SU($3$) cases. In particular, we compare our results for the corresponding transition temperatures to both lattice and non-perturbative continuum results. We also investigate the order parameter in various representations and test the so-called Casimir scaling hypothesis. Finally, we discuss some open questions concerning the thermodynamical properties which, in fact, go beyond the mere framework of the Curci-Ferrari model and affect most, if not all, present continuum approaches.

\section{Landau-deWitt gauge}
Since the Landau-deWitt gauge has only been mentioned formally so far, let us derive the associated gauge-fixed action, together with the relevant Curci-Ferrari completion.

\subsection{Faddeev-Popov action}
The Landau-deWitt gauge corresponds to the following choice of gauge-fixing functional
\begin{equation}
 F^a[A]=\bar D_\mu(A_\mu^a-\bar A_\mu^a)\,,
\end{equation} 
with $\bar D_\mu\varphi^a\equiv\partial_\mu\varphi^a+f^{abc}\bar A_\mu^b\varphi^c$. 

Applying the general formula derived in Chap.~\ref{chap:intro}, we find that the Faddeev-Popov gauge-fixing term reads here
\begin{equation}\label{eq:LdW_FP}
 \delta S_{\mbox{\tiny FP}}=\int d^dx\,\Big\{\bar D_\mu\bar c^a(x)\,D_\mu c^a(x)+ih^a(x)\,\bar D_\mu a_\mu^a(x)\Big\}\,,
\end{equation}
with $\smash{a_\mu^a(x)\equiv A_\mu^a(x)-\bar A_\mu^a(x)}$ and where, for convenience, we have used an integration by parts in the ghost term. Of course, all the general properties derived in Chap.~\ref{chap:intro} apply here. In particular, the gauge-fixed action is BRST invariant and both the partition function and the expectation value of any gauge-invariant observable are independent of the choice of the background $\bar A_\mu^a(x)$, seen as an infinite collection of gauge-fixing parameters.

The gauge-fixing term (\ref{eq:LdW_FP}) just looks like the corresponding gauge-fixing term in the Landau gauge upon making the replacement $\partial_\mu\to \bar D_\mu$ and $A_\mu^a\to a_\mu^a$. Moreover, it is easily checked that
\begin{equation}\label{eq:FFbar}
 F_{\mu\nu}^a=\bar F_{\mu\nu}^a+\bar D_\mu a_\nu^a-\bar D_\nu a_\mu^a+f^{abc}a_\mu^ba_\nu^c\,,
\end{equation}
where $\bar F_{\mu\nu}^a$ is the field-strength tensor associated to the background. Therefore, for constant, temporal and diagonal backgrounds such as those considered in this work, the above replacement rule applies also to the original Yang-Mills contribution. This means in particular that the background components can be interpreted as imaginary chemical potentials associated to the (commuting) color charges \cite{Reinosa:2016iml}. Together with the fact that the gauge-fixed action boils down to the one in the Landau gauge in the limit of vanishing background, this ensures the renormalizability of the Landau-deWitt gauge-fixed action and the fact that the background is not renormalized, as any other chemical potential.\footnote{This can be seen as the consequence of the abelian symmetry associated to the color charges.}

\subsection{Curci-Ferrari completion}
Similarly to the Landau gauge, the Landau-deWitt gauge-fixing is hampered by the Gribov ambiguity which means that the Faddeev-Popov action (\ref{eq:LdW_FP}) should be considered as an approximation, valid at best at high energies, and that it should be extended at lower energies. Since the Curci-Ferrari model has proven to be a good candidate for a completion of the gauge-fixing in the case of the Landau gauge, here we complete the Faddeev-Popov action using a background extension of the Curci-Ferrari model.

The two candidates to construct such a background extension are
\begin{equation}
 \int d^dx\,\frac{1}{2}m^2 A_\mu^a A_\mu^a \quad \mbox{or} \quad \int d^dx\,\frac{1}{2}m^2 a_\mu^a a_\mu^a\,.
\end{equation}
However, if we want to maintain the order parameter interpretation of the background, it is crucial that the symmetry (\ref{eq:inv}) is preserved, which leaves us only with the second possibility. The background extension of the Curci-Ferrari model that we should consider in the rest of this manuscript is therefore
\begin{equation}\label{eq:LdW_CF}
 \delta S_{\mbox{\tiny CF}}=\int d^dx\,\Big\{\bar D_\mu\bar c^a(x)\,D_\mu c^a(x)+ih^a(x)\,\bar D_\mu a_\mu^a(x)+\frac{1}{2}m^2 a_\mu^a(x)a_\mu^a(x)\Big\}\,.
\end{equation}
Following the same steps as in Chap.~\ref{chap:intro}, it is easily seen that this action is invariant under a modified BRST symmetry
\begin{equation}
 s_mA_\mu^a=D_\mu c^a, \quad s_m c^a=\frac{1}{2}f^{abc}c^bc^c, \quad s_m\bar c^a=ih^a, \quad s_mih^a=m^2 c^a\,,
\end{equation}
which ensures in particular the renormalizability of the model. Equivalently one can again use the interpretation of the background as a chemical potential.

\subsection{Order parameter interpretation}
As we saw in Chap.~\ref{chap:bg}, the order parameter interpretation of the background relied not only on the symmetry property (\ref{eq:inv}) but also on the convexity of the generating functional $W[J;\bar A]$ and the background independence of the limit $W[0,\bar A]$.

In the case of the Faddeev-Popov action (\ref{eq:LdW_FP}) the partition function is background independent as we have shown in Chap.~\ref{chap:intro}, but the measure is not positive definite, which can seriously jeopardize the convexity of $W[J;\bar A]$. The presence of the Curci-Ferrari mass in (\ref{eq:LdW_CF}) reduces the positivity violation and, a not so large mass could be enough to ensure the convexity of the generating functional $W[J;\bar A]$. However, the presence of the Curci-Ferrari mass violates the background independence of the $W[0,\bar A]$ since we now find
\begin{equation}
 \frac{\delta\ln W[0,\bar A]}{\delta\bar A_\mu^a(x)}=m^2 \big(\bar A_\mu^a(x)-A_{\mbox{\tiny min},\mu}^a[\bar A](x)\big)\,.
\end{equation}
The right-hand side vanishes if the background is self-consistent. Generalizing the arguments given in Chap.~\ref{chap:bg}, this is enough to show that the self-consistent backgrounds are extrema of the background field effective action $\tilde\Gamma[\bar A]$. Thus, even though, in this case it is not clear which extrema represent the self-consistent backgrounds, the order parameter interpretation of the latter survives.

One can try to restore the background independence of the partition function, and thus the identification of the self-consistent backgrounds with the absolute minima of $\tilde\Gamma[\bar A]$, by introducing a background dependence of the Curci-Ferrari mass. It remains to be investigated, however, whether this leads to a consistent system of equations and also whether this is enough to ensure the background independence of other observables. These questions are beyond the scope of the present discussion and, in a first approximation, we should neglect the possible background dependence of the mass. 

Moreover, it should be kept in mind that the actual action beyond the Faddeev-Popov prescription may contain other operators that help restoring the background independence of the partition function. For this reason, in what follows, we will assume that the state of the system is indeed described by the minima of $\tilde\Gamma[\bar A]$. In fact, the previous difficulties of principle are present in most continuum approaches and there is always an implicit assumption made that the correct recipe is that of minimizing $\tilde\Gamma[\bar A]$.\footnote{This is not just a formal discussion. In Chap.~\ref{chap:quarks}, we will see that the non-positivity of the measure due to the presence of quarks at finite density, modifies this recipe substantially.}

\section{Background field effective potential}
\label{sec:1loop}
The evaluation of the background field effective potential $V(r)$ at one-loop order requires only the quadratic part of the gauge-fixed action (\ref{eq:LdW_CF}) with respect to the fields $a_\mu=A_\mu-\bar A_\mu$, $h$, $c$ and $\bar c$.\footnote{More generally, the effective action $\Gamma[\tilde A,\bar A]$ is obtained by expanding the action in the fields $A-\tilde A$, $h$, $c$ and $\bar c$. We shall compute $\Gamma[\tilde A,\bar A]$ to one-loop order in Chapter \ref{chap:correlators}.} It reads
\begin{equation}\label{eq:S0}
 S_0\equiv\int d^dx\,\left\{\frac{1}{2}\big(\bar D_\mu a_\nu;\bar D_\mu a_\nu\big)+\frac{m^2_0}{2}\big(a_\mu;a_\mu\big)+\big(ih;\bar D_\mu a_\mu\big)+\big(\bar D_\mu\bar c;\bar D_\mu c\big)\!\right\},\nonumber
\end{equation}
where, for later purpose, we have rescaled all fields by $g_0$, including the background. In particular, the background covariant derivative reads now $\bar D_\mu\,\_\,\equiv \partial_\mu\,\_\,-g_0\,[\bar A_\mu,\,\_\,]$. We have also used the intrinsic notation introduced in the first chapter, in order to facilitate the change from Cartesian color bases to Cartan-Weyl color bases that we shall consider below.\\

The presence of a preferred color direction, as provided by the background $\beta g_0\bar A_\mu(x)=ir_jH_j\delta_{\mu0}$, renders the color structure a bit more complicated than usual. Indeed, the background covariant derivative in a Cartesian basis writes 
\begin{equation}
 \bar D_\mu^{ab}=\partial_\mu\delta_{ab}+g_0 f^{acb}\bar A_\mu^c\,,
\end{equation} 
which is not diagonal in color space for $\mu=0$. This calls for the introduction of bases that diagonalize the action of the background covariant derivative in color space. Since the background is assumed to lie in the Cartan sub-algebra, these bases are nothing but the Cartan-Weyl bases introduced in Chapter \ref{chap:symmetries}.

\subsection{Notational convention}
In order to fully exploit the properties of Cartan-Weyl bases, we introduce the following notation. Given an element $X$ of the algebra, we write its decomposition into a Cartan-Weyl basis $\{iH_j,iE_\alpha\}$ as $X=i(X_jH_j+X_\alpha E_\alpha)\equiv iX_\kappa t_\kappa$, where the label $\kappa$ can take two types of values, $\kappa=0^{(j)}$ or $\kappa=\alpha$, with $t_{0^{(j)}}\equiv H_j$ and $t_\alpha\equiv E_\alpha$. The label $0^{(j)}$ is referred to as a {\it zero} and should be taken literally as corresponding to the nul vector in the same space as the one containing the roots $\alpha$ (the dual background space in the terminology of the previous chapter). There is of course only one such nul vector. However, there are as many color labels associated to zero as there are elements in the Cartan sub-algebra, hence the label $j$ used to denote the various zeros.

Given these notational conventions, the definition of a Cartan-Weyl basis can be written compactly as
\begin{equation}
 [t_j,t_\kappa]=\kappa_j t_\kappa\,,
\end{equation}
where, of course, $0^{(k)}_j=0$, no matter the values of $j$ and $k$. The vectors $\kappa$ appear therefore as the weights of the adjoint representation. Now, since the background is assumed to lie in the Cartan sub-algebra, the background covariant derivative acting on $X$ writes
\begin{eqnarray}
 \bar D_\mu X & \!\!\!\!\!=\!\!\!\!\! & \partial_\mu X-g_0[\bar A_\mu,X]\\
& \!\!\!\!\!=\!\!\!\!\! & i\partial_\mu X_\kappa t_\kappa+T\,r_j X_\kappa [t_j,t_\kappa]\,\delta_{\mu0}\nonumber\\
& \!\!\!\!\!=\!\!\!\!\! & i\partial_\mu X_\kappa t_\kappa+T\,r_j \kappa_j\,X_\kappa t_\kappa\,\delta_{\mu0}=i\bar D_\mu^\kappa X_\kappa t_\kappa\,,\nonumber
\end{eqnarray}
with $\bar D_\mu^\kappa\equiv\partial_\mu-iT\,r\cdot\kappa\,\delta_{\mu0}$ and $r\cdot\kappa\equiv r_j\,\kappa_j$.

As announced, the background covariant derivative is diagonal in color space when expressed in a Cartan-Weyl basis. This simplifies considerably the evaluation of the background field effective potential. Similarly, in order to evaluate the background-dependent Polyakov loop, it will be convenient to diagonalize the defining action of the $H_j$ as $H_j|\rho\rangle=\rho_j|\rho\rangle$, with $\rho$ the weights of the defining representation.

\subsection{General one-loop expression}
Coming back to the quadratic action (\ref{eq:S0}) and choosing the Cartan-Weyl basis such that $(it_\kappa;it_\lambda)=\delta_{\kappa,-\lambda}$, see App.~\ref{app:sun}, we find
\begin{equation}
 S_0\equiv \int d^dx\,\left\{\frac{1}{2}\bar D_\mu^{-\kappa} a_\nu^{-\kappa}\bar D_\mu^\kappa a_\nu^\kappa+\frac{m^2_0}{2} a_\mu^{-\kappa}a_\mu^\kappa+ih^{-\kappa}\bar D_\mu^\kappa a_\mu^\kappa+\bar D_\mu^{-\kappa}\bar c^{-\kappa}\bar D_\mu^\kappa c^\kappa\!\right\}.
\end{equation}
In Fourier space, using that $\varphi^{-\kappa}(x)=\varphi^\kappa(x)^*$ for bosonic fields, see App.~\ref{app:sun}, this reads
\begin{equation}\label{eq:SF}
 S_0\equiv \int_Q^T\,\left\{\frac{1}{2}a_\mu^{\kappa}(Q)^*(Q_\kappa^2+m^2_0)\,a_\mu^\kappa(Q)+h^{\kappa}(Q)^*Q_\mu^\kappa a_\mu^\kappa(Q)+\bar c^{-\kappa}(-Q)\,Q_\kappa^2 c^\kappa(Q)\!\right\},
\end{equation}
where $Q_\kappa\equiv Q+T\,r\cdot\kappa\,n$ is referred to as the {\it generalized momentum}, with $Q=(\omega_n,q)$, $\omega_n=2\pi nT$ a bosonic Matsubara frequency, and $n\equiv (1,\vec{0})$. We have also introduced the notation
\begin{equation}
 \int_Q^Tf(Q)\equiv T\sum_{n\in\mathds{Z}} \int\frac{d^{d-1}q}{(2\pi)^{d-1}}f(\omega_n,q)
\end{equation}
for the Matsubara sum-integrals.\\

From the quadratic action (\ref{eq:SF}), a standard Gaussian integration, leads to the one-loop potential
\begin{equation}\label{eq:V_1loop}
 V_{\mbox{\scriptsize 1loop}}(r;T)=\sum_\kappa\left[\frac{d-1}{2} \int_Q^T \ln\,\big(Q^2_\kappa+m^2\big)-\frac{1}{2}\int_Q^T \ln\,Q^2_\kappa\right],
\end{equation}
where we have set the bare mass $m_0$ equal to the renormalized mass $m$, since we are computing to one-loop accuracy and the mass enters already a one-loop integral. For later purpose, it is important to emphasize that the massive contribution originates from the transverse gluonic modes whereas the massless contribution results from an over-cancellation between the one massless longitudinal mode and two massless ghost modes. It is also worth noting that the zero temperature limit of the above expression at fixed $\hat r\equiv T r$ does not depend on $\hat r$. Indeed, in this limit, the Matsubara sum-integral becomes a $d$-dimensional integral and the frequency shift $\hat r\cdot\kappa$ hidden in $Q_\kappa$ can be absorbed using a simple change of variables in the frequency integral.\footnote{As we show in the next section, this property generalizes in fact to any loop order.} If we disregard this background-independent, zero-temperature limit, we can set $d=4$ and use an integration by parts (with respect to the variable $q$) to find
\begin{equation}
 V_{\mbox{\scriptsize 1loop}}(r;T)=\sum_\kappa \left[-\int_Q^T \frac{q^2}{Q_\kappa^2+m^2_0}+\frac{1}{3}\int_Q^T \frac{q^2}{Q^2_\kappa}\right],
\end{equation}
which involves convergent Matsubara sums unlike expression (\ref{eq:V_1loop}).\\

Using  standard contour integration techniques, see for instance \cite{Reinosa:2014zta}, we arrive at (again, we disregard the zero-temperature limit)
\begin{equation}\label{eq:pot_n}
 V_{\mbox{\scriptsize 1loop}}(r;T)=\frac{1}{2\pi^2}\sum_\kappa\left[-\int_0^\infty dq\,\frac{q^4}{\varepsilon_q}\,\mbox{Re}\,n_{\varepsilon_q-i\hat r\cdot\kappa}+\,\frac{1}{3}\int_0^\infty dq\,q^3\,\mbox{Re}\,n_{q-i\hat r\cdot\kappa}\right],
\end{equation}
with $n(x)\equiv 1/(e^{\beta x}-1)$ the Bose-Einstein distribution function and $\varepsilon_q\equiv\sqrt{q^2+m^2}$. Using the integration by parts backwards, this can also be rewritten as
\begin{equation}\label{eq:V222}
 V_{\mbox{\scriptsize 1loop}}(r;T)=\frac{T}{2\pi^2}\sum_\kappa\int_0^\infty dq\,q^2\,\mbox{Re}\,\ln\frac{\big(1-e^{-\beta\varepsilon_q+ir\cdot\kappa}\big)^3}{1-e^{-\beta q+ir\cdot\kappa}}\,.
\end{equation}
Moreover, given that the root diagram is invariant under $\alpha\to -\alpha$, we can remove the explicit real parts in the previous expressions. Alternatively, we can evaluate these real parts to find
\begin{eqnarray}
V_{\mbox{\scriptsize 1loop}}(r;T) & \!\!\!\!\!=\!\!\!\!\! & \frac{1}{2\pi^2}\sum_\kappa\left[-\int_0^\infty dq\,\frac{q^4}{\varepsilon_q}\,\frac{e^{\beta\varepsilon_q}\cos(r\cdot\kappa)-1}{e^{2\beta\varepsilon_q}-2e^{\beta\varepsilon_q}\cos(r\cdot\kappa)+1}\right.\\
& & \hspace{1.5cm}\left.+\,\frac{1}{3}\int_0^\infty dq\,q^3\,\,\frac{e^{\beta q}\cos(r\cdot\kappa)-1}{e^{2\beta q}-2e^{\beta q}\cos(r\cdot\kappa)+1}\right],\nonumber
\end{eqnarray}
or, equivalently,
\begin{equation}\label{eq:V3}
V_{\mbox{\scriptsize 1loop}}(r;T)=\frac{T}{4\pi^2}\sum_\kappa\int_0^\infty dq\,q^2\,\ln\,\frac{\big(e^{-2\beta\varepsilon_q}-2e^{-\beta\varepsilon_q}\cos(r\cdot\kappa)+1\big)^3}{e^{-2\beta q}-2e^{-\beta q}\cos(r\cdot\kappa)+1}\,.
\end{equation}

\subsection{Checking the symmetries}
As a crosscheck, let us verify that the above formulas fulfil all the symmetries that we identified in Chapter \ref{chap:symmetries}.\\

First of all, the potential should be invariant under winding transformations which appear in the restricted background space as translations along any of the vectors $\bar\alpha^{(k)}$ that generate the lattice dual to the one generated by the roots:
\begin{equation}\label{eq:V_winding}
 V\big(r+\bar\alpha^{(k)}\big)=V(r)\,,\,\,\forall\bar\alpha^{(k)}\,.
\end{equation}
This identity is trivially fulfilled by Eq.~(\ref{eq:V_1loop}) because, under a winding transformation, the generalized momentum $Q_\kappa$ becomes $Q_\kappa+T\bar\alpha^{(k)}\cdot\kappa$ and the shift $T\bar\alpha^{(k)}\cdot\kappa\in 2\pi T\mathds{Z}$ can be absorbed using a change of variables in the Matsubara sum. Similarly, $e^{ir\cdot\kappa}$ and $\cos(r\cdot\kappa)$, which appear for instance in Eqs.~(\ref{eq:V222}) and (\ref{eq:V3}) respectively, remain invariant under winding transformations.

The potential should also be invariant under Weyl transformations which appear as reflections with respect to hyperplanes orthogonal to the roots:
\begin{equation}\label{eq:V_Weyl}
 V\left(r-2\frac{r\cdot\alpha}{\alpha^2}\alpha\right)=V(r)\,,\,\,\forall\alpha\,.
\end{equation}
This identity is indeed fulfilled by Eq.~(\ref{eq:V_1loop}) because, under a Weyl transformation, the momentum $Q_\kappa$ becomes $Q+T\left(r\cdot\kappa-2\frac{r\cdot\alpha \alpha\cdot\kappa}{\alpha^2}\right)=Q_{\kappa-2\frac{\kappa\cdot\alpha}{\alpha^2}\alpha}$ and this change of $\kappa$ can be absorbed using a redefinition of the sum over $\kappa$ for it is generally true that the root diagram is invariant under Weyl transfomations. The same remark applies to $e^{ir\cdot\kappa}$ and $\cos(r\cdot\kappa)$.

Finally, the potential should be invariant under charge conjugation which appears as the reflection about the origin of the restricted background space:
\begin{equation}\label{eq:V_C}
 V(-r)=V(r)\,.
\end{equation}
This identity is fulfilled by Eq.~(\ref{eq:V_1loop}) because, under charge conjugation, the momentum $Q_\kappa$ is changed to $Q_{-\kappa}$ and this change of $\kappa$ can again be absorbed using a redefinition of the sum over $\kappa$ since the root diagram is invariant under $\alpha\to -\alpha$. The same remark applies to $e^{;\cdot\kappa}$ and $\cos(r\cdot\kappa)$.\\

As a consequence of the symmetry identity (\ref{eq:V_winding}), we can show that the background-independence of the zero-temperature limit of the potential, that we observed at one-loop, generalizes in fact to any loop order. To see this, apply this identity $n$ times along the direction $\bar\alpha^{(k)}$ and take the zero-temperature limit in the form $T=u/n$, with $u\in\mathds{R}$. Writing the potential as a function of $\hat r$ and $T$ (rather than $r$ and $T$), we find
\begin{equation}
 V(\hat r+u\bar\alpha^{(k)};T\to 0)=V(\hat r;T\to 0)\,.
\end{equation}
Now, since the vectors $\bar\alpha^{(k)}$ form a basis of the restricted background space, we have
\begin{equation}
 V(\hat r+\delta\hat r;T\to 0)=V(\hat r;T\to 0)\,,\,\,\forall\delta\hat r\,,
\end{equation}
as announced.

Another consequence of the symmetry identity (\ref{eq:V_winding}) is that the analysis of the background field effective potential can be restricted to the fundamental parallelepiped defined by the vectors $\bar\alpha^{(j)}$. In fact, it is convenient to decompose the restricted background as $r=x_j\bar\alpha^{(j)}$, with $2\pi\,x_j=r\cdot\alpha^{(j)}$, and to express the potential as a function of the $x_i\in [0,1]$. We shall use this decomposition in the various plots to be shown in what follows.

\section{SU($2$) and SU($3$) gauge groups}
We now use the formulas derived in the previous section to investigate the deconfinement transition in the SU($2$) and SU($3$) cases.

\begin{figure}[t]
\begin{center}
\includegraphics[height=0.21\textheight]{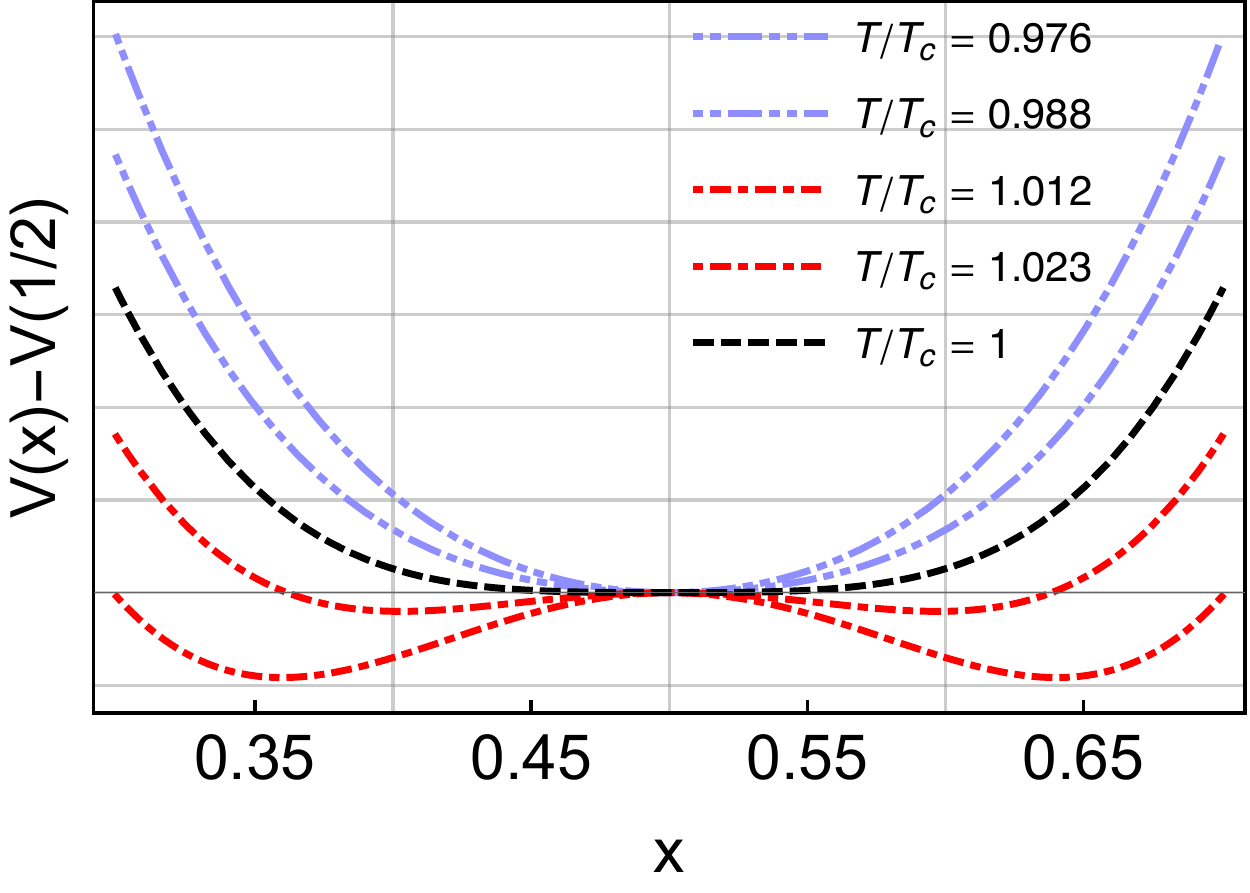}\hspace{1.0cm}\includegraphics[height=0.21\textheight]{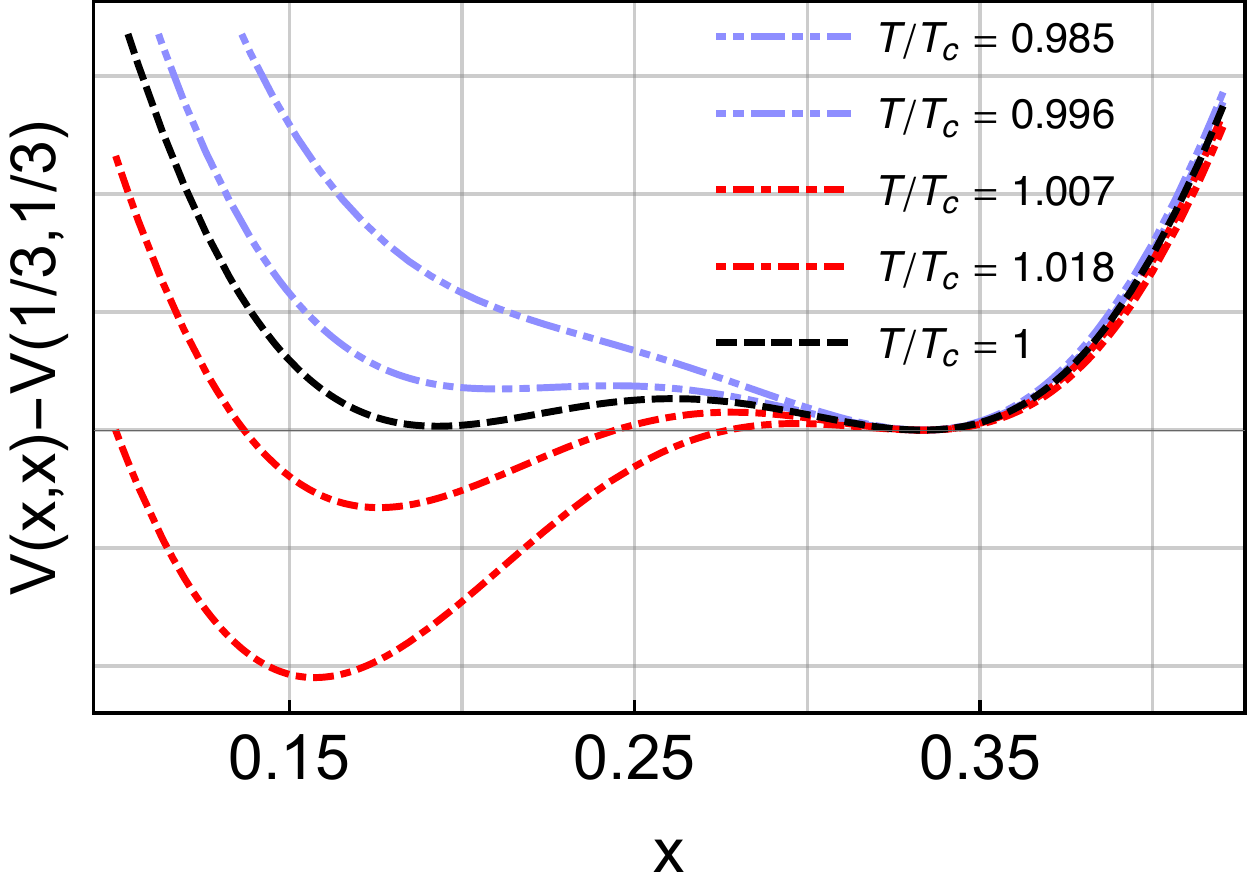}
\mycaption[SU($2$) transition]{Left: SU($2$) background field effective potential in the fundamental parallelepiped $r=x\bar\alpha^{(1)}$, with $\bar\alpha_1=2\pi$ and $0<x<1$, which is also the fundamental Weyl chamber. The center-symmetric or confining point is located at $x=1/2$. Right: SU($3$) background field effective potential along the line $x_1=x_2\equiv x$ of the fundamental parallelepiped. This fundamental parallelepiped contains two Weyl chamber and we show the potential in the fundamental Weyl chamber only. The center-symmetric or confining point is located at $x=1/3$. In both cases, we have subtracted the value of the potential at the confining point, for more readability.}
\label{fig:pot_1loop_su2}
\end{center}
\end{figure}

\subsection{Deconfinement transition}
In the SU($2$) case, we can work over the fundamental Weyl chamber $r\in [0,2\pi]$ whose only center-symmetric point, or confining state, is at $r=\pi$. Using that $\kappa\in\{-1,0,+1\}$ and writing only the background dependent contributions to the potential, we find
\begin{equation}
 V_{\mbox{\scriptsize 1loop}}^{\mbox{\scriptsize SU(2)}}(r;T)=\frac{T}{2\pi^2}\int_0^\infty dq\,q^2\,\ln\,\frac{\big(e^{-2\beta\varepsilon_q}-2e^{-\beta\varepsilon_q}\cos(r)+1\big)^3}{e^{-2\beta q}-2e^{-\beta q}\cos(r)+1}\,.
\end{equation}
This potential is plotted in Fig.~\ref{fig:pot_1loop_su2} (left) for various temperatures. As the temperature is increased, the minimum of the potential moves continuously from the confining, center-symmetric state at $r=\pi$ to a pair of degenerate states $(r,2\pi-r)$ connected by center symmetry. The transition is then of the second order type, in agreement with the results of lattice simulations \cite{Svetitsky:1985ye,Kaczmarek:2002mc,Lucini:2005vg,Greensite:2012dy,Smith:2013msa} or continuum non-perturbative approaches \cite{Marhauser:2008fz,Braun:2007bx,Braun:2010cy,Fister:2013bh,Epple:2006hv,Alkofer:2006gz,Fischer:2009wc,Fischer:2009gk,Reinhardt:2012qe,Reinhardt:2013iia,Fukushima:2012qa,Quandt:2016ykm}.

The transition temperature $T_c$ can be determined in terms of $m$ by requiring that the curvature of the potential vanishes for $r=\pi$. This condition reads
\begin{equation}
0=\frac{T}{\pi^2}\int_0^\infty dq\,q^2\,\left[\frac{e^{\beta q}}{(e^{\beta q}+1)^2}-\frac{3e^{\beta\varepsilon_q}}{(e^{\beta\varepsilon_q}+1)^2}\right],
\end{equation}
from which we find $T_{\mbox{\scriptsize c,1loop}}^{\mbox{\scriptsize SU($2$)}}\simeq 0.336m$.\\

\begin{figure}[t]
\begin{center}
\hspace{0.4cm}\includegraphics[height=0.2\textheight]{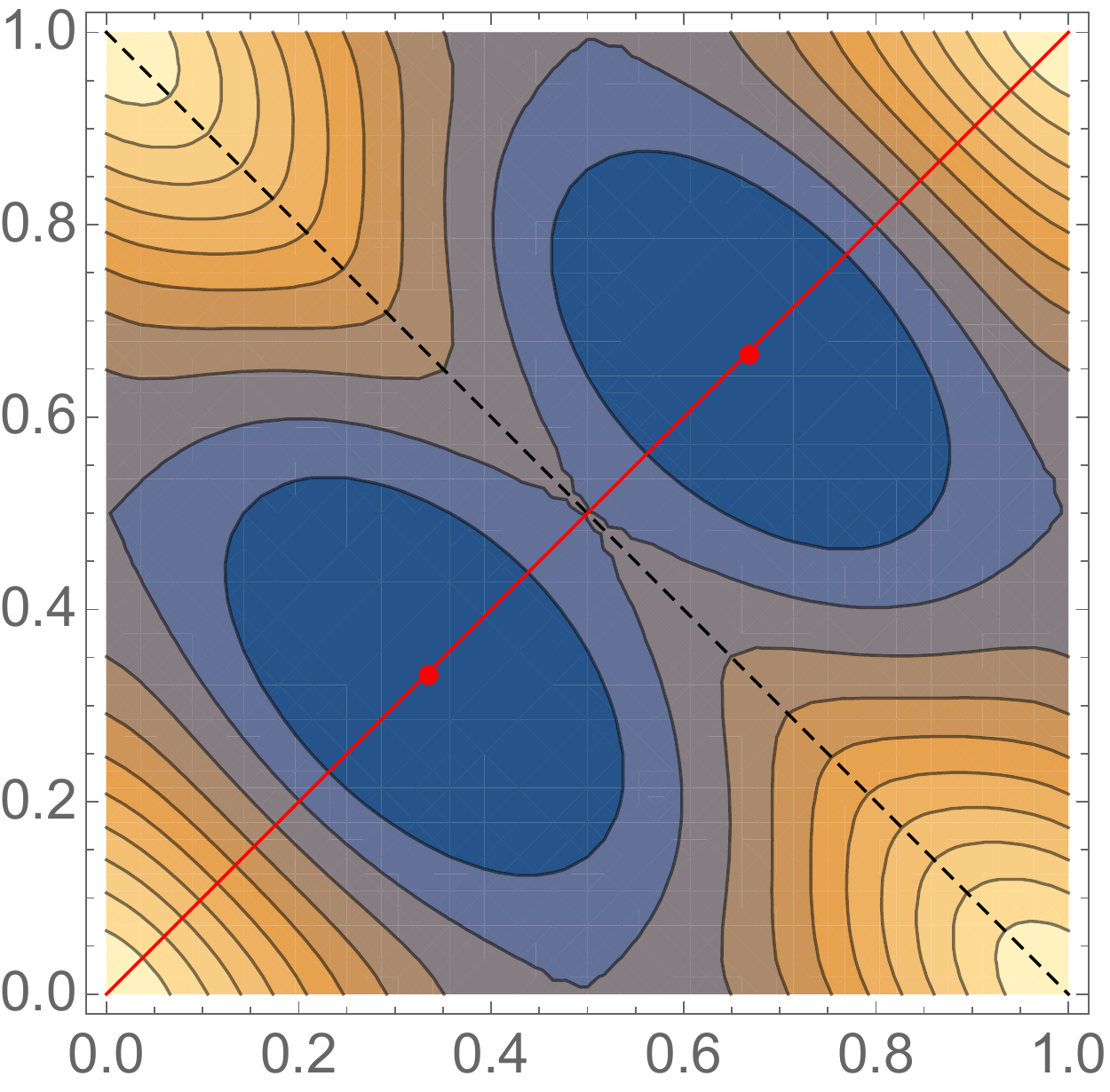}\hspace{0.5cm}\includegraphics[height=0.2\textheight]{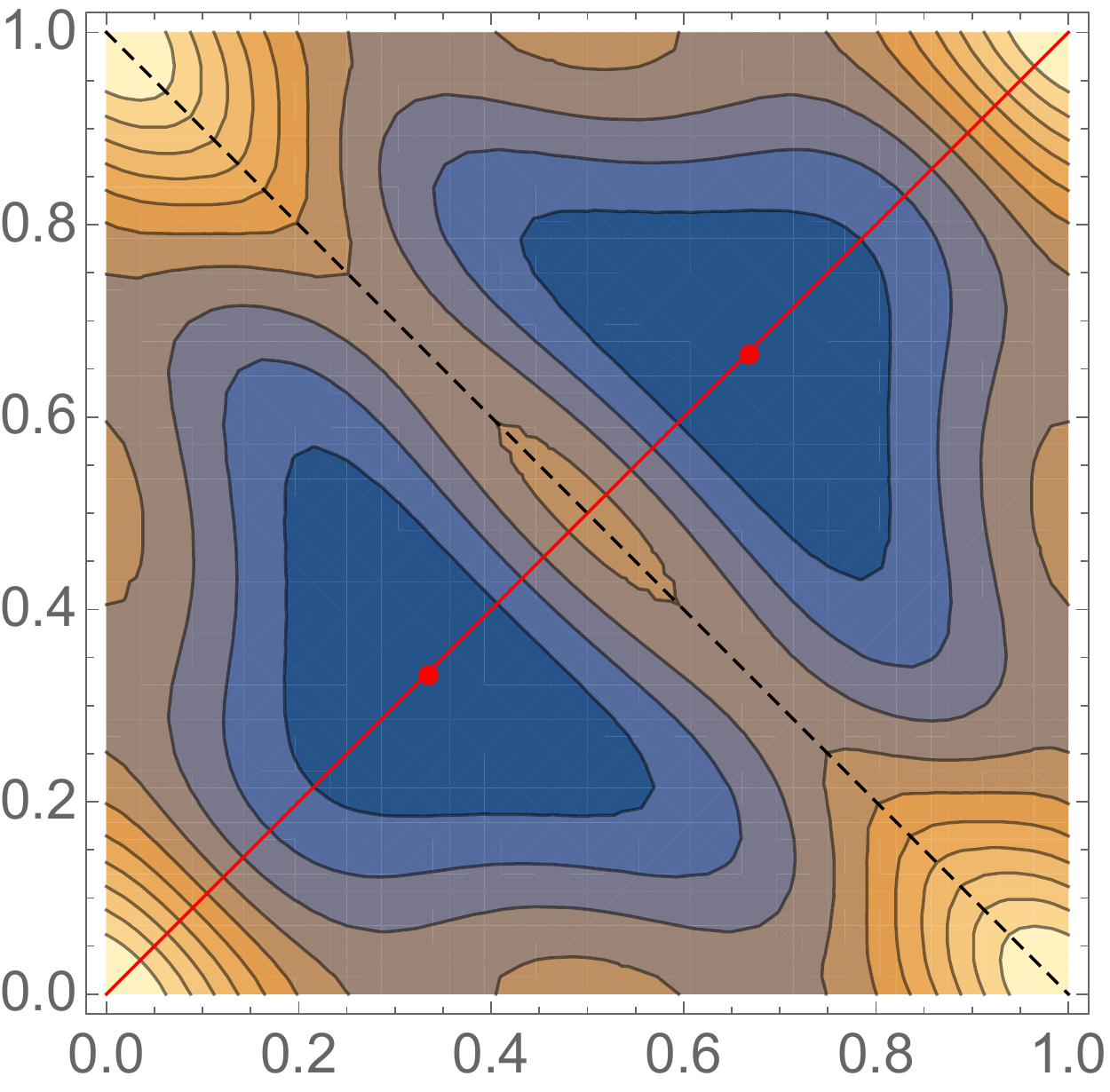}\hspace{0.5cm}\includegraphics[height=0.2\textheight]{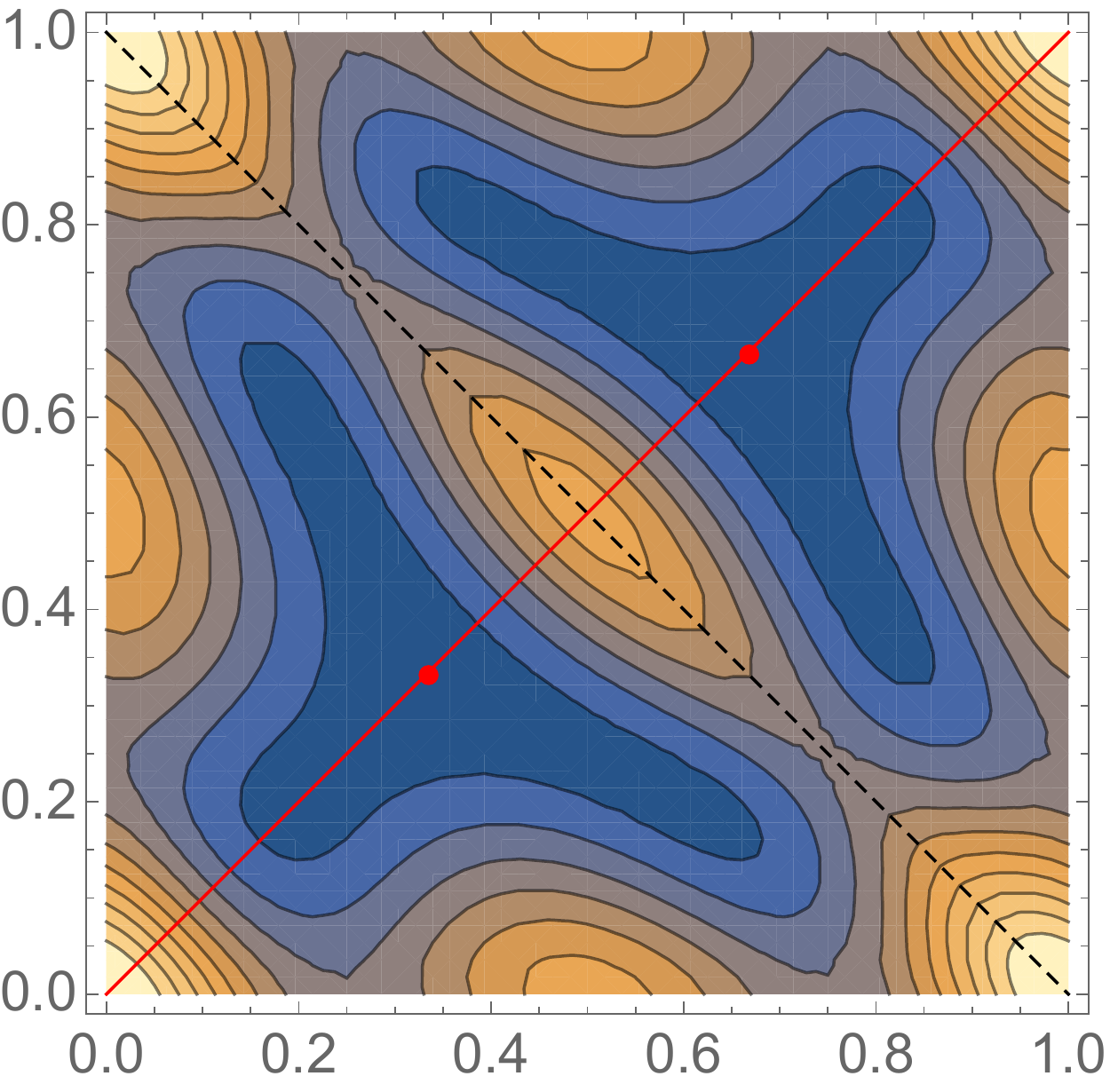}\\
\vspace{0.7cm}
\hspace{0.4cm}\includegraphics[height=0.2\textheight]{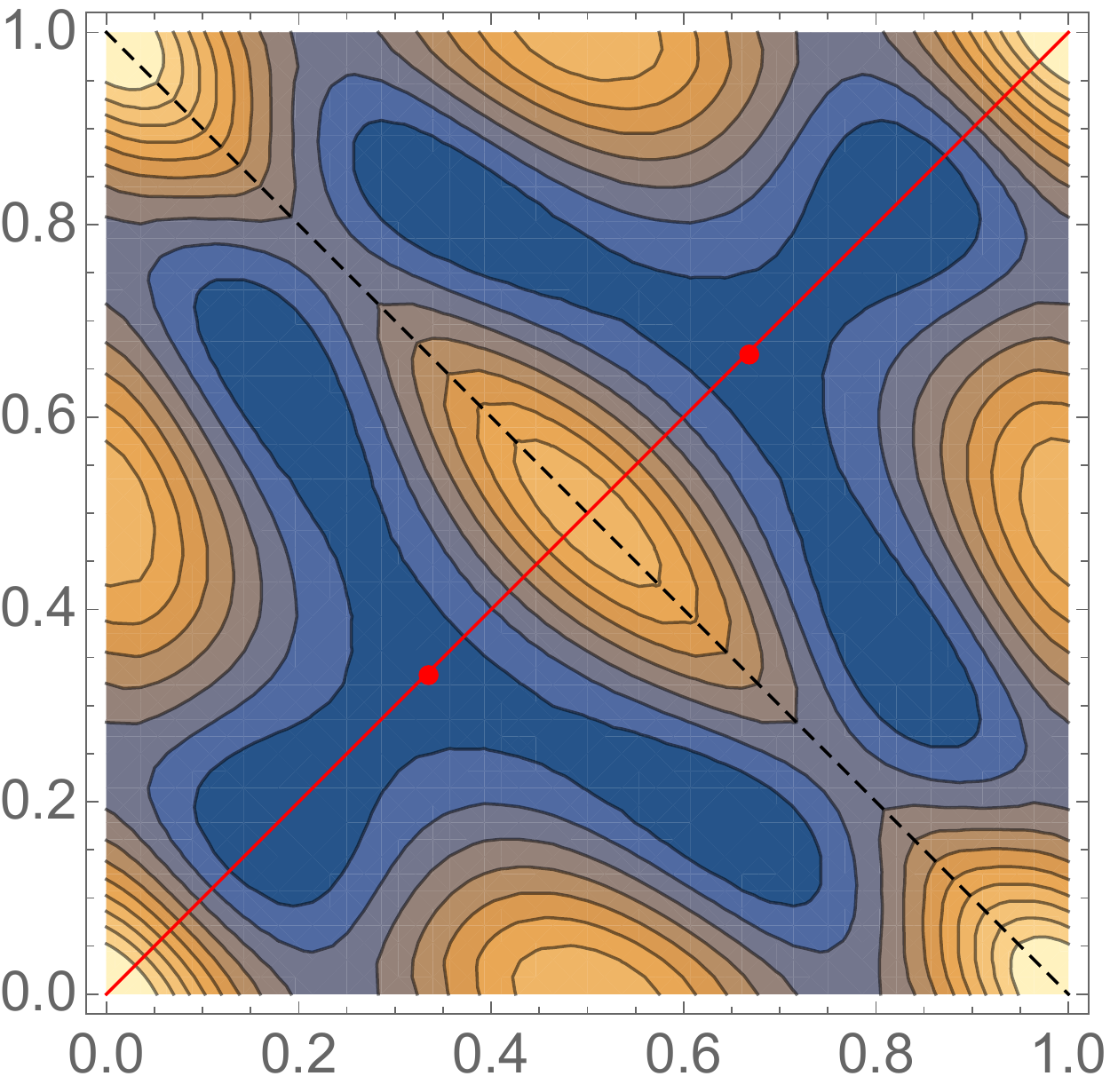}\hspace{0.5cm}\includegraphics[height=0.2\textheight]{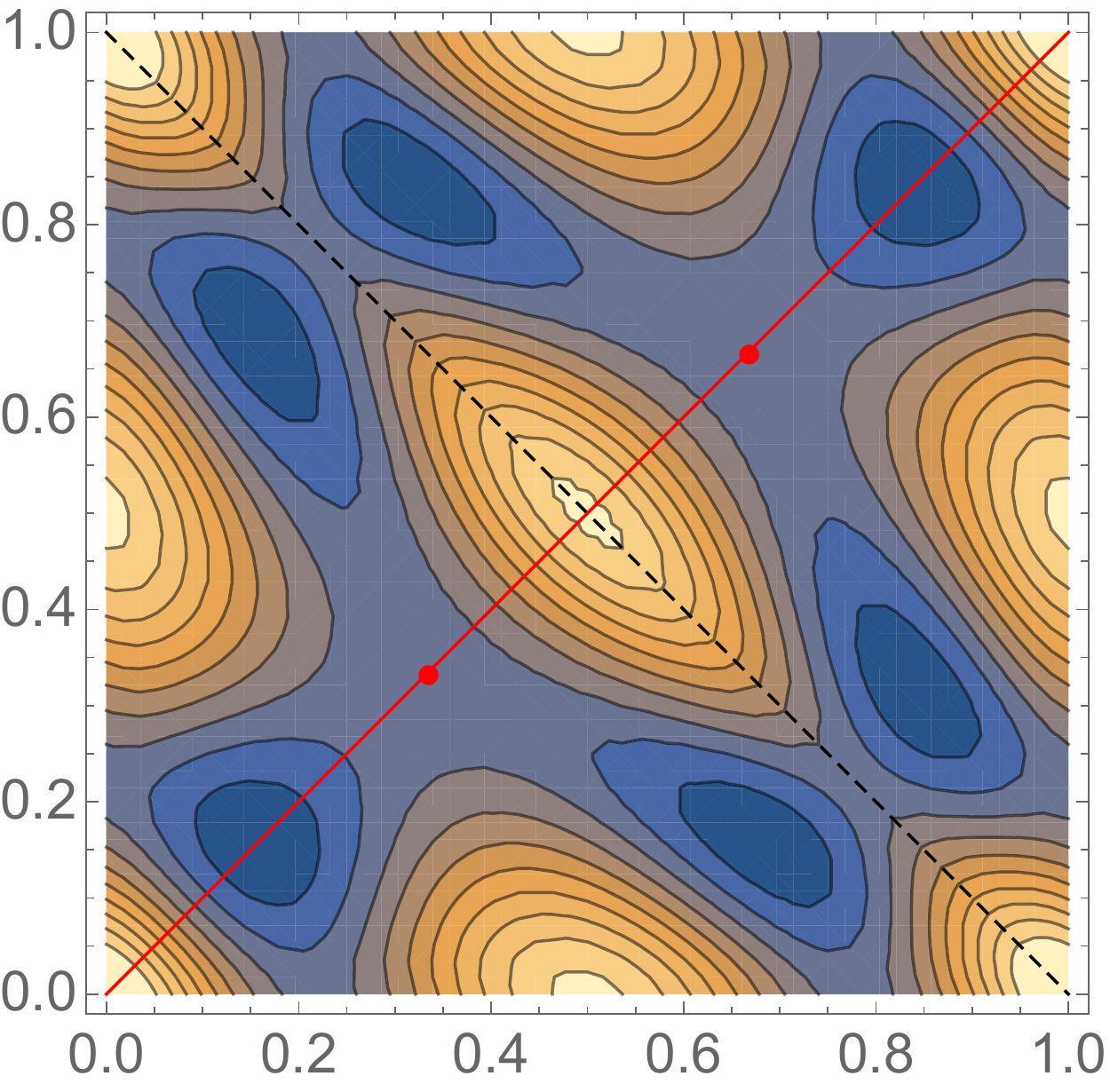}\hspace{0.5cm}\includegraphics[height=0.2\textheight]{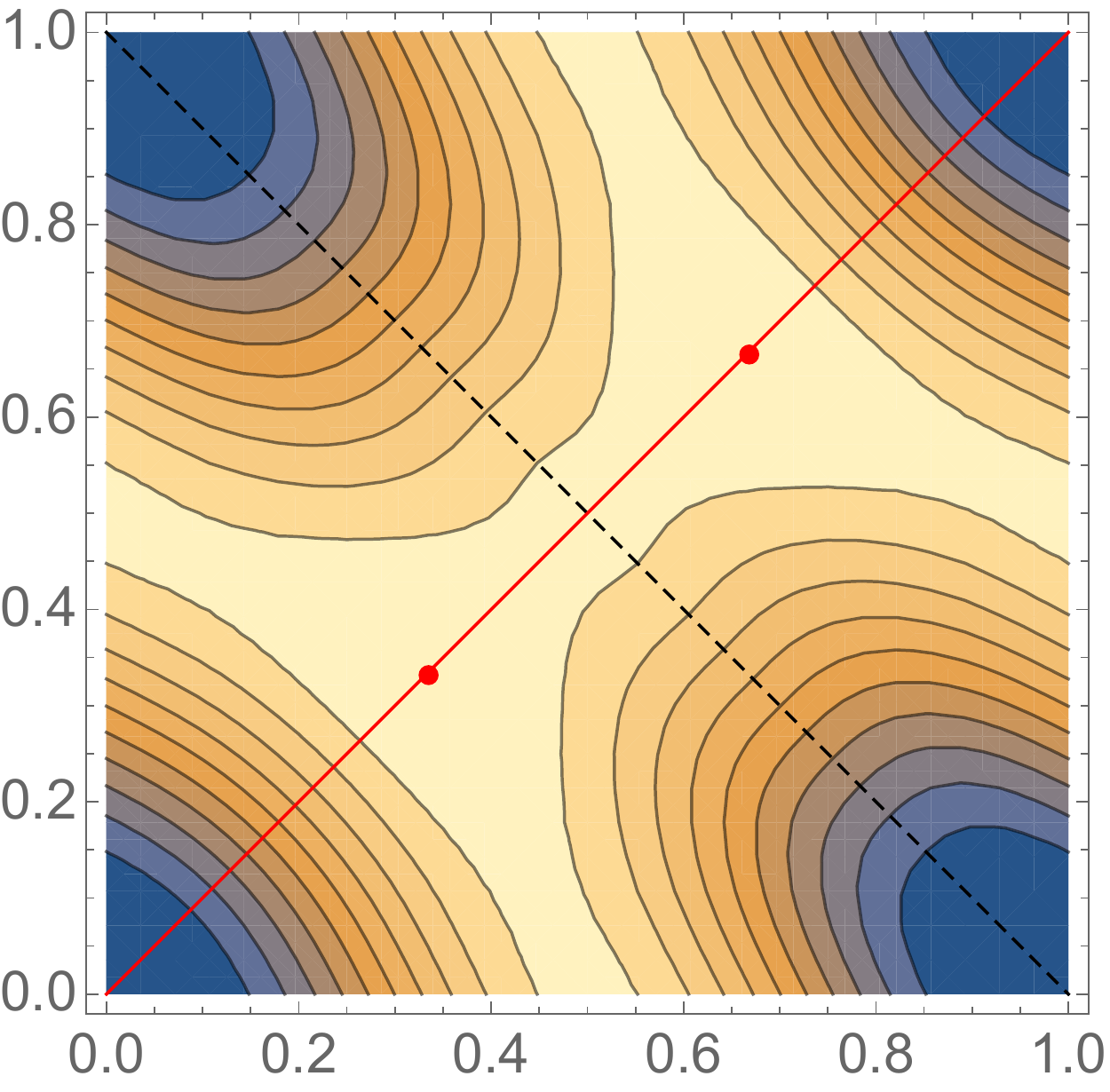}
\mycaption[SU($3$) transition]{SU($3$) background field effective potential in the fundamental parallelepiped $r=x_1\bar\alpha^{(1)}+x_2\bar\alpha^{(2)}$, with $\bar\alpha_1=2\pi\,(1,-1/\sqrt{3})$, $\bar\alpha_2=2\pi\,(1,1/\sqrt{3})$ and $0<x_i<1$. The fundamental parallelepiped is made of two Weyl chambers (separated by the dashed line) which contain each one center-symmetric or confining point, located $(1/3,1/3)$ and $(2/3,2/3)$ respectively (red dots). Finally, the charge conjugation invariant states are indicated by a red line.}
\label{fig:pot_1loop_su3}
\end{center}
\end{figure}

The corresponding one-loop expression for the potential in the SU($3$) case can be obtained by observing that each pair of roots is associated to a SU($2$) subgroup. It follows that (again we keep only the background dependent contributions)
\begin{equation}
V_{\mbox{\scriptsize 1loop}}^{\mbox{\scriptsize SU(3)}}(r;T)=V_{\mbox{\scriptsize 1loop}}^{\mbox{\scriptsize SU(2)}}(r_3;T)+V_{\mbox{\scriptsize 1loop}}^{\mbox{\scriptsize SU(2)}}\left(\frac{r_3-r_8\sqrt{3}}{2};T\right)+V_{\mbox{\scriptsize 1loop}}^{\mbox{\scriptsize SU(2)}}\left(\frac{r_3+r_8\sqrt{3}}{2};T\right)\,.
\end{equation}
In Fig.~\ref{fig:pot_1loop_su3}, we show contour graphs of the potential in the fundamental Weyl chamber for various temperatures. We observe that, as the temperature is increased, the minimum of the potential jumps eventually from the confining, center-symmetric state at $r=(4\pi/3,0)$ to a triplet of degenerate states connected by center symmetry. We find that one of the absolute minima lies always along the $r_8=0$ axis. As we have explained in the previous chapter, this is a consequence of charge conjugation invariance.\footnote{The presence of two other directions along which the minima move is in accordance with center symmetry and just means that, as long as center symmetry is not explicitly broken, there are three equivalent ways to define charge conjugation symmetry.} For us, this means that we can restrict the analysis to the $r_8=0$ axis, along which the potential reads more simply
\begin{equation}
 V_{\mbox{\scriptsize 1loop}}^{\mbox{\scriptsize SU(3)}}(r;T)=V_{\mbox{\scriptsize 1loop}}^{\mbox{\scriptsize SU(2)}}(r_3;T)+2V_{\mbox{\scriptsize 1loop}}^{\mbox{\scriptsize SU(2)}}\left(\frac{r_3}{2};T\right)\,.
\end{equation}
 In Fig.~\ref{fig:pot_1loop_su2}, we show the potential along this axis, in the fundamental Weyl chamber, for various temperatures, with a clear signal of a first order type transition, in agreement with other approaches.

From center-symmetry, it follows that the confining point is always an extremum, so it is possible to locate the highest spinodal by requiring the curvature of the potential to vanish at this point. To this purpose, we notice first that
\begin{equation}
\left.\frac{\partial^2}{\partial r^2}V_{\mbox{\scriptsize 1loop}}^{\mbox{\scriptsize SU(2)}}(r;T)\right|_{r=4\pi/3}=\left.\frac{\partial^2}{\partial r^2}V_{\mbox{\scriptsize 1loop}}^{\mbox{\scriptsize SU(2)}}(r;T)\right|_{r=2\pi/3}\,.
\end{equation}
Using this identity, the equation for a vanishing curvature at the confining point is found to be
\begin{equation}
0=\frac{T}{2\pi^2}\int_0^\infty dq\,q^2\,\left[\frac{e^{-3\beta q}+4e^{-2\beta q}+e^{-\beta q}}{(e^{-2\beta q}+e^{-\beta q}+1)^2}-3\frac{e^{-3\beta\varepsilon_q}+4e^{-2\beta\varepsilon_q}+e^{-\beta\varepsilon_q}}{(e^{-2\beta\varepsilon_q}+e^{-\beta\varepsilon_q}+1)^2}\right],
\end{equation}
from which we deduce that $T_{\mbox{\scriptsize s,1loop}}^{\mbox{\scriptsize SU($3$)}}\simeq 0.382m$, while the transition temperature is found to be $T_{\mbox{\scriptsize c,1loop}}^{\mbox{\scriptsize SU($3$)}}\simeq 0.364m$.\\

Finally, given that the mass scale $m$ can be fixed by fitting lattice Landau-gauge propagators with the corresponding one-loop expressions obtained within the Curci-Ferrari model,\footnote{We fix the parameters of the model at zero temperature, in which case the background $\hat r=rT$ vanishes and the Landau-deWitt gauge becomes the Landau gauge.} we can estimate the values for the transition temperatures. Table \ref{tab:1loop} summarizes our results and compares them to other approaches. The obtained values for $T_c$ are already quite good given the simplicity of the one-loop approximation that we have considered. In the next chapter, we shall see that the two-loop corrections considerably improve these values.

\begin{table}[h]
$$\begin{tabular}{|c||c||c||c||c|}
\hline
$T_c$ (MeV) & lattice \cite{Lucini:2012gg} & fRG \cite{Fister:2013bh} & Variational \cite{Quandt:2016ykm} & CF model \cite{Reinosa:2014ooa}\\
\hline\hline
SU(2) & 295 & 230 & 239 & 238\\
SU(3) & 270 & 275 & 245 &185\\
\hline
\end{tabular}$$
\caption{Transition temperatures for the SU($2$) and SU($3$) deconfinement transitions as computed from various approaches.}\label{tab:1loop}
\end{table}

\subsection{Inversion of the Weiss potential}
The previous results can be understood in a simple way by comparing the high and low temperature forms of the background field effective potential.

For high temperatures ($T\gg m$), the mass can be neglected and the gluonic contributions combine with the massless contributions to yield
\begin{equation}
 V_{\mbox{\scriptsize 1loop}}(r;T\gg m)\simeq T^4 v_{\mbox{\scriptsize Weiss}}(r)\,,
\end{equation}
where
\begin{eqnarray}
v_{\mbox{\scriptsize Weiss}}(r) &\!\!\!\!\!\equiv\!\!\!\!\! & \frac{1}{2\pi^2}\sum_\kappa\int_0^\infty dq\,q^2\,\ln\,\big(e^{-2q}-2e^{-q}\cos(r\cdot\kappa)+1\big)\,,\\
& \!\!\!\!\!=\!\!\!\!\! & -\frac{1}{3\pi^2}\sum_\kappa\int_0^\infty dq\,q^3\,\frac{e^q\cos(r\cdot\kappa)-1}{e^{2q}-2e^q\cos(r\cdot\kappa)+1}
\end{eqnarray}
is known as the {\it Weiss potential} \cite{Weiss:1980rj,Gross:1980br}. This potential rewrites
\begin{equation}
v_{\mbox{\scriptsize Weiss}}(r)=-\frac{1}{3\pi^2}\sum_\kappa P_4(\underline{r\cdot\kappa})\,,
\end{equation}
where
\begin{equation}
P_4(r)\equiv-\frac{(\pi-r)^4}{8}+\frac{\pi^2(\pi-r)^2}{4}-\frac{7\pi^4}{120}\,,
\end{equation}
is related to the Bernouilli polynomial of degree $4$ and the notation $\underline{r\cdot\kappa}$ stands for the remainder of the Euclidean division of $r\cdot\kappa$ by $2\pi$.

In contrast, for low temperatures ($T\ll m$), the gluonic contributions can be neglected and one obtains again the Weiss potential, but inverted with respect to the high temperature regime:
\begin{equation}
 V_{\mbox{\scriptsize 1loop}}(r;T\ll m)\simeq -\frac{T^4}{2} v_{\mbox{\scriptsize Weiss}}(r)\,.
\end{equation}
This inversion of the Weiss potential (due to the presence of the scale $m$) triggers the deconfinement transition if the center-symmetric state of the system corresponds to the absolute maximum of the Weiss potential. It is a simple exercice to check that this is so in the SU($2$) and SU($3$) cases.\\

The inversion mechanism is present in other continuum approaches and was in fact originally proposed in \cite{Braun:2007bx}. Let us emphasize that it relies on the dominance of ghost degrees of freedom at low temperatures, which obviously poses some conceptual questions such as how this temperature regime could be {\it a priori} dominated by unphysical degrees of freedom, while maintaining basic principles of thermodynamical consistency. We shall analyse this question below and try to answer it, at least partially.

\subsection{Polyakov loops}
We can also evaluate the background-dependent Polyakov loop $\ell(r)$ which becomes the physical order parameter when evaluated at the minimum of $V(r)$.

Let us recall that $\ell(r)$ is nothing but the expectation value of $\Phi_A(\vec{x})$ in the presence of a source $J[\bar A]$ such that $\langle A\rangle_{\bar A}=\bar A$. Therefore, to evaluate $\ell(r)$, we expand $\Phi_A(\vec{x})$ in powers of $a_\mu(x)=A_\mu(x)-\bar A_\mu(x)$ and substitute any one-point function $\langle a_\mu(x)\rangle_{\bar A}$ by $0$ in the evaluation of the expectation value. At leading order, we write $\Phi_A(\vec{x})\simeq \Phi_{\bar A}(\vec{x})=\mbox{tr}\,e^{i\beta g\bar A}/N$, where we have used that the background is constant to remove the time-ordering. We find
\begin{eqnarray}
 \ell(r) & \!\!\!\!\!=\!\!\!\!\! & \frac{1}{N}\mbox{tr}\,e^{i r_j H_j}=\frac{1}{N}\sum_\rho e^{ir\cdot\rho}\,,\\
 \bar\ell(r) & \!\!\!\!\!=\!\!\!\!\! & \frac{1}{N}\mbox{tr}\,e^{-i r_j H_j}=\frac{1}{N}\sum_\rho e^{-ir\cdot\rho}\,.
\end{eqnarray}
where $\rho$ are the weights of the defining representation, and $-\rho$ those of the corresponding contragredient representation.

These formulas obey the expected symmetries. Under a winding transformation $r\to r+\bar\alpha^{(k)}$ (representing a center transformation, with center element $e^{i\bar\alpha^{(k)}\cdot\rho}$), the phase $e^{i\rho\cdot r}$ becomes $e^{i\rho\cdot r}e^{i\rho\cdot\bar\alpha^{(k)}}$ with $e^{i\rho\cdot\bar\alpha^{(k)}}$ not depending on the chosen weight $\rho$, as we have seen in the Chapter \ref{chap:symmetries}. Therefore
\begin{equation}\label{eq:l_winding}
 \ell\big(r+\bar\alpha^{(k)}\big)=e^{i\bar\alpha^{(k)}\cdot\rho}\ell(r)\,,\,\,\forall\bar\alpha^{(k)}\,,
\end{equation}
as it should for a winding transformation. Under a Weyl transformation, $e^{ir\cdot\rho}$ is transformed into $e^{ir\cdot\rho}e^{-2i\frac{r\cdot\alpha \alpha\cdot\rho}{\alpha^2}}=e^{ir\cdot\left(\rho-\frac{\alpha\cdot\rho}{\alpha^2}\alpha\right)}$ and the change of $\rho$ can be absorbed in a redefinition of the sum over $\rho$ for it is generally true that the weight diagram is globally invariant under Weyl transformations. It follows that
\begin{equation}\label{eq:l_Weyl}
 \ell\left(r-2\frac{r\cdot\alpha}{\alpha^2}\alpha\right)=\ell(r)\,,
\end{equation}
as it should for a Weyl transformations. Finally, we observe that $\ell(-r)=\bar\ell(r)$, in line with charge conjugation invariance, see Chapter \ref{chap:symmetries}. As we also saw there, we have $\ell(-r)=\ell^*(r)$ and, therefore, $\bar\ell(r)=\ell^*(r)$.\\

In the SU($2$) case, we find
\begin{equation}
 \ell(r)=\bar\ell(r)=\cos(r/2)\,,
\end{equation}
whereas in the SU($3$) case,
\begin{eqnarray}
 \ell(r) & \!\!\!\!\!=\!\!\!\!\! &  \frac{e^{-i\frac{r_8}{\sqrt{3}}}+2e^{i\frac{r_8}{2\sqrt{3}}}\cos(r_3/2)}{3}\,,\\
 \bar\ell(r) & \!\!\!\!\!=\!\!\!\!\! &  \frac{e^{i\frac{r_8}{\sqrt{3}}}+2e^{-i\frac{r_8}{2\sqrt{3}}}\cos(r_3/2)}{3}\,.
\end{eqnarray}
We mention that for the SU($2$) gauge group, the constraint from charge conjugation invariance can also be seen as arising from Weyl symmetry, and therefore, as we mentioned already, there is no real constraint from charge conjugation invariance in this case. The reason is that there exists a Weyl transformation $W=i\sigma_2$ such that $(i\sigma_2)^\dagger A_0(i\sigma_2)=A_0^{\cal C}$. From this, it is easily deduced that $\ell(-r)=\bar\ell(r)$. Since Weyl transformations also imply $\ell(-r)=\ell(r)$, it follows that $\ell(r)=\bar\ell(r)$. This last property is in general not true in the SU($3$) case, except for a charge conjugation invariant state $r=(r_3,0)$ for which
\begin{equation}
 \ell(r)=\bar\ell(r)=\frac{1+2\cos(r_3/2)}{3}\,.
\end{equation}
That $\ell$ and $\bar\ell$ become equal and real in this case follows from hermiticity and charge conjugation invariance, as we discuss more generally in Chap.~\ref{chap:bfg_quarks}.

\begin{figure}[t]
\begin{center}
\hspace{0.75cm}\includegraphics[height=0.32\textheight]{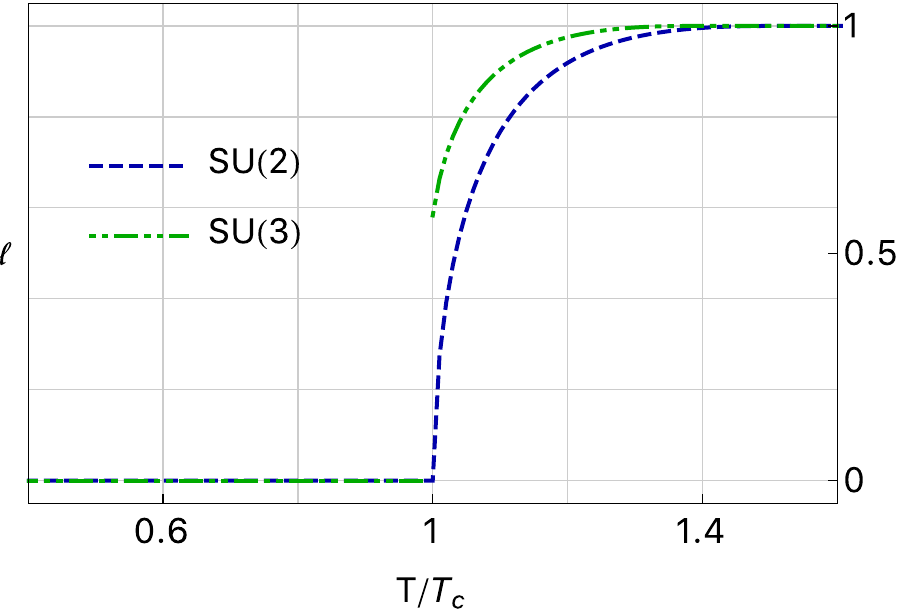}
\mycaption[Polyakov loops]{Polyakov loops for the SU($2$) and SU($3$) defining representations, as functions of $T/T_c$.}
\label{fig:polys_1loop}
\end{center}
\end{figure}

The physical Polyakov loops, as obtained from the one-loop potential derived above, are shown in Fig.~\ref{fig:polys_1loop}. They display of course the same behavior at the transition that the one observed by following the minimum of the potential. We mention that both Polyakov loops saturate to $1$ at a temperature $T_\star\simeq 1.45m$, corresponding to the background reaching $0$ at this temperature. This occurs with a discontinuity in the second derivative of the order parameter which leads to artificial singularities in the thermodynamical observables. A similar feature is observed in other approaches, see for instance \cite{Quandt:2016ykm}. We will see in the next chapter that these artefacts disappear at two-loop order.

\section{Thermodynamics}
\label{sec:thermo}

In principle, the free-energy density of the system is obtained as minus the background field effective potential evaluated at its minimum $r(T)$:
\begin{equation}
 f(T)=V(r(T);T)\,.
\end{equation}
From this, all other thermodynamical observables can be derived. In particular the pressure and the entropy density are given respectively by
\begin{eqnarray}
p(T) & \!\!\!\!\!=\!\!\!\!\! & -f(T)=-V(r(T);T)\,,\label{eq:pressure}\\
s(T) & \!\!\!\!\!=\!\!\!\!\! & p'(T)=-\left.\frac{\partial V(r;T)}{\partial T}\right|_{r=r(T)}\,.
\end{eqnarray}
The energy density can then be computed as
\begin{equation}
e(T)=f(T)+Ts(T)=T^2(p/T)'\,,
\end{equation}
and the so-called interaction measure is
\begin{equation}
i(T)=\frac{e(T)-3p(T)}{T^4}=T(p/T^4)'\,.
\end{equation}

\subsection{High and low temperature behavior}
In the high temperature limit, all modes contributing to the potential (\ref{eq:V_1loop}) become massless and the resulting potential counts effectively two degrees of freedom per color mode. This is the result of the usual cancellation between the two unphysical components of the gauge field $A_\mu$ and the two ghost degrees of freedom which, due to their Grasmannian nature, contribute to the potential with an extra minus sign.

In the low temperature phase, the situation is in general more delicate and still open to debate \cite{Reinosa:2014zta,Reinosa:2015gxn,Quandt:2017poi}. In the present approach, since three of the four components of the gauge field become massive while the ghosts remain massless, the potential is essentially dominated by ghost degrees of freedom. While this {\it ghost domination} is not specific to the present approach and is in fact a key ingredient in the Weiss potential inversion mechanism described above, one may fear the presence of inconsistencies in the behavior of the thermodynamical observables, such as a negative (thermal) pressure or a negative entropy density. As we now explain, this unphysical behavior is not present.

The point is that, although the low temperature phase is indeed dominated by the one uncanceled ghost degree of freedom, the latter is surrounded by a background $\bar A_0$. Instead of cancelling each other, these two unphysical degrees of freedom are subtlety combined in such a way that the thermodynamics remains consistent at low temperatures. More precisely, the confining value of the background $\bar A_0$ operates a transmutation of the thermal distribution function of the ghost field from a negative distribution to a positive distribution.

Let us first illustrate this transmutation mechanism in the SU($2$) case where it is particularly compelling. From Eqs.~(\ref{eq:pressure}) and (\ref{eq:pot_n}), the contribution to the pressure at low temperatures writes
\begin{equation}
 p(T)=\frac{1}{6\pi^2}\int_0^\infty dq\,q^3\,\big[-n_q-n_{q-i\hat r(T)}-n_{q+i\hat r(T)}\big]\,.
\end{equation}
The bracket contains the three color mode contributions, corresponding to $\kappa\in\{-1,0,+1\}$. The neutral mode $\kappa=0$ is blind to the background and contributes negatively to the thermal pressure, as we would expect from a ghost degree of freedom. However, in the presence of a confining background $\hat r(T)=\pi\,T$, the charge modes $\kappa=\pm 1$ contribute with the transmuted distribution functions 
\begin{equation}
 -n_{q-i\hat r(T)}=-\frac{1}{e^{\beta q-i\pi}-1}=\frac{1}{e^{\beta q}+1}\equiv f_q>0\,,
\end{equation}
and thus behave effectively as true fermions! Finally, using $f_q=n_q-2n_{2q}$ and a simple change of variables, on obtains the total contribution to the thermal pressure as
\begin{equation}
 p(T)=\frac{3}{24\pi^2}\int_0^\infty dq\,q^3\,n_q\,,
\end{equation}
which is positive. Since $s(T)/T^3\sim 4p(T)/T^4$, the same conclusion is reached for the entropy density.

We can proceed identically in the SU($3$) case. The thermal pressure reads
\begin{equation}
p(T)=\frac{1}{6\pi^2}\int_0^\infty dq\,q^3\,\big[-2n_q-2\mbox{Re}\,n_{q-i\hat r(T)}-4\mbox{Re}\,n_{q-i\hat r(T)/2}\big]\,.
\end{equation}
In the confining phase $r(T)=4\pi/3$ and
\begin{equation}
-\mbox{Re}\,n_{q-i\hat r(T)}=-\mbox{Re}\,n_{q-i\hat r(T)/2}=\frac{e^{\beta q}/2+1}{e^{2\beta q}+e^{\beta q}+1}>0\,.
\end{equation}
To check that the total contribution to the pressure is positive, we can use the identity\footnote{This is simply derived by writing
$$\sum_{k=0}^{N-1}n_{q-i\frac{2\pi}{N}k}=\sum_{k=0}^{N-1}\sum_{p=1}^\infty e^{\beta q p}e^{-i\frac{2\pi}{N}kp}=\sum_{p=1}^\infty e^{\beta q p}\sum_{k=0}^{N-1}e^{-i\frac{2\pi}{N}kp}\,,$$
where we have inverted the order of the summations. Now, the inner sum over $k$ vanishes unless $p$ is a multiple of $N$, in which case it equals $N$. Therefore, the sum becomes $N\sum_{p=1}^\infty e^{N\beta q p}=Nn_{Nq}$, as announced. This result is in fact more general. Given a Matsubara sum $S(T)\equiv T\sum_{n\in\mathds{Z}} f(i\omega_n)$, we may need to evaluate the related double sum $$\sum_{k=0}^{N-1} T\sum_{n\in\mathds{Z}} f\left(i\omega_n+iT\frac{2\pi}{N}k\right)=N\frac{T}{N}\sum_{k=0}^{N-1}\sum_{n\in\mathds{Z}} f\left(i2\pi\frac{T}{N}(Nn+k)\right).$$ With the exception of the factor of $N$, we can interprete the double sum in the RHS as a single Matsubara sum at a reduced temperature $T/N$. Therefore, the result of the sum in the LHS is $NS(T/N)$.}
\begin{equation}
\sum_{k=0}^{N-1}n_{q-i\frac{2\pi}{N}k}=Nn_{Nq}\,,
\end{equation}
After a simple change of variables, we find
\begin{equation}
 p(T)=\frac{8}{54\pi^2}\int_0^\infty dq\,q^3\,n_q\,.
\end{equation}
We conclude therefore that, despite the ghost dominance at low temperatures, the thermodynamical properties seem consistent.

\subsection{Vicinity of the transition}
The transmutation mechanism identified in the previous section could lead to inconsistencies in the vicinity of the transition. Indeed, as one approaches the transition temperature from below, the gluonic degrees of freedom start playing a role, while the confining background remains confining. Then, because the gluonic distribution functions are positive in the absence of background, those orthogonal to the background transmute into negative distribution functions in the confining phase and can potentially turn the entropy density or the thermal pressure negative. A calculation reveals that a very tiny violation occurs at leading order over a narrow range of temperatures below $T_c$, see \cite{Reinosa:2014zta,Reinosa:2015gxn}. In the next chapter, we investigate whether higher loop corrections can cure this behavior.\\

In the SU(3) case, the first order transition is characterized by a discontinuity of the entropy density, associated to a latent heat $\Delta\epsilon/T_c^4=\Delta s/T_c^3$. At one-loop order in the Curci-Ferrari approach, we find $0.41$ that is only one third of the lattice value \cite{Beinlich:1996xg}. It remains to be investigated whether this result improves when including higher order corrections.

\clearemptydoublepage
%
%
%
%
\let\textcircled=\pgftextcircled
\chapter{YM deconfinement transition at next-to-leading order}
\label{chap:YM_2loop}

\initial{W}e have seen in the previous chapter that the background extension of the Curci-Ferrari model captures the physics of the confinement/deconfinement transition in Yang-Mills theories already at leading order. In particular, we have obtained rather good estimates for the SU($2$) and SU($3$) transition temperatures, given the simplicity of the considered approximation.

In this chapter, we investigate the next-to-leading order corrections to the background field effective potential and to the background-dependent Polyakov loop. Not only are these corrections necessary to assess the convergence properties of the approach but they are also needed to try to cure some of the problematic features identified at leading order.

We shall see that the two loop corrections help improving the values for the transition temperatures and cure some of the problematic features. There will remain some open questions, however. As we argue, these question are not restricted to the mere Curci-Ferrari approach but pose in fact challenges to most, if not all, present continuum approaches.

\section{Feynman rules and color conservation}
\label{sec:feynman_rules}

In order to evaluate the two-loop corrections to the background field effective potential, we first need to determine the Feynman rules in the Landau-deWitt gauge. Due to our previous experience with the one-loop calculation, we suspect these rules to become simpler when expressed in a Cartan-Weyl basis. We will check below that this is indeed the case. But let us first see how this result can be anticipated using the symmetries of the gauge-fixed action.

\subsection{Color conservation}
In fact, there is still one invariance that we have not yet exploited, namely the invariance under background gauge transformations
\begin{equation}
S_{\mbox{\scriptsize CF}}[A^U,h^U,c^U,\bar c^U;\bar A^U]=S_{\mbox{\scriptsize CF}}[A,h,c,\bar c;\bar A]\,,
\end{equation} 
in the case where the background is left invariant, $\bar A^U=\bar A$. Indeed, since only the dynamical fields are transformed in this case, this invariance implies certain conservation rules that should appear explicitly at the level of the action provided the fields are decomposed into the eigenmodes of the corresponding Noether charges.

In the case where the background is constant, temporal and diagonal, the transformations that leave the background invariant are color rotations of the form $e^{i\theta_j H_j}$, with $\theta_j$ constant. The corresponding conserved charges on the Lie algebra are therefore the operators $\mbox{ad}_{H_j}$, whose eigenstates are nothing but the elements $t_\kappa$ of a Cartan-Weyl basis, with associated charges combined into the vector $\kappa$.

Using (\ref{eq:FFbar}) and (\ref{eq:LdW_CF}) and the Killing form introduced in Chapter 2, the background extension of the Curci-Ferrari model writes
\begin{eqnarray}
 S_{\mbox{\scriptsize CF}} & \!\!\!\!\!=\!\!\!\!\! & \int d^dx\,\left\{\frac{1}{2}\big(\bar D_\mu a_\nu;\bar D_\mu a_\nu\big)+\frac{m^2_0}{2}\big(a_\mu;a_\mu\big)+\big(ih;\bar D_\mu a_\mu\big)+\big(\bar D_\mu\bar c;\bar D_\mu c\big)\!\right.\nonumber\\
& & \hspace{1.5cm}\left.-\,\left(\bar D_\mu\bar c;[a_\mu,c]\right)-\left(\bar D_\mu a_\nu;[a_\mu,a_\nu]\right)+\frac{1}{4}\left([a_\mu,a_\nu];[a_\mu,a_\nu]\right)\right\},
\end{eqnarray}
and decompose each field into a Cartan-Weyl basis, we find
\begin{eqnarray}
 S_{\mbox{\scriptsize CF}} & \!\!\!\!\!=\!\!\!\!\! &  \int d^dx\,\left\{\frac{1}{2}\bar D_\mu^{-\kappa} a_\nu^{-\kappa}\bar D_\mu^\kappa a_\nu^\kappa+\frac{m^2_0}{2} a_\mu^{-\kappa}a_\mu^\kappa+ih^{-\kappa}\bar D_\mu^\kappa a_\mu^\kappa+\bar D_\mu^{-\kappa}\bar c^{-\kappa}\bar D_\mu^\kappa c^\kappa\!\right.\nonumber\\
& & \hspace{0.5cm}\left.-\,if_{\kappa\lambda\tau}\big(\bar D^\kappa_\mu\bar c^\kappa\big)\,a^\lambda_\mu c^\tau-if_{\kappa\lambda\tau}\big(\bar D^\kappa_\mu a^\kappa_\nu\big)\,a^\lambda_\mu a^\tau_\nu-\frac{1}{4}f_{\kappa\lambda\xi}f_{(-\xi)\tau\sigma}\,a^\kappa_\mu a^\lambda_\nu a^\tau_\mu a^\sigma_\nu\right\},\label{eq:S_kappa}
\end{eqnarray}
where we have used $(t_\kappa;t_\lambda)=-\delta_{\kappa,-\lambda}$ and we have introduced the notation $(t_\kappa;[t_\lambda,t_\tau])\equiv -f_{\kappa\lambda\tau}$. This tensor is totally antisymmetric, owing to the antisymmetry of the bracket and the cyclicity of the Killing form. In fact, it plays the role of the structure constant tensor in the Cartan-Weyl basis since, as it is easily shown, $[t_\lambda,t_\tau]=f_{\kappa\lambda\tau}t_{-\kappa}$. This has been used in particular to write the four gluon interaction.

We may now ask how is the conservation of color manifest at the level of the action (\ref{eq:S_kappa}). To answer this question, let us first notice that $f_{\kappa\lambda\tau}$ vanishes if two or three labels are zeros since $[t_{0^{(j)}},t_{0^{(k)}}]=0$. Moreover, since $[t_{0^{{j}}},t_\alpha]=\alpha_jt_\alpha$, $f_{0^{(j)\alpha(-\alpha)}}=\alpha_j$. Finally, let us evaluate how a given $t_\alpha$ acts on $t_\beta$ in the adjoint representation. Since we know already how the $t_{0^{(j)}}$ act, we write
\begin{equation}\label{eq:74}
 \mbox{ad}_{t_{0^{(j)}}}\mbox{ad}_{t_\alpha}t_\beta=\mbox{ad}_{t_\alpha}\mbox{ad}_{t_{0^{(j)}}}t_\beta+\mbox{ad}_{[t_{0^{(j)}},t_\alpha]} t_\beta\,,
\end{equation} 
where we have used that $[\mbox{ad}_X,\mbox{ad}_Y]=\mbox{ad}_{[X,Y]}$. This rewrites
\begin{equation}\label{eq:75}
 \mbox{ad}_{t_{0^{(j)}}}\mbox{ad}_{t_\alpha}t_\beta=(\alpha+\beta)_j\mbox{ad}_{t_\alpha}t_\beta\,.
\end{equation}
So either $\mbox{ad}_{t_\alpha}t_\beta=[t_\alpha,t_\beta]=0$ or it has to be collinear to $t_{\alpha+\beta}$ (in which case $\alpha+\beta$ has to be a root). In all cases, we notice that $f_{\kappa\lambda\tau}=0$ if $\kappa+\lambda+\tau\neq 0$, which ensures color conservation at the level of (\ref{eq:S_kappa}). 

\begin{figure}[h]
\begin{center}
\includegraphics[height=0.16\textheight]{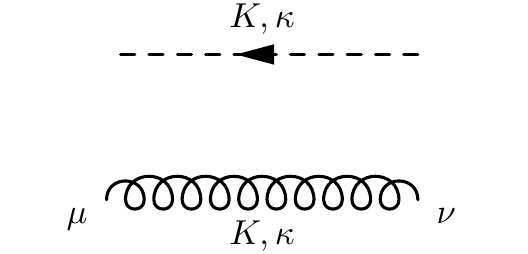}\\
\mycaption[Representation of the propagators]{Diagrammatic representation of the propagators in the Landau-deWitt gauge.}
\label{fig:props}
\end{center}
\end{figure}

Let us mention finally that upon complex conjugation in the complexified algebra, see Appendix \ref{app:sun}, and using that it is possible to choose the $t_\kappa$ such that $\bar t_\kappa=-t_{-\kappa}$ (where the bar denotes the conjugation in the complexified algebra) we find $[t_\lambda,t_\tau]=-f^*_{(-\kappa)(-\lambda)(-\tau)}t_{-\kappa}$ which implies 
\begin{equation}\label{eq:f_complex}
f_{\kappa\lambda\tau}=-f^*_{(-\kappa)(-\lambda)(-\tau)}\,.
\end{equation}

\subsection{Feynman rules}
The action (\ref{eq:S_kappa}) is pretty similar to the action in the Landau gauge, up to the missing conventional factor of $i$ in the structure constants and the presence of background covariant derivatives that depend on the color charge of the field they act upon. In Fourier space, these covariant derivatives become generalized momenta, the color charge of the field entering as a shift $Tr\cdot\kappa$ of the frequencies, similar to an imaginary chemical potential for the color charge. This means that the Feynman rules for the Landau-deWitt gauge in a Cartan-Weyl basis are pretty similar to those for the Landau gauge in a conventional Cartesian basis. For the gluon and ghost propagators, we find respectively
\begin{equation}
 P^\perp_{\mu\nu}(Q_\kappa)\,G_{m_0}(Q_\kappa)
\end{equation}
and
\begin{equation}
 G_0(Q_\kappa)\,,
\end{equation}
where we have introduced
\begin{equation}
 G_{m_0}(Q)\equiv\frac{1}{Q^2+m^2_0}\,,
\end{equation}
as well as the transverse projector $P^\perp_{\mu\nu}(Q)\equiv\delta_{\mu\nu}-Q_\mu Q_\nu/Q^2$. These propagator are represented diagrammatically in Fig.~\ref{fig:props}. Each line carries both a momentum $Q$ and a color charge $\kappa$. We note the property $Q_\kappa=-(-Q)_{-\kappa}$ which implies $G_{m_0}(Q_\kappa)=G_{m_0}((-Q)_{-\kappa})$ and $P^\perp_{\mu\nu}(Q_\kappa)=P^\perp_{\mu\nu}((-Q)_{-\kappa})$ and thus that the common orientation of momentum and charge can be chosen arbitrarily.

\begin{figure}[t]
\begin{center}
\includegraphics[height=0.24\textheight]{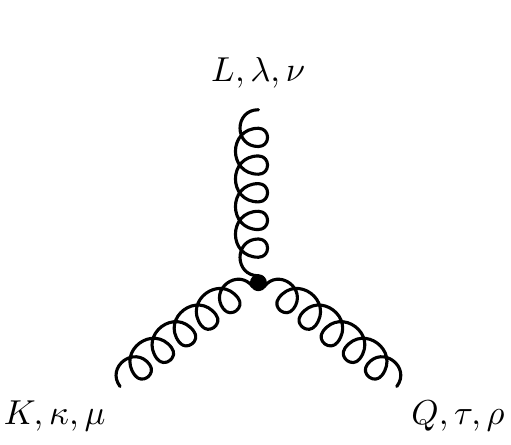}\hspace{0.4cm}\includegraphics[height=0.24\textheight]{./fig07/vertex-3g.pdf}\\
\mycaption[Representation of the derivative vertices.]{Diagrammatic representation of the derivative vertices.}
\label{fig:der}
\end{center}
\end{figure}

\begin{figure}[t]
\begin{center}
\includegraphics[height=0.24\textheight]{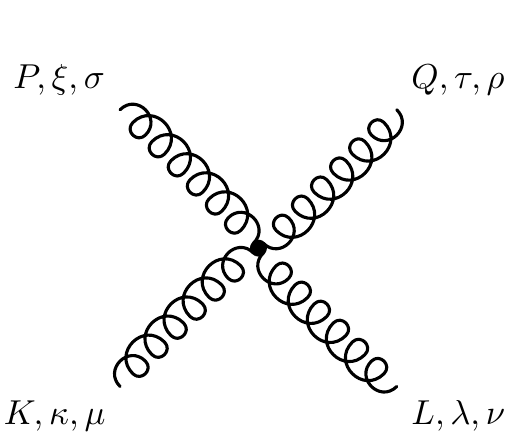}\\
\mycaption[Representation of the four-gluon vertex.]{Diagrammatic representation of the four-gluon vertex.}
\label{fig:four}
\end{center}
\end{figure}

As for the vertices, the ghost-antighost-gluon vertex is given by
\begin{equation}\label{eq:sdyd}
 g_0f_{\kappa\lambda\tau} K_\nu^\kappa,
\end{equation}
where $K$ is the momentum of the outgoing antighost and $\kappa$ is the corresponding outgoing color charge, and similarly, $\lambda$ and $\tau$ are the charges of the outgoing gluon and ghost. Similarly, the three-gluon vertex is given by
\begin{equation}\label{eq:sdysfd}
 \frac{g_0}{6}f_{\kappa\lambda\tau}\Big\{\delta_{\mu\rho}(K^\kappa_\nu-Q^\tau_\nu)+\delta_{\nu\mu}(L^\lambda_\rho-K^\kappa_\rho)+\delta_{\rho\nu}(Q^\tau_\mu-L^\lambda_\mu)\Big\},
\end{equation}
where $Q_\kappa$, $K_\lambda$ and $L_\tau$ denote the shifted outgoing momenta associated to the indices $\mu$, $\nu$ and $\rho$. These derivative vertices are represented in Fig.~\ref{fig:der}. Finally, the four-gluon vertex is given by
\begin{eqnarray}
 & & \frac{g^2_0}{24}\sum_\eta \Big\{f_{\kappa\lambda\eta}f_{\tau\xi(-\eta)}(\delta_{\mu\rho}\delta_{\nu\sigma}-\delta_{\mu\sigma}\delta_{\nu\rho})\nonumber\\
& & \hspace{1.0cm}+f_{\kappa\tau\eta}f_{\lambda\xi(-\eta)}(\delta_{\mu\nu}\delta_{\rho\sigma}-\delta_{\mu\sigma}\delta_{\nu\rho})\nonumber\\
& & \hspace{1.0cm}+f_{\kappa\xi\eta}f_{\tau\lambda(-\eta)}(\delta_{\mu\rho}\delta_{\nu\sigma}-\delta_{\mu\nu}\delta_{\sigma\rho})\Big\}\,,\label{eq:kjdg}
\end{eqnarray}
where $\kappa$, $\lambda$, $\tau$ and $\xi$ represent the outgoing charges. This vertex is represented in Fig.~(\ref{fig:four}). Using Eq.~(\ref{eq:f_complex}), the momenta/charges in the vertices of Figs.~\ref{fig:der} and \ref{fig:four} can also all be considered incoming, provided one replaces Eqs.~(\ref{eq:sdyd}), (\ref{eq:sdysfd}), and (\ref{eq:kjdg}) by the complex-conjugate expressions.

\section{Two-loop effective potential}
\label{sec:pot_2loop}

There are three diagrams contributing to the background field effective potential at two-loop order. We can label them according to the number of gluon propagators, $\delta V_{\mbox{\scriptsize 2loop}}^{\mbox{\scriptsize 1gl}}$, $\delta V_{\mbox{\scriptsize 2loop}}^{\mbox{\scriptsize 2gl}}$ and $\delta V_{\mbox{\scriptsize 2loop}}^{\mbox{\scriptsize 3gl}}$ respectively. Applying the Feynman rules listed above we find
\begin{eqnarray}\label{eq:V_feynman}
 \delta V_{\mbox{\scriptsize 2loop}}^{\mbox{\scriptsize 1gl}} & \!\!\!\!\!=\!\!\!\!\! & \sum_{\kappa\lambda\tau}{\cal C}_{\kappa\lambda\tau}\Big[-\frac{g^2}{2}\int_Q^T\int_K^T Q_\kappa\cdot P^\perp(L_\tau)\cdot K_\lambda\,G_0(Q_\kappa)\,G_0(K_\lambda)\,G_m(L_\tau)\Big],\nonumber\\
\delta V_{\mbox{\scriptsize 2loop}}^{\mbox{\scriptsize 2gl}} & \!\!\!\!\!=\!\!\!\!\! &  \frac{g^2}{4}\sum_{\kappa\lambda\tau}{\cal C}_{\kappa\lambda\tau} \int_Q^T\int_K^T\Big[\mbox{tr}\,P^\perp(Q_\kappa)\,\mbox{tr}P^\perp(K_\lambda)\nonumber\\
& & \hspace{4.0cm}-\,\mbox{tr}\,P^\perp(Q_\kappa)\,P^\perp(K_\lambda)\Big]G_m(Q_\kappa)\,G_m(K_\lambda)\,,\nonumber\\
\delta V_{\mbox{\scriptsize 2loop}}^{\mbox{\scriptsize 3gl}} & \!\!\!\!\!=\!\!\!\!\! &  -g^2\sum_{\kappa\lambda\tau}{\cal C}_{\kappa\lambda\tau}\int_Q^T\int_K^T {\cal B}(Q_\kappa,K_\lambda,L_\tau)\,G_m(Q_\kappa)\,G_m(K_\lambda)\,G_m(L_\tau)\,,\nonumber\\
\end{eqnarray}
where we have defined the tensor ${\cal C}_{\kappa\lambda\tau}\equiv|f_{\kappa\lambda\tau}|^2$. Since $f_{\kappa\lambda\tau}$ vanishes if $\kappa+\lambda+\tau\neq 0$, the same is true for ${\cal C}_{\kappa\lambda\tau}$ and we can assume that the color sums in Eq.~(\ref{eq:V_feynman}) are constrained by color conservation, meaning that $\kappa+\lambda+\tau=0$. Since we have chosen the momenta such that $Q+K+L=0$, the generalized momenta are also conserved in the sense that 
\begin{equation}\label{eq:mom_conserv}
 Q_\kappa+K_\lambda+L_\tau=0\,.
\end{equation} 
We shall repeatedly make use of this property in what follows.

Let us also mention that, for the last diagram, we have introduced the function
\begin{eqnarray}
 {\cal B}(Q_\kappa,K_\lambda,L_\tau) & \!\!\!\!\!\equiv\!\!\!\!\! & -\left[Q_\kappa\cdot P^\perp(L_\tau)\cdot K_\lambda\right]\mbox{tr}\,\left[P^\perp(K_\lambda)\,P^\perp (Q_\kappa)\right]\\
& & -\,2\,L_\tau\cdot P^\perp(Q_\kappa)\cdot P^\perp(L_\tau)\cdot P^\perp(K_\lambda)\cdot L_\tau\,.\nonumber\
\end{eqnarray}
Owing to the symmetry of the corresponding summand/integrand in Eq.~(\ref{eq:V_feynman}) with respect to permutations of the triplet $(Q_\kappa,K_\lambda,L_\tau)$, we can replace this function by its symmetrized version. A straightforward calculation shows that
\begin{eqnarray}\label{eq:Bsym}
  {\cal B}_{\mbox{\tiny sym}}(Q_\kappa,K_\lambda,L_\tau) & \!\!\!\!\!=\!\!\!\!\! & \frac{1}{3}\left\{\left(d-\frac{3}{2}\right)\left(\frac{1}{Q^2_\kappa}+\frac{1}{K^2_\lambda}+\frac{1}{L^2_\tau}\right)\right.\\
& & \hspace{0.7cm}\left.+\,\frac{Q^4_\kappa+K^4_\lambda+L^4_\tau}{4Q^2_\kappa K^2_\lambda L^2_\tau}\right\}\left[Q^2_\kappa K^2_\lambda-\left(Q_\kappa\cdot K_\lambda\right)^2\right],\nonumber
\end{eqnarray}
where we note that the combination $Q^2_\kappa K^2_\lambda-\left(Q_\kappa\cdot K_\lambda\right)^2$ is also totally symmetric, although not explicitly. This is easily seen by writing
\begin{eqnarray}\label{eq:trans_sym}
 Q^2_\kappa K^2_\lambda-(Q_\kappa\cdot K_\lambda)^2 & \!\!\!\!\!=\!\!\!\!\! & Q_\kappa \cdot P^\perp(K_\lambda)\cdot Q_\kappa\,K^2_\lambda\\
& \!\!\!\!\!=\!\!\!\!\! & L_\tau \cdot P^\perp(K_\lambda)\cdot L_\tau\,K^2_\lambda=L^2_\tau K^2_\lambda-(L_\tau\cdot K_\lambda)^2\,,\nonumber
\end{eqnarray}
where we have used Eq.~(\ref{eq:mom_conserv}). Again, we shall make use of these identities below.

\subsection{Reduction to scalar sum-integrals}
It is convenient to reduce the sum-integrals in Eq.~(\ref{eq:V_feynman}) to a set of simpler sum-integrals. To this purpose, we proceed in two steps. 

We first reduce everything to the set
\begin{equation}
 J_{\alpha}^\kappa\equiv \int_Q^T G_\alpha(Q_\kappa)\,, \quad J_{\mu\nu}^\kappa\equiv \int_Q^T \frac{Q^\kappa_\mu Q^\kappa_\nu}{Q^2_\kappa}G_m(Q_\kappa)\,.
\end{equation}
and
\begin{equation}
 I^{\kappa\lambda\tau}_{\alpha\beta\gamma}\equiv \int_Q^T\int_K^T \!\!\big[Q^2_\kappa K_\lambda^2-(Q_\kappa\cdot K_\lambda)^2\big]G_\alpha(Q_\kappa)\,G_\beta(K_\lambda)\,G_\gamma(L_\tau)\,.
\end{equation}
For instance, using 
\begin{equation}
 \mbox{tr}\,P^\perp(Q_\kappa)=d-1 \quad \mbox{and} \quad  \mbox{tr}\,P^\perp(Q_\kappa)\,P^\perp(K_\lambda)=d-2+\frac{(Q_\kappa\cdot K_\lambda)^2}{Q^2_\kappa K^2_\lambda}\,,
\end{equation} 
we find
\begin{equation}
 \delta V_{\mbox{\scriptsize 2loop}}^{\mbox{\scriptsize 2gl}}=\frac{g^2}{4}\sum_{\kappa,\lambda,\tau}{\cal C}_{\kappa\lambda\tau}\left[(d^2-3d+3)J_m^\kappa J_m^\lambda-J^\kappa_{\mu\nu}J^\lambda_{\mu\nu}\right],
\end{equation}
for the diagram with two gluon propagators. The diagram with one gluon propagator can be treated using Eqs.~(\ref{eq:mom_conserv}) and (\ref{eq:trans_sym}), together with the identity
\begin{equation}
 G_0(L_\tau)\,G_m(L_\tau)=\frac{1}{m^2}\Big[G_0(L_\tau)-G_m(L_\tau)\Big]\,.
\end{equation}
We find
\begin{equation}
  \delta V_{\mbox{\scriptsize 2loop}}^{\mbox{\scriptsize 1gl}}=\frac{g^2}{2m^2}\sum_{\kappa,\lambda,\tau}{\cal C}_{\kappa\lambda\tau}\Big[I^{\kappa\lambda\tau}_{000}-I^{\kappa\lambda\tau}_{00m}\Big].
\end{equation}
As for the diagram with three gluon propagators, similar considerations using the symmetrized function (\ref{eq:Bsym}) lead to
\begin{eqnarray}
 \delta V_{\mbox{\scriptsize 2loop}}^{\mbox{\scriptsize 3gl}} & \!\!\!\!\!=\!\!\!\!\! & -\frac{g^2}{4} \sum_{\kappa,\lambda,\tau}{\cal C}_{\kappa\lambda\tau}\Bigg\{J_m^\kappa J_m^\lambda-J^\kappa_{\mu\nu}J^\lambda_{\mu\nu}\nonumber\\
& & \hspace{3.0cm}-\,\frac{4}{m^2}\left[\left(d-\frac{5}{4}\right)I^{\kappa\lambda\tau}_{mmm}-(d-1)I^{\kappa\lambda\tau}_{mm0}+\frac{1}{4}I^{\kappa\lambda\tau}_{m00}\right]\Bigg\}\,.
\end{eqnarray}
Adding up the three diagrams, we arrive finally at
\begin{eqnarray}\label{eq:V_red1}
 \delta V_{\mbox{\scriptsize 2loop}} & \!\!\!\!\!=\!\!\!\!\! & g^2\sum_{\kappa,\lambda,\tau}{\cal C}_{\kappa\lambda\tau}\left\{\frac{(d-1)(d-2)}{4}J_m^\kappa J_m^\lambda+\frac{1}{2m^2}I^{\kappa\lambda\tau}_{000}\right.\nonumber\\
& & \hspace{2.5cm}+\,\left.\frac{1}{m^2}\left[\left(d-\frac{5}{4}\right)I^{\kappa\lambda\tau}_{mmm}-(d-1)I^{\kappa\lambda\tau}_{mm0}-\frac{1}{4}I^{\kappa\lambda\tau}_{m00}\right]\right\},
\end{eqnarray}
where we note that the contributions involving $J^\kappa_{\mu\nu}$ have cancelled out.

In a second step, we reduce all the sum-integrals to the set $J^\kappa_\alpha$,
\begin{equation}
 \tilde J_{\alpha}^\kappa\equiv \int_Q^T Q_\kappa^0 G_\alpha(Q_\kappa) \quad \mbox{and} \quad S^{\kappa\lambda\tau}_{\alpha\beta\gamma}\equiv \int_Q^T\int_K^T \!\!G_\alpha(Q_\kappa)\,G_\beta(K_\lambda)\,G_\gamma(L_\tau)\,.
\end{equation}
To this purpose, we use
\begin{equation}
 Q_\kappa\cdot K_\lambda=\frac{1}{2}\Big[\alpha^2+\beta^2-\gamma^2+L_\tau^2+\gamma^2-(Q_\kappa^2+\alpha^2)-(K_\lambda^2+\beta^2)\Big]
\end{equation}
to obtain, after a lengthy but straightforward calculation,
\begin{eqnarray}
 I^{\kappa\lambda\tau}_{\alpha\beta\gamma} & \!\!\!\!\!=\!\!\!\!\! &  \frac{1}{4}\left[\left(\gamma^2-\alpha^2-\beta^2\right)J_\alpha^\kappa J_\beta^\lambda+\left(\beta^2-\alpha^2-\gamma^2\right)J_\alpha^\kappa J_\gamma^\tau +\left(\alpha^2-\beta^2-\gamma^2\right)J_\beta^\lambda J_\gamma^\tau\right]\nonumber\\
& &  - \frac{1}{4}\left(\alpha^4+\beta^4+\gamma^4-2\alpha^2\beta^2-2\alpha^2\gamma^2-2\beta^2\gamma^2\right) S^{\kappa\lambda\tau}_{\alpha\beta\gamma}\nonumber\\
& & -\,\frac{1}{2}\Big[\tilde J_{\alpha}^\kappa \tilde J_{\beta}^\lambda+\tilde J_{\alpha}^\kappa \tilde J_{\gamma}^\tau+\tilde J_{\beta}^\lambda \tilde J_{\gamma}^\tau\Big].
\end{eqnarray}
Plugging this identity back into Eq.~(\ref{eq:V_red1}), we arrive finally at
\begin{eqnarray}\label{eq:V_red2}
 \delta V_{\mbox{\scriptsize 2loop}} & \!\!\!\!\!=\!\!\!\!\! & g^2{ \sum_{\kappa,\lambda,\tau}{\cal C}_{\kappa\lambda\tau}\Bigg\{\frac{1}{4}\left(d^2-4d+\frac{15}{4}\right)J_m^\kappa J_m^\lambda+\frac{1}{8}J_0^\kappa J_m^\lambda-\frac{1}{16}J_0^\kappa J_0^\lambda}\nonumber\\
& & \hspace{2.5cm}-\,\frac{1}{m^2} \left(d-\frac{11}{8}\right)\tilde J_m^\kappa \tilde J_m^\lambda+\frac{1}{m^2} \left(d-\frac{3}{4}\right)\tilde J_0^\kappa \tilde J_m^\lambda-\frac{5}{8m^2}\tilde J_0^\kappa \tilde J_0^\lambda\nonumber\\
& & \hspace{2.5cm}+\,\frac{3m^2}{4}\left(d-\frac{5}{4}\right) S_{mmm}^{\kappa\lambda\tau}+\frac{m^2}{16} S^{\kappa\lambda\tau}_{m00}\Bigg\}\,.
\end{eqnarray}
This expression is convenient because the Matsubara sum-integrals $J_\alpha^\kappa$, $\tilde J_\alpha^\kappa$ and $S_{\alpha\beta\gamma}^{\kappa\lambda\tau}$ are essentially those found in a scalar theory and can be easily evaluated using standard techniques, see \cite{Reinosa:2014zta}.

\subsection{Thermal decomposition and renormalization}
After evaluating the elementary sum-integrals identified above, one can typically decompose them as
\begin{eqnarray}
 J_\alpha^\kappa & \!\!\!\!\!=\!\!\!\!\! & J_\alpha(0n)+J_\alpha^\kappa(1n)\,,\\
 \tilde J_\alpha^\kappa & \!\!\!\!\!=\!\!\!\!\! & \tilde J_\alpha(0n)+\tilde J_\alpha^\kappa(1n)\,,\\
 S_{\alpha\beta\gamma}^{\kappa\lambda\tau} & \!\!\!\!\!=\!\!\!\!\! & S_{\alpha\beta\gamma}(0n)+S_{\alpha\beta\gamma}^{\kappa\lambda\tau}(1n)+S_{\alpha\beta\gamma}^{\kappa\lambda\tau}(2n)\,,
\end{eqnarray}
where $\mbox{($i$n)}$ refers to the number of loops that are cut off by the presence of thermal factors $n$.\footnote{More precisely, these should be factors of $n$ with an argument whose real part is positive, so that the loop momentum is indeed cut off.} 

Plugging these decompositions into Eq.~(\ref{eq:V_red2}), we arrive at a similar decomposition for the two-loop correction to the background field effective potential:
\begin{equation}
 \delta V_{\mbox{\scriptsize 2loop}}=\delta V_{\mbox{\scriptsize 2loop}}(0n)+\delta V_{\mbox{\scriptsize 2loop}}(1n)+\delta V_{\mbox{\scriptsize 2loop}}(2n)\,.
\end{equation}
As we have shown in the previous chapter, the zero-temperature contribution $\delta V_{\mbox{\scriptsize 2loop}}(0n)$ does not depend on the background and can be ignored in the following analysis. 

The contribution $\delta V_{\mbox{\scriptsize 2loop}}(1n)$ involves products of the form $J_\alpha(0n)J_\beta^\lambda(1n)$ or $\tilde J_\alpha(0n)\tilde J_\beta^\lambda(1n)$, as well as $S_{\alpha\beta\gamma}^{\kappa\lambda\tau}(1n)$. This latter quantity can be expressed in terms of $J_\alpha^\kappa(1n)$ as 
\begin{eqnarray}
 S_{\alpha\beta\gamma}^{\kappa\lambda\tau}(1n) & \!\!\!\!\!=\!\!\!\!\! & J_\alpha^\kappa(1n)\,\,\mbox{Re}\,\tilde I_{\beta\gamma}(\varepsilon_{\alpha,q}+i0^+;q)\\
& & J_\beta^\lambda(1n)\,\,\mbox{Re}\,\tilde I_{\gamma\alpha}(\varepsilon_{\beta,k}+i0^+;k)\nonumber\\
& & J_\gamma^\tau(1n)\,\,\mbox{Re}\,\tilde I_{\alpha\beta}(\varepsilon_{\gamma,l}+i0^+;l)\,,\nonumber
\end{eqnarray}
where $\mbox{Re}\,\tilde I_{\beta\gamma}(\varepsilon_{\alpha,q}+i0^+;q)$ is defined as the real part of the vacuum bubble integral
\begin{equation}
 I_{\beta\gamma}(Q)\equiv \int_K^{T=0} G_\beta(K)\,G_\gamma(L)\,,
\end{equation}
analytically continued to real frequencies, $Q=(\omega_n,q)\to (-iq_0+\epsilon,q)$ and evaluated on the mass shell $q_0^2-q^2=\alpha^2$, see \cite{Reinosa:2014zta} for more details. For the cases of interest here, we find
\begin{eqnarray}
  \mbox{Re}\,\tilde I_{00}(\varepsilon_{q}\!+\!i0^+;q) & \!\!\!\!\!=\!\!\!\!\! & \frac{1}{16\pi^2}\!\!\left[\frac{1}{\epsilon}+\ln\frac{\bar\mu^2}{m^2}+2\right],\\
  \mbox{Re}\,\tilde I_{m0}({q}\!+\!i0^+;q) & \!\!\!\!\!=\!\!\!\!\! & \frac{1}{16\pi^2}\!\!\left[\frac{1}{\epsilon}+\ln\frac{\bar\mu^2}{m^2}+1\right],\\
  \mbox{Re}\,\tilde I_{mm}(\varepsilon_{q}\!+\!i0^+;q) & \!\!\!\!\!=\!\!\!\!\! & \frac{1}{16\pi^2}\!\!\left[\frac{1}{\epsilon}+\ln\frac{\bar\mu^2}{m^2}+2-\frac{\pi}{\sqrt{3}}\right],
\end{eqnarray}
where we note that the RHS do not depend on $q$.\footnote{This follows from the definition of $\mbox{Re}\,\tilde I_{\beta\gamma}(\varepsilon_{\alpha,q}+i0^+;q)$ and Lorentz invariance.}
From these relations, together with the standard results
\begin{equation}
 J_\alpha(0n)=-\frac{\alpha^2}{16\pi^2}\left[\frac{1}{\epsilon}+\ln\frac{\bar\mu^2}{\alpha^2}+1\right],
\end{equation}
and $\tilde J_\alpha(0n)=0$, one arrives at
\begin{equation}\label{eq:V1}
 \delta V_{\mbox{\scriptsize 2loop}}(1n)=\frac{g^2C_{\mbox{\scriptsize ad}}}{128\pi^2}m^2\left(\frac{35}{\epsilon}+35\ln\frac{\bar\mu^2}{m^2}+\frac{313}{3}-\frac{99\pi}{2\sqrt{3}}\right)\sum_\kappa J_m^\kappa(1n)\,,
\end{equation}
where we have made use of the symmetry of ${\cal C}_{\kappa\lambda\tau}$ together with the Casimir identity $\sum_\tau {\cal C}_{\kappa\lambda\tau}=C_{\mbox{\scriptsize ad}}\delta_{\kappa,-\lambda}$ \cite{Reinosa:2015gxn}.

Finally, the contribution $\delta V_{\mbox{\scriptsize 2loop}}(2n)$ with two thermal factors is UV finite by construction and one can evaluate it directly in $d=4$ dimensions. One finds
\begin{eqnarray}\label{eq:V2}
 \delta V_{\mbox{\scriptsize 2loop}}(2n) & \!\!\!\!\!=\!\!\!\!\! & \frac{3g^2}{8}\sum_{\kappa,\lambda,\tau}{\cal C}_{\kappa\lambda\tau}\left[\frac{5}{2}U^\kappa V^\lambda- \frac{7}{m^2}\tilde U^\kappa\tilde V^\lambda\right]\\
& & +\,\frac{g^2m^2}{16}\sum_{\kappa,\lambda,\tau}{\cal C}_{\kappa\lambda\tau}\left[33S_{mmm}^{\kappa\lambda\tau}(2n)+S_{m00}^{\kappa\lambda\tau}(2n)\right],\nonumber
\end{eqnarray}
with $U^\kappa\equiv J_m^\kappa(1n)+J^\kappa_0(1n)/3$, $V^\kappa\equiv J_m^\kappa(1n)-J^\kappa_0(1n)/5$, $\tilde U^\kappa\equiv\tilde J_{m}^\kappa(1n)-\tilde J_{0}^\kappa(1n)$ and $\tilde V^\kappa\equiv\tilde J_{m}^\kappa(1n)-5\tilde J_{0}^\kappa(1n)/21$.\\

One could be worried by the presence of a temperature-dependent divergence in $\delta V_{\mbox{\scriptsize 2loop}}(1n)$. However, at two-loop order, one should also consider one-loop diagrams involving counterterms.\footnote{The counterterms should be evaluated at one-loop order, so that the counterterm diagram counts effectively as a two-loop order diagram.} These are obtained by first rewriting the action in terms of renormalized fields and parameters as
\begin{equation}
 a_\mu\to Z_a^{1/2}a_\mu\,, \quad c\to Z_c^{1/2}c\,, \quad m^2\to Z_{m^2} m^2\,, \quad g\to Z_g g\,,
\end{equation}
writing $Z=1+\delta Z$ and treating the various $\delta Z$ as new interaction vertices. At the considered order, we find the counterterm contribution
\begin{eqnarray}
{\color{red} \checkmark } \delta V_{\mbox{\scriptsize 2loop}}^{\mbox{\scriptsize ct}} & \!\!\!\!\!=\!\!\!\!\! & -\delta Z_c\sum_\kappa\int_Q^T  Q^2_\kappa G_0(Q_\kappa)\\
& & +\,\frac{1}{2}\sum_\kappa\int_Q^T\Big(\delta Z_a(Q^2_\kappa+m^2)+m^2\delta Z_{m^2}\Big) \mbox{tr}\,P^\perp(Q_\kappa)\,G(Q_\kappa)\,.\nonumber
\end{eqnarray}
Using $\mbox{tr}\,P^\perp(Q_\kappa)=d-1$ together with the fact that $\int_Q^T 1=0$ in dimensional regularization, this rewrites
\begin{equation}\label{eq:Vct}
 \delta V_{\mbox{\scriptsize 2loop}}^{\mbox{\scriptsize ct}}=\frac{d-1}{2}m^2\delta Z_{m^2}\sum_\kappa J_m^\kappa\,.
\end{equation}
We note that this is precisely of the same form as the divergence that needs to be cancelled out in Eq.~(\ref{eq:V1}). In fact, the value of $\delta Z_{m^2}$ is fixed from the renormalization of the gluon two-point function at one-loop order, which we discuss in Chapter \ref{chap:correlators}. In the zero-momentum scheme considered there, one finds
\begin{equation}\label{eq:Zm}
 \delta Z_{m^2}=\frac{g^2N}{192\pi^2}\left[-\frac{35}{\epsilon}+z_f\right],
\end{equation}
and
\begin{eqnarray}
 z_f & \!\!\!\!\!\!=\!\!\!\!\! & -\frac{1}{s^2}+\frac{111}{2s}-\frac{287}{6}-35\ln \left(\bar s \right)-\frac{1}{2}\left(s^2-2\right) \ln (s)\nonumber\\
& & +\left(s^2-10 s+1\right) \left(\frac{1}{s}+1\right)^3 \ln (s+1)\nonumber\\
& & +\frac{1}{2}(s^2-20s+12)\left(\frac{4}{s}+1\right)^{3/2}  \ln \left(\frac{\sqrt{4/s+1}-1}{\sqrt{4/s+1}+1}\right),
\end{eqnarray}
with $s\equiv \mu^2/m^2$ and $\bar s\equiv \bar\mu^2/m^2$. The pole in Eq.~(\ref{eq:Zm}) is universal and exactly cancels the one in Eq.~(\ref{eq:V1}), as expected in a renormalizable theory such as the Curci-Ferrari model. Consequently, in what follows, we redefine $\delta V_{\mbox{\scriptsize 2loop}}(1n)$ to be the sum of (\ref{eq:V1}) and (\ref{eq:Vct}):
\begin{equation}\label{eq:V12}
 \delta V_{\mbox{\scriptsize 2loop}}(1n)=\frac{g^2C_{\mbox{\scriptsize ad}}}{128\pi^2}m^2\left(z_f+35\ln \bar s^2+\frac{313}{3}-\frac{99\pi}{2\sqrt{3}}\right)\sum_\kappa J_m^\kappa(1n)\,.
\end{equation}

\subsection{UV finite contributions}
To complete our evaluation of the two-loop corrections, we just need to recall the expressions for the UV finite contributions $J_\alpha^\kappa(1n)$, $\tilde J_\alpha^\kappa(1n)$ and $S_{\alpha\beta\gamma}^{\kappa\lambda\tau}(2n)$ \cite{Reinosa:2015gxn}. Setting $\hat r\equiv r\,T$
\begin{eqnarray}
 J_\alpha^\kappa(1n) & \!\!\!\!\!=\!\!\!\!\! & \frac{1}{2\pi^2}\int_0^\infty dq\, \frac{q^2} {\varepsilon_{\alpha,q}} \mbox{Re}\,n_{\varepsilon_{\alpha,q}-i\hat r\cdot\kappa}\,,\\
 \tilde J_\alpha^\kappa(1n) & \!\!\!\!\!=\!\!\!\!\! & \frac{1}{2\pi^2}\int_0^\infty dq\, q^2 \mbox{Im}\,n_{\varepsilon_{\alpha,q}-i\hat r\cdot\kappa}\,,
\end{eqnarray}
and
\begin{eqnarray}
 S_{\alpha\beta\gamma}^{\kappa\lambda\tau}(2n) & \!\!\!\!\!=\!\!\!\!\! & \frac{1}{32\pi^4}\int_0^\infty \!\!dq\,q\int_0^\infty \!\!dk\,k\, \mbox{Re}\,\frac{n_{\varepsilon_{\alpha,q}-i\hat r\cdot\kappa}\,n_{\varepsilon_{\beta,k}-i\hat r\cdot\lambda}}{\varepsilon_{\alpha,q}\,\varepsilon_{\beta,k}}\nonumber\\
& & \hspace{4.0cm}\times\,\mbox{Re}\,\ln\frac{(\varepsilon_{\alpha,q}+\varepsilon_{\beta,k}+i0^+)^2-(\varepsilon_{\gamma,k+q})^2}{(\varepsilon_{\alpha,q}+\varepsilon_{\beta,k}+i0^+)^2-(\varepsilon_{\gamma,k-q})^2}\nonumber\\
& \!\!\!\!\!+\!\!\!\!\! &\,\frac{1}{32\pi^4}\int_0^\infty \!\!dq\,q\int_0^\infty \!\!dk\,k \,\mbox{Re}\,\frac{n_{\varepsilon_{\alpha,q}-i\hat r\cdot\kappa}\,n_{\varepsilon_{\beta,k}+i\hat r\cdot\lambda}}{\varepsilon_{\alpha,q}\,\varepsilon_{\beta,k}}\nonumber\\
& & \hspace{4.0cm}\times\,\mbox{Re}\,\ln\frac{(\varepsilon_{\alpha,q}-\varepsilon_{\beta,k}+i0^+)^2-(\varepsilon_{\gamma,k+q})^2}{(\varepsilon_{\alpha,q}-\varepsilon_{\beta,k}+i0^+)^2-(\varepsilon_{\gamma,k-q})^2}\nonumber\\
& \!\!\!\!\!+\!\!\!\!\! & \mbox{cyclic permutations of $(\alpha,\kappa)$, $(\beta,\lambda)$ and $(\gamma,\tau)$.}
\end{eqnarray}
These formulas, together with Eqs.~(\ref{eq:V12}) and (\ref{eq:V2}) form the basis for the evaluation of the two-loop corrections to the background field effective potential.

\section{Next-to-leading order Polyakov loop}
Corresponding to the two-loop background field effective potential, we can evaluate the background-dependent Polyakov loop at next-to-leading order. When evaluated at the minimum of the two-loop background field effective potential, this gives access to the physical Polyakov loop at next-to-leading order. As already emphasized, we do not really need to evaluate the physical Polyakov loop in order to assess the phase transition since the background that minimizes the background field effective potential is an equally relevant order parameter for center symmetry, simpler to compute in practice. Computing the Polyakov loop is nevertheless interesting for it is a gauge independent quantity and can therefore be compared, in principle, with non-gauge fixed lattice simulations. Another motivation for evaluating the next-to-leading order corrections to the Polyakov loop is to assess the role of higher order corrections in curing some of the artificial behaviors observed at leading order. 

\subsection{Setting up the expansion}
As emphasized in Chapter \ref{chap:bg}, the background-dependent Polyakov loop is defined in the presence of a source that forces the background to be self-consistent, that is $\bar A=\langle A\rangle_{\bar A}$ or $\langle a\rangle_{\bar A}=0$. We thus need to expand the action and the Wilson line in powers of $a\equiv A-\bar A$, and take an average where we set $\langle a\rangle_{\bar A}=0$. 

Due to the presence of time-ordering, the expansion of the Wilson line $L_A(\vec{x})$ in $a=A-\bar A$ (and not $A$) is cumbersome. We can however use the following trick.\footnote{We have checked that the brute force calculation leads to the same result.} Consider the gauge transformation
\begin{equation}
 U(\tau)=e^{-g_0\bar A\tau}\,.
\end{equation}
It is such that $\bar A^U=0$ and therefore
\begin{equation}
 L_{\bar A+a}(\vec{x})=U^\dagger(\beta)\,L_{(\bar A+a)^U}(\vec{x})\,U(0)=U^\dagger(\beta)\,L_{a^U}(\vec{x})\,,
\end{equation}
with $a^U\equiv U\,a\,U^\dagger$. The expansion in $a$ is easily done, despite the presence of the time ordering. To order $g^2$, we find
\begin{equation}
 L_{\bar A+a}(\vec{x})=U^\dagger(\beta)\left[\mathds{1}+g\int_0^\beta d\tau\,a^U_0(\tau,\vec{x})+\frac{g^2}{2}\int_0^\beta d\tau \int_0^\beta d\sigma\,{\cal P}\,a^U_0(\tau,\vec{x})\,a^U_0(\sigma,\vec{x})\right].
\end{equation}
Upon taking the trace and averaging with $\langle a\rangle_{\bar A}=0$, we find
\begin{eqnarray}
\ell(\bar A)=\ell_0(\bar A) & \!\!\!\!\!-\!\!\!\!\! & \frac{g^2}{2N}\int_0^\beta d\tau \int_0^{\tau} d\sigma\,G^{\kappa\lambda}_{00}(\tau-\sigma,\vec{0})\,\,\mbox{tr}\left[U^\dagger(\beta-\tau+\sigma)\,t^\kappa\,U^\dagger(\tau-\sigma)\,t^\lambda\right]\nonumber\\
& \!\!\!\!\!-\!\!\!\!\! & \frac{g^2}{2N}\int_0^\beta d\tau \int_\tau^\beta d\sigma\,G^{\kappa\lambda}_{00}(\sigma-\tau,\vec{0})\,\,\mbox{tr}\left[U^\dagger(\beta-\sigma+\tau)\,t^\lambda\,U^\dagger(\sigma-\tau)\,t^\kappa\right]\,,
\end{eqnarray}
where we have made use of the cyclic property of the trace and where
\begin{equation}
G^{\kappa\lambda}_{00}(\tau-\sigma,\vec{0})\equiv \langle {\cal P}\,a^\kappa_0(\tau,\vec{x})\,a^\lambda_0(\sigma,\vec{x})\rangle_{\bar A}
\end{equation}
denotes the Euclidean time component of the gluon propagator, computed at tree-level. We can make the changes of variables $\sigma\to\beta-\sigma$ and $\sigma\to\beta+\tau-\sigma$ in each of the $\sigma$-integrals respectively. We arrive at
\begin{eqnarray}
\ell(\bar A)=\ell_0(\bar A) & \!\!\!\!\!-\!\!\!\!\! & \frac{g^2}{2N}\int_0^\beta d\tau \int_0^{\tau} d\sigma\,G^{\kappa\lambda}_{00}(\sigma,\vec{0})\,\,\mbox{tr}\left[U^\dagger(\beta-\sigma)\,t^\kappa\,U^\dagger(\sigma)\,t^\lambda\right]\nonumber\\
& \!\!\!\!\!-\!\!\!\!\! & \frac{g^2}{2N}\int_0^\beta d\tau \int_\tau^{\beta} d\sigma\,G^{\kappa\lambda}_{00}(\sigma-\beta,\vec{0})\,\,\mbox{tr}\left[U^\dagger(\beta-\sigma)\,t^\kappa\,U^\dagger(\sigma)\,t^\lambda\right]\,,
\end{eqnarray}
where we have again used the cyclic property of the trace. Using the periodicity of the propagator, this rewrites eventually as
\begin{equation}
\ell(\bar A)=\ell_0(\bar A)-\frac{g^2\beta}{2N} \int_0^\beta d\sigma\,G^{\kappa\lambda}_{00}(\sigma,\vec{0})\,\,\mbox{tr}\left[U^\dagger(\beta-\sigma)\,t^\kappa\,U^\dagger(\sigma)\,t^\lambda\right]\,,
\end{equation}
where the $\tau$-integral has become trivial, producing a factor $\beta$.

\subsection{Using the weights}
To continue the calculation, we need to evaluate the trace. Before doing so, let us notice that, with little modifications, the previous formula applies to the Polyakov loop in any representation. We have indeed
\begin{equation}
\ell_R(\bar A)=\ell_{R,0}(\bar A)-\frac{g^2\beta}{2d_R} \int_0^\beta d\sigma\,G^{\kappa\lambda}_{00}(\sigma,\vec{0})\,\,\mbox{tr}\left[U^\dagger(\beta-\sigma)\,t^\kappa_R\,U^\dagger(\sigma)\,t^\lambda_R\right]\,.
\end{equation}
We shall continue with this more generic formula from now on. Now, the color trace is easily evaluated using the weights $\rho$ of the representation, defined such that $t^{0^{(j)}}|\rho\rangle=\rho_j|\rho\rangle$. Since the background is taken along the Cartan sub-algebra, we have $U^\dagger(\sigma)|\rho\rangle=e^{i\frac{\sigma}{\beta}r\cdot\rho}|\rho\rangle$ and then
\begin{equation}
\mbox{tr}\left[U^\dagger(\beta-\sigma)\,t^\kappa\,U^\dagger(\sigma)\,t^\lambda\right]=\sum_{\rho\rho'}\,e^{i\frac{\beta-\sigma}{\beta}r\cdot\rho+i\frac{\sigma}{\beta}r\cdot\rho'}\langle\rho |t^\kappa_R|\rho'\rangle\langle\rho'|t^\lambda|\rho\rangle\,.
\end{equation}
Since the free gluon propagator has non-zero components only for $\kappa=\lambda=0^{(j)}$ and $\kappa=-\lambda=-\alpha$, the color trace has to be considered in those cases only. In the first case, from the definition of the weights, the color trace rewrites
\begin{equation}\label{eq:758}
\mbox{tr}\left[U^\dagger(\beta-\sigma)\,t^{0^{(j)}}\,U^\dagger(\sigma)\,t^{0^{(j)}}\right]=\sum_{\rho}\,\rho_j^2\,e^{ir\cdot\rho}\,.
\end{equation}
In the second case, similarly to what we did for the adjoint representation in Eqs.~(\ref{eq:74}) and (\ref{eq:75}), it is easily argued that $t^{\alpha}|\rho\rangle$ is either zero or has a definite weight equal to $\rho+\alpha$ (a necessary condition is that $\rho+\alpha$ belongs to the weight diagram). Therefore, in the second case the color trace rewrites
\begin{equation}\label{eq:759}
\mbox{tr}\left[U^\dagger(\beta-\sigma)\,t^{-\alpha}\,U^\dagger(\sigma)\,t^{\alpha}\right]=\sum_{\rho}\,e^{ir\cdot\rho+i\frac{\sigma}{\beta}r\cdot\alpha}|\langle\rho+\alpha |t^\alpha_R|\rho\rangle|^2\,,
\end{equation}
where it is understood that $|\rho-\alpha\rangle=0$ if $\rho-\alpha$ is not a weight. In fact, Eqs.~(\ref{eq:758}) and (\ref{eq:759}) can be written as a single result. One has just to replace $\alpha$ by $\kappa$ in Eq.~(\ref{eq:759}). The next-to-leading order correction to the background-dependent Polyakov loop reads, therefore,
\begin{equation}\label{eq:DlR}
\Delta\ell_R(\bar A)=-\frac{g^2\beta}{2d_R} \sum_\rho e^{ir\cdot\rho}\sum_\kappa |\langle\rho+\kappa |t^\kappa_R|\rho\rangle|^2\,\int_0^\beta d\sigma\,e^{i\frac{\sigma}{\beta}r\cdot\kappa}\,G^{(-\kappa)\kappa}_{00}(\sigma,\vec{0})\,.
\end{equation}

\subsection{Completing the calculation}
We finally note that
\begin{eqnarray}
G^{(-\kappa)\kappa}_{00}(\sigma,\vec{0}) & \!\!\!\!\!=\!\!\!\!\! & \int_Q^T \frac{q^2}{(\omega_n+Tr\cdot\kappa)^2+q^2}\frac{e^{i\omega_n\sigma}}{(\omega_n+Tr\cdot\kappa)^2+q^2+m^2}\nonumber\\
& \!\!\!\!\!=\!\!\!\!\! & \frac{1}{m^2}\left[\int_Q^T \frac{q^2\,e^{i\omega_n\sigma}}{(\omega_n+Tr\cdot\kappa)^2+q^2}-\int_Q^T\frac{q^2\,e^{i\omega_n\sigma}}{(\omega_n+Tr\cdot\kappa)^2+q^2+m^2}\right].
\end{eqnarray}
Standard contour techniques for the evaluation of Matsubara sums lead to (recall that $0<\sigma<\beta$)
\begin{eqnarray}
& & \int_Q^T \frac{q^2\,e^{i\omega_n\sigma}}{(\omega_n+Tr\cdot\kappa)^2+q^2+m^2}\\
& & \hspace{1.0cm}=\,\int\frac{d^3q}{(2\pi)^3}\frac{q^2}{2\varepsilon_q}\Big[e^{(\varepsilon_q-iTr\cdot\kappa)\sigma}n_{\varepsilon_q-iTr\cdot\kappa}-e^{(-\varepsilon_q-iTr\cdot\kappa)\sigma}n_{-\varepsilon_q-iTr\cdot\kappa}\Big]\,,\nonumber
\end{eqnarray}
which generalizes the well known result for $\kappa=0$. It follows that
\begin{eqnarray}
& & \int_0^\beta d\sigma\,e^{i\frac{\sigma}{\beta}r\cdot\kappa}\,G^{(-\kappa)\kappa}_{00}(\sigma,\vec{0})\nonumber\\
& & \hspace{0.5cm}=\,\frac{1}{2m^2}\int\frac{d^3q}{(2\pi)^3}\Bigg\{\Big[(e^{\beta q}-1)n_{q-iTr\cdot\kappa}+\,(e^{-\beta q}-1)n_{-q-iTr\cdot\kappa}\big]\nonumber\\
& & \hspace{4.0cm}-\frac{q^2}{q^2+m^2}\Big[(e^{\beta\varepsilon_q}-1)n_{\varepsilon_q-iTr\cdot\kappa}+\,(e^{-\beta\varepsilon_q}-1)n_{-\varepsilon_q-iTr\cdot\kappa}\Big]\Bigg\}\,.
\end{eqnarray}
Finally, we write
\begin{equation}
(e^{\beta \varepsilon_q}-1)n_{\varepsilon_q-iTr\cdot\kappa}=e^{ir\cdot\kappa}+(e^{ir\cdot\kappa}-1)n_{\varepsilon_q-iTr\cdot\kappa}\nonumber
\end{equation}
from which it follows that
\begin{eqnarray}
& & (e^{\beta \varepsilon_q}-1)n_{\varepsilon_q-iTr\cdot\kappa}+(e^{-\beta \varepsilon_q}-1)n_{-\varepsilon_q-iTr\cdot\kappa}\\
& & \hspace{0.5cm}=\,2e^{ir\cdot\kappa}+(e^{ir\cdot\kappa}-1)(n_{\varepsilon_q-iTr\cdot\kappa}+n_{-\varepsilon_q-iTr\cdot\kappa})\nonumber\\
& & \hspace{0.5cm}=\,e^{ir\cdot\kappa}+1+(e^{ir\cdot\kappa}-1)(n_{\varepsilon_q-iTr\cdot\kappa}-n_{\varepsilon_q+iTr\cdot\kappa})\,,\nonumber
\end{eqnarray}
and thus that
\begin{eqnarray}
& & \int_0^\beta d\sigma\,e^{i\frac{\sigma}{\beta}r\cdot\kappa}\,G^{(-\kappa)\kappa}_{00}(\sigma,\vec{0})\\
& & \hspace{0.5cm}=\,\frac{e^{ir\cdot\kappa}+1}{2}\int\frac{d^3q}{(2\pi)^3}\frac{1}{q^2+m^2}\nonumber\\
& & \hspace{0.5cm}+\,i\frac{e^{ir\cdot\kappa}-1}{m^2}\int\frac{d^3q}{(2\pi)^3}\Bigg\{\mbox{Im}\,n_{q-iTr\cdot\kappa}-\frac{q^2}{q^2+m^2}\mbox{Im}\,n_{\varepsilon_q-iTr\cdot\kappa}\Bigg\}\,.\nonumber
\end{eqnarray}
which we conveniently rewrite as
\begin{eqnarray}
& & \int_0^\beta d\sigma\,e^{i\frac{\sigma}{\beta}r\cdot\kappa}\,G^{(-\kappa)\kappa}_{00}(\sigma,\vec{0})\\
& & \hspace{0.5cm}=\,\frac{e^{ir\cdot\kappa}+1}{2}\left[\int\frac{d^3q}{(2\pi)^3}\frac{1}{q^2+m^2}+\frac{2}{m^2}\sin^2\left(\frac{r\cdot\kappa}{2}\right)\right.\nonumber\\
& & \hspace{2.5cm}\left.\times\,\int\frac{d^3q}{(2\pi)^3}\Bigg\{\frac{q^2}{\varepsilon_q^2}\frac{1}{\cosh(\varepsilon_q/T)-\cos(r\cdot\kappa)}-(m\to 0)\Bigg\}\right]\,.\nonumber
\end{eqnarray}
We can now plug this expression back into Eq.~(\ref{eq:DlR}) by noticing that
\begin{eqnarray}
\sum_\rho e^{ir\cdot\rho}\sum_\kappa |\langle\rho+\kappa |t^\kappa_R|\rho\rangle|^2\,e^{ir\cdot\kappa}\,e_\kappa & \!\!\!\!\!=\!\!\!\!\! & \sum_\rho e^{ir\cdot\rho}\sum_\kappa |\langle\rho+\kappa |t^\kappa_R|\rho\rangle|^2\,e_\kappa\,,\nonumber\\
\end{eqnarray}
for any $e_\kappa$ such that $e_{-\kappa}=e_\kappa$. We arrive at
\begin{eqnarray}
\Delta\ell_R(\bar A) & = & \frac{g^2}{8\pi} \frac{C_R}{d_R}\frac{m}{T}\sum_\rho e^{ir\cdot\rho}\nonumber\\
& & +\,\frac{g^2}{2\pi^2}\frac{1}{d_R}\frac{m}{T}\sum_\rho e^{ir\cdot\rho}\sum_\kappa |\langle\rho+\kappa |t^\kappa_R|\rho\rangle|a(T,r\cdot\kappa)\sin^2\left(\frac{r\cdot\kappa}{2}\right),
\end{eqnarray}
where we have used that $\sum_\kappa |\langle\rho+\kappa |t^\kappa_R|\rho\rangle|^2=C_R$ is the Casimir of the representation,
\begin{equation}
\int\frac{d^3q}{(2\pi)^3}\frac{1}{q^2+m^2}=-\frac{m}{4\pi}
\end{equation}
in dimensional regularization, and we have introduced the function
\begin{equation}
a(T,r\cdot\kappa)\equiv\int_0^\infty \frac{dq\,q^2}{m^3}\left(\frac{1}{\cosh(q/T)-\cos(r\cdot\kappa)}-\frac{q^2}{\varepsilon_q^2}\frac{1}{\cosh(\varepsilon_q/T)-\cos(r\cdot\kappa)}\right).
\end{equation}

\section{Results}
\label{sec:results}

We use a renormalization scheme at zero-temperature. In this limit, the Landau-deWitt gauge boils down to the Landau gauge. We can therefore adjust the values of $m$ and $g$ to those for which the Curci-Ferrari propagators (computed within the same renormalization scheme) best fit the Landau gauge lattice propagators at zero temperature, namely $g=4.9$ and $m=540$ MeV, see \cite{Tissier:2011ey}.

\subsection{SU($2$) and SU($3$) transitions}

We find once more that the transition is second order in the SU(2) case and first order in the SU(3) case. Our updated values for the transition temperatures are shown in Table \ref{tab:2loop} and have neatly improved as compared to the leading order estimations.
\begin{table}[h]
$$\begin{tabular}{|c||c||c||c||c|}
\hline
$T_c$ (MeV) & lattice \cite{Lucini:2012gg} & fRG \cite{Fister:2013bh} & Variational \cite{Quandt:2016ykm} & $2$-loop CF \cite{Reinosa:2015gxn}\\
\hline\hline
SU(2) & 295 & 230 & 239 & 284\\
SU(3) & 270 & 275 & 245 & 254\\
\hline
\end{tabular}$$
\caption{Transition temperatures for the SU($2$) and SU($3$) deconfinement transitions as computed from various approaches, compared to the two-loop Curci-Ferrari model.}\label{tab:2loop}
\end{table}

\subsection{Polyakov loops}
\begin{figure}[t]
\begin{center}
\hspace{0.75cm}\includegraphics[height=0.32\textheight]{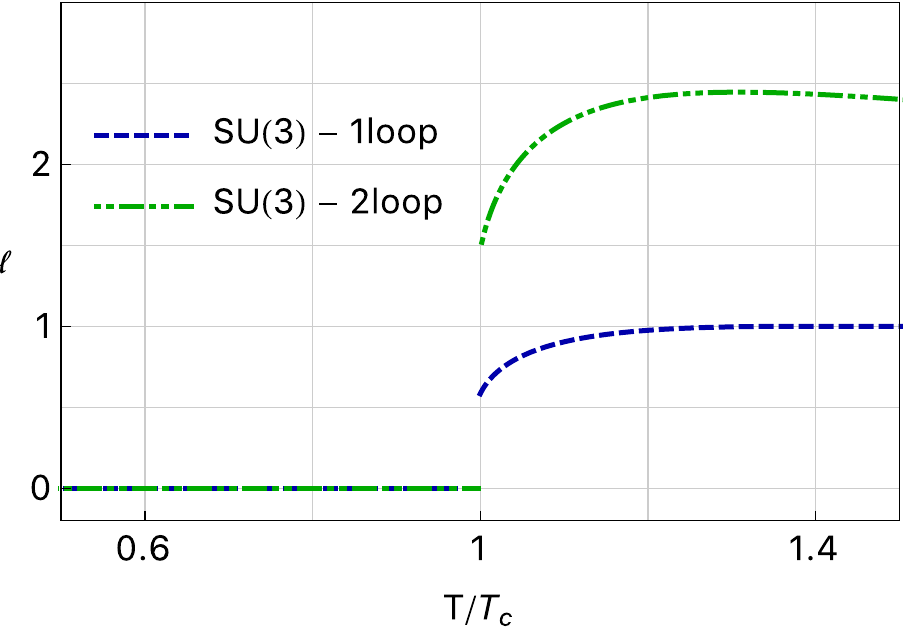}
\mycaption[Polyakov loops]{Leading and next-to-leading order Polyakov loops for the SU($3$) defining representation, as functions of $T/T_c$.}
\label{fig:polys_2loop}
\end{center}
\end{figure}

The Polyakov loop is shown in Fig.~\ref{fig:polys_2loop} for both the SU(2) and SU(3) groups. Let us first mention that the singularity that we found at leading order (in addition to the one at $T_c$) seems to have disappeared. Interestingly, the Polyakov displays a similar non-monotonous behavior as the Polyakov loop evaluated on the lattice, approaching its large temperature limit from above. However, the rise of the Polyakov loop above the transition is faster than on the lattice, which seems to be a common feature of many calculations in the continuum. 

We mention that the Polyakov loop computed here is the bare Polyakov loop. The latter is known not to have a proper continuum limit even though, in dimensional regularization, and at the present level of accuracy, it is UV finite. A direct comparison to the lattice results requires the computation of the renormalized Polyakov, including the running of parameters with the temperature. This calculation is beyond the scope of the present manuscript. For an evaluation within the functional renormalization group, see \cite{Herbst:2015ona}.

\subsection{Vicinity of the transition}
\begin{figure}[t]
\begin{center}
\hspace{0.75cm}\includegraphics[height=0.32\textheight]{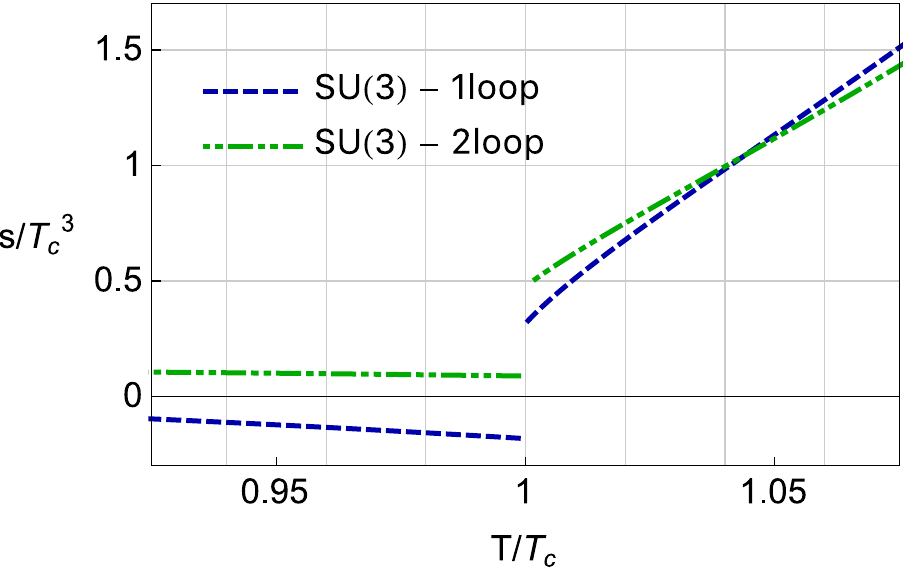}
\mycaption[Polyakov loops]{SU($3$) entropy density at one- and two-loop order, in the vicinity of $T_c$.}
\label{fig:entropy_2loop}
\end{center}
\end{figure}

Now that we have the two-loop corrections to the potential, we can revisit the vicinity of the transition. We find that the slight violation of the entropy density that was found at one-loop has disappeared, see Fig.~\ref{fig:entropy_2loop}. However, we find a slight change of the latent which remains roughly $1/3$ of the lattice value.

\subsection{Low temperature behavior}
A simple analysis \cite{} shows that, in the absence of RG improvement at least, the two-loop corrections are sub-leading with respect to the one-loop ones. This seems a good news since, as we saw, the latter are thermodynamically consistent despite being dominated by the ghosts. However, one important problem remains since these contributions are massless and contribute to the thermodynamical observables in powers of $T$, a behavior which is not observed on the lattice.

This problem is not specific to the Curci-Ferrari model and occurs in fact in most continuum approaches in the Landau gauge (that do not throw the ghost degrees of freedom by hand), see for instance \cite{Quandt:2017poi} and illustrates their inability, so far, to properly exclude the massless degrees of freedom from the thermodynamical quantities.

\clearemptydoublepage
%
%
%
%
\let\textcircled=\pgftextcircled
\chapter{More on the relation between the center symmetry group and the deconfinement transition}
\label{chap:more}

\initial{T}he confinement/deconfinement transition in pure Yang-Mills theories has been studied for a variety of gauge groups using lattice simulations \cite{Moriarty:1981wc,Green:1984ks,Batrouni:1984vd} or non-perturbative continuum methods \cite{Braun:2010cy}. Not only is this useful as a benchmark for the various theoretical approaches to non-abelian gauge theories, but it also helps refining the connection between the confinement/deconfinement transition, the Polyakov loop, and center symmetry breaking. Here, we provide a general discussion of these questions in the case of the SU($N$) gauge group and investigate them in more detail in the case of the SU($4$) gauge group using the background extended Curci-Ferrari model. 

In Chapter \ref{chap:generalities}, we saw that, if center symmetry is realized in the Wigner-Weyl sense, the Polyakov loop associated to the defining representation needs to vanish. We also assumed the reciprocal property to hold true. This is far from obvious, however, because one could imagine scenarios where only a subgroup of the center symmetry group is broken, in which case the defining Polyakov loop would still be equal to zero. Although this is not a possibility for any prime value of $N$ since $\mathds{Z}_N$ does not have any non-trivial subgroup in this case,\footnote{This includes the cases $N=2$ and $N=3$ treated so far.} for other values of $N$, the question needs to be posed how to characterize the Wigner-Weyl realization of center symmetry in terms of order parameters. To this purpose, it is convenient to consider Polyakov loops in representations other than the defining representation, and introduce the concept of $N$-ality. We shall then start with a brief review of these notions (see \cite{Greensite:2011zz,Cohen:2014swa} for a more thorough discussion) before summarizing the results obtained within the Curci-Ferrari model for the SU($4$) gauge group.

\section{Polyakov loops in other representations}
To a given representation $R:t^a\mapsto t^a_R$ of dimension $d_R$, one associates a temporal Wilson line as
\begin{equation}
 L_{A^R}(\vec{x})\equiv {\cal P}\,\exp\left\{i\int_0^\beta d\tau\, A_0^a(\tau,\vec{x})\,t^a_R\right\}.
\end{equation}
The corresponding Polyakov loop is defined as
\begin{equation}
 \ell_R\equiv \big\langle \Phi_{A^R}(\vec{x})\big\rangle_{YM}\equiv\frac{\int {\cal D}A\,{\cal \Phi}_{A^R}(\vec{x})\,e^{-S_{YM}[A]}}{\int {\cal D}A\,e^{-S_{YM}[A]}}\,,
\end{equation}
with $\Phi_{A^R}(\vec{x})\equiv \mbox{tr}\,L_{A^R}(\vec{x})\,/\,d_R$.

\subsection{$N$-ality of a representation}
To any representation, one can associate its $N$-ality that characterizes the way the corresponding Polyakov loop transforms under center transformations. In general, a given representation $R$ can be obtained by decomposing tensor products of the form $N\otimes N\otimes\dots\bar N\otimes\dots$ involving a certain number $n$ of defining representations and a certain number $\bar n$ of contragredient representations. Using the two properties
\begin{eqnarray}\label{eq:char}
L_{A_{R_1\oplus R_2}} & \!\!\!\!\!\!=\!\!\!\! & L_{A_{R_1}}\oplus L_{A_{R_2}}\,,\\
L_{A_{R_1\otimes R_2}} & \!\!\!\!\!\!=\!\!\!\! & L_{A_{R_1}}\otimes L_{A_{R_2}}\,,
\end{eqnarray}
it is then easily seen that all representations extracted from the same tensor product transform in the same way under center transformations. In place of the transformation rule (\ref{eq:l_transfo}), we now have
\begin{equation}\label{eq:lR_transfo}
\ell^R_{\alpha-\frac{2\pi}{N}k}=e^{i\frac{2\pi}{N}\nu_R k}\ell^R_\alpha\,,
\end{equation}
where $\nu_R\equiv n-\bar n$ is known as the $N$-ality of the representation. In fact, the $N$-ality is defined only modulo multiples of $N$ and, for convenience, we shall make the choice $1\leq\nu_R\leq N$ in what follows. 

Of particular interest are those representations whose $N$-ality divides $N$. Indeed, it is easily seen in this case that the values of $k$ that make the phase equal to $1$ in Eq.~(\ref{eq:lR_transfo}) are multiples of $N/\nu_R$. The corresponding elements of $\mathds{Z}_{N}$ are of the form $e^{i\frac{2\pi}{\nu_R}k'}$ and constitute, therefore, the $\mathds{Z}_{\nu_R}$ subgroup of $\mathds{Z}_N$.\footnote{For a generic $N$-ality $\nu_R$, the phase in Eq.~(\ref{eq:lR_transfo}) is equal to $1$ for any $k$ multiple of $\mbox{lcm}\,(N,\nu_R)\,/\,\nu_R$, where $\mbox{lcm}\,(p,q)$ stands for the least common multiple of $p$ and $q$. This corresponds to elements of $\mathds{Z}_N$ in the subgroup $\mathds{Z}_{N\nu_R\,/\,\mbox{\scriptsize lcm}\,(N,\nu_R)}=\mathds{Z}_{\mbox{\scriptsize gcd}\,(N,\nu_R)}$, where $\mbox{gcd}\,(p,q)$ stands for the greatest common divisor of $p$ and $q$.} It follows that the vanishing of $\ell_R$ does not tell us anything about the way center symmetry is realized in this subgroup. However, and interestingly, it tells us that center symmetry is necessarily realized in the Wigner-Weyl sense by some elements of $\mathds{Z}_N$ not included in $\mathds{Z}_{\nu_R}$.\footnote{We are here implicitly excluding the presence of symmetries other than the center that would make the Polyakov loops vanish. We mention also that Polyakov loops could vanish accidentally, without the need of a symmetry. However, if this happens, it most probably does for isolated values of the external parameters (temperature, \dots) and we should not consider these exceptional cases in the characterization of center symmetry to be given below.}

\subsection{Center symmetry characterization}
Let us now use the previous notions in order to find a characterization of center symmetry.\\

Recall that the subgroups of $\mathds{Z}_N$ are the groups $\mathds{Z}_\nu$ with $1\leq \nu\leq N$ a divisor of $N$, and that, for each $\nu$ dividing $N$, there is only one such subgroup in $\mathds{Z}_N$, with $\mathds Z_1=\{1\}$ and $\mathds{Z}_{\nu=N}=\mathds{Z}_N$. In order to characterize the Wigner-Weyl realization of center symmetry, it is then enough to consider a collection of representations that exhaust all $N$-alities $1\leq\nu<N$ that divide $N$. Indeed the vanishing of the Polyakov loops in all these representations implies that center symmetry needs to be realized for elements that do not belong to any of the strict subgroups of $\mathds{Z}_N$. Since these elements generate $\mathds{Z}_N$, any element of $\mathds{Z}_N$ will then realize the symmetry in the Wigner-Weyl sense. In summary:\\

\vglue-11mm

\noindent{$\_\_\_\_\_\_\_\_\_\_\_\_\_\_\_\_\_\_\_\_\_\_\_\_\_\_\_\_\_\_\_\_\_\_\_\_\_\_\_\_\_\_\_\_\_\_\_\_\_\_\_\_\_\_\_\_\_\_\_\_\_\_\_\_\_\_\_$}\\
Center symmetry is manifest in the Wigner-Weyl sense iff all Polyakov loops vanish within a collection of representations covering all possible $N$-alities dividing $N$ but not equal to $N$ (i.e. $\nu\,|\,N$ and $\nu\neq N$).\\
\vglue-12mm
\noindent{$\_\_\_\_\_\_\_\_\_\_\_\_\_\_\_\_\_\_\_\_\_\_\_\_\_\_\_\_\_\_\_\_\_\_\_\_\_\_\_\_\_\_\_\_\_\_\_\_\_\_\_\_\_\_\_\_\_\_\_\_\_\_\_\_\_\_\_$}\\

\vglue-7mm

\noindent{We stress that this characterization does not include representations with $\nu_R=N$, also known as representations with {\it vanishing $N$-ality} (since $N$ equals $0$ modulo $N$). The Polyakov loops in such representations are not transformed under any center transformation and, therefore, are not constrained to vanish.\footnote{In particular, the present formalism does not allow to address the confinement of colored objects in representations with vanishing $N$-ality, such as gluons.} It is then no surprise that they do not appear in the characterization of center symmetry. Let us also mention that, once the criterion applies, the Polyakov loops in any representation with non-vanishing $N$-ality need to vanish.\\

In fact, the characterization of the center can be further simplified as follows. The set of divisors of $N$ strictly smaller than $N$ is naturally equipped with a partial ordering corresponding to the relation {\it being a divisor of.} We call {\it maximal $N$-alities,} the maximal elements for this partial ordering over this set.\footnote{For instance, if $N=24=2^3\times 3$, then the maximal $N$-alities are $8=2^3$ and $12=2^2\times 3$. If $N=3$, the only maximal $N$-ality is $1$.} We can now state that}\\

\vglue-11mm

\noindent{$\_\_\_\_\_\_\_\_\_\_\_\_\_\_\_\_\_\_\_\_\_\_\_\_\_\_\_\_\_\_\_\_\_\_\_\_\_\_\_\_\_\_\_\_\_\_\_\_\_\_\_\_\_\_\_\_\_\_\_\_\_\_\_\_\_\_\_$}\\
Center symmetry is manifest in the Wigner-Weyl sense iff all Polyakov loops vanish within in a collection of representations covering all possible maximal $N$-alities.\\
\vglue-12mm
\noindent{$\_\_\_\_\_\_\_\_\_\_\_\_\_\_\_\_\_\_\_\_\_\_\_\_\_\_\_\_\_\_\_\_\_\_\_\_\_\_\_\_\_\_\_\_\_\_\_\_\_\_\_\_\_\_\_\_\_\_\_\_\_\_\_\_\_\_\_$}\\

\vglue-7mm

\noindent{Indeed, the union of the corresponding subgroups is equal to the union of all the strict subgroups of $Z_N$. Then, the vanishing of the Polyakov loops with maximal $N$-alities leads to the same consequences than the vanishing of the Polyakov loops with non-vanishing $N$-alities dividing $N$.\\

Finally, we mention that the characterization of center symmetry can also be formulated in terms of the confinement of color charges: the Wigner-Weyl realization of center symmetry is equivalent to the confinement of color charges within a collection of representations covering all possible maximal $N$-alities. Again, once the criterion is obeyed, all types of color charges with non-vanishing $N$-ality are confined.

\subsection{Fundamental representations}
The prototypes for representations covering all possible $N$-alities are the so-called {\it fundamental representations} obtained by anti-symmetrizing successive tensor products of the defining representation.\footnote{Another, equivalent possibility, followed by \cite{Dumitru:2012fw}, is to consider expectation values of traced powers of the defining Wilson line, $\langle\mbox{tr}\,L_A^k\rangle/N^k$.} These correspond to the one-column Young tableaux with a number of boxes equal to the $N$-ality. From those, we need only to consider those with maximal $N$-alities.\footnote{We mention that the fundamental representations are naturally organized in pairs of $N$-alities $(\nu,N-\nu)$. The representations in each pair are contragredient of each other which, in the case of pure Yang-Mills theory, implies that the corresponding Polyakov loops are equal to each other, see Chapter \ref{chap:bg}. This is in line with the fact that, in the above characterization of center symmetry, at most one of the representations in each pair is considered, since when $\nu$ divides $N$, $N-\nu$ does in general not divide $N$ (the only exception is when $N$ is even and $\nu=N/2=N-\nu$ but, in this case, there is anyway only one representation in the pair).}

To take an example consider SU($4$). There are three fundamental representations $4$, $6$ and $\bar 4$, of respective $N$-alities $1$, $2$ and $3$. According to the previous criterion, in order to fully characterize the center, we need only to consider the representation $6$ whose $N$-ality $2$ is the only maximal divisor of $4$ (within the set of divisors of $4$ not equal to $4$). The subgroup which is not probed by $\ell_6$ is $\mathds{Z}_2$. Since this is the only non-trivial subgroup of $\mathds{Z}_4$, the vanishing of $\ell_6$ characterizes entirely the Wigner-Weyl realization of $\mathds{Z}_4$. In contrast, the vanishing of $\ell_4$ is not enough to characterize the centre since this could also mean a partial breaking of the symmetry via the $\mathds{Z}_2$ subgroup.

In practice, the breaking pattern depends on the dynamics and, therefore, on the considered model and approximation. Its determination requires an explicit calculation. In Sec.~\ref{sec:pattern}, we shall investigate the breaking pattern in the SU($4$) case using the Curci-Ferrari model in the presence of a background. Before doing so, in Sec.~\ref{sec:su4_weyl}, we review the properties of the Weyl chambers in this case, in particular the location of the center invariant states. We recall that, in this framework, center symmetry is characterized by the fact that the background is found at particular points in the Weyl chambers, those points that are left invariant under the action of the center symmetry group on the Weyl chambers. In fact, it is enough to look for points in the Weyl chamber that are invariant under a center element that generates the complete center symmetry group, in other words, a center element that does not belong to any of the strict subgroups of the center symmetry group.  Such transformations are precisely those that act non-trivially on the Polyakov loops with maximal $N$-ality, and, therefore, if the background is located at such invariant points, the corresponding background dependent Polyakov loops vanish, which in turn is enough to ensure that the center symmetry is fully manifest.

\section{SU($4$) Weyl chambers}\label{sec:su4_weyl}
We saw in Chapter \ref{chap:symmetries} that, in the reduced background space ($\bar r\equiv r/4\pi$), the $SU(N)$ Weyl chambers are sub-pavings of the parallelepipeds generated by $N-1$ weights $\rho^{(j)}$ of the defining representation, with $1\leq j\leq N-1$, dual to the roots $\alpha^{(j)}=\rho^{(j)}-\rho^{(N)}$, with $1\leq j\leq N-1$, such that $\alpha^{(j)}\cdot\rho^{(k)}=\delta^{jk}/2$. To construct the Weyl chambers, one needs to study how these parallelepipeds are subdivided by each network of hyperplanes orthogonal to a given of the remaining roots, $\rho^{(j)}-\rho^{(j')}$ with $1\leq j<j'\leq N-1$, and translated by multiples of half that root. The corresponding hyperplane passing through the origin contains $\rho^{(k)}$ for $k\neq j$ and $k\neq j'$, as well as $\rho^{(j)}+\rho^{(j')}$.

The defining weights of SU($N$) can be found in App.~\ref{app:sun}. For SU($4$), we may choose
\begin{equation}\label{eq:we}
\rho^{(1)}=\frac{1}{2}\left(
\begin{array}{c}
-1\\
\sqrt{1/3}\\
\sqrt{1/6}
\end{array}\right)\,, \quad
\rho^{(2)}=\frac{1}{2}\left(
\begin{array}{c}
0\\
-\sqrt{4/3}\\
\sqrt{1/6}
\end{array}\right)\,,\quad
\rho^{(3)}=\frac{1}{2}\left(
\begin{array}{c}
0\\
0\\
-\sqrt{3/2}
\end{array}\right)\,,
\end{equation}
the remaining weight being $\rho^{(4)}=-(\rho^{(1)}+\rho^{(2)}+\rho^{(3)})$. Consider the fundamental parallelepiped generated by the $\rho^{(j)}$ with $1\leq j\leq 3$. It is further divided in Weyl chambers by the planes containing $\rho^{(3)}$ and $\rho^{(1)}+\rho^{(2)}$, $\rho^{(1)}$ and $\rho^{(2)}+\rho^{(3)}$, and finally $\rho^{(2)}$ and $\rho^{(3)}+\rho^{(1)}$, as shown in Fig.~\ref{fig:SU4_center}. Altogether they divide the fundamental parallelepiped into six tetrahedra, see Fig.~\ref{fig:SU4_center}. A simple calculation shows that these tetrahedra have two nonadjacent edges that are longer than the other four, by a factor of $2/\sqrt{3}$.\\

\begin{figure}[h]
\begin{center}
\includegraphics[height=0.25\textheight]{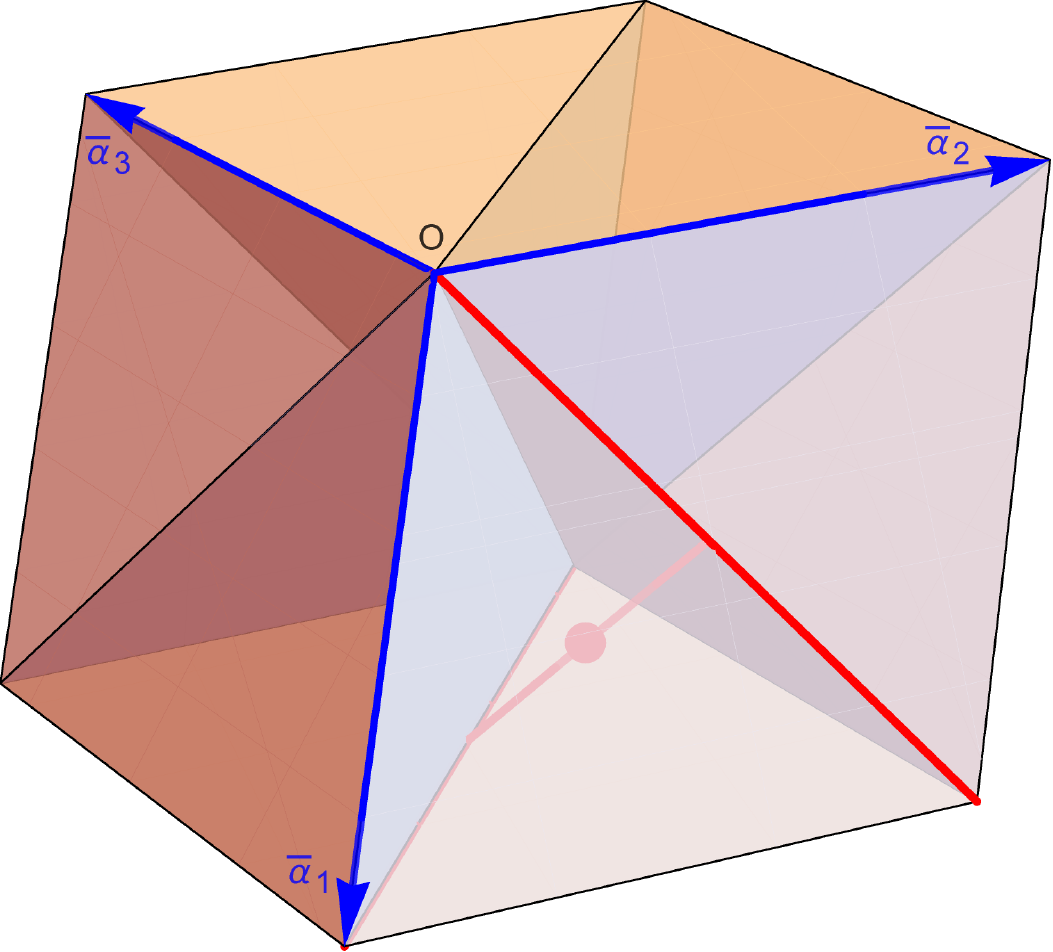}\\
\mycaption[SU($4$) Weyl chambers and center-symmetric states.]{The fundamental parallelepiped and its division into SU($4$) Weyl chambers in the restricted background space. The vectors $\bar\alpha^{(1)}$, $\bar\alpha^{(2)}$, $\bar\alpha^{(3)}$ (blue) define the fundamental parallelepiped. The latter is composed of six tetrahedral Weyl chambers:  $\{O,\bar\alpha^{(1)},\bar\alpha^{(1)}+\bar\alpha^{(2)},\bar\alpha^{(1)}+\bar\alpha^{(2)}+\bar\alpha^{(3)}\}$, $\{O,\bar\alpha^{(2)},\bar\alpha^{(1)}+\bar\alpha^{(2)},\bar\alpha^{(1)}+\bar\alpha^{(2)}+\bar\alpha^{(3)}\}$ (blue), $\{O,\bar\alpha^{(1)},\bar\alpha^{(1)}+\bar\alpha^{(3)},\bar\alpha^{(1)}+\bar\alpha^{(2)}+\bar\alpha^{(3)}\}$, $\{O,\bar\alpha^{(3)},\bar\alpha^{(1)}+\bar\alpha^{(3)},\bar\alpha^{(1)}+\bar\alpha^{(2)}+\bar\alpha^{(3)}\}$ (red), $\{O,\bar\alpha^{(2)},\bar\alpha^{(2)}+\bar\alpha^{(3)},\bar\alpha^{(1)}+\bar\alpha^{(2)}+\bar\alpha^{(3)}\}$, $\{O,\bar\alpha^{(3)},\bar\alpha^{(2)}+\bar\alpha^{(3)},\bar\alpha^{(1)}+\bar\alpha^{(2)}+\bar\alpha^{(3)}\}$ (yellow). Each chamber has two edges longer than the other four, namely the one connecting the origin to the sum of two $\bar\alpha$'s and the one connecting one $\bar\alpha$ to the sum of the three $\bar\alpha$'s, as illustrated in the figure (red lines).}
\label{fig:SU4_center}
\end{center}
\end{figure}

\subsection{Symmetries}
Such an irregular tetrahedron, made of four identical isosceles triangles, is called a tetragonal disphenoid, whose symmetry group is the dihedral group $D_{2d}$, with eight elements. These are the identity, the three rotations by an angle $\pi$ around any of the axes which relate the midpoints of nonadjacent edges, the reflections about the two planes perpendicular to one of the long edges and containing the other, and two other elements that can be obtained by combining the former. Using the method described in Chapter \ref{chap:symmetries}, it is easily seen that these transformations correspond to the center transformations, charge conjugation, or any combination of these, as we now discuss in more details. 

Let us first consider center transformations. As we have seen in Chapter \ref{chap:symmetries}, any translation along the defining weights $\rho^{(j)}$ corresponds to a center transformation with center element $e^{-i2\pi/N}=-i$. In order to generate the other non-trivial transformations, we consider sums of two or three weights. For a given Weyl chamber connected to the origin, these correspond to translations along the three edges connected to the origin. It is easily seen that among those, the translation along the longer edge corresponds to a center element $-1$ and is associated to a rotation by an angle $\pi$ around the axis that connects the midpoints of the two long edges of the Weyl chamber. The other two center transformations, corresponding to $\pm i$, are obtained by combining the two other rotations with any of the reflection planes described above. Following the same method, one can see that charge conjugation corresponds to one of the reflection planes---the other one being a combination of charge conjugation and the center transformation associated to $-1$.

Let us mention finally that one very convenient way to guess the various geometrical transformations of the Weyl chamber associated to physical transformations is to notice that the values taken by the leading order background-dependent Polyakov loop in the defining representation at the nodes of the Weyl chamber span the center of the group (this is because it equals $1$ at the origin and translations along the edges of the Weyl chamber span all possible center transformations). Upon a given physical transformation, these values are permuted in a certain way, which in turn allows one to infer the corresponding geometrical transformation of the Weyl chamber, together with its invariant points.

\subsection{Invariant states}
As mentioned above, center symmetry is characterized once the invariant states for a center transformations that generates the whole center symmetry group have been identified. Any such center transformation does the job and here these correspond to the center elements $i$ or $-i$. These transformations have only one fixed point, the barycenter of the tetrahedron,\footnote{In general, for SU($N$), it is pretty obvious that the barycenter of any Weyl chamber is invariant under any center transformation \cite{Dumitru:2012fw} and in fact under any physical transformation (commuting with the periodic gauge transformations in the sense of Eq.~(\ref{eq:toto})). A less trivial question is to identify all possible invariant states associated to a given physical transformation. Here we shall limit our analysis to the case of SU($4$).} which is then the only point compatible with center symmetry. 

In contrast, the center transformation with center element $-1$ has a whole line of fixed points, the one connecting the midpoints of the two long edges, which contains in particular the barycenter of the tetrahedron. For any point on this line, except for the barycenter, the Polyakov loop $\ell_4$ vanishes but not $\ell_6$, corresponding to a scenario where center symmetry is partially broken down to the subgroup $\mathds{Z}_2$. The complete breaking of the symmetry corresponds to points away from this line.

\section{One-loop results}\label{sec:pattern}
It is convenient to locate the states in the fundamental parallelepiped in terms of their coordinates in the basis $\rho^{(j)}$. For a restricted background $r$, we write $r=x_j\bar\alpha^{(j)}$, where the $0\leq x_j\leq 1$ can be obtained as $2\pi x_j=r\cdot\alpha^{(j)}$. In this basis, the center-symmetric point represented in Fig.~\ref{fig:SU4_center}  (dot in the figure)  is located at the coordinates $(3/4,1/2,1/4)$. The $\mathds{Z}_2$ invariant line in the same Weyl chamber is defined by the equations $x_1=x_2+x_3$ and $x_2=1/2$. To simplify the discussion, we note that we can restrict to charge conjugation invariant states. In the considered Weyl chamber these are the points in the plane $x_1=x_2+x_3$. In fact, it is even more convenient to work in a basis of this plane $r=y_2(\bar\alpha^{(1)}+\bar\alpha^{(2)})+y_3(\bar\alpha^{(1)}+\bar\alpha^{(3)})$. The confining point is located at $(1/2,1/4)$ and the $\mathds{Z}_2$ line is corresponds to $y_2=1/2$.
 
\subsection{Deconfinement transition}
Figure~\ref{fig:su4} shows contour plots of the potential in this plane. At low temperatures, the minimum of the potential is at the center-symmetric point (upper panels), whereas we find a $Z_4$ quadruplet of degenerate minima at high temperatures (lower panels), located pairwise in the two reflection-symmetry planes of the tetrahedron but away from the line of $Z_2$ symmetry. The breaking of center symmetry is thus complete. We have checked that the transition is first order with $T_c^{LO}/m\approx 0.367$, close to the value obtained for SU($3$) at LO.

\begin{figure}[t!]  
\begin{center}
\includegraphics[height=0.2\textheight]{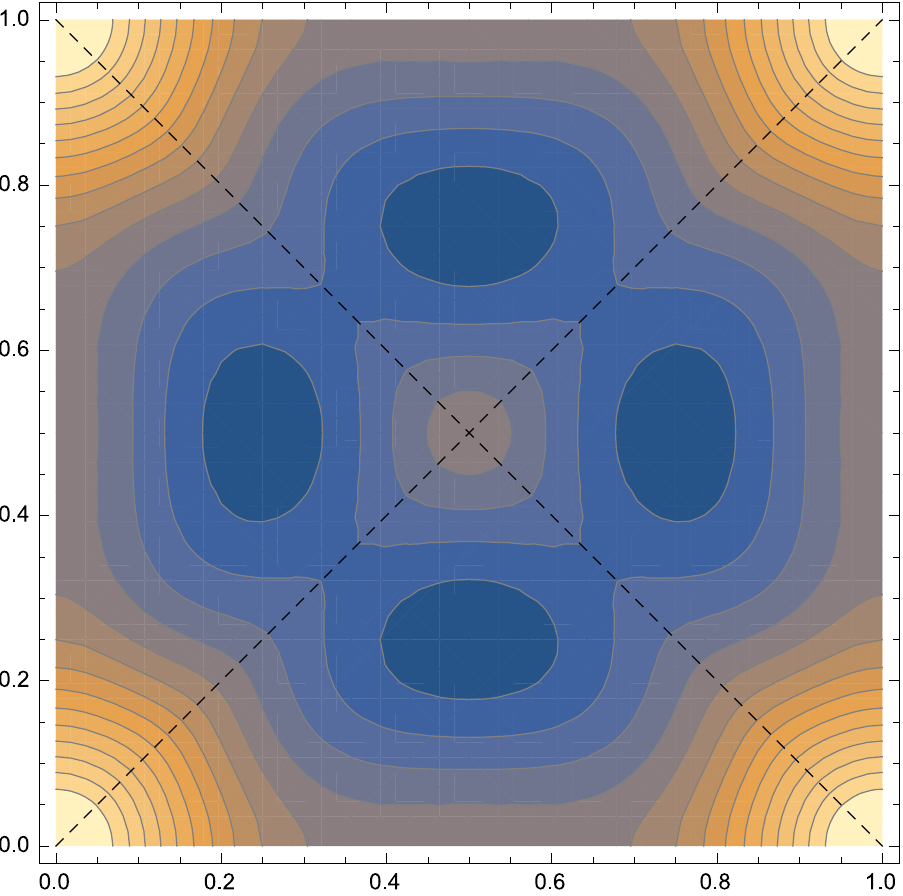}\quad\includegraphics[height=0.2\textheight]{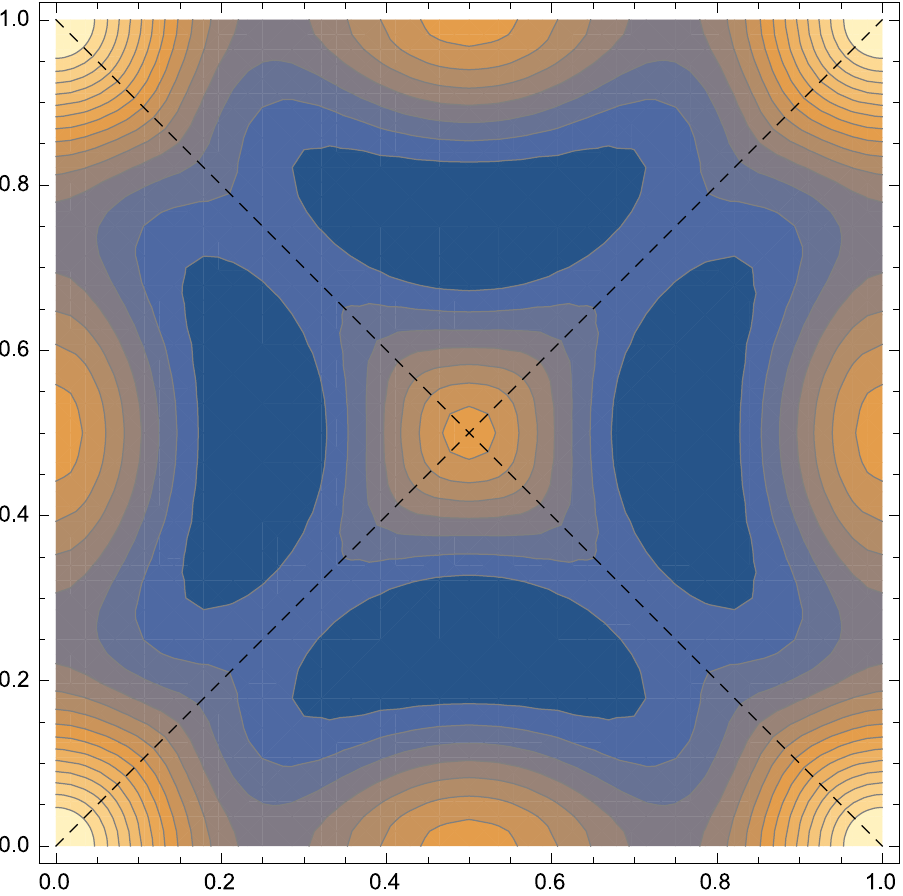}\quad\includegraphics[height=0.2\textheight]{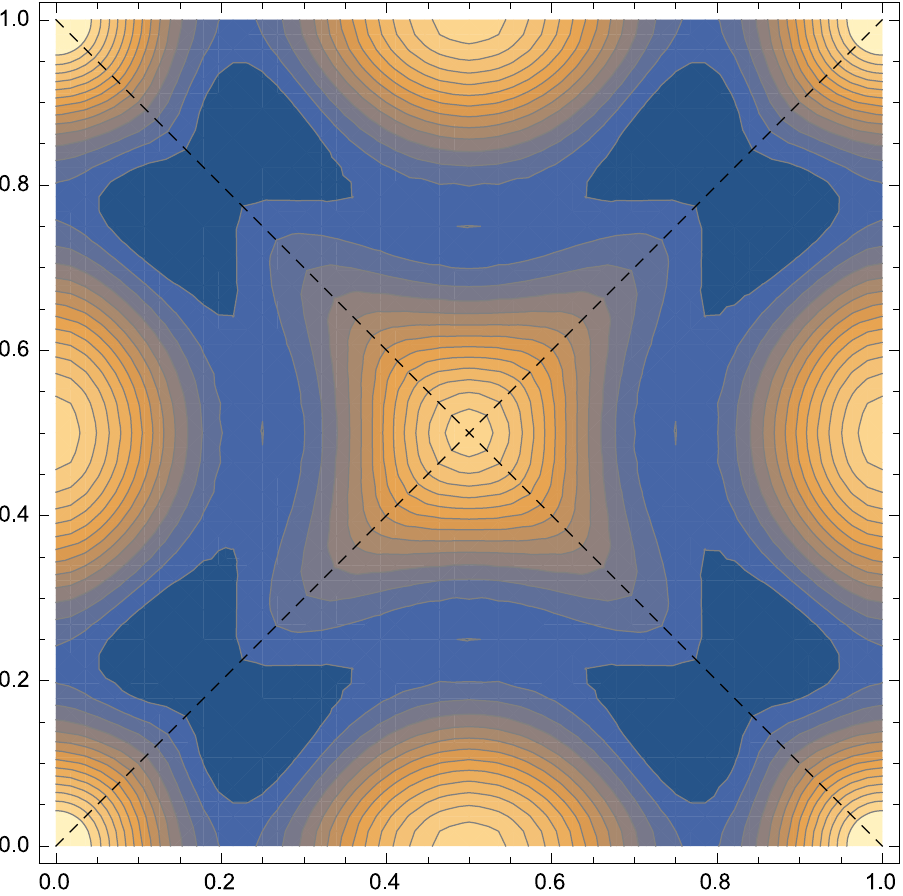}\\
\vspace{0.6cm}
\includegraphics[height=0.2\textheight]{./fig08/su4_pot_37.pdf}\quad\includegraphics[height=0.2\textheight]{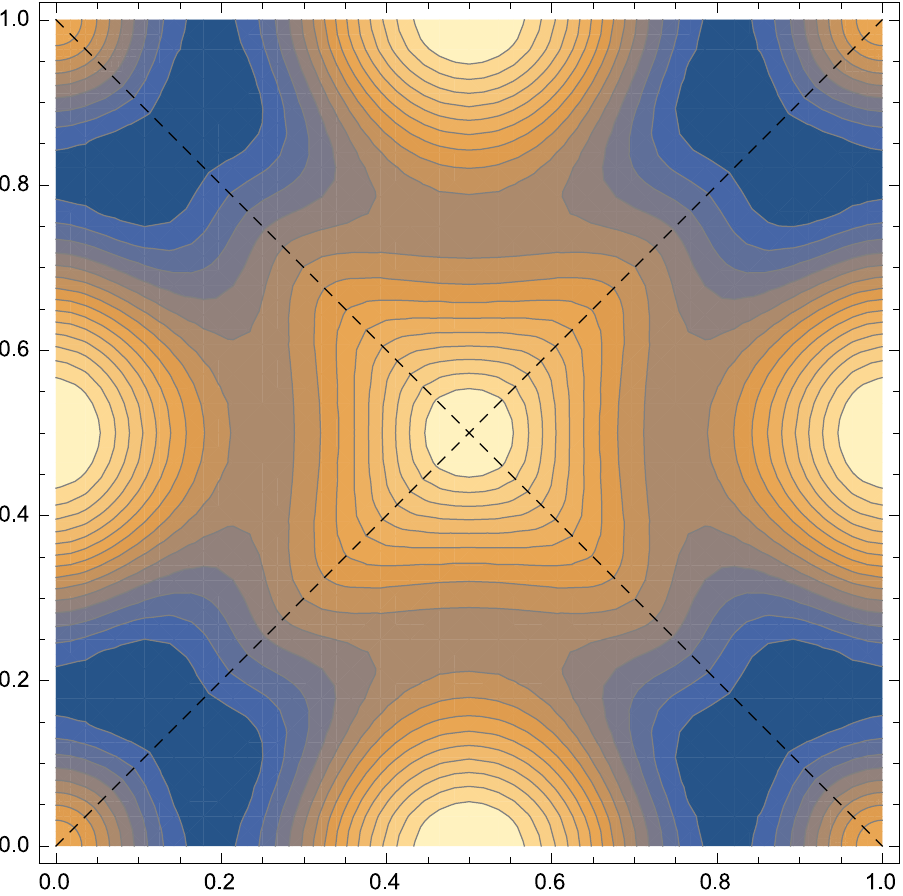}\quad\includegraphics[height=0.2\textheight]{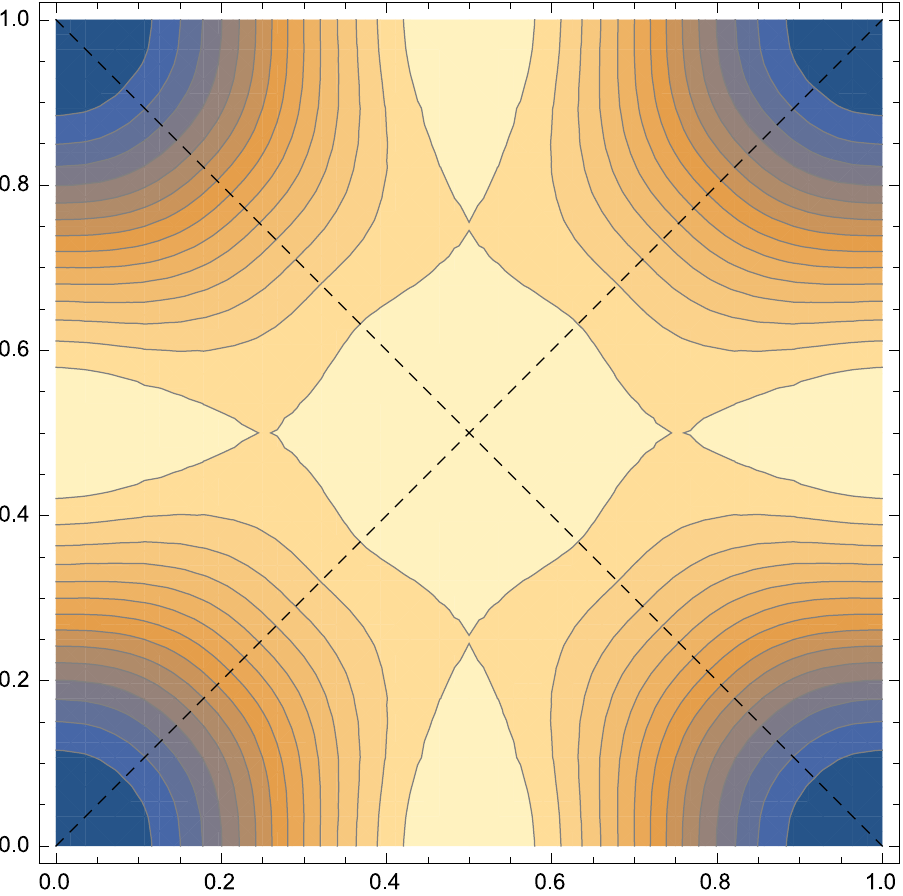}
 \caption{Contour plots of the SU($4$) potential at LO in a charge-conjugation-invariant plane. The latter intersects four Weyl chambers separated by dashed lines. The $\mathds{Z}_2$ invariant lines are the vertical and horizontal lines (not shown) passing through the origin. The upper (respectively lower) plots correspond to temperatures below (respectively above) the transition temperature.}\label{fig:su4}
\end{center}
\end{figure}

\subsection{Background-dependent Polyakov loops}
It is interesting to evaluate the (background-dependent) fundamental Polyakov loops. The one in the defining representation writes
\begin{equation}
\ell(r)=\frac{1}{4}\Big[e^{i\rho^{(1)}\cdot r}+e^{i\rho^{(2)}\cdot r}+e^{i\rho^{(3)}\cdot r}+e^{i\rho^{(4)}\cdot r}\Big]\,.
\end{equation}
Using the fact that $\alpha^{(j)}=\rho^{(j)}-\rho^{(4)}$ and that $\alpha^{(1)}+\alpha^{(2)}+\alpha^{(3)}=-4\rho^{(4)}$ this rewrites
\begin{equation}
\ell(r)=\frac{e^{-i\frac{\pi}{2}(x_1+x_2+x_3)}}{4}\Big[1+e^{2\pi i x_1\cdot r}+e^{2\pi i x_2\cdot r}+e^{2\pi i x_3\cdot r}\Big]\,,
\end{equation}
in terms of the coordinates in the fundamental parallelepiped. Then, we can interpret the vanishing of $\ell(r)$ as the closure of a rhombus with external angles $2\pi x_1$, $2\pi(x_2-x_1)$, $2\pi(x_3-x_2)$ and $2\pi(x_1-x_3)$ modulo $2\pi$ and up to permutations of $x_1$, $x_2$, and $x_3$. This implies $2\pi x_1+2\pi(x_2-x_1)=\pi$ and $2\pi x_1=2\pi(x_3-x_2)$ modulo $2\pi$, that is,
\begin{equation}
\label{eq_zero_pol_1}
x_2=\frac{1}{2}\,\,\mbox{mod}\,1\quad \mbox{and} \quad x_3=x_1-\frac{1}{2}\,\,\mbox{mod}\,1\,.
\end{equation}
In a given Weyl chamber, this corresponds to the segment joining the centers of the long edges of the tetrahedron, to which belong, in particular, the center-symmetric point. The corresponding segments in the other chambers in the fundamental parallelepiped are obtained by permutations of $x_1$, $x_2$ and $x_3$. This are precisely the states where the defining Polyakov loop needs to vanish if the subgroup $\mathds{Z}_2$ of center transformations is realized in the Wigner Weyl sense.\\

Let us now evaluate the Polyakov loop associated to the representation $6$. Since fundamental representations are constructed by anti-symmetrizing tensor products of defining representations, the weights of the fundamental representation with $N$-ality $\nu$ are obtained by considering all the possible sums of $\nu$ different weights. There are $C_N^\nu$ such weights, which is precisely the dimension of the considered fundamental representation. It follows that, in general
\begin{equation}
\ell_{C_N^\nu}(r)=\frac{1}{C_N^\nu}\sum_{j_1<\cdots<j_\nu}e^{i\big(\rho^{(j_1)}+\cdots+\rho^{(j_\nu)}\big)\cdot r}\,.
\end{equation}
Here we have
\begin{eqnarray}
\ell_6(r) & \!\!\!\!\!=\!\!\!\!\! & \frac{1}{6}\Big[e^{i(\rho^{(1)}+\rho^{(2)})\cdot r}+e^{i(\rho^{(1)}+\rho^{(3)})\cdot r}+e^{i(\rho^{(1)}+\rho^{(4)})\cdot r}\\
& & \hspace{1.5cm}+\,e^{i(\rho^{(2)}+\rho^{(3)})\cdot r}+e^{i(\rho^{(2)}+\rho^{(4)})\cdot r}+e^{i(\rho^{(3)}+\rho^{(4)})\cdot r}\Big].\nonumber
\end{eqnarray}
Using again that $\rho^{(1)}+\rho^{(2)}+\rho^{(3)}+\rho^{(4)}=0$, this rewrites
\begin{equation}
\ell_6(r)=\frac{1}{3}\mbox{Re}\,\Big[e^{i(\rho^{(1)}+\rho^{(4)})\cdot r}+e^{i(\rho^{(2)}+\rho^{(4)})\cdot r}+e^{i(\rho^{(1)}+\rho^{(4)})\cdot r}\Big]
\end{equation}
which is not surprising since $6=\bar 6$. Using the same remarks as before this also rewrites as
\begin{eqnarray}
\ell_6(r) & \!\!\!\!\!=\!\!\!\!\! & \frac{1}{3}\mbox{Re}\,e^{-i\pi(x_1+x_2+x_3)}\Big[e^{2\pi ix_1}+e^{2\pi ix_2}+e^{2\pi ix_3}\Big]\\
& \!\!\!\!\!=\!\!\!\!\! & \frac{1}{3}\Big[\cos(\pi(x_1-x_2-x_3))\nonumber\\
&  & \hspace{0.5cm}+\,\cos(\pi(x_2-x_3-x_1))\nonumber\\
& & \hspace{0.5cm}+\,\cos(\pi(x_3-x_1-x_2))\Big]\nonumber
\end{eqnarray}
or, equivalently
\begin{eqnarray}
\ell_6(r) & \!\!\!\!\!=\!\!\!\!\! & \frac{1}{6}\mbox{Re}\,\Big[(e^{\pi i x_1}+e^{-\pi i x_1})(e^{\pi i x_2}+e^{-\pi i x_2})(e^{\pi i x_3}+e^{-\pi i x_3})\\
& & \hspace{1.2cm}-\,e^{\pi i(x_1+x_2+x_3)}-e^{-\pi i(x_1+x_2+x_3)}\Big]\nonumber\\
& \!\!\!\!\!=\!\!\!\!\! & \frac{1}{3}\Big[4\cos(\pi x_1)\cos(\pi x_2)\cos(\pi x_3)-\cos(\pi(x_1+x_2+x_3))\Big].\nonumber
\end{eqnarray}
This defines a surface in the Weyl chamber, containing in particular the center-symmetric point.

We mention that the fact that the Polyakov loop $\ell_6$ can vanish away from the center-symmetric point seems in contradiction with the discussion given above. However, this discussion concerned the physical Polyakov loop, whereas here we are discussing the background dependent Polyakov loop which corresponds to a situation where one artificially imposes the state of the system to correspond to a given background. The system does not necessarily explores these states and, if it does, there is no symmetry principle that would enforce it to remain in such a state for a certain range of temperatures. We can disregard these accidental vanishings for the present discussion. We also expect the location of these accidental vanishing points in the Weyl chamber to change, or even disappear, at next-to-leading order.

\section{Casimir scaling}
Another application of Polyakov loops in higher representations is the so-called Casimir scaling. It has been observed that the Polyakov loops in different representations obey the following scaling law \cite{Gupta:2007ax}
\begin{equation}
\ell_R^{1/C_R}=\ell_{R'}^{1/C_{R'}}\,,
\end{equation}
where $C_R$ denotes the Casimir of the representation $R$. This in turn implies that the free-energy cost for bringing a colored object into the medium scales like the square of the corresponding color charge.

To investigate Casimir scaling at leading order in the SU($3$) case, we use (\ref{eq:char}). Taking the traces and normalizing by the dimensions of the representations, we arrive at\footnote{These are similar to the identities obeyed by the characters of the corresponding representations.}
\begin{eqnarray}
 (d_1+d_2)\,\Phi_{A_{R_1\oplus R_2}} & \!\!\!\!\!=\!\!\!\!\! & d_1\Phi_{A_{R_1}}+ d_2 L_{A_{R_2}}\,,\\
  \Phi_{A_{R_1\otimes R_2}} & \!\!\!\!\!=\!\!\!\!\! & \Phi_{A_{R_1}}\Phi_{A_{R_2}}\,.
\end{eqnarray}
At leading order, these properties are transferred to the corresponding Polyakov loops. Moreover, if $\bar R$ denotes the contragredient of a given representation $R$, we can once again show that the Polyakov loops are equal and real (on a charge conjugation invariant state). To take a few examples, we know that $3\otimes 3=\bar 3\oplus 6$. From this it follows that $9\ell_3^2=3\ell_3+6\ell_6$ and then $6\ell_6=9\ell_3^2-3\ell_3$. Similarly, one finds
\begin{eqnarray}
8\ell_8 & \!\!\!\!\!=\!\!\!\!\! & 9\ell_3^2-1\,,\\
10\ell_{10} & \!\!\!\!\!=\!\!\!\!\! & 18\ell_3\ell_6-8\ell_8\,,\\
15\ell_{15} & \!\!\!\!\!=\!\!\!\!\! & 18\ell_3\ell_6-3\ell_3\,,\\
15\ell_{15'} & \!\!\!\!\!=\!\!\!\!\! & 30\ell_3\ell_{10}-15\ell_{15}\,,\\
24\ell_{24} & \!\!\!\!\!=\!\!\!\!\! & 30\ell_3\ell_{10}-6\ell_6\,,\\
27\ell_{27} & \!\!\!\!\!=\!\!\!\!\! & 36\ell_6^2-8\ell_8-1\,,
\end{eqnarray}
see \cite{Gupta:2007ax} for more details.

In Fig.~\ref{fig:casimir}, we show how Casimir scaling is satisfied in our approach. The scaling is well satisfied down to $\simeq 1.1T_c$. For temperatures below, we observe partial Casimir scaling, with certain groups of representations obeying scaling ($6$ and $8$, $10$ and $15$, $15'$, $24$ and $27$, \dots).

\begin{figure}[t]
\begin{center}
\hspace{0.8cm}\includegraphics[height=0.32\textheight]{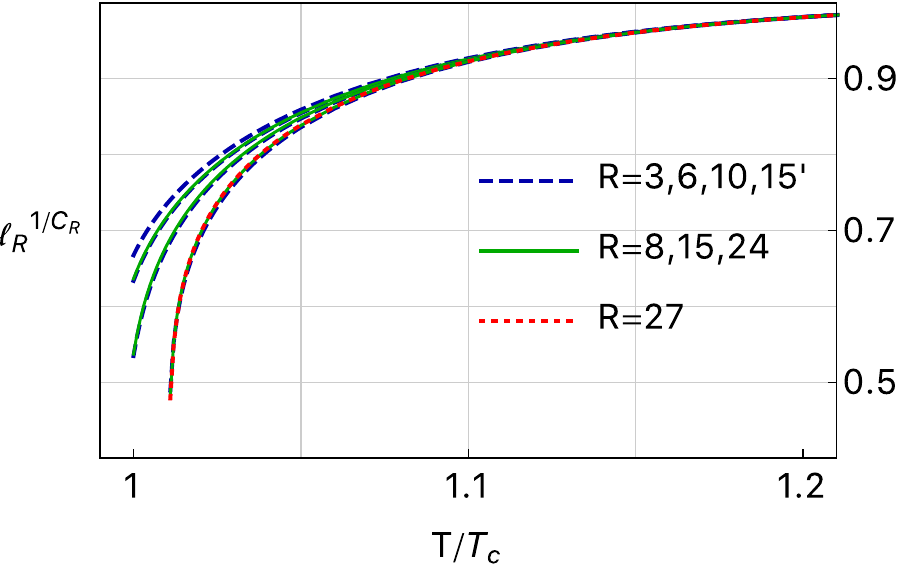}
\mycaption[Casimir scaling]{Testing the Casimir scaling at leading order.}
\label{fig:casimir}
\end{center}
\end{figure}

\clearemptydoublepage
%
%
%
%
\let\textcircled=\pgftextcircled
\chapter{Background field gauges:\\ Adding quarks and density}
\label{chap:bfg_quarks}

\initial{I}n the next two chapters, we pursue our investigation of the deconfinement transition by complementing the background-extended Curci-Ferrari model with quark degrees of freedom. Including quarks is in principle straightforward since one has simply to add the usual matter contribution of the QCD action. In practice, however, the presence of quarks leads to new difficulties that require one to revisit the very foundations of the background field method.

These difficulties concern in fact any continuum approach that relies, in one way or another, on the use of effective actions. They appear at finite density, as the functional integral measure becomes complex, and are of two different types: those that relate to the integration measure not being real-valued and those that relate to its non-positivity. We show how the first type can be cured by exploiting the symmetries of the system at hand. In contrast, the second type is more subtle as it connects directly to the sign problem of lattice QCD \cite{deForcrand:2010ys,Philipsen:2010gj}. A simple recipe will be postulated to handle this second type of difficulty, but its justification will remain an open question beyond the scope of the present manuscript.

\section{General considerations}
\label{sec:general}

In this section, we derive some general properties that will be used throughout the chapter. We do so by using the QCD action in its non gauge-fixed version. We shall later extend the discussion to the Landau-deWitt gauge and to the corresponding Curci-Ferrari model.\\

The QCD action is obtained by adding the usual matter contribution to the Yang-Mills action, $S_{YM}\to S_{QCD}=S_{YM}+\delta S$, with
\begin{equation}\label{eq:QCD_corr}
 \delta S[A,\psi,\bar\psi;\mu]\equiv\int d^dx\,\sum_{f=1}^{N_f}\bar\psi_f(x)\,\big( {\cal D}\!\!\!\!\slash +M_f-\mu\gamma_0\big)\,\psi_f(x)\,.
\end{equation}
Here, the quarks are taken in the fundamental representation and ${\cal D}_\mu=\partial_\mu-igA^a_\mu t^a$ is the corresponding covariant derivative. We have also introduced the usual notation ${\cal D}\!\!\!\!\slash\equiv \gamma_\mu{\cal D}_\mu$ with $\gamma_\mu$ the Euclidean Dirac matrices \cite{Laine:2016hma}. These are related to the standard Minkowski matrices $\gamma^\mu_M$ as $\gamma_0=\gamma_M^0$ and $\gamma_i=-i\gamma_M^i$. They are thus Hermitian and satisfy the anti-commutation relations $\{\gamma_\mu,\gamma_\nu\}=2\delta_{\mu\nu}$. In the following, we work in the Weyl  basis, where $\gamma_{0,2}^*=\gamma_{0,2}^t=\gamma_{0,2}$ and $\gamma_{1,3}^*=\gamma_{1,3}^t=-\gamma_{1,3}$. Within this basis, one easily derives the two identities \cite{Reinosa:2015oua}
\begin{equation}
 \gamma_2\gamma_0\gamma_\mu\gamma_0\gamma_2=-\gamma_\mu^{\mbox{\tiny t}} \quad \mbox{and} \quad \gamma_3\gamma_1\gamma_\mu\gamma_1\gamma_3=\gamma_\mu^*\,,
\end{equation}
which will play a major role in the following discussion.

The conjugated field $\bar\psi$ is defined as usual as $\bar\psi\equiv\psi^\dagger\gamma_0$. It is convenient to extend such conjugation to complex numbers and matrices as $\bar z\equiv z^*$ and $\overline{M}\equiv \gamma_0M^\dagger\gamma_0$.  In particular, it is readily checked that $\overline{\gamma}_0=\gamma_0$ and $\overline{\gamma}_i=-\gamma_i$. Moreover, given two matrices $M$ and $N$, one finds $\smash{\overline{MN}=\overline{N}\,\overline{M}}$. These properties allow for an easy determination of the conjugate of any expression, similar to the determination of the bra associated to a ket in quantum mechanics. We also introduce $\gamma_5\equiv\gamma_0\gamma_1\gamma_2\gamma_3$. It obeys the properties $\{\gamma_5,\gamma_\mu\}=0$, $\gamma_5^2=1$, $\gamma_5^\dagger=\gamma_5$ and $\overline{\gamma}_5=-\gamma_5$.\\

Finally, let us recall that the last term in Eq.~(\ref{eq:QCD_corr}) corresponds to the baryonic charge of the system. It is proportional to the associated chemical potential $\mu_B\equiv 3\mu$.\footnote{Note that we use the opposite convention for the sign of $\mu$ than in Ref.~\cite{Reinosa:2015oua}.} For reasons that shall become clear below, we allow for both real and imaginary chemical potentials. As usual, the temperature $T\equiv 1/\beta$ enters via the boundary conditions of the fields, and, unlike gluons, quarks obey anti-periodic boundary conditions: $\psi(\tau+\beta,\vec{x})=-\psi(\tau,\vec{x})$.

\subsection{Polyakov loops}
Our investigation of the deconfinement transition will again be achieved by means of the Polyakov loops associated to the defining and contragredient representations,\footnote{This is because we are considering the SU($3$) case. As we discuss in a future version of the manuscript, the case of SU($N$) requires one to introduce a priori all the fundamental loops.}
\begin{equation}
 \ell\equiv\left\langle\frac{1}{N}\,\mbox{tr}\,{\cal P}\exp\left\{i\int_0^\beta d\tau\, A^a_0(\tau,\vec{x})\,t^a\right\}\right\rangle
\end{equation}
and
\begin{equation}
 \bar\ell\equiv\left\langle\frac{1}{N}\,\mbox{tr}\,{\cal P}\exp\left\{-i\int_0^\beta d\tau\, A^a_0(\tau,\vec{x})\,t_a^{\mbox{\tiny t}}\right\}\right\rangle.
\end{equation}
Those are directly related to the free-energy cost for bringing a quark or an anti-quark into an equilibrated bath of quarks and gluons \cite{Svetitsky:1985ye,Philipsen:2010gj}:
\begin{equation}
 \ell\propto e^{-\beta\Delta F} \quad \mbox{and} \quad  \bar\ell\propto e^{-\beta\Delta \bar F}\,.
\end{equation} 
We mention that this interpretation has been questioned in the presence of a chemical potential due to the non-monotonous behavior of the Polyakov loops as the chemical potential is varied \cite{Dumitru:2003hp}. We will argue instead in the next chapter that the non-monotony as a function of $\mu$ can be seen precisely as a consequence of the free-energy interpretation.

\subsection{Symmetries}
The other crucial ingredients in the analysis to follow are of course the symmetries of the problem. As we now recall, those are notably modified by the presence of quarks in the fundamental representation.\\

First of all, center symmetry is explicitly broken. This is because the fundamental quarks transform as $\psi^U(x)=U(x)\,\psi(x)$. So, even though the action remains invariant, $S[A^U,\psi^U,\bar\psi^U;\mu]=S[A,\psi,\bar\psi;\mu]$, the anti-periodic boundary conditions of the quarks are changed into anti-periodic boundary conditions modulo a phase, $\psi^U(\tau+\beta,\vec{x})=-e^{i\frac{2\pi}{N}k}\psi^U(\tau,\vec{x})$, and the symmetry is not manifest at the level of the quantum/thermal expectation values. As a consequence, the Polyakov loops have no reason to vanish anymore. We will see nonetheless that the Polyakov loops possess a singular behavior in the heavy quark regime which qualifies them still as good order parameters.

We expect quarks not to break charge conjugation symmetry as long as the chemical potential is equal to zero. This is easily checked by evaluating the action with charge conjugated fields\footnote{The transformation of the field $\bar\psi$ is easily obtained from the conjugation rule described above: $ \psi^{\cal C}=\gamma_0\gamma_2\bar\psi^{\mbox{\tiny t}}\to \bar\psi^{\cal C}=\overline{\bar\psi^{\mbox{\tiny t}}} \overline{\gamma}_2\overline{\gamma}_0=-\psi^{\mbox{\tiny t}}\gamma_2\gamma_0$, where we have used that $\overline{\gamma}_0=\gamma_0$, $\overline{\gamma}_2=-\gamma_2$ and $\overline{\bar\psi^{\mbox{\tiny t}}}=\psi^{\mbox{\tiny t}}$.}
\begin{equation}
 A^{\cal C}\equiv -A^{\mbox{\tiny t}}\,, \quad \psi^{\cal C}\equiv\gamma_0\gamma_2\bar\psi^{\mbox{\tiny t}}\,, \quad \bar\psi^{\cal C}\equiv-\psi^{\mbox{\tiny t}}\gamma_2\gamma_0\,,
\end{equation}
and by using the property $\gamma_2\gamma_0\gamma_\mu\gamma_0\gamma_2=-\gamma_\mu^{\mbox{\tiny t}}$. More generally, it is easily verified that charge conjugation changes the sign of the baryonic charge term in the action, and thus that 
\begin{equation}
 S[A^{\cal C},\psi^{\cal C},\bar\psi^{\cal C};\mu]=S[A,\psi,\bar\psi;-\mu]\,.
\end{equation} 
Since charge conjugation does not alter the boundary conditions of the fields, the symmetry applies also to expectation values. In particular, it is easily shown that, as a consequence of charge conjugation, $\ell(\mu)=\bar\ell(-\mu)$. For $\mu=0$, we recover the result $\bar\ell=\ell$ obtained in Chapter \ref{chap:bg}.\footnote{Here, we assume that there is no spontanously broken symmetry (as we have seen above, center symmetry is explicitly broken in the presence of fundamental quarks). A counterexample is provided by Yang-Mills theory where center symmetry can be spontanously broken. In that case, $\ell=\bar\ell$ does not apply to all possible states of the system.} 

Another useful transformation is
\begin{equation}
 A^{\cal K}\equiv A^*\,, \quad \psi^{\cal K}\equiv\gamma_1\gamma_3\psi, \quad \bar\psi^{\cal K}\equiv\bar\psi\gamma_3\gamma_1\,.
\end{equation}
Indeed, owing to the property $\gamma_3\gamma_1\gamma_\mu\gamma_1\gamma_3=\gamma_\mu^*$ and assuming that the fermionic fields are real,\footnote{The Grasmannian fields $\psi$ and $\bar\psi$ are a formal device to write the determinant of an operator, $\mbox{det}\,M=\int {\cal D}[\psi,\bar\psi]\,\exp\left\{\int d^dx\int d^d\!y\,\bar\psi(x) M(x,y)\psi(y)\right\}$. Since $(\mbox{det}\,M)^*=\mbox{det}\,M^*$, it is perfectly consistent to assume that $\psi^*=\psi$ and $\bar\psi^*=\bar\psi$, see also \cite{Reinosa:2015oua}.} it is easily checked that 
\begin{equation}
 S[A^{\cal K},\psi^{\cal K},\bar\psi^{\cal K};\mu]=S[A,\psi,\bar\psi;\mu^*]^*\,.
\end{equation} 
Although this is strictly speaking not a symmetry for it requires the complex conjugation of the action, it plays a major role in what follows. It implies for instance that $\ell(\mu)=\ell(\mu^*)^*$ and $\bar\ell(\mu)=\bar\ell(\mu^*)^*$ and, thus, that $\ell$ and $\bar\ell$ are both real if the chemical potential is real. 

For completeness, let us finally combine the previous two transformations into the transformation
\begin{equation}
 A^{\cal K \cal C}=A\,, \quad \psi^{\cal K\cal C}=-\gamma_5\bar\psi^{\mbox{\tiny t}}\,, \quad \bar\psi^{\cal K\cal C}= \psi^{\mbox{\tiny t}}\gamma_5\,.\label{eq:g5t}
\end{equation}
Using  $\{\gamma_5,\gamma_\mu\}=0$ together with $\gamma_\mu^\dagger=\gamma_\mu$, one finds 
\begin{equation}
 S[A,\psi^{\cal K \cal C},\bar\psi^{\cal K\cal C};\mu]=S[A,\psi,\bar\psi;-\mu^*]^*\,.
\end{equation} 
In particular, it follows that $\ell(\mu)=\bar\ell(-\mu^*)^*$ and, thus, $\ell$ and $\bar\ell$ become complex conjugate of each other if the chemical potential is imaginary. The transformation (\ref{eq:g5t}) should not be mistaken with $\psi\to\gamma_5\psi$, $\bar\psi\to-\bar\psi\gamma_5$ which leaves the chemical potential unchanged, while flipping the sign of all quark masses. We shall also make use of this property below.

\subsection{Fermion determinant}
Let us end this section by recalling that the fermionic part of the action being quadratic, the fermionic fields can be integrated out exactly. This means that any observable can be written formally as the corresponding integral in the pure Yang-Mills theory, but with a functional measure modified by the so-called fermion determinant. For instance, the partition function reads
\begin{equation}\label{eq:Z}
 Z=\int {\cal D}A\,\Delta[A;\mu]\,e^{-S_{YM}[A]}\,,
\end{equation}
with
\begin{equation}
 \Delta[A;\mu]\equiv \int {\cal D}[\psi,\bar\psi]\,e^{-\delta S[\psi,\bar\psi,A;\mu]}=\mbox{det}\,{\cal M}[A;\mu]
\end{equation} 
the determinant of the Dirac operator in the presence of the gauge field $A$ and a chemical potential $\mu$:
\begin{equation}
 {\cal M}[A;\mu]=\bigotimes_{f=1}^{N_f}\,(\partial\!\!\!\!\slash-gA\!\!\!\!\slash +M_f-\mu\gamma_0)\,.
\end{equation} 
A similar rewriting applies of course to the Polyakov loops.\\

Now, using similar arguments as in the previous section, one finds
\begin{eqnarray}
 \gamma_2\gamma_0\,{\cal M}[A;\mu]\,\gamma_0\gamma_2 & \!\!\!\!\!=\!\!\!\!\! & {\cal M}[A^{\cal C};-\mu]^{\mbox{\scriptsize t}}\,,\\
 \gamma_3\gamma_1\,{\cal M}[A;\mu]\,\gamma_1\gamma_3 & \!\!\!\!\!=\!\!\!\!\! & {\cal M}[A^{\cal K};\mu^*]^*,\\
 \gamma_5\,{\cal M}[A;\mu]\,\gamma_5 & \!\!\!\!\!=\!\!\!\!\! & {\cal M}[A;-\mu^*]^\dagger\,.\label{eq:Dirac3}
\end{eqnarray}
Again, the last identity was obtained  by combining the previous two. If we use instead the chiral transformation $\psi\to\gamma_5\psi$, $\bar\psi\to-\bar\psi\gamma_5$, we obtain that $-\gamma_5\,{\cal M}[A;\mu]\,\gamma_5$ is the original Dirac operator but with a sign flip of all fermion masses:
\begin{equation}\label{eq:gamma5}
 -\gamma_5\,{\cal M}[A;\mu]\,\gamma_5=\bigotimes_{f=1}^{N_f}\,(\partial\!\!\!\!\slash-gA\!\!\!\!\slash-M_f-\mu\gamma_0)\,.
\end{equation}
For the fermion determinant, we obtain, correspondingly,
\begin{eqnarray}
 \Delta[A;\mu] & \!\!\!\!\!=\!\!\!\!\! & \Delta[A^{\cal C};-\mu]\,,\label{eq:det1}\\
 \Delta[A;\mu] & \!\!\!\!\!=\!\!\!\!\! & \big(\Delta[A^{\cal K};\mu^*]\big)^*\,,\label{eq:det2}\\
 \Delta[A;\mu] & \!\!\!\!\!=\!\!\!\!\! & \big(\Delta[A;-\mu^*]\big)^*\,.\label{eq:det3}
\end{eqnarray}

\section{Continuum sign problem(s)}\label{sec:72}
As it can be seen from Eq.~(\ref{eq:det2}), the fermion determinant becomes a priori complex when the chemical potential is taken real. This is the source of various difficulties, usually collected under the generic name of {\it continuum sign problems} \cite{Fukushima:2006uv,Roessner:2006xn,Rossner:2007ik,Mintz:2012mz,Stiele:2016cfs,Kovacs:2016juc,Folkestad:2018psc} and which we now discuss in some detail.\\

We put special care into discriminating between true sign problems that emanate from the non-positivity of the fermion determinant, and related difficulties that emanate from its non-real-valuedness. To make the distinction clear, we label the first type as (Pn) and the second type as (Rn). As we argue, this latter type of difficulties are easily handled after one acknowledges the fact that, in the presence of a complex integration measure, a given observable (with real spectrum) does not necessarily lead to a real expectation value. In contrast, the former type are more difficult to handle since they relate directly to the lattice sign problem.\\

In order to better appreciate the various possible difficulties, we first consider the case of an imaginary chemical potential where these difficulties are absent.\footnote{As we discuss in the next chapter, the case of imaginary chemical potential possesses further interesting features that make it a case worth of study.} Also, we first discuss the various problems in a non gauge-fixed setting, in a way similar to \cite{Dumitru:2005ng}, and then extend the discussion to background field gauges.

\pagebreak

\subsection{Imaginary chemical potential}
In the case of an imaginary chemical potential, the fermion determinant is positive. To see this \cite{Fukushima:2006uv}, we note first that the massless Dirac operator is anti-hermitian. It is thus diagonalizable with a purely imaginary spectrum. Moreover, the mass operator being proportional to the identity, the massive Dirac operator remains diagonalizable. Finally, it follows from Eq.~(\ref{eq:gamma5}) and $\gamma_5^2=1$ that, for each eigenstate of eigenvalue $\mu+i\lambda$, there is another eigenstate of eigenvalue $\mu-i\lambda$. The fermion determinant is then the product of positive factors of the form $\mu^2+\lambda^2$.\footnote{For completeness, we mention that the real-valuedness of the fermion determinant (but not its positivity) follows more directly from Eq.~(\ref{eq:det3}).}

The positivity of the fermion determinant is a welcome feature at various levels. On a fundamental level, it ensures the positivity of the partition function (\ref{eq:Z}) and thus the real-valuedness of any thermodynamical observable derived from it (since $\ln Z$ is real).\footnote{We shall not try, for the moment, to give a physical interpretation to the case of imaginary chemical potential. We shall later show that, for certain values of the (imaginary) chemical potential, there is a simple physical interpretation.} On a practical level, it makes it possible to evaluate the partition function with, discrete, importance sampling Monte-Carlo techniques. As we now discuss, it has also its importance within any continuum approach based on the use of effective actions.

Suppose indeed that we introduce sources $\bar\eta$ and $\eta$ coupled respectively to the operators $\Phi_A$ and $\bar\Phi_A$, thus defining the generating functional $W[\eta,\bar\eta] \equiv\ln Z[\eta,\bar\eta]$, with
\begin{equation}
Z[\eta,\bar\eta] \equiv\int {\cal D}A\,\Delta[A;\mu]\,e^{-S_{\mbox{\tiny YM}}[A]}\exp\left\{\int d\vec{x}\,\Big(\bar\eta(\vec{x})\,\Phi_A(\vec{x})+\bar\Phi_A(\vec{x})\,\eta(\vec{x})\Big)\right\}.
\end{equation}
The sources are a practical way to generate the expectation values of the Polyakov loops or more generally correlations between various Polyakov loops. In the limit of zero sources, one can extract their values from the extremization of the Polyakov loop effective action, defined as the Legendre transform of $W[\eta,\bar\eta]$:
\begin{equation}
 \Gamma[\ell,\bar\ell]\equiv -W[\eta,\bar\eta]+\int d\vec{x}\,\Big(\bar\eta(\vec{x})\,\ell(\vec{x})+\bar\ell(\vec{x})\,\eta(\vec{x})\Big)\,,
\end{equation}
where the conjugated variables $\ell$ and $\bar\ell$ are defined as
\begin{equation}
 \ell(\vec{x})\equiv \frac{\delta W}{\delta\bar\eta(\vec{x})}\,, \quad \bar\ell(\vec{x})\equiv\frac{\delta W}{\delta\eta(\vec{x})}\,,
\end{equation} 
and are nothing but the Polyakov and anti-Polyakov loops in the presence of the sources $\eta$ and $\bar\eta$.

For the purpose of generating correlation functions, one could consider the sources to be two independent complex numbers, that is $(\eta,\bar\eta)\in\mathds{C}\times\mathds{C}$. Correspondingly, the variables $\ell$ and $\bar\ell$ that enter the effective action are complex and independent (possibly within some region of $\mathds{C}\times\mathds{C}$). This is not the most convenient choice, however, because the effective action becomes generically complex, obscuring the fact that it should be real in the limit of zero sources where it corresponds to $-W[0,0]=-\ln Z$. Moreover, a complex effective action makes it difficult to devise a definite criterion to identify the physical extremum.\footnote{In principle, one expects that the limit of zero sources corresponds to one extremum only, or to a collection of degenerated extrema. However, due to the approximations inherent to any approach, additional extrema can appear and one needs a criterion to identify the correct one.} 

As we now explain, we can avoid these difficulties by restricting $(\eta,\bar\eta)$ to a subspace of $\mathds{C}\times\mathds{C}$ such that $\Gamma[\ell,\bar\ell]$ is real. This restriction is of course allowed as long as the considered subspace contains the limit of zero sources.\footnote{We assume again that the limit of zero sources does not depend on the way it is taken, which is legitimate here since there is no spontanously broken symmetry.}  However, one problem with restricting the space of sources is that it is not always easy to identify the target space where the conjugated variables $\ell$ and $\bar\ell$ should vary. Interestingly, we will show that it is possible to find a subspace of $\mathds{C}\times\mathds{C}$ which is both stabilized by the Legendre transformation and over which the effective action is real.\\

Suppose that we restrict the sources such that $\bar\eta=\eta^*$, that is the pair $(\eta,\bar\eta)$ is taken in the subspace $\Sigma\equiv\left\{(\eta,\bar\eta)\in\mathds{C}\times\mathds{C}\,|\,\bar\eta=\eta^*\right\}$ of complex conjugated sources. This choice seems natural here since the quantities the sources are coupled to are also complex conjugate of each other, $\bar\Phi_A=\Phi_A^*$.  We emphasize, however, that this very choice will not be the natural one in the case of a real chemical potential to be treated in the next section. In the present case, this choice of sources, together with the real-valuedness of the fermion determinant, implies that $Z[\eta,\bar\eta]$ is real: 
\begin{equation}
 \boxed{\mbox{(R1)} \quad \forall (\eta,\bar\eta)\in\Sigma\,, Z[\eta,\bar\eta]\in\mathds{R}}\,\,. 
\end{equation} 
A stronger result is obtained from the positivity of the fermion determinant: $Z[\eta,\bar\eta]$ is in fact positive over $\Sigma$. It follows that the generating functional $W[\eta,\bar\eta]$ is real over $\Sigma$: 
\begin{equation}
 \boxed{\mbox{(P1)} \quad \forall (\eta,\bar\eta)\in\Sigma\,, W[\eta,\bar\eta]\in\mathds{R}}\,\,.
\end{equation}
\vglue4mm

Next, we consider the effective action $\Gamma[\ell,\bar\ell]$. It is easily seen that, if the sources are taken in the subspace $\Sigma$, the conjugated variables belong to the same space: 
\begin{equation}
 \boxed{\mbox{(R2)} \quad (\eta,\bar\eta)\in\Sigma \Rightarrow (\ell,\bar\ell)\in\Sigma}\,\,.
\end{equation} 
This result relies crucially on the real-valuedness of the fermion determinant. Indeed, in the presence of a real-valued integration measure, the expectation values of two complex conjugated quantities, such as $\Phi_A$ and $\bar\Phi_A$, are themselves complex conjugate of each other. One can also see (R2) as a consequence of (R1) and the identities $\ell=(\delta Z/\delta\bar\eta)/Z$ and $\bar\ell=(\delta Z/\delta\eta)/Z$. Moreover, by combining the properties (P1) and (R2), we deduce that $\Gamma[\eta,\bar\eta]$ is real over the space $\Sigma$: 
\begin{equation}
 \boxed{\mbox{(P2)} \quad \forall (\ell,\bar\ell)\in\Sigma\,, \Gamma[\ell,\bar\ell]\in\mathds{R}}\,\,.
\end{equation} 

\pagebreak

Finally, from the positivity of the integration measure, we can use similar arguments as those given in Appendix \ref{app:convexity} and in Chap.~\ref{chap:bg} to show that $W[\eta,\bar\eta]$ is convex over $\Sigma$ and that the the limit of zero sources corresponds to the minimization of $\Gamma[\ell,\bar\ell]$ over this space: 
\begin{equation}
 \boxed{\mbox{(P3)} \quad (\eta,\bar\eta)\in\Sigma \to (0,0)\Leftrightarrow \mbox{min}_\Sigma\,\Gamma[\ell,\bar\ell]}\,\,.
\end{equation}

In summary, we have just shown that, in the case of an imaginary chemical potential, the real-valuedness of the fermion determinant ensures the existence of a subspace over which the functional $Z[\eta,\bar\eta]$ is real (R1) and, thus, such it is stabilized by Legendre transformation (R2). On the other hand, the positivity of the fermion determinant ensures that  the functionals $W[\eta,\bar\eta]$ and $\Gamma[\ell,\bar\ell]$ are real over this subspace (P1/P2) and that there exists an unambiguous characterization of the limit of zero sources that allows to select the correct extremum of the effective action over this subspace (P3). We mention that varying the effective action over $\Sigma$ is also convenient because it makes sure that the property of the physical Polyakov loops being complex conjugate of each other in the case of an imaginary chemical potential, see above, is automatically fulfilled.

\subsection{Real chemical potential}
In the case of a real chemical potential, both the reality and the positivity of the fermion determinant are lost and we expect the properties (R1), (R2), (P1), (P2) and (P3) not to hold true anymore, at least not in such a simple way as above. As we now recall, the properties (R1) and (R2) can be shown to still hold true with however a new invariant subspace. In contrast, to the best of our knowledge, the properties (P1), (P2) and (P3), for they quite crucially rely on the positivity of the fermion determinant, have not been extended so far to the case of a real chemical potential.\\

(R1/R2) For a real chemical potential, $Z[\eta,\bar\eta]$ is not real in general, even when restricted to $\Sigma$. Correspondingly, the Legendre transformation does not stabilize $\Sigma$. This problem, however, should not be qualified as a sign problem for it does not originate in the integration measure not being positive, but only in the integration measure not being real. Moreover, it has a simple solution, as originally suggested in the matrix model analysis of \cite{Dumitru:2005ng}. In fact, using the identity (\ref{eq:det2}) as well as $\Phi_{A^{\cal K}}=\Phi_A^*$, it is easily checked that $Z[\eta,\bar\eta]$ is real over the subspace $\mathds{R}\times\mathds{R}$ and thus that this subspace is stable under Legendre transformation. Moreover, this is the natural subspace where to consider the effective action, since as we have seen above, the physical Polyakov loops should be real in this case.

The above considerations give a formal basis to the standard rule on how to analyse the Polyakov loop effective action/potential (see for instance \cite{Roessner:2006xn,Rossner:2007ik,Kovacs:2016juc}), when changing from the case of an imaginary chemical potential to that of a real chemical potential: in contrast to the former case where $\ell$ and $\bar\ell$ are taken complex conjugate of each other, in the latter case, one chooses instead $\ell$ and $\bar\ell$ real and independent. This change of subspace is illustrated in Fig.~\ref{fig:mu_im_re}.\\

\begin{figure}[t]
\begin{center}
\includegraphics[width=.75\linewidth]{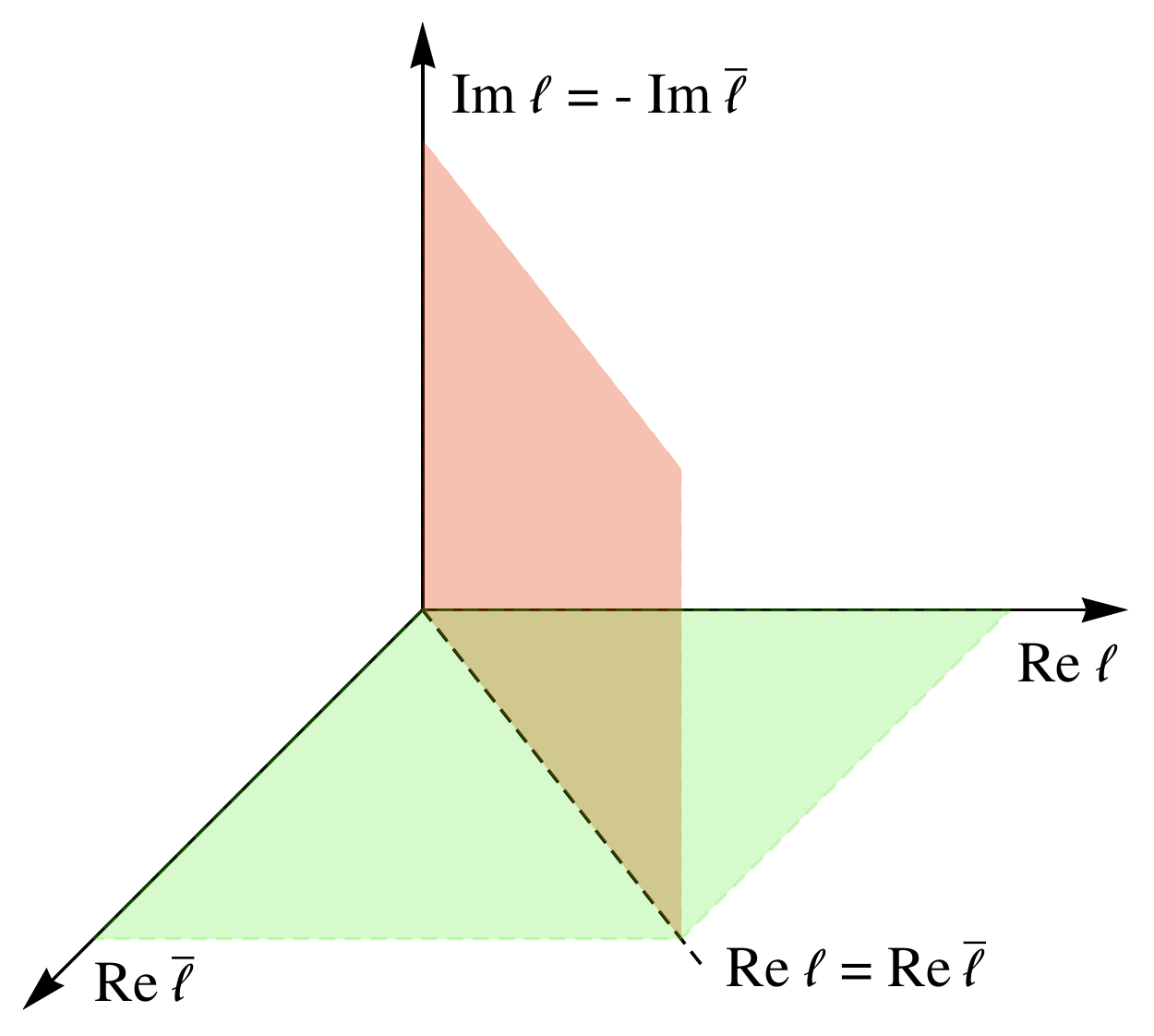}
\caption{The two different planes where the Polyakov loop effective action should be considered depending on whether $\mu\in i\mathds{R}$ or $\mu\in\mathds{R}$. These two planes are sub-manifolds of $\mathds{C}\times\mathds{C}$, but for the sake of representing them in a three dimensional figure, we have glued together the axes $\mbox{Im}\,\ell$ and $-\mbox{Im}\,\bar\ell$.}
\label{fig:mu_im_re}
\end{center}
\end{figure}

(P1/P2) As we have already emphasized, the question of the reality of $Z[\eta,\bar\eta]$, and in particular of the partition function $Z\equiv Z[\eta=0,\bar\eta=0]$, has not much to do with the sign problem and can be addressed using the identity (\ref{eq:det2}). The true sign problem appears when one tries to argue that the partition function $Z$ is positive, as it should be the case if $-T\ln Z$ is to represent the free-energy of the system. The fact that the fermion determinant is not positive-definite makes the proof of the positivity of $Z$ difficult a priori and, to our knowledge, this question has not been clarified yet. From the reality of $Z[\eta,\bar\eta]$, the (assumed) positivity of $Z$ and the expectation that $W[\eta,\bar\eta]$ should not diverge at any value of the sources, we also expect $Z[\eta,\bar\eta]$ to be always positive. Again, the non-positivity of the fermion determinant makes this statement difficult to prove. These issues are minor, however, in the sense that, within any continuum approach that gives access to an explicit expression for $\Gamma[\ell,\bar\ell]$, one can always check whether $\Gamma[\ell,\bar\ell]$ is real over $\mathds{R}\times\mathds{R}$. For instance, in the next chapter, we check that this is the case for the one-loop Polyakov loop potential computed within the Curci-Ferrari model.\\

(P3) A slightly more serious sign problem is that, even after checking that $\Gamma[\ell,\bar\ell]$ is real over $\mathds{R}\times\mathds{R}$, it is not at all obvious which extremum one should choose in this subspace, in the case where various extrema are present. The criterion that we identified in the case of an imaginary chemical potential, because it crucially relied on the positivity of the fermion determinant, is not valid anymore. In fact, it is easily seen that the physical point cannot correspond to the absolute minimum of the effective action. This is because, for a vanishing chemical potential, one can choose to work either over $\Sigma$ or over $\mathds{R}\times\mathds{R}$. Due to charge conjugation invariance, the physical point is such that $\ell=\bar\ell$ and lies therefore at the intersection of these two subspaces. Seen from the perspective of the subspace $\Sigma$, the physical point appears as the absolute minimum, but seen from the perspective of the subspace $\mathds{R}\times\mathds{R}$ it appears as a saddle-point.\\

We mention that this version of the sign problem is not as dramatic as the one on the lattice, because one can always try to locate the different extrema (assuming that there is a finite number of them) and select the correct one based on some additional physical insight. However, this entails inevitably a loss of prediction power of the continuum approach.  We can try to circumvent the problem by inferring a general rule from the case of zero chemical potential. In that case, the physical point in the subspace $\mathds{R}\times\mathds{R}$ is not only a saddle-point, it is in fact the deepest saddle point.\footnote{At least among those compatible with the charge conjugation invariance present in the $\mu=0$ case.} In what follows, we assume that this rule applies even at non-zero chemical potential. We stress however that we found no way to justify this rule from first principles. The best we can do is to provide a posteriori justifications, based on the relevance of the physical results that we obtain with this rule, in comparison to other strategies.

\section{Background field gauges}
The previous considerations extend of course also to a gauge-fixed setting. In particular, as we now discuss, they have a strong imprint on the way the background field method should be implemented at finite density.

\subsection{Complex self-consistent backgrounds}
To understand this point, recall that the background field approach at finite temperature relies crucially on the use of self-consistent backgrounds defined by the condition $\bar A=\langle A\rangle_{\bar A}$. Making the expectation value explicit and projecting along the generators $t^a$, this condition writes
\begin{equation}
 \bar A_\nu^a=\frac{\int {\cal D}_{\mbox{\tiny gf}}[A;\bar A]\,\Delta[A;\mu]\,A_\nu^a\,e^{-S_{YM}[A]}}{\int {\cal D}_{\mbox{\tiny gf}}[A;\bar A]\,\Delta[A;\mu]\,e^{-S_{YM}[A]}}\equiv\langle A_\nu^a\rangle_{\bar A}\,.
\end{equation}
Since the background is taken constant, temporal and along the Cartan sub-algebra, this identity is trivially valid for $\nu\neq 0$ or for $a\neq 3$ and $a\neq 8$. Self-consistent backgrounds appear, therefore, as fixed points of the mapping 
\begin{equation}
 (\bar A_0^3,\bar A_0^8)\mapsto(\langle A_0^3\rangle_{\bar A},\langle A_0^8\rangle_{\bar A})\,.\label{eq:mapping}
\end{equation}
Of course, the existence of fixed points depends crucially on which space the fixed points are searched for. If one views the background as a particular configuration of the gauge field $A_\mu^a$, the natural choice would be to look for fixed points in the space of real background components. As we now discuss, however, this choice is not always the appropriate one. The right choice depends on the considered situation and is again intimately related to the properties of the measure under the functional integral.\\

Let us consider the case of an imaginary chemical potential first. In this case, the fermion determinant is real. Then, if we choose the background components to be real, the measure under the functional integral is real. The expectation values $\langle A_0^3\rangle_{\bar A}$ and $\langle A_0^8\rangle_{\bar A}$ are also real and (\ref{eq:mapping}) maps $\mathds{R}\times\mathds{R}$ into itself. This is certainly a favourable condition for the existence of a fixed point solution, and, thus, of self-consistent backgrounds. 

The problem occurs when changing to a real chemical potential. Indeed, the fermion determinant being complex in this case, the expectation values $\langle A_0^3\rangle_{\bar A}$ and $\langle A_0^8\rangle_{\bar A}$ are not real anymore if we insist in keep the background components real, and the previous favourable condition for the existence of a (real) self-consistent background is not met. The way out is to allow for complex components of the background. This may look surprising at first sight if one insists in viewing the background as a particular configuration of the gauge-field. However, from the perspective of the gauge-fixing procedure the background should be rather seen as an infinite collection of gauge-fixing parameters that characterizes the particular gauge that one is considering.\footnote{Complex backgrounds have also been considered in \cite{Nishimura:2014rxa,Nishimura:2014kla}. There, the search for saddle points in the presence of a complex action forces one to continue the original real gauge field to complex configurations.} From this perspective, nothing prevents the background components to be taken complex.

To see how such an extended background allows us to solve the problem, we need the counterpart of the property (\ref{eq:det2}) regarding the gauge-fixed measure. One finds
\begin{equation}
 {\cal D}_{\mbox{\tiny gf}}[A;\bar A]=\big({\cal D}_{\mbox{\tiny gf}}[A^{\cal K};\bar A^{\cal K}]\big)^*\,.
\end{equation}
We note that, since we leave open the possibility of complex background components, we have $\bar A^{\cal K}\neq \bar A^{\cal C}$. More precisely, for our constant, temporal and diagonal background, we have $(\bar A^{\cal K}_0)^{3,8}=-(\bar A_0^{3,8})^*$, whereas $(\bar A^{\cal C}_0)^{3,8}=-(\bar A_0^{3,8})$. The background is thus invariant under ${\cal K}$ if its components are taken purely imaginary. In this case, it is readily checked that
\begin{equation}
 \big\langle A_0^{3,8}\big\rangle_{\bar A}=-\big\langle A_0^{3,8}\big\rangle_{\bar A}^*\,,
\end{equation}
and, therefore (\ref{eq:mapping}) maps $i\mathds{R}\times i\mathds{R}$ into itself. Thus, there is again some chance to find self-consistent backgrounds, but with purely imaginary components this time.

Another possibility is to consider $\bar A_0^3$ real and $\bar A_0^8$ imaginary. Indeed, in this case, the background is invariant under a combination of ${\cal K}$ and the Weyl transformation that flips the sign of $\bar A_0^3$. Since the Weyl transformation is a symmetry of the problem, it is readily checked in this case that
\begin{eqnarray}
 \big\langle A_0^{3}\big\rangle_{\bar A}=\big\langle A_0^{3}\big\rangle_{\bar A}^*\,,\\
 \big\langle A_0^{8}\big\rangle_{\bar A}=-\big\langle A_0^{8}\big\rangle_{\bar A}^*\,.
\end{eqnarray}
It follows that (\ref{eq:mapping}) maps $\mathds{R}\times i\mathds{R}$ into itself, opening the possibility for the existence of self-consistent backgrounds with $\bar A_0^3$ real and $\bar A_0^8$ imaginary.

\subsection{Background field effective potential}
The previous discussion extends to the background field effective potential, from which the self-consistent backgrounds should be obtained in principle. It is convenient to introduce the generating functional
\begin{equation}
 Z[J;\bar A_0,\mu]\equiv \int {\cal D}_{\mbox{\tiny gf}}[A;\bar A_0]\,\Delta[A;\mu]\,\exp\left\{-S_{YM}[A]+\int d^dx \,J^j(x)A_0^j(x)\right\}\,,
\end{equation}
with $J\equiv (J^3,J^8)$ and $\bar A_0\equiv (\bar A_0^3,\bar A_0^8)$. We recall that the background field effective action $\tilde\Gamma[\bar A_0]$ is constructed from this functional by first Legendre transforming $W[J;\bar A_0,\mu]\equiv-\ln Z[A_0;\bar A_0,\mu]$ with respect to the sources
\begin{equation}
 \Gamma[A_0;\bar A_0,\mu]=-W[J;\bar A_0,\mu]+\int d^dx\,J^j(x)A^j_0(x)\,,
\end{equation}
with
\begin{equation}\label{eq:A0def}
 A_0^j(x)\equiv \frac{\delta W}{\delta J^j(x)}=\frac{1}{Z}\frac{\delta Z}{\delta J^j(x)}\,,
\end{equation}
and then evaluating $\tilde\Gamma[\bar A_0,\mu]=\Gamma[\bar A_0;\bar A_0,\mu]$. We can now follow a similar discussion as the one presented in Sec.~\ref{sec:72}.

By using the identities (\ref{eq:det1})-(\ref{eq:det3}), it is found that
\begin{eqnarray}
 Z[J;\bar A_0,\mu] & \!\!\!\!\!=\!\!\!\!\! & Z[-J;-\bar A_0,-\mu]\label{eq:Z1}\\
& \!\!\!\!\!=\!\!\!\!\! & Z[-J^*;-\bar A_0^*,\mu^*]^*\label{eq:Z2}\\
& \!\!\!\!\!=\!\!\!\!\! & Z[J^*;\bar A_0^*,-\mu^*]^*\,,\label{eq:Z3}
\end{eqnarray}
\begin{figure}[t]
\begin{center}
\includegraphics[width=.6\linewidth]{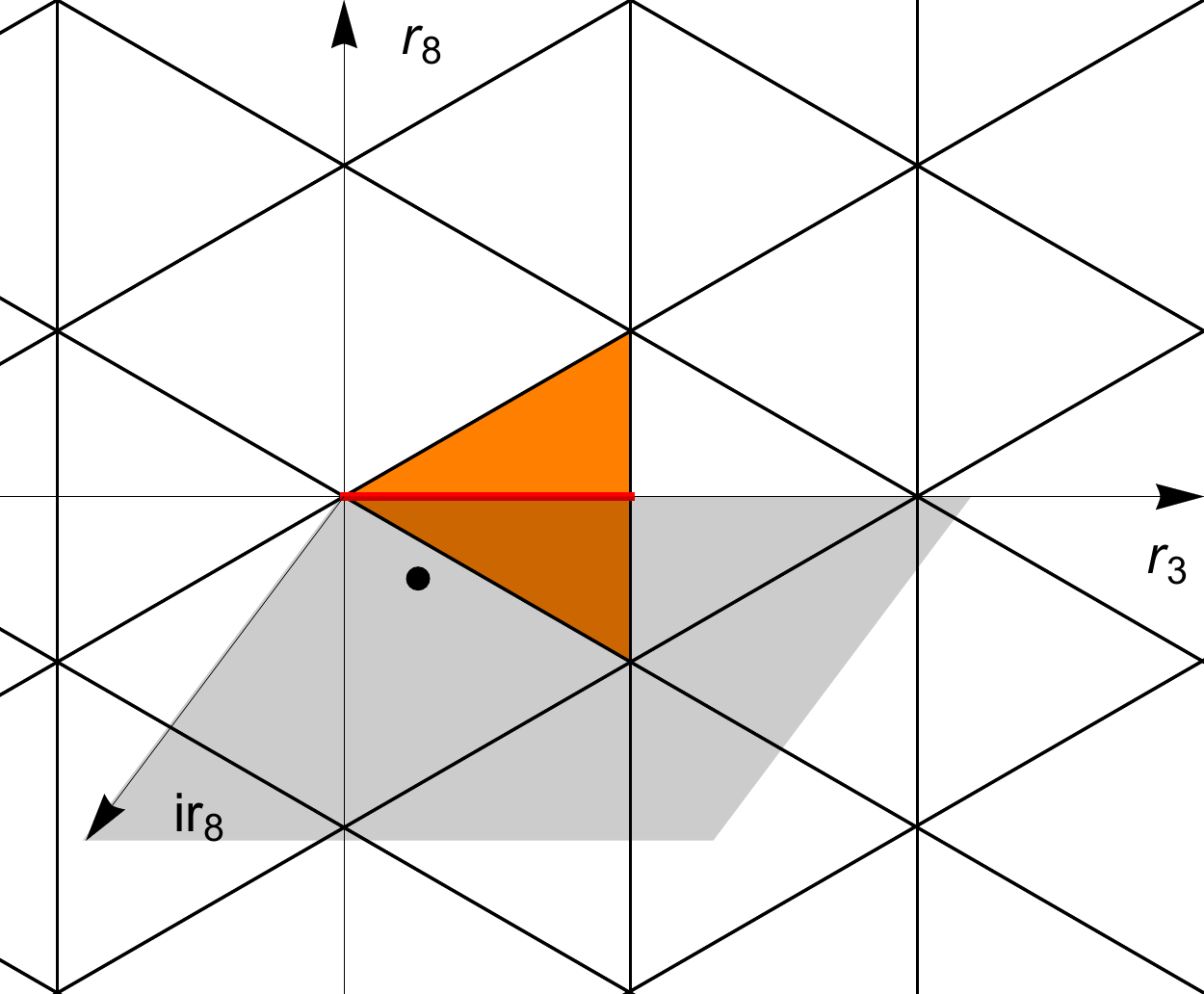}
\caption{The two different planes where the background field effective action should be considered depending on whether $\mu\in i\mathds{R}$ or $\mu\in\mathds{R}$. This figure is the counterpart of Fig.~\ref{fig:mu_im_re}.}\label{fig:away}
\label{fig:away}
\end{center}
\end{figure}

where the last identity is a combination of the previous two. One can also use a Weyl transformation to flip the sign of both $J^3$ and $\bar A_0^3$ in Eq.~(\ref{eq:Z2}). Using these identities, it is easily checked that, in the case of an imaginary chemical potential, $Z[J;\bar A_0,\mu]$ is real if $J\in\mathds{R}\times\mathds{R}$ and $\bar A_0\in\mathds{R}\times\mathds{R}$. Similarly, in the case of a real chemical potential, $Z[J;\bar A_0,\mu]$ is real if $J\in i\mathds{R}\times i\mathds{R}$ and $\bar A_0\in i\mathds{R}\times i\mathds{R}$, or, if $J\in \mathds{R}\times i\mathds{R}$ and $\bar A_0\in \mathds{R}\times i\mathds{R}$. From (\ref{eq:A0def}), it then follows that these subspaces are stabilized by the Legendre transform and are thus the natural subspaces where to study the background effective action $\tilde\Gamma[\bar A_0]$. In fact, we will see in the next chapter that the effective action is somewhat ill-defined over $i\mathds{R}\times i\mathds{R}$, so we shall restrict to $\mathds{R}\times i\mathds{R}$ from now on. The change of subspace for the background components, as one changes from the case of imaginary chemical potential to the case of a real chemical potential is illustrated in Fig.~\ref{fig:away}.\\

The properties that we have just discussed are the counterpart of properties (R1) and (R2) discussed in the previous section. The equivalent of the properties (P1), (P2) and (P3) can again be shown in the case of an imaginary chemical potential using the positivity of the fermion determinant. In particular, the background field effective action $\tilde\Gamma [\bar A_0]$ is real if $(\bar A_0^3,\bar A_0^8)\in\mathds{R}\times\mathds{R}$, and self-consistent backgrounds are obtained by minimizing the background field effective action over this subspace. In the case of a real chemical potential, the proof is jeopardized by the non-positivity of the fermion determinant but it is reasonable to admit that the background field effective action $\tilde\Gamma[\bar A_0]$ is again real if $\bar A_0\in\mathds{R}\times i\mathds{R}$, which we shall check on explicit examples in the next chapter. Again, we shall also choose the physical point as deepest saddle point in this subspace but here we lack a derivation of this recipe from first principles.

\subsection{Background dependent Polyakov loop}
The connection with the general discussion of Sec.~\ref{sec:72} can be done using the background dependent Polyakov loops. From the identities (\ref{eq:det1})-(\ref{eq:det3}), we find that
\begin{eqnarray}
 \ell[\bar A_0,\mu] & \!\!\!\!=\!\!\!\! & \bar\ell[-\bar A_0,-\mu]\label{eq:l1}\\
& \!\!\!\!=\!\!\!\! & \ell[-\bar A_0^*,\mu^*]^*\label{eq:l2}\\
& \!\!\!\!=\!\!\!\! & \bar\ell[\bar A_0^*,-\mu^*]^*\,.\label{eq:l3}
\end{eqnarray}
In the case of an imaginary chemical potential, choosing $\bar A_0\in\mathds{R}\times\mathds{R}$, one finds from Eq.~(\ref{eq:l3}) that $\ell[\bar A_0,\mu]$ and $\bar\ell[\bar A_0,\mu]$ are complex conjugate of each other, in agreement with the previous discussion. In the case of a real chemical potential, choosing $\bar A_0\in\mathds{R}\times i\mathds{R}$, one finds from Eq.~(\ref{eq:l2}) that $\ell[\bar A_0,\mu]$ and $\bar\ell[\bar A_0,\mu]$ are both real.

In particular, at leading order, the background dependent Polyakov loops are not modified by the quark content and one finds
\begin{equation}
 \ell(r_3,r_8,\mu)=\frac{e^{-i\frac{r_8}{\sqrt{3}}}+2\cos\left(\frac{r_3}{2}\right)\,e^{i\frac{r_8}{2\sqrt{3}}}}{3}=\bar\ell(r_3,-r_8,\mu)\,,
\end{equation}
which obey the above mentioned properties.
 
\subsection{Other approaches}

Some works propose instead to restrict to $r_8=0$ as a way to ensure that the background potential remains real. This is done either directly \cite{Fischer:2013eca,Fischer:2014vxa,Folkestad:2018psc} or effectively by first dropping the imaginary part of the potential and then realizing that the real part has a minimum such that $r_8=0$ \cite{Mintz:2012mz}. However, this is at odds with the fact that a non-vanishing chemical potential breaks charge conjugation symmetry and should then correspond to a non-zero $r_8$. We will see that approaches that artificially set $r_8=0$ miss part of the physical picture by not reproducing the expected behavior of the Polayakov loops as a function of the chemical potential. However, for some other aspects, the $r_8=0$ and $r_8\in i\mathds{R}$ prescriptions give quantitatively similar results. In particular, we evaluated the thermodynamical observables in the Polyakov-extended Quark-Meson model of \cite{Folkestad:2018psc}. Figure \ref{fig:r8} shows both the deviation of the pressure $p$ and the trace anomaly $e-3p$ with respect to the zero-density case, using both prescriptions for $r_8$.

The prescription $r_8=0$ has also been used in \cite{Stiele:2016cfs} to obtain bubble nucleation rates by computing the barrier between two minima of the potential in the case of a first order phase transition. We stress again that the analysis should in principle be carried out with a non-zero, imaginary $r_8$. In this case however, the very method for extracting nucleation rates has to be revisited before a comparison such as the one in Fig.~\ref{fig:r8} can be even considered.

\begin{figure}[t]
\begin{center}
\includegraphics[width=.8\linewidth]{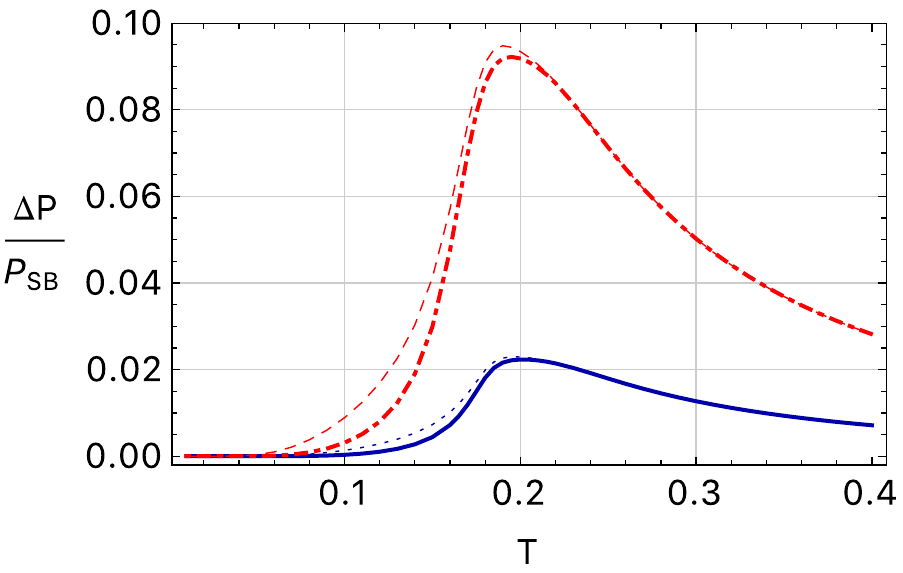}\\
\vglue4mm
\includegraphics[width=.8\linewidth]{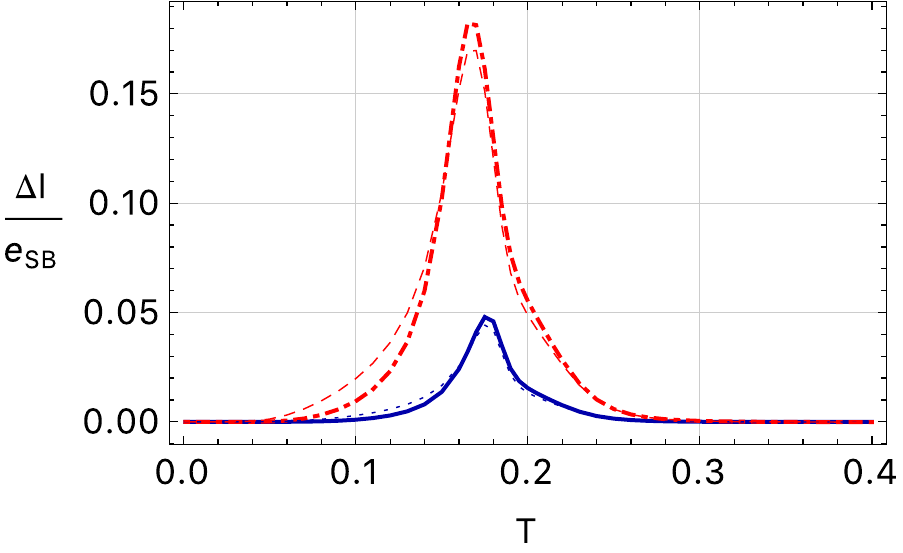}
\caption{Pressure and trace anomaly deviation with respect to the zero-density case (divided respectively by the Stefan-Boltzmann pressure and energy density), as functions of the temperature, for various values of the baryonic chemical potential $\mu_B=3\mu$, $\mu_B=200$ MeV (blue, plain and dotted) and $\mu_B=400$ MeV (red, dashed and dashed-dotted). The thin lines correspond to the results obtained with the $r_8=0$ prescription, as in Ref.~\cite{Folkestad:2018psc}, whereas the thick lines shows the results obtained with the $r_8\in i\mathds{R}$ presciption.}\label{fig:r8}
\end{center}
\end{figure}

\clearemptydoublepage
%
%
%
%
\let\textcircled=\pgftextcircled
\chapter{QCD deconfinement transition\\ in the heavy quark regime}
\label{chap:quarks}

\initial{W}e now apply the general considerations of the previous chapter to the study of the deconfinement transition in the presence of quark degrees of freedom. To keep the picture relatively simple, we consider a formal regime where the up, down and strange quarks are considered heavy, and the other, even heavier quarks, are neglected. This departure from the physical QCD case has been largely explored in the literature, from lattice simulations \cite{Fromm:2011qi}, first principle continuum methods \cite{Fischer:2014vxa} or models \cite{Kashiwa:2012wa}, and this for several reasons.

First, it allows to assess the impact of dynamical quarks on the deconfinement transition without the contamination from the breaking of chiral symmetry, the other relevant transition at play in physical QCD.\footnote{It is interesting, of course, to study the interplay between these two transitions in the physical case. This lies, however, beyond the scope of the present manuscript.} Moreover, in this formal regime, QCD presents a rich phase structure that allows for comparison and benchmarking of various approaches. For us, it will serve as a further testing ground of the Curci-Ferrari model and the related hypothesis that some of the low energy properties of QCD can be described with perturbative methods.

\section{Background effective potential}
\label{sec:1_loop}

As it was already the case in Yang-Mills theory, the evaluation of the one-loop background effective potential in the presence of quarks requires only the quadratic part of the action. Thus, for the matter part we only need
\begin{equation}
 \delta S_0[\psi,\bar\psi;\mu]\equiv\int d^dx\,\sum_{f=1}^{N_f}\bar\psi_f(x)\big(\bar {\cal D}\!\!\!\!\slash+M_f-\mu\gamma_0\big)\,\psi_f(x)\,,
\end{equation}
with $\bar{\cal D}_\nu=\partial_\nu-iT\delta_{\nu0}\,r^jt^j$. Decomposing the quark fields into a Cartan-Weyl basis $\{|\rho\rangle\}$ that diagonalizes simultanously the generators $t^j$, $\psi=\sum_\rho\psi_\rho|\rho\rangle$, with $t^j|\rho\rangle=\rho_j|\rho\rangle$ and $\langle\rho|\sigma\rangle=\delta_{\rho\sigma}$, one finds
\begin{equation}
 \delta S_0[\psi,\bar\psi;\mu]\equiv\int d^dx\,\sum_{f=1}^{N_f}\bar\psi_{f,\rho}(x)\big( \bar {\cal D}^\rho\!\!\!\!\!\!\!\!\slash\,\,+M_f-\mu\gamma_0\big)\,\psi_{f,\rho}(x)\,,
\end{equation}
with ${\cal D}_\nu^\rho=\partial_\nu-iTr\cdot\rho\,\delta_{\nu0}$. In Fourier space, this becomes
\begin{equation}
 \delta S_0[\psi,\bar\psi;\mu]\equiv \int_P^T\,\sum_{f=1}^{N_f}\bar\psi_{f,\rho}(P)\,(-iP^\rho\!\!\!\!\!\!\!\!\!\slash\,\,\,+M_f-\mu\gamma_0)\,\psi_{f,\rho}(P)\,,
\end{equation}
where  $P^\rho\equiv P+T\,r\cdot\rho\,n$ is the generalized momentum in the presence of the background, with $P=(\omega,\vec{p})$, $\omega$ a fermionic Matsubara frequency, and $n=(1,\vec{0})$. 

\subsection{General one-loop expression}
We mention that each shift $T\,r\cdot\rho\,n$ of momentum can be interpreted as an imaginary shift of the chemical potential $\mu$: $\mu\to\mu+i\,r\cdot\rho\,T$. This illustrates once more the interpretation of the background as an imaginary chemical potential and allows us to derive the one-loop matter contribution to the background effective potential using the well known one-loop expression for the free-energy density of a colorless fermionic field of flavour $f$ in the absence of background:
\begin{equation}
 V_f(T,\mu)=-\frac{T}{\pi^2}\!\int_0^\infty \!\!dq\,q^2\Big\{\ln\Big[1+e^{-\beta\,(\varepsilon_{q,f}-\mu)}\Big]+\ln\Big[1+e^{-\beta\,(\varepsilon_{q,f}+\mu)}\Big]\Big\}.
\end{equation}
We find
\begin{equation}\label{eq:dV}
   \delta V(r;T,\mu)=\sum_{f\!,\,\rho}V_f\big(T,\mu+iT\,r\cdot\rho\big)\,,
\end{equation}
where $\rho$ runs over the weights (colors) of the defining representation.

\subsection{Real-valuedness in the SU(3) case}
In the SU(3) case, we recall that $r=(r_3,r_8)$ and the weights are $(0,-1/\sqrt{3})$, $(1,1/\sqrt{3})/2$ and $(-1,1/\sqrt{3})/2$. Therefore
\begin{eqnarray}\label{eq:86}
& &  \delta V_{\mbox{\scriptsize 1loop}}^{\mbox{\scriptsize SU(3)}}(r;T,\mu)\nonumber\\
& & \hspace{0.7cm}=\,-\frac{T}{\pi^2}\!\int_0^\infty \!\!dq\,q^2\Bigg\{\ln\Big[1+e^{-\beta\,(\varepsilon_{q,f}-\mu)-i\frac{r_8}{\sqrt{3}}}\Big]+\ln\Big[1+e^{-\beta\,(\varepsilon_{q,f}+\mu)+i\frac{r_8}{\sqrt{3}}}\Big]\nonumber\\
& & \hspace{2.95cm}+\,\ln\Big[1+e^{-\beta\,(\varepsilon_{q,f}-\mu)+\frac{i}{2}\left(r_3+\frac{r_8}{\sqrt{3}}\right)}\Big]+\ln\Big[1+e^{-\beta\,(\varepsilon_{q,f}+\mu)-\frac{i}{2}\left(r_3+\frac{r_8}{\sqrt{3}}\right)}\Big]\nonumber\\
& & \hspace{2.75cm}+\,\ln\Big[1+e^{-\beta\,(\varepsilon_{q,f}-\mu)+\frac{i}{2}\left(-r_3+\frac{r_8}{\sqrt{3}}\right)}\Big]+\ln\Big[1+e^{-\beta\,(\varepsilon_{q,f}+\mu)-\frac{i}{2}\left(-r_3+\frac{r_8}{\sqrt{3}}\right)}\Big]\Bigg\}.
\end{eqnarray}
For completeness, we recall that the pure glue part of the potential reads
\begin{eqnarray}\label{eq:87}
& &  V_{\mbox{\scriptsize 1loop}}^{\mbox{\scriptsize SU(3)}}(r;T)\nonumber\\
& & \hspace{0.7cm}=\,\frac{T}{2\pi^2}\int_0^\infty dq\,q^2\,\left\{\ln\frac{\Big(1-e^{-\beta\varepsilon_q+ir_3}\Big)^3}{1-e^{-\beta q+ir_3}}+\ln\frac{\Big(1-e^{-\beta\varepsilon_q-ir_3}\Big)^3}{1-e^{-\beta q-ir_3}}\right.\nonumber\\
& & \hspace{3.1cm}+\,\ln\frac{\Big(1-e^{-\beta\varepsilon_q+i\frac{r_3+r_8\sqrt{3}}{2}}\Big)^3}{1-e^{-\beta q+i\frac{r_3+r_8\sqrt{3}}{2}}}+\ln\frac{\Big(1-e^{-\beta\varepsilon_q-i\frac{r_3+r_8\sqrt{3}}{2}}\Big)^3}{1-e^{-\beta q-i\frac{r_3+r_8\sqrt{3}}{2}}}\nonumber\\
& & \hspace{3.1cm}\left.+\,\ln\frac{\Big(1-e^{-\beta\varepsilon_q+i\frac{r_3-r_8\sqrt{3}}{2}}\Big)^3}{1-e^{-\beta q+i\frac{r_3-r_8\sqrt{3}}{2}}}+\ln\frac{\Big(1-e^{-\beta\varepsilon_q-i\frac{r_3-r_8\sqrt{3}}{2}}\Big)^3}{1-e^{-\beta q-i\frac{r_3-r_8\sqrt{3}}{2}}}\right\}.
\end{eqnarray}
It is readily checked that the above expressions are real if $(\mu,r_3,r_8)\in i\mathds{R}\times\mathds{R}\times\mathds{R}$, $(\mu,r_3,r_8)\in \mathds{R}\times i\mathds{R}\times i\mathds{R}$ or $(\mu,r_3,r_8)\in \mathds{R}\times\mathds{R}\times i\mathds{R}$, as anticipated in the previous chapter. We mention however that the case $(\mu,r_3,r_8)\in \mathds{R}\times i\mathds{R}\times i\mathds{R}$ is problematic since some of the bosonic integrals become ill-defined. For this reason, in the case of a real chemical potential, we shall only consider the case $(\mu,r_3,r_8)\in \mathds{R}\times\mathds{R}\times i\mathds{R}$.

\subsection{Polyakov loop potential}
As we have already mentioned in the previous chapter, at leading order, the background dependent Polyakov loops take the same expressions as in the pure YM case, namely
\begin{equation}
 \ell(r)=\frac{1}{N}\sum_\rho e^{ir\cdot\rho} \quad \mbox{and} \quad  \bar\ell(r)=\frac{1}{N}\sum_\rho e^{-ir\cdot\rho}\,.
\end{equation}
In the SU(3) case, these two Polyakov loops are in one-to-one correspondance with the background components $r_3$ and $r_8$ and one can consider expressing the latter in terms of the former. Doing so, the background field effective potential becomes an effective potential for the Polyakov loop.\footnote{That this corresponds to the Polyakov loop potential, as it would be obtained from a Legendre transformation with respect to sources coupled to the Polyakov loops, can be shown up to two-loop order, see \cite{Reinosa:2015gxn}.}

Expressing the background components in terms of the Polyakov loops is not particularly simple. However, expressing the background field effective potential in terms of the Polyakov loops can be done as follows. First, by combining the various terms in the sum (\ref{eq:dV}), one arrives at the following logarithms
\begin{eqnarray}
& &  \ln\Big[1+e^{-\beta\,(\varepsilon_{q,f}\mp\mu)\pm ir\cdot\rho_1}\Big]\Big[1+e^{-\beta\,(\varepsilon_{q,f}\mp\mu)\pm ir\cdot\rho_2}\Big]\Big[1+e^{-\beta\,(\varepsilon_{q,f}\mp\mu)\pm ir\cdot\rho_3}\Big]\\
& & \hspace{1.5cm}=\,\ln\Big[1+A_\pm e^{-\beta\,(\varepsilon_{q,f}\mp\mu)}+B_\pm e^{-2\beta\,(\varepsilon_{q,f}\mp\mu)}+C_\pm e^{-3\beta\,(\varepsilon_{q,f}\mp\mu)}\Big],\nonumber
\end{eqnarray}
where
\begin{eqnarray}
 A_\pm & \!\!\!\!\!=\!\!\!\!\! & e^{\pm ir\cdot\rho_1}+e^{\pm ir\cdot\rho_2}+e^{\pm ir\cdot\rho_3}\,,\\
 B_\pm & \!\!\!\!\!=\!\!\!\!\! & e^{\pm ir\cdot(\rho_1+\rho_2)}+e^{\pm ir\cdot(\rho_2+\rho_3)}+e^{\pm ir\cdot(\rho_3+\rho_1)}\,,\\
 C_\pm & \!\!\!\!\!=\!\!\!\!\! & e^{\pm ir\cdot(\rho_1+\rho_2+\rho_3)}\,.
\end{eqnarray}
Using that $\rho_1+\rho_2+\rho_3=0$ leads to $A_+=3\ell=B_-$, $B_+=3\bar\ell=A_-$ and $C_+=1=C_-$, and one arrives eventually at
\begin{eqnarray}\label{eq:123}
& &  \delta V_{\mbox{\scriptsize 1loop}}^{\mbox{\scriptsize SU(3)}}(\ell,\bar\ell;T,\mu)\nonumber\\
& & \hspace{0.7cm}=\,-\frac{T}{\pi^2}\!\int_0^\infty \!\!dq\,q^2\Bigg\{\ln\Big[1+3\ell e^{-\beta\,(\varepsilon_{q,f}-\mu)}+3\bar\ell e^{-2\beta\,(\varepsilon_{q,f}-\mu)}+e^{-3\beta\,(\varepsilon_{q,f}-\mu)}\Big]\nonumber\\
& & \hspace{3.5cm}+\ln\Big[1+3\bar\ell e^{-\beta\,(\varepsilon_{q,f}+\mu)}\ell+3\ell e^{-2\beta\,(\varepsilon_{q,f}+\mu)}+e^{-3\beta\,(\varepsilon_{q,f}+\mu)}\Big]\Bigg\}.
\end{eqnarray}
This formula can be extended to SU(N) but in this case one needs to introduce $N-1$ Polyakov loops to be mapped to the $N-1$ background components. These are the Polyakov loops associated to the $N-1$ fundamental SU(N) representations \cite{Zuber}. More details can be found in App.~\ref{app:fund}. There we also show how to express the one-loop glue potential in terms of the Polyakov loops, in the case of the SU($3$) gauge group. One obtains
\begin{eqnarray}\label{eq:456}
& & V_{\mbox{\scriptsize 1loop}}^{\mbox{\scriptsize SU(3)}}(\ell,\bar\ell;T)\nonumber\\
& & \hspace{0.7cm}=\,\frac{3T}{2\pi^2}\!\!\int_0^\infty dq\,q^2\ln\Big[1+e^{-8\beta\varepsilon_q}-\big(9\ell\bar\ell-1\big)\big(e^{-\beta\varepsilon_q}+e^{-7\beta\varepsilon_q}\big)\nonumber\\
& & \hspace{2.5cm} -\,\big(81\ell^2\bar\ell^2-27\ell\bar\ell+2\big)\big(e^{-3\beta\varepsilon_q}+e^{-5\beta\varepsilon_q}\big)\nonumber\\
& & \hspace{2.5cm} +\,\big(27\ell^3+27\bar\ell^3-27\ell\bar\ell+1\big)\big(e^{-2\beta\varepsilon_q}+e^{-6\beta\varepsilon_q}\big)\nonumber\\
& & \hspace{2.5cm} +\,\big(162\ell^2\bar\ell^2-54\ell^3-54\bar\ell^3+18\ell\bar\ell-2\big)e^{-4\beta\varepsilon_q}\Big],\nonumber\\
& & \hspace{0.7cm}-\,\frac{T}{2\pi^2}\!\!\int_0^\infty dq\,q^2\ln\Big[1+e^{-8\beta q}-\big(9\ell\bar\ell-1\big)\big(e^{-\beta q}+e^{-7\beta q}\big)\nonumber\\
& & \hspace{2.5cm}-\,\big(81\ell^2\bar\ell^2-27\ell\bar\ell+2\big)\big(e^{-3\beta q}+e^{-5\beta q}\big)\nonumber\\
& & \hspace{2.5cm}+\,\big(27\ell^3+27\bar\ell^3-27\ell\bar\ell+1\big)\big(e^{-2\beta q}+e^{-6\beta q}\big)\nonumber\\
& & \hspace{2.5cm}+\,\big(162\ell^2\bar\ell^2-54\ell^3-54\bar\ell^3+18\ell\bar\ell-2\big)e^{-4\beta q}\Big].
\end{eqnarray}
In line with the discussion of the previous chapter, these expressions should be considered for complex conjugated variables $\bar\ell=\ell^*$ in the case of an imaginary chemical potential and for real and independent variables $\ell$ and $\bar\ell$ in the case of a real chemical potential.\footnote{Being the average values of traced unitary matrix, these variables are further constrained to lie in some subregion of $\Sigma$ or $\mathds{R}\times\mathds{R}$ respectively.}

\section{Phase structure at $\mu=0$}

Let us now combine Eqs.~(\ref{eq:86}) and (\ref{eq:87}) to study the phase structure of the model as a function of the (heavy) quark masses. This dependence has been studied in various approaches, including non gauge-fixed lattice QCD, and offers, therefore, a valuable benchmark for the Curci-Ferrari model. For simplicity, we consider the case of two degenerate flavors with mass $M_u=M_d$ and a third flavor with mass $M_s$.

We first analyze the phase diagram at $\mu=0$. As we have discussed above, in this case, the analysis of the background effective potential can be done either over $(r_3,r_8)\in\mathds{R}\times\mathds{R}$ or over $(r_3,r_8)\in\mathds{R}\times i\mathds{R}$. This is related to the fact that, because charge conjugation is not broken, we expect the physical point to lie along the axis $r_8=0$ in the fundamental Weyl chamber. This is indeed what we find numerically. Correspondingly, we set $r_8=0$ in what follows. Along this axis, the background dependent Polyakov loops $\ell(r)$ and $\bar\ell(r)$, are guaranteed to be equal and real, in line with their standard interpretation in terms of the free energy of a static quark or antiquark \cite{Svetitsky:1985ye,Philipsen:2010gj} and the fact that there should be no distinction between the free energy of a quark and that of an antiquark at $\mu=0$.\\

\begin{figure}[t]
  \centering
  \includegraphics[width=.43\linewidth]{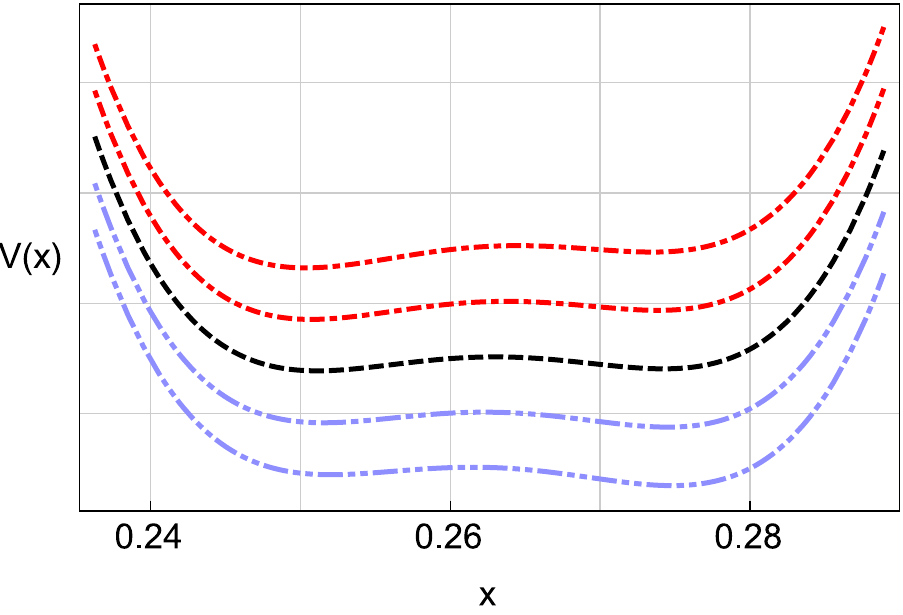}
  \hglue4mm\includegraphics[width=.43\linewidth]{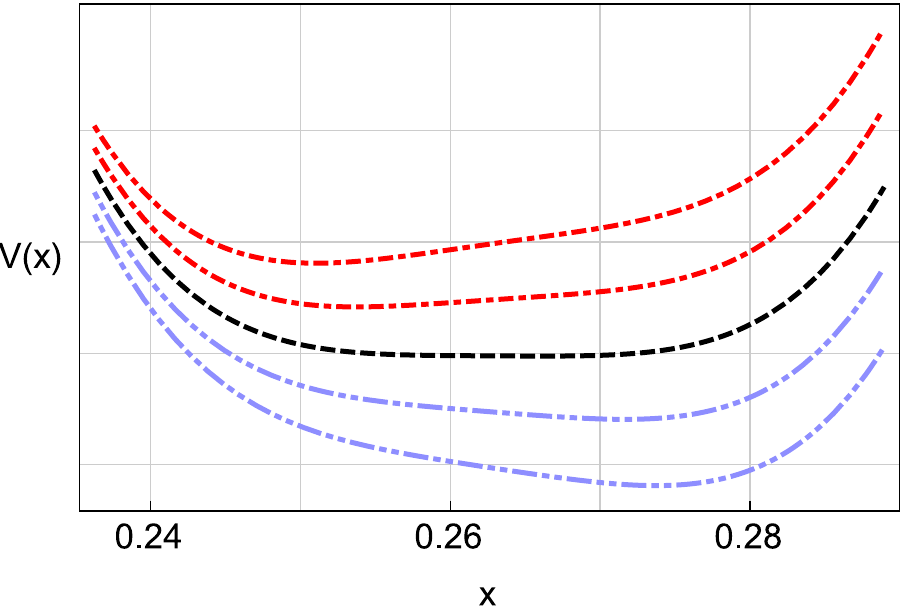}\\
  \vglue4mm
  \includegraphics[width=.43\linewidth]{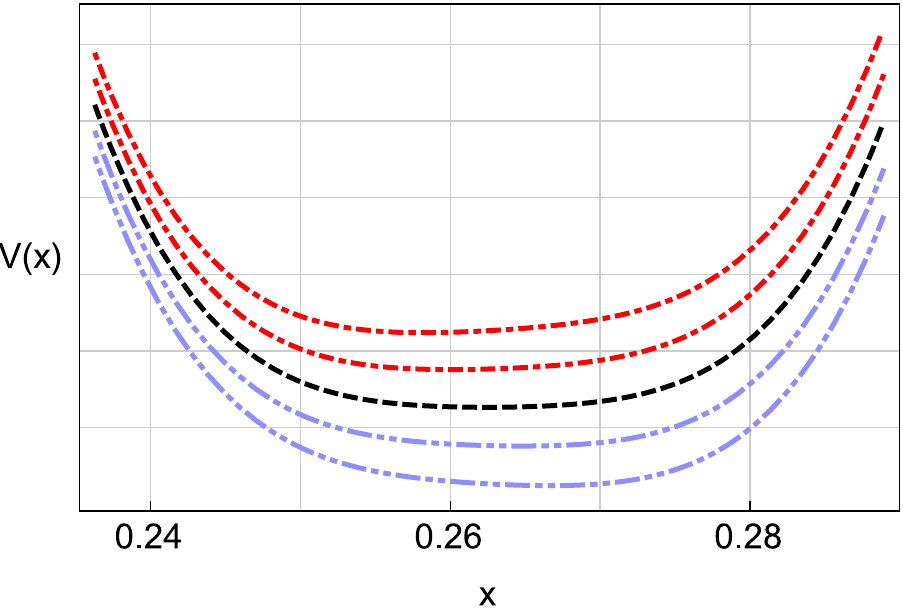}
  \caption{The background field potential (in arbitrary units) for $\mu=0$ and $r_8=0$, as a function of $x=r_3/(2\pi)$ for different temperatures in the degenerate case $M_u=M_s=M$. Top
    figure: $M$ slightly larger than the critical value $M_c$; see
    Fig.~\ref{fig_columbia_mu_O}. Temperature increases for curves from bottom to top. The dashed line corresponds to the
    transition temperature. Middle
    figure: $M=M_c$ with the same conventions as
    for the top figure. Bottom figure: $M<M_c$. The dashed line is for the crossover
    temperature, where the curvature of the potential at the minimum is the smallest.}
  \label{fig_pot}
\end{figure}

Depending on the values of the quark masses, we find different types
of behaviors as the temperature is varied, see Fig.~\ref{fig_pot}.  For large masses, the absolute minimum presents a finite jump at some transition temperature, signalling a first-order transition. Instead, for small masses, there is
always a unique minimum, whose location rapidly changes with temperature in some crossover regime. At the common boundary of these two mass regions, the system presents a critical
behavior: there exists a unique minimum of the potential for all
temperatures, which however behaves as a power-law around some
critical temperature. At this critical point, the curvature of the potential at the minimum needs to vanish, and, to distinguish it from a mere spinodal in the first order transition region, we need to require that the third derivative vanishes as well (which is the condition for the merging of spinodals in the first order transition region). The three conditions
\begin{equation}
0=V'(r_3)=V''(r_3)=V'''(r_3)
\end{equation}
thus determine the critical values $T_c$ and $r_3^c$ for the temperature and the background, together with a critical line in the $(M_u,M_s)$ plane, the so-called Columbia plot. 

Our result for the Columbia plot are shown in Fig.~\ref{fig_columbia_mu_O}. In the degenerate case $M_u=M_s$ we get, for the critical mass, $M_c/m=2.867$ and, for the critical temperature, $T_c/m=0.355$. We thus have $M_c/T_c=8.07$. This dimensionless ratio does not depend on the value of $m$ and can be directly compared to lattice results. For instance, the calculation of Ref.~\cite{Fromm:2011qi} yields, for $3$ degenerate quarks, $(M_c/T_c)^{\mbox{\tiny latt.}}=8.32$. We obtain similar good agreement for different numbers of degenerate quark flavors, as summarized in Table~\ref{tab:mu0}.\\

\begin{figure}[t]
  \centering
  \includegraphics[width=.84\linewidth]{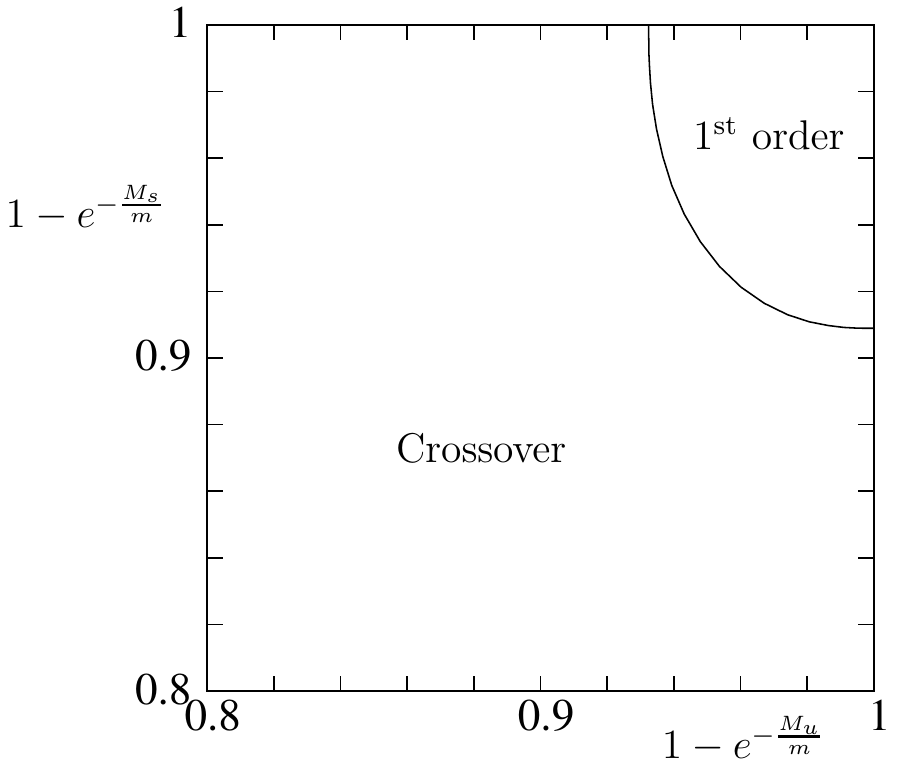}
  \caption{Columbia plot at $\mu=0$. In the upper
    right corner of the plane $(M_u,M_s)$, the phase transition is
    of the first-order type. In the lower left corner, the system presents a
    crossover. On the plain line, the system has a critical behavior. }
  \label{fig_columbia_mu_O}
\end{figure}

\begin{table}[h!]
  \centering
  \begin{tabular}{|c|c|c|c|c|c|}
\hline
$\,N_f\,$&$\,M_c/m\,$&$\,T_c/m\,$&$\,M_c/T_c\,$&$\left(M_c/T_c\right)^{\mbox{\tiny latt.}}$&$\left(M_c/T_c\right)^{\mbox{\tiny matr.}}$\\
\hline
1&2.395&0.355&6.74&7.22(5)&8.04\\
\hline
2&2.695&0.355&7.59&7.91(5)&8.85\\
\hline
3&2.867&0.355&8.07&8.32(5)&9.33\\
\hline
  \end{tabular}
  \caption{Values of the critical quark mass and temperature for $N_f=1,2,3$ degenerate quark flavors from the present one-loop calculation. The values of the dimensionless and parameter independent ratio $M_c/T_c$ are compared to the lattice results of Ref.~\cite{Fromm:2011qi} (previous to last column). { For comparison we also show the values from the matrix model of Ref.~\cite{Kashiwa:2012wa} (last column).}}\label{tab:mu0}
\end{table}

We observe that the critical temperature is essentially unaffected by the presence of quarks. It is actually close to the one obtained in the present approach for the pure gauge SU($3$) theory \cite{Reinosa:2014ooa}. This is due to the fact that, for the typical values of $M/T$ near the critical line, the quark contribution to the potential is Boltzmann suppressed as compared to that of the gauge sector \cite{Reinosa:2015oua}.

We also mention that recent calculations in the Dyson-Schwinger approach \cite{Fischer:2014vxa} yield values of the ratio $M_c/T_c$ that are systematically smaller than the ones obtained on the lattice. The origin of this discrepancy lies in the fact that in these studies a certain renormalized quark mass is used to compute the ratios $M_c/T_c$, whereas on the lattice and in any one-loop approach such as the one considered here, the bare quark mass is used.

Two-loop corrections to the previous results within the Curci-Ferrari model have been evaluated in \cite{Maelger:2017amh} and improve the agreement with lattice results. More recently, another perturbative approach inspired by the Gribov-Zwanziger approach has been considered in \cite{Kroff:2018ncl} and supports once more the idea that the phase structure of QCD in the top-left corner of the Columbia plot is akin to perturbative methods.

\section{Phase structure for $\mu\in i\mathds{R}$}

The case of imaginary chemical potential is interesting in many respects. First, as already mentioned in the previous chapter, the sign problem is under control, allowing for the use of lattice simulations. But more importantly, the corresponding phase structure is quite rich and offers a new source for comparison between the various approaches.

\subsection{Roberge-Weiss symmetry}
The richness of the phase structure has to do with the fact that, despite the explicit breaking of center-symmetry by the fundamental quarks, a symmetry exists for particular values of the chemical potential. 

To see this, consider a center transformation such that $U(\tau+\beta)=e^{-2\pi/3}U(\tau)$. We have seen that this transformation modifies the boundary conditions of the quark field to $\psi(\tau+\beta,\vec{x})=-e^{-2\pi/3}\psi(\tau,\vec{x})$ which explicitly breaks center symmetry at the quantum level. However, the usual anti-periodic boundary conditions can be restored by means of an abelian transformation $e^{i2\pi/3T\tau}$, with the effect of shifting the chemical potential by $-i2\pi/3T$. It follows that
\begin{equation}
 S[A^U,e^{i2\pi/3T\tau}U\psi,e^{-i2\pi/3T\tau}U^\dagger\bar\psi;\mu+i2\pi/3T]=S[A,\psi,\bar\psi;\mu]\,.
\end{equation}
Now, had we first applied charge conjugation to the system we would have ended up with the identity
\begin{equation}
 S[(A^{\cal C})^U,e^{i2\pi/3T\tau}U\psi^{\cal C},e^{-i2\pi/3T\tau}U^\dagger\bar\psi^{\cal C};i2\pi/3T-\mu]=S[A,\psi,\bar\psi;\mu]\,.
\end{equation}
In particular, if we choose $\mu=i\pi/3T$, we have a symmetry that survives at the quantum level since the boundary conditions are unaffected. This is the so-called Roberge-Weiss symmetry, a subtle combination of center, charge conjugation and abelian transformations \cite{Roberge:1986mm}.

The Roberge-Weiss symmetry imposes constraints on certain observables that one can then use as order parameters testing the possible spontanous breaking of the symmetry. In particular, if the Roberge-Weiss symmetry is not spontanously broken, we must have (at $\mu=i\pi/3 T$)
\begin{equation}
 \ell=e^{-i2\pi/3}\bar\ell=e^{-i2\pi/3}\ell^*\,,
\end{equation}
where we used that $\bar\ell=\ell^*$ when the chemical potential is imaginary. The argument of the Polyakov loop is then such that
\begin{equation}
 \mbox{Arg}\,\ell=-i\frac{2\pi}{3}-\mbox{Arg}\,\ell\,,
\end{equation}
and thus $\mbox{Arg}\,\ell=-i\pi/3 T$. Any departure from this value signals the breaking of the Roberge-Weiss symmetry.

We mention finally that, as a consequence of the above properties, the system is invariant under $\mu\to \mu+i2\pi/3T$ and $\mu\to i2\pi/3T-\mu$. We shall thus consider chemical potentials $\mu=ixT$ with $x\in[0,\pi/3]$.

\subsection{Results}

\begin{figure}[t]
  \centering
  \includegraphics[width=.7\linewidth]{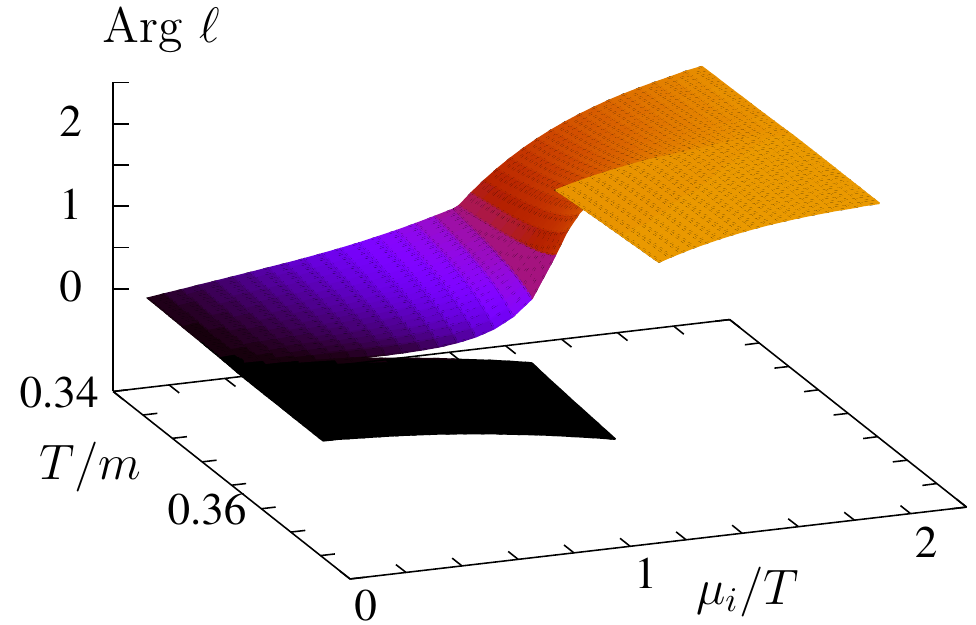}
  \caption{Argument of the Polyakov loop in the $(\mu_i/T,T)$ plane for a degenerate quark mass $M$.}
  \label{fig_arg_pol}
\end{figure}

We consider for simplicity the case of three degenerate quarks with $M_u=M_s=M$. For a large enough value of $M$, we find indeed that the Roberge-Weiss symmetry can be spontanously broken, with a discontinuity of $\mbox{Arg}\,\ell$ as $\mu_i/T$ crosses $\pi/3$, if the temperature is large enough, see Fig.~\ref{fig_arg_pol}. It is interesting to follow the evolution of the $(\mu_i/T,T)$ phase diagram as the degenerate quark mass $M$ is varied. 

For any mass larger than $M_c(\mu=0)\simeq 2.8m$, such that the transition at vanishing chemical potential is first-order (see Fig.~\ref{fig_columbia_mu_O}), we find that the transition persists at non-vanishing $\mu_i$, as depicted in the top panel of Fig.~\ref{fig_phase_diag}. This line of first order transitions, its mirror image by the symmetry $\mu_i/T\to2\pi/3-\mu_i/T$ and the line of first order transitions associated to the Roberge-Weiss symmetry merge into a triple point at $\mu_i/T=\pi/3$ . 

\begin{figure}[h]
  \centering
  \includegraphics[width=.47\linewidth]{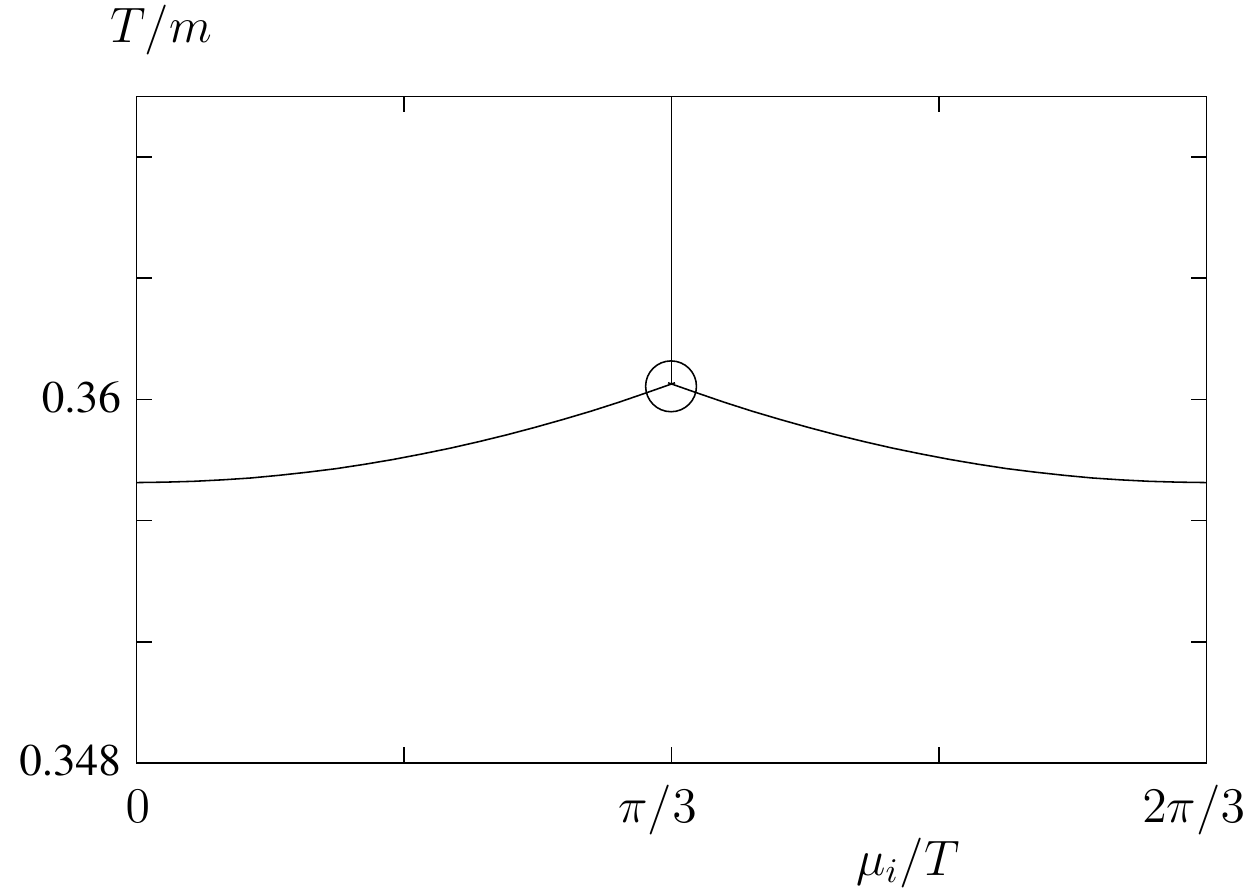}
  \hspace{0.5cm}\vspace{-0.1cm}\includegraphics[width=.45\linewidth]{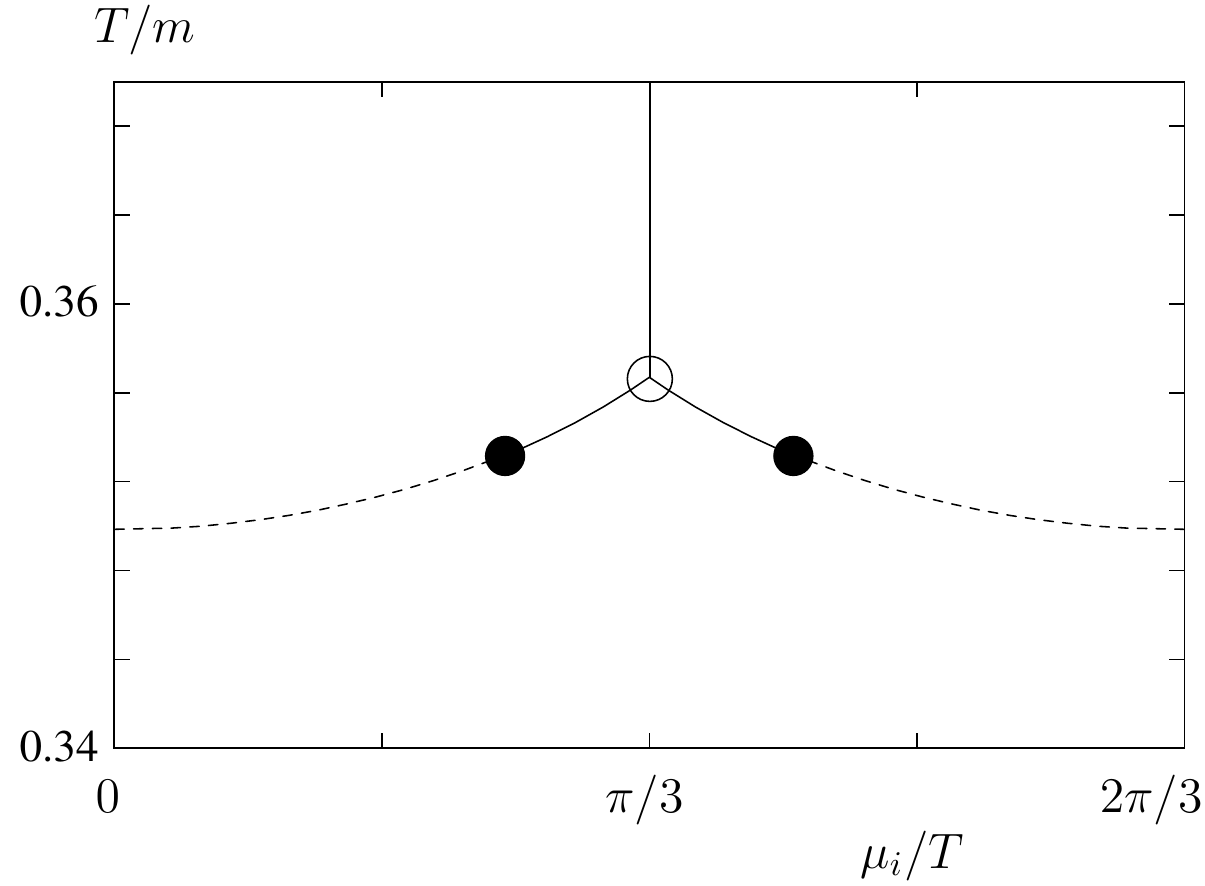}
  \includegraphics[width=.45\linewidth]{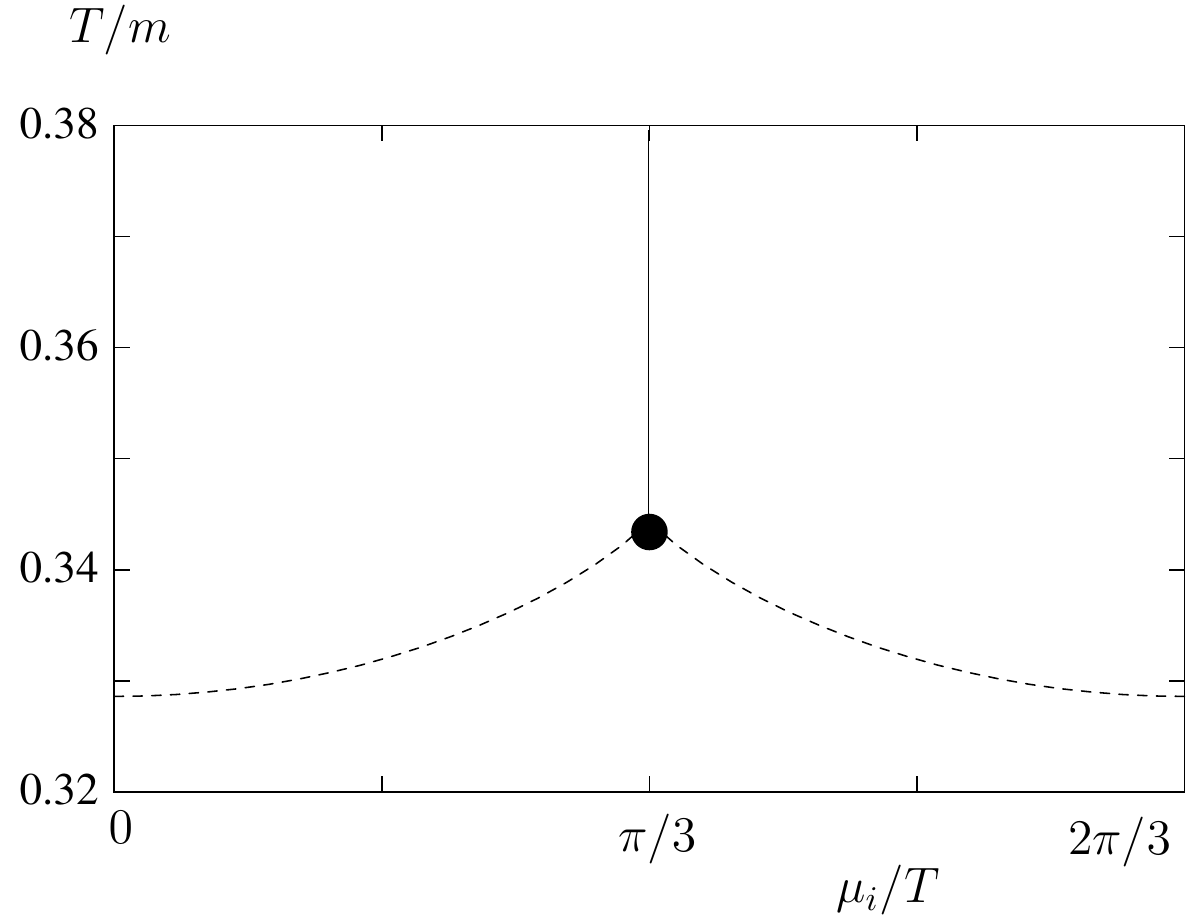}
  \caption{ Phase diagram in the plane $(\mu_i/T,T)$ for different values of the degenerate quark mass $M$. Top panel: $M=3m$ is larger that the critical mass at vanishing chemical potential $M_c(0)\simeq2.8m$. Middle panel: $M=2.6m$ is below $M_c(0)$ but larger than the second critical mass $M_c(i\pi T/3)\simeq2.2m$; see text. Bottom panel: $M=2m$ is smaller that $M_c(i\pi T/3)$. Plain lines correspond to first-order phase transitions,
    dashed lines to crossovers, black dots represent $Z_2$ critical
    points, and the empty circles are triple points.}
  \label{fig_phase_diag}
\end{figure}

For $M=M_c(\mu=0)$, the transition at vanishing chemical potential is second order and there appears, in the $(\mu_i/T,T)$ phase diagram, a couple of $Z_2$ critical points which terminate the lines of first-order transitions described above at $\mu_i/T=0$ and $\mu_i/T=2\pi/3$. Decreasing the mass $M$ further, these critical points penetrate deeper in the phase diagram towards $\mu_i/T=\pi/3$, as shown in the middle panel of  Fig.~\ref{fig_phase_diag}. The critical points are located by generalizing the approach at $\mu=0$. Since we have one extra variable $r_8$, we need of course one additional condition. We require that both equations of motion for $r_3$ and $r_8$ are fulfilled:
\begin{equation}
 0=\frac{\partial V}{\partial r_3} \quad \mbox{and} \quad 0=\frac{\partial V}{\partial r_8}\,,\label{eq:eq1}
\end{equation}
that the determinant of the Hessian vanishes, meaning that the curvature of the potential vanishes in a certain direction,
\begin{equation}
 0=\frac{\partial^2 V}{\partial r_3^2}\frac{\partial^2 V}{\partial r_8^2}-\left(\frac{\partial^2 V}{\partial r_3\partial r_8}\right)^2\,,\label{eq:eq2}
\end{equation}
and that the third derivative in this direction vanishes as well
\begin{equation}
 0=a^3\frac{\partial^3 V}{\partial r_3^3}+3a^2 b\frac{\partial^3 V}{\partial r_3^2\partial r_8}+3a b^2\frac{\partial^3 V}{\partial r_3\partial r_8^3}+b^3\frac{\partial^3 V}{\partial r_3^3}\,,\label{eq:eq3}
\end{equation}
with $a=-\partial^2 V/\partial r_3\partial r_8$ and $b=\partial^2 V/\partial r_3^2$.

At a critical value $M_c(\mu=i\pi T/3)\simeq 2.2m$, the  two $Z_2$ critical points merge at the symmetric point $\mu_i/T=\pi/3$ and give rise to a tricritical point which terminates the vertical line of first-order transition \cite{deForcrand:2010he}. The horizontal lines of first-order transitions for $\mu_i/T\neq\pi/3$ have completely disappeared and are replaced by crossovers. For $M<M_c(i\pi T/3)$, the picture is the same with, however, the tricritical point ending the first-order transition line at $\mu_i/T=\pi/3$ replaced by a $Z_2$ critical point, as shown in the lower panel of Fig.~\ref{fig_phase_diag}. 

The approach to tricriticality is controlled by the scaling behavior \cite{deForcrand:2010he}
\begin{equation}
\label{eq:tricsca}
  \frac{M_c(\mu)}{T_c(\mu)}= \frac{M_{\mbox{\scriptsize tric.}}}{T_{\mbox{\scriptsize tric.}}}+K\left[\left(\frac{\pi}{3}\right)^2+\left(\frac{\mu}{T}\right)^2\right]^{2/5}.
\end{equation}
A fit of our results at $\mu=i\mu_i$ yields $M_{\mbox{\scriptsize tric.}}/T_{\mbox{\scriptsize tric.}}=6.15$ and $K=1.85$,\footnote{We mention that the tricritical point can be obtained by setting directly $\mu_i/T=\pi/3$ and using the Roberge-Weiss symmetry, see \cite{Maelger:2018vow}.} to be compared with the lattice result of Ref.~\cite{Fromm:2011qi}, $(M_{\mbox{\scriptsize tric.}}/T_{\mbox{\scriptsize tric.}})^{\mbox{\scriptsize latt.}}=6.66$ and $K^{\mbox{\scriptsize latt.}}=1.55$ for $3$ degenerate quark flavors.

\section{Phase structure for $\mu\in\mathds{R}$}

\subsection{Columbia plot}
Using Eqs.~(\ref{eq:eq1})-(\ref{eq:eq3}), we can follow the critical line in the Columbia plot for increasing values of $\mu^2>0$. The only subtlety is that we need to solve these equations for $(r_3,r_8)\in\mathds{R}\times i\mathds{R}$. We mention however that there is no ambiguity here on the choice of the saddle point since, on the critical line there is typically only one saddle point. Our result is shown in Fig.~\ref{fig_columbia_mu_real} and shows that the critical line moves towards the Yang-Mills point, in line with the observations made on the lattice. We can also compare our result at real chemical potential with the extrapolation of the tricritical scaling law. We observe that tricritical scaling survives deep in the $\mu^2>0$ region \cite{Reinosa:2015oua}, as also observed in other approaches.

\begin{figure}[t]
  \centering
  \includegraphics[width=.7 \linewidth]{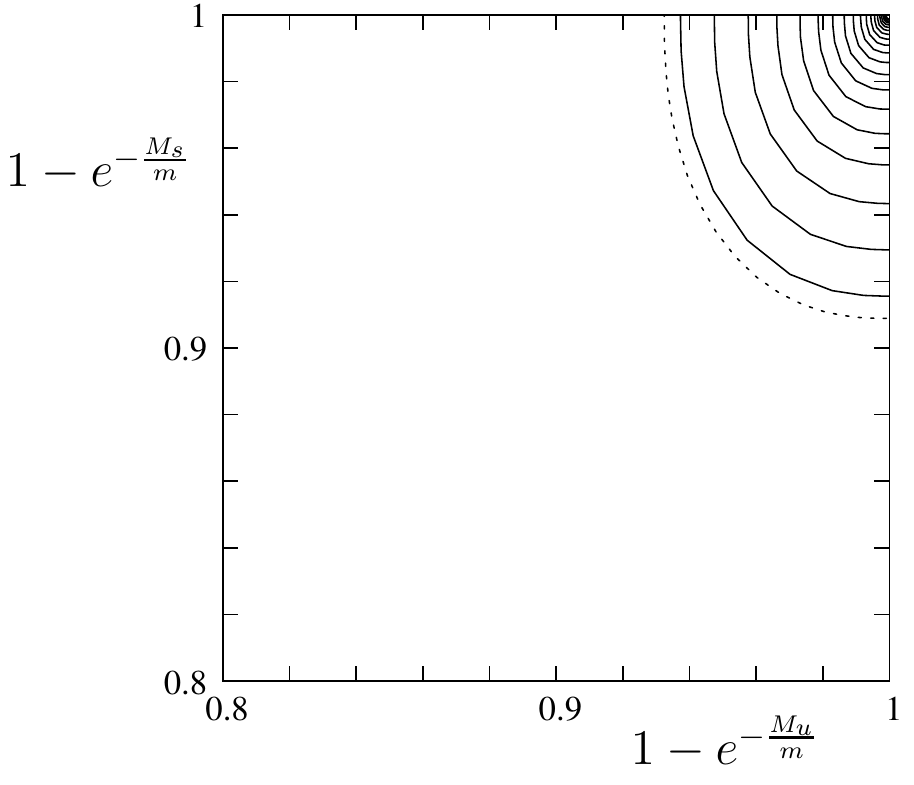}
  \caption{Columbia plot for real chemical potential. The first-order
    region contracts with increasing $\mu$. The dotted
    line corresponds to $\mu=0$. Successive plain lines correspond to a chemical
    potential increased by steps $\delta\mu=0.1$~GeV.}
  \label{fig_columbia_mu_real}
\end{figure}

 \begin{figure}[t]
	\centering
	\includegraphics[width=0.67\textwidth]{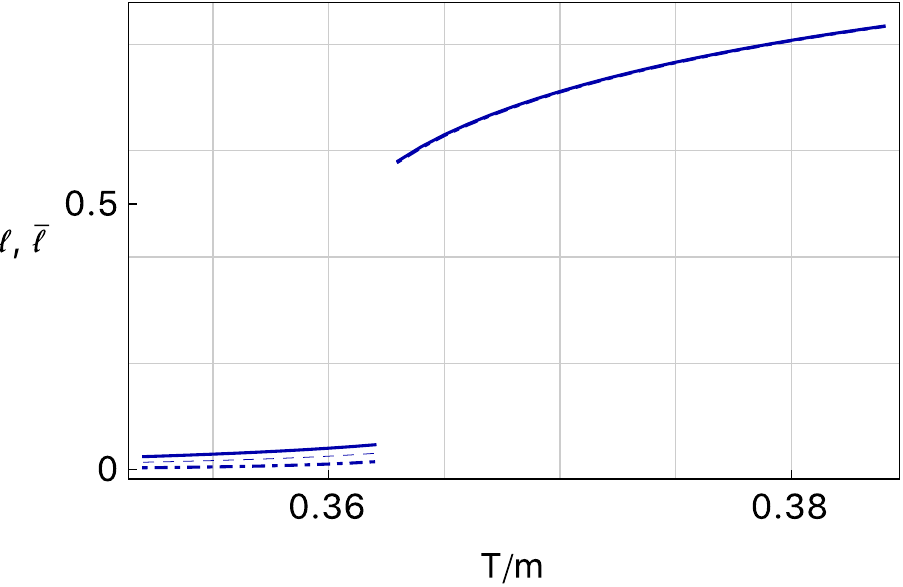}\\
	\vglue8mm
	\hglue6mm\includegraphics[width=0.64\textwidth]{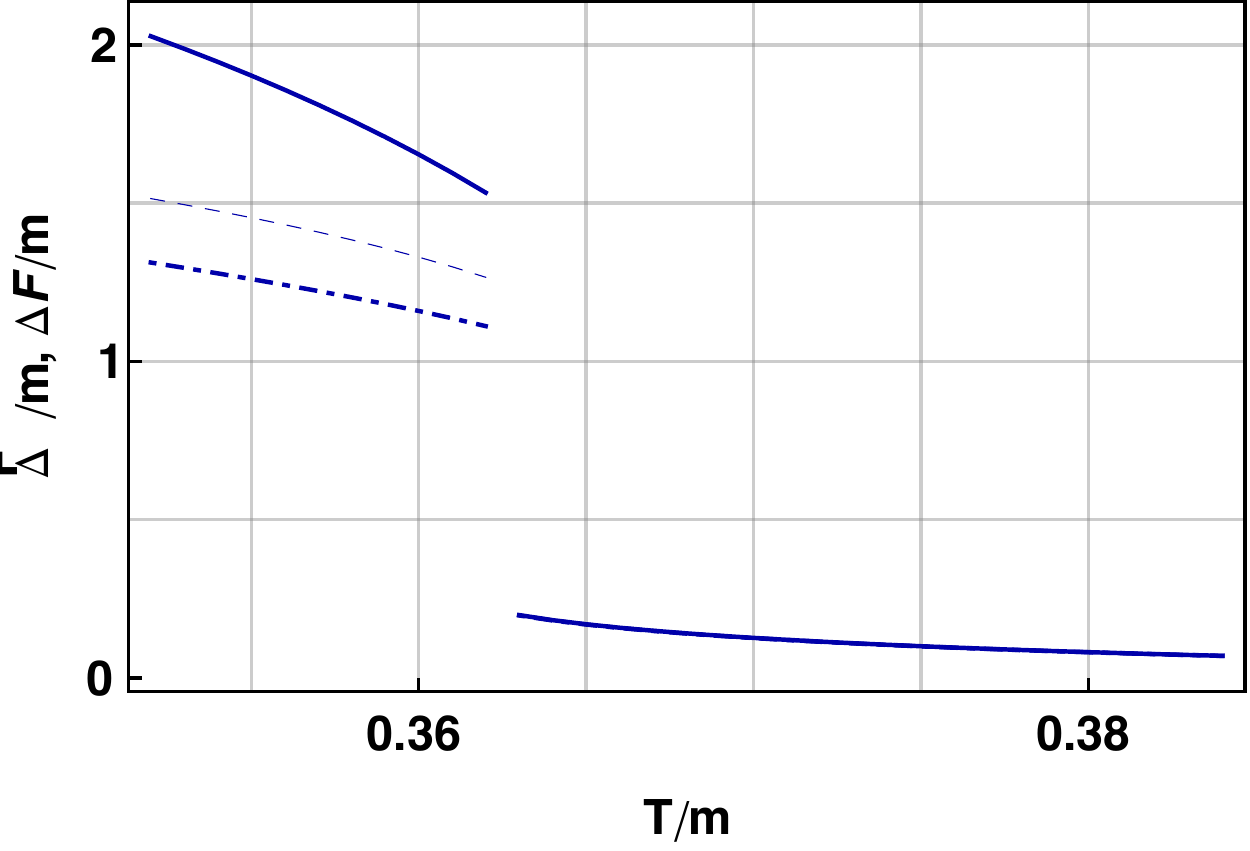}
	\caption{\label{fig:PressureEntropy} {The Polyakov loops $\ell$ (dot-dashed) and $\bar\ell$ (plain) and the corresponding quark and anti-quark free energies computed with a nonzero imaginary background along the $r_8$ direction, in the case of a positive chemical potential ($\mu=0.6m$). We also show the same quantities computed with $r_8 =0$ (dashed) and $r_3$ at the minimum of the  potential along this axis.}}
\end{figure}

\subsection{$T$-dependence of the Polyakov loops}
Our analysis reveals that the location of the saddle point is typically at $r_8\neq0$, in line with the fact that charge conjugation invariance is explicitly broken by the presence of a finite chemical potential. A non-vanishing $r_8$ induces a difference between $\ell(\mu)$ and $\bar\ell(\mu)$, and therefore between the associated free-energies for quarks and anti-quarks. This is illustrated in Fig.~\ref{fig:PressureEntropy}, which shows the temperature dependence of the averaged Polyakov loops in the region of first-order transition. We observe a significant difference between $\ell(\mu)$ and $\bar\ell(\mu)$ below the transition temperature, whereas they essentially agree in the high-temperature phase. In other words, the energetic price to pay for a static quark is much higher than that for an anti-quark (at $\mu>0$) in the quasi-confined, low temperature phase, where the Polyakov loops are small, whereas it is essentially the same in the high-temperature deconfined phase. In this latter case, the explanation is that, in a deconfined phase, quarks and anti-quarks are free to roam around and therefore the energetic cost is essentially the same for bringing a quark or an anti-quark. In contrast, in the quasi-confined phase, the capacity of the medium to confine a test color charge depends more notably on the properties of the medium. That anti-quarks are more easily confined can be interpreted in terms of screening of a quark by the anti-quarks of the thermal bath \cite{Kaczmarek:2002mc}.

We mention that previous studies of the phase diagram with background field methods in the functional renormalization group and Dyson-Schwinger approaches of Refs.~\cite{Fischer:2013eca,Fischer:2014vxa,Folkestad:2018psc} have employed another criterion than the one used here to determine the physical properties of the system. Instead of searching for a saddle point of the function $V(r_3,ir_8,\mu)$ as we propose here, the authors of Refs.~\cite{Fischer:2013eca,Fischer:2014vxa} define the physical point as the absolute minimum of the function $V(r_3,0,\mu)$ as a function of $r_3$. We have repeated our analysis using this procedure for comparison. A clear artefact of this procedure is that on the axis $r_8=0$, the tree-level expressions of the Polyakov loops $\ell(\mu)$ and $\bar\ell(\mu)$ are equal, as already mentioned. However, we have found that both criteria give essentially the same critical temperatures in our calculation. This is illustrated in Fig.~\ref{fig:PressureEntropy}. 

That the critical temperatures are not significantly modified in these two prescriptions can be traced back to the relative smallness of the values of $r_8$ obtained by following the saddle points in our procedure. This, in turn, originates from the strong Boltzmann suppression of the (heavy) quark contribution to the potential, which is responsible for the departure of the saddle point from the axis $r_8=0$. We point out that the situation might be very different in the case of light quarks \cite{Fischer:2013eca,Fischer:2014vxa,Folkestad:2018psc} and that different procedures for identifying the relevant extremum of the potential may have more dramatic consequences. This needs to be investigated further.

\subsection{$\mu$-dependence of the Polyakov loops}
Finally, in Fig.~\ref{fig:PressureEntropy2}, we show the Polyakov loops and the corresponding free energies as functions of $\mu$ for fixed $T/m=0.33$ and $M/m=2.22$, for $N_f=3$ degenerate flavors. We observe that the Polyakov loops have a different monotony at small $\mu$ but then increase together towards one, in line with the observations of Ref.~\cite{Dumitru:2003hp}. In this reference, the different monotony at small $\mu$ was used to question the interpretation of the logarithms of the Polyakov loops as free energies. Here, we show instead that this behavior is perfectly in line with the free-energy interpretation.

The point is that, even though the average charge $Q$ of the thermal bath should vanish for a vanishing chemical potential, this is not so for the average charge $Q_q$ and $\bar Q_q$ of the thermal bath in the presence of a test quark/anti-quark. Therefore, we must have
\begin{equation}
\left.\frac{\partial\Delta F}{\partial\mu}\right|_{\mu=0}=-\left.\frac{\partial\Delta \bar F}{\partial\mu}\right|_{\mu=0}\neq 0\,,
\end{equation}
which explains why the two Polyakov loops have a different monotony for small $\mu$. In fact we find that 
\begin{equation}
\Delta Q_{\mu=0}\equiv-\left.\frac{\partial\Delta F}{\partial\mu}\right|_{\mu=0}<0 \quad \mbox{and} \quad \Delta \bar Q_{\mu=0}\equiv-\left.\frac{\partial\bar\Delta F}{\partial\mu}\right|_{\mu=0}>0\,,
\end{equation}
so that, in the presence of a test quark (anti-quark) at $\mu=0$, the thermal bath charges negatively (positively), see Fig.~\ref{fig:PressureEntropy2}. We also show $\Delta Q=Q_q-Q$ and $\Delta\bar Q=\bar Q_q-Q$ for increasing values of $\mu$. The fact that they both approach $0$ is expected since both $Q_q$ and $\bar Q_q$ should approach $Q$ at large $\mu$.

  \begin{figure}[t]
	\centering
	\includegraphics[width=0.64\textwidth]{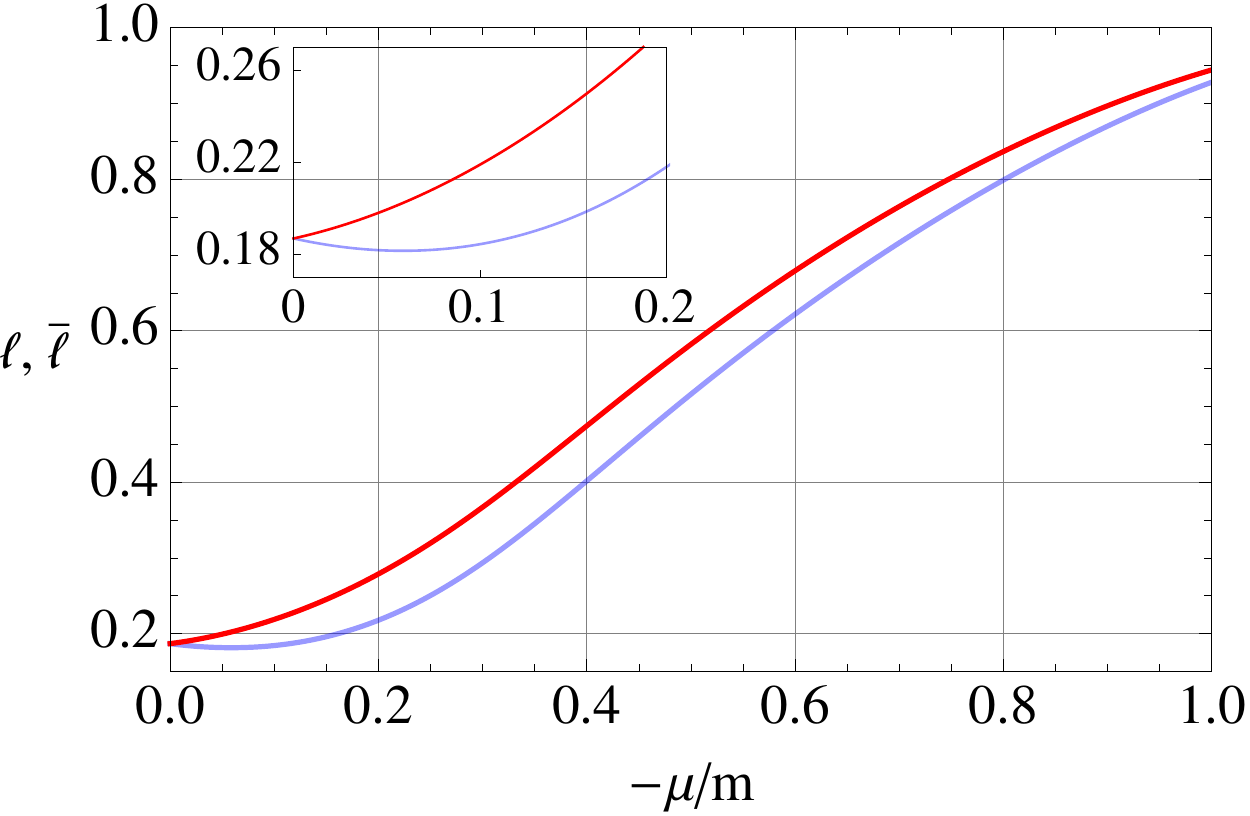}\\
	\vglue5mm
	\includegraphics[width=0.64\textwidth]{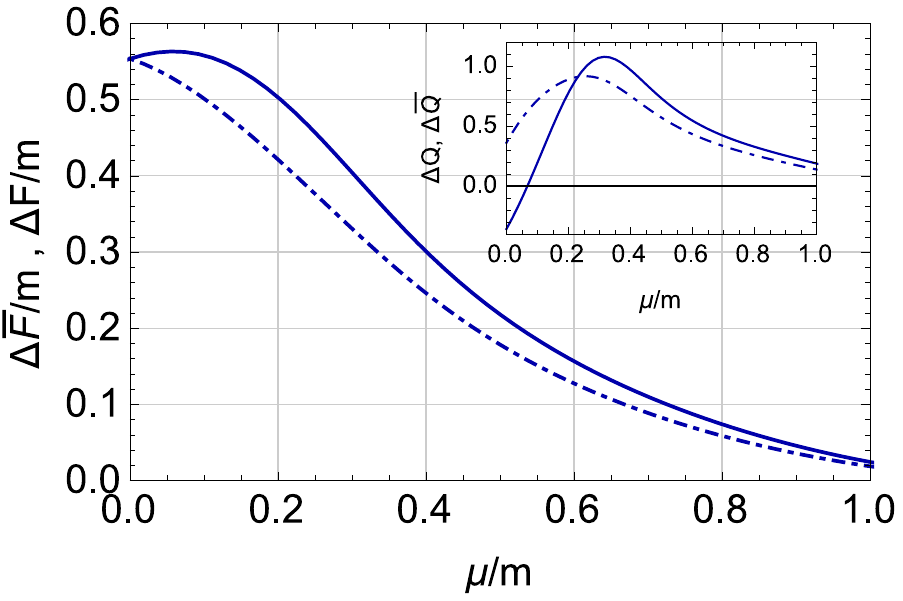}
	\caption{\label{fig:PressureEntropy2} The Polyakov loops $\ell$ (plain) and $\bar\ell$ (dot-dashed) and the corresponding free energies as functions of $\mu/m$. The inset in the first plot shows a close-up view on the small $\mu$ region where the change of monotony for $\ell$ { (and therefore $\Delta F_q$) occurs. The second inset shows the difference of average baryonic charge of the bath in the presence of a test quark (plain) or a test anti-quark (dot-dashed), with respect to the average charge in the absence of test charges.}}
\end{figure}

\clearemptydoublepage
%
%
%
%
\let\textcircled=\pgftextcircled
\chapter{Two-point correlators in the Landau-deWitt gauge}
\label{chap:correlators}

\initial{I}n this final chapter, and as an echo to the discussion at the beginning of the manuscript, we revisit the two-point correlation functions at finite temperature from the perspective of the Landau-deWitt gauge. We have argued that the order parameter for the deconfinement transition is properly accounted for in this gauge and appears as a shift of momentum in the Feynman rules. Therefore, we expect that the corresponding correlation functions are more sensitive to the deconfinement transition than the corresponding correlation functions in the Landau gauge. We shall test precisely these expectations by computing the various components of the ghost and gluon propagators in the massive extension of the Landau-deWitt gauge.

The main interest of these calculations is that they should be easily comparable with lattice simulations. In particular, in the low temperature phase, the value of the order parameter is fixed to zero, corresponding to a confining background configuration, and the lattice simulations are similar to those in the Landau gauge but with, definite, twisted boundary conditions.

This chapter will also be the opportunity to clarify further aspects of the background field method at finite temperature. In particular, we shall investigate to which extent the violation of the underlying assumptions described in Chap.~\ref{chap:bg} on the generating functional $W[J;\bar A]$, as it is the rule in most practical implementations of the gauge-fixing, impacts on the very use of the background field method. Based on this discussion, we shall also conjecture a specific behavior of the gluon susceptibility at the transition in any implementation of the gauge-fixing that preserves these basic assumptions.

\section{Propagators in the LdW gauge}
We have already given many details in previous chapters on how perturbative calculations are conveniently carried out in the Landau-deWitt gauge in a way that pretty much mimics the calculations in the Landau gauge. Here, we shall be briefer, highlighting only those aspects specific to the evaluation of two-point functions.
 
In contrast to what happens in the Landau gauge, the two-point correlators are not proportional to the identity in color space because color invariance is explicitly broken by the presence of the background field. There remains nevertheless a residual invariance under color rotations of the form $\exp\{i\theta^jt^j\}$, with $t^j$ the generators of the Cartan sub-algebra. Under these transformations, the fields transform as \cite{Reinosa:2015gxn}
\begin{equation}
  \varphi^\kappa\to e^{i\kappa\cdot \theta}\varphi^\kappa\,,
\end{equation}
where we note that neutral modes $\kappa=0^{(j)}$ are invariant. It follows that that two-point correlators are block diagonal, with blocks in the neutral and charged sectors, but no mixing between the neutral and charged sectors. Moreover the block in the charged sector is diagonal. The block in the neutral sector does not need to be diagonal but this question is irrelevant in the SU(2) case considered here, see \cite{Reinosa:2016iml} for a discussion in the SU(3) case.

We define the color components of the ghost and gluon propagators as 
\begin{equation}
\label{eq:propertyiy}
   {\cal G}^{\kappa\lambda}(K)=\delta^{-\kappa,\lambda}{\cal G}^\lambda(K)\,\,,\quad{\cal G}_{\mu\nu}^{\kappa\lambda}(K)=\delta^{-\kappa,\lambda}{\cal G}_{\mu\nu}^\lambda(K)\,,
\end{equation}
 and those of the corresponding self-energies as,
\begin{equation}
   \Sigma^{\kappa\lambda}(K)=\delta^{-\kappa,\lambda}\Sigma^\kappa(K)\,\,,\quad\Pi_{\mu\nu}^{\kappa\lambda}(K)=\delta^{-\kappa,\lambda}\Pi_{\mu\nu}^\kappa(K)\,.
\end{equation}
With these conventions, we have, for the ghost propagator
\begin{equation}
\label{eq:lefantome}
  {\cal G}^\lambda(K)=\frac{1}{\left(K^\lambda\right)^2+\Sigma^\lambda(K)}\,.
\end{equation}
As for the gluon propagator, it is transverse with respect to the generalized momentum: $\smash{K_\mu^\lambda {\cal G}^\lambda_{\mu \nu}(K)= {\cal G}^\lambda_{\mu \nu}(K)K_\nu^\lambda=0}$ \cite{Reinosa:2016iml}. At finite temperature, it then admits the following tensorial decomposition
\begin{equation}\label{eq:decomp}
  {\cal G}^\lambda_{\mu \nu}(K)= {\cal G}^\lambda_T(K) P_{\mu \nu}^T(K^\lambda)+{\cal G}^\lambda_L(K) P_{\mu \nu}^L(K^\lambda)\,,
\end{equation}
where $P_{\mu \nu}^T(K)$ and $P_{\mu \nu}^L(K)$ are the transverse and longitudinal projectors with respect to the frame of the thermal bath, defined as [we write $K_\mu=(\omega,{ k})$ and $k^2={ k}^2$]
\begin{equation}\label{eq_projT}
  P^T_{\mu \nu}(K)=\left(1-\delta_{\mu 0} \right)\left(1-\delta_{\nu 0} \right) \left(\delta_{\mu \nu}-\frac{K_\mu K_\nu}{k^2}\right)
\end{equation}
and
\begin{equation}\label{eq_projL}
  P^L_{\mu \nu}(K)+P_{\mu \nu}^T(K)=P_{\mu \nu}^\perp(K)= \delta_{\mu \nu}-\frac{K_\mu K_\nu}{K^2}.
\end{equation}
{ It follows in particular that ${\cal G}^\lambda_{\mu\nu}(K)={\cal G}^\lambda_{\nu\mu}(K)$. In terms of the projected self-energies
\begin{eqnarray}
\label{eq_decompo_propag-2}
  \Pi^\lambda_{T}(K) & = & \frac{P^T_{\mu \nu}(K^\lambda)\Pi_{\mu\nu}^\lambda(K)}{d-2}\,,\\
\label{eq_decompo_propag-3}
  \Pi^\lambda_{L}(K) & = & P^L_{\mu \nu}(K^\lambda)\Pi_{\mu\nu}^\lambda(K)\,,
\end{eqnarray}
the scalar components of the gluon propagator read 
\begin{equation}\label{eq_decompo_propag}
  {\cal G}^\lambda_{T/L}(K)=\frac{1}{\left(K^\lambda\right)^2+m^2_0+\Pi^\lambda_{T/L}(K)}.
\end{equation}}
In the present case, one can also show that \cite{Reinosa:2016iml}
\begin{align}
\label{eq:prop1}
   {\cal G}^\lambda(K)&={\cal G}^{-\lambda}(-K)\,,\\ 
\label{eq:LT}
  {\cal G}^\lambda_{L,T}(K)&={\cal G}^{-\lambda}_{L,T}(-K)\,,
\end{align}
and similarly for the self-energies.

\subsection{Ghost propagator}
The ghost self-energy involves only one diagram, see Fig.~\ref{fig:ccb}. Using similar reduction techniques as the ones employed for the diagrams contributing to the background effective potential, we arrive at
\begin{eqnarray}
 \label{eq:ghost_self}
 \Sigma^\lambda(K) & = &  \sum_{\kappa,\tau}{\cal C}_{\kappa \lambda \tau} \left[\frac{K_\lambda^2-m^2}{4m^2} \left(J_m^\kappa-J_0^\kappa \right)-\frac{\omega^\lambda}{2m^2}(\tilde J_m^\kappa - \tilde J_0^\kappa)\right.\nonumber\\
& &  \hspace{2.0cm}\left.+\,\frac{K_\lambda^4}{4m^2}I_{00}^{\kappa\tau}(K)-\frac{\left(K_\lambda^2+m^2\right)^2}{4m^2} I_{m0}^{\kappa\tau}(K)\right],
\end{eqnarray}
where the tensor ${\cal C}_{\kappa\lambda\tau}$ and the sum-integrals $J_\kappa$ and $\tilde J_\kappa$ were introduced in Chap.~\ref{chap:YM_2loop} and 
\begin{equation}
\label{eq_integralsss}
  I_{m_1m_2}^{\kappa\tau}(K)\equiv\int_Q^T G_{m_1}(Q^\kappa)\,G_{m_2}(L^\tau)\,,
\end{equation}
with $\smash{K_\lambda+Q_\kappa+L_\tau=0}$. The scalar-type Matsubara sum (\ref{eq_integralsss}) can be easily evaluated using standard techniques \cite{Reinosa:2016iml}. 
 \begin{figure}[t]
  \centering
  \includegraphics[width=.4\linewidth]{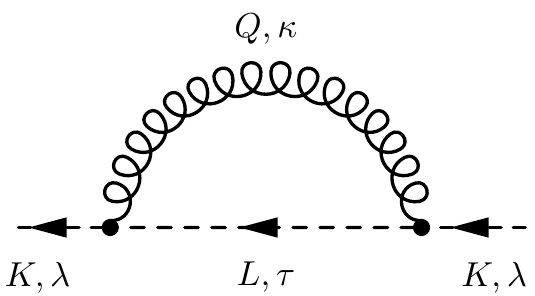}
   \caption{One-loop contribution to the ghost self-energy. Momenta and color charges are all either entering or outgoing at the vertices.}
  \label{fig:ccb}
\end{figure}

\subsection{Gluon propagator}
As for the gluon self-energy, it requires the evaluation of the diagrams depicted in Fig.~\ref{fig:AA}. Using similar techniques as above we arrive at
\begin{eqnarray}
 \Pi_{T/L}^\lambda(K) & \!\!\!\!\!=\!\!\!\!\! &  \sum_{\kappa,\tau}{\cal C}_{\lambda\kappa\tau}\left\{\left(1-\frac{K^4_\lambda}{2m^4}\right)\!\{I_{T/L}^\lambda\}_{00}^{\kappa\tau}(K)+\left(1+\frac{K^2_\lambda}{m^2}\right)^{\!\!2}\!\{I_{T/L}^\lambda\}_{m0}^{\kappa\tau}(K)\right.\nonumber\\
& & \hspace{1.6cm}-\,2\!\left[d-2+\left(1+\frac{K^2_\lambda}{2m^2}\right)^{\!\!2}\right]\!\{I_{T/L}^\lambda\}_{mm}^{\kappa\tau}(K)\nonumber\\
& & \hspace{1.6cm}+\,(d-2)J_m^\kappa+\frac{K_\lambda^2+m^2}{m^2}\left(J_m^\kappa-J_0^\kappa\right)-\frac{2\omega^\lambda}{m^2}(\tilde J^\kappa_m-\tilde J_0^\kappa)\nonumber\\
& & \hspace{1.6cm}\left.+\,\frac{(K^2_\lambda+m^2)^2}{m^2}I_{m0}^{\kappa\tau}(K)-K^2_\lambda\left(4+\frac{K^2_\lambda}{m^2}\right)I_{mm}^{ \kappa\tau}(K)\right\},
\end{eqnarray}
where we have introduced the following integrals
\begin{align} \label{eq:integrals2}
 \{I_{T}^\lambda\}_{m_1m_2}^{\kappa\tau}(K) & =  \frac{P^T_{\mu \nu}(K^\lambda)\{I_{\mu\nu}\}_{m_1m_2}^{\kappa\tau}(K)}{d-2}\,,\\
 \{I_L^\lambda\}_{m_1m_2}^{\kappa\tau}(K) & =  P^L_{\mu \nu}(K^\lambda)\{I_{\mu\nu}\}_{m_1m_2}^{\kappa\tau}(K)\,,  \label{eq:integrals3}
\end{align}
with
\begin{equation}
\label{eq:hehehe}
 \{I_{\mu\nu}\}_{m_1m_2}^{\kappa\tau}(K)\equiv\int_Q { Q_\mu^\kappa Q_\nu^\kappa}\,G_{m_1}(Q^\kappa)\,G_{m_2}(L^\tau)\,,
\end{equation}
and $\smash{K_\lambda+Q_\kappa+L_\tau=0}$. Again, these sum-integrals can be evaluated using standard techniques and final expressions can be found in \cite{Reinosa:2016iml}.

\begin{figure}[t]
\centering
    \includegraphics[width=0.95\linewidth]{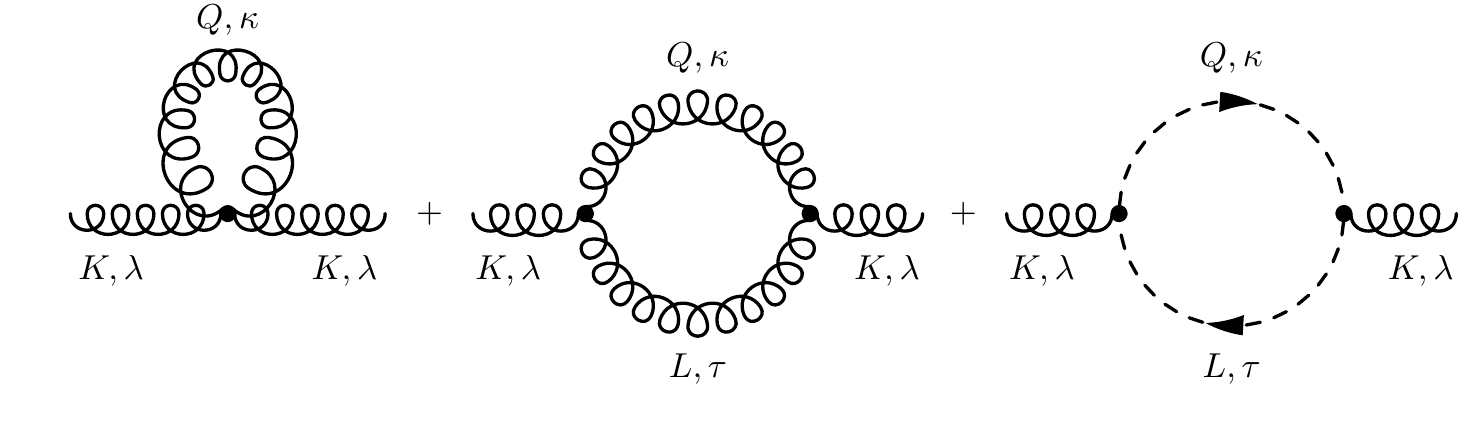}
  \caption{One-loop diagrams for the gluon self-energy.}
  \label{fig:AA}
\end{figure}

\subsection{Discussion}
For a thorough discussion of the propagators, we refer to \cite{Reinosa:2016iml}. Here, we concentrate on the longitudinal susceptibility $\chi$, defined as the zero-momentum value of the longitudinal propagator, more precisely its neutral component. We find
\begin{eqnarray}
\chi^{-1}=m^2 & \!\!\!\!\!-\!\!\!\!\! & {g^2\over\pi^2}\int_0^\infty\!\!dq\,\mbox{Re}\,n_{q-irT}\,q\left(\frac{1}{2}+{q^2\over m^2}\right)\nonumber\\
& \!\!\!\!\!+\!\!\!\!\! & {g^2m^2\over\pi^2}\int_0^\infty\!\!dq\,\mbox{Re}\,\frac{n_{\varepsilon_{m,q}-irT}}{\varepsilon_{m,q}}\left(3+6{q^2\over m^2}+{q^4\over m^4}\right).\label{eq:AAA}
\end{eqnarray}
Here, it is understood that $r$ is to be taken at the minimum $r_{\mbox{\scriptsize min}}(T)$ of the background effective potential. For a generic $r$, the same expression corresponds to the inverse susceptibility computed in the presence of a source that forces the background $r$ to be self-consistent, the same source used to define the background dependent Polyakov loop in Chap.~\ref{chap:bg}. The inverse susceptibility in the Landau gauge is easily obtained by setting $r=0$ in Eq.~(\ref{eq:AAA}).

The temperature dependence of the susceptibility across the phase transition is shown in Fig.~\ref{fig_k0masses} for the choice of parameters $m=0.71$GeV and $g=7.5$. We also show the corresponding result in the Landau gauge to quantify the effect of the nontrivial background. We observe that the susceptibility in the Landau-deWitt gauge presents a cusp at the transition, in sharp contrast with the results in the Landau gauge. The cusp reflects the non-analytic behavior of the order parameter $r_{\mbox{\scriptsize min}}(T)$ across the transition and is in fact present in all the correlator components. However, it is only in the neutral longitudinal component where it appears in such a pronounced manner, turning the slight non-monotonous behavior in the Landau gauge into a rather identifiable peak.

This rise of the susceptibility as one approaches the deconfinement transition could be easily studied on the lattice. Indeed, in the low temperature phase, the background is fixed to its confining value $r=\pi$. Using a center transformation, it is then possible to map the Landau-deWitt gauge-fixing action into the Landau gauge-fixing action, the only difference with respect to the standard lattice simulations being that the fields should obey anti-periodic boundary conditions.

 \begin{figure}[t]
  \centering
  \includegraphics[width=0.8\linewidth]{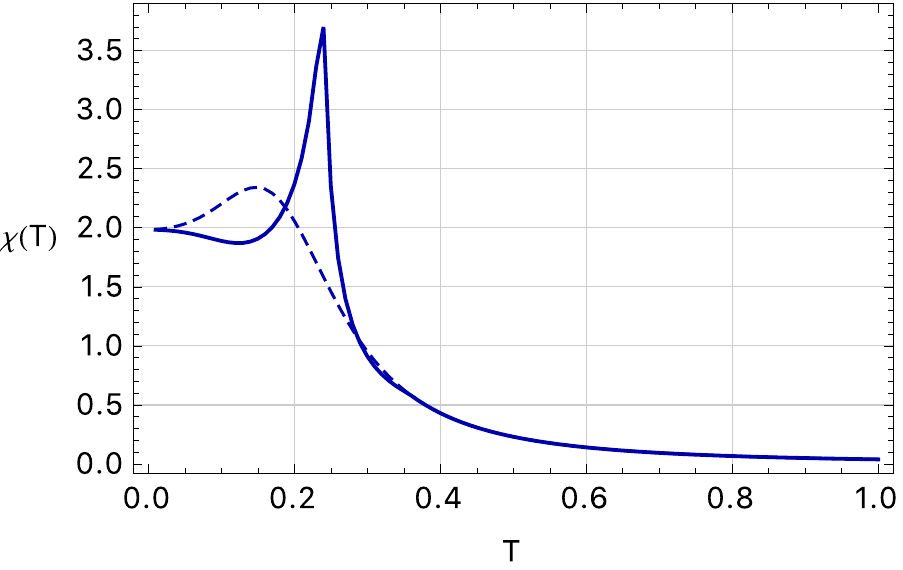}
   \caption{Temperature dependence of the susceptibility in the longitudinal and neutral sector of the Landau-deWitt gauge, compared to the corresponding susceptibility in the Landau gauge. The parameters used are $m=0.71$GeV and $g=7.5$ and $r_{\mbox{\scriptsize}}(T)$ is determined from the one-loop background field effective potential.}
  \label{fig_k0masses}
\end{figure}

\section{Back to the effective action $\Gamma[A,\bar A]$}

The previous calculation also sheds considerable light on the use of the background field method and the relation between the functionals $\Gamma[A,\bar A]$ and $\tilde\Gamma[\bar A]=\Gamma[\bar A,\bar A]$. In particular, we argued in Chap.~\ref{chap:bg} that self-consistent backgrounds defined by the condition $A_{\mbox{\scriptsize min}}(\bar A)=\bar A$ correspond to the absolute minima of $\tilde\Gamma[\bar A]$. This result is at the basis of the identification of the self-consistent backgrounds as order parameters for the deconfinement transition. It requires however working with an ``ideal'' gauge-fixing such that the generating functional $W[J,\bar A]$ is convex with respect to $J$ and its zero-source limit $W[0,\bar A]$ is background independent. These two properties are usually not fulfilled in practice, leading to potential artefacts.

Here, we can investigate this question in the case of the Curci-Ferrari model at one-loop order. We restrict the effective action $\Gamma[A,\bar A]$ to field configurations $A$ that are constant, temporal and diagonal, just like the background $\bar A$. The discussion then boils down to that of a function of two variables $V(\hat r,r)$ with $\hat r=g A$ and $r=g\bar A$. The background field effective potential is obtained by evaluating $V(\hat r,r)$ along the diagonal $\hat r=r$, $V(r)=V(r,r)$. Another important identity is
\begin{equation}\label{eq:susc}
 \frac{\chi^{-1}}{g^2}=\left.\frac{\partial^2 V}{\partial \hat r^2}\right|_{\hat r=r}\,,
\end{equation}
which connects the susceptibility with the existence of self-consistent backgrounds, the presence of a self-consistent background at $r_s$ implying necessarily $\chi(r_s)>0$.

We shall find that there is some tension in the Curci-Ferrari model between the expected and the actual characterization of the absolute minima of $V(r)$ as self-consistent backgrounds such that $\hat r_{\mbox{\scriptsize min}}(r)=r$ where $\hat r_{\mbox{\scriptsize min}}(r)$ denotes the absolute minimum of $V(\hat r,r)$ with respect to $\hat r$ for a fixed $r$. In particular, in some temperature interval above $T_c$, we shall find self-consistent backgrounds that are not anymore absolute minima of $V(r)$. We shall interpret this tension as resulting from the violation of the basic assumptions referred to above. Reverting the argument, we shall then conjecture a certain behavior of the susceptibility at the deconfinement transition in the case where the basic assumptions are fulfilled.

\subsection{One-loop expression for $V(\hat r,r)$}
To obtain $V(\hat r,r)$ at one-loop order, we expand the classical action (\ref{eq:S0}) to quadratic order around a generic gluon configuration: $A\to A+a$. In Fourier space, the quadratic part in the ghost sector reads
\begin{equation}
 \int_Q^T\,\bar c^{-\kappa}(-Q)\,Q_\kappa\cdot \hat{Q}_\kappa c^\kappa(Q)\,,
\end{equation}
with $Q_\kappa=Q+\kappa r\,n$ and ${\hat Q}_\kappa=Q+\kappa \hat r\,n$. Therefore
\begin{equation}
 \delta V_{\mbox{\scriptsize gh}}(\hat r,r)=-\sum_\kappa \int_Q^T \ln\big[Q_\kappa\cdot\hat Q_\kappa\big].
\end{equation}
In the gluonic sector, the quadratic part reads
\begin{equation}
  \int_Q^T\,\left\{\frac{1}{2}a_\mu^{\kappa}(Q)^*({\hat Q}_\kappa^2P^\perp_{\mu\nu}(\hat{Q}_\kappa)+m^2_0\delta_{\mu\nu})\,a_\nu^\kappa(Q)+h^{\kappa}(Q)^* Q_\mu^\kappa a_\mu^\kappa(Q)\right\}.
\end{equation}
The corresponding determinant can be evaluated using Schur's formula. We find $({\hat Q}^2_\kappa+m^2_0)^{d-2}\left((Q_\kappa\cdot{\hat Q}_\kappa)^2+m_0^2Q_\kappa^2\right)$. It follows that
\begin{equation}
\delta V_{\mbox{\scriptsize gl}}(\hat r,r)=\frac{d-2}{2}\sum_\kappa\int_Q^T\ln\big[{\hat Q}_\kappa^2+m_0^2\big]+\frac{1}{2}\sum_\kappa\int_Q^T\ln\big[(Q_\kappa\cdot{\hat Q}_\kappa)^2+m_0^2Q_\kappa^2\big].
\end{equation}
Putting all one-loop contributions together and adding the tree-level contribution, we arrive at
\begin{eqnarray}
 V(\hat r,r) & \!\!\!\!\!=\!\!\!\!\! & \frac{m^2_0}{2g^2}(\hat r-r)^2+\frac{d-2}{2}\sum_\kappa\int_Q^T\ln\big[{\hat Q}_\kappa^2+m^2\big]\nonumber\\
& \!\!\!\!\!+\!\!\!\!\! & \frac{1}{2}\sum_\kappa\int_Q^T\ln\left[1+\frac{m^2Q_\kappa^2}{(Q_\kappa\cdot{\hat Q}_\kappa)^2}\right],
\end{eqnarray}
where we have replaced the bare mass $m_0$ by the renormalized mass $m$ in the one-loop terms. We check that for $\hat r=r$, we recover the one-loop expression for the background field effective potential
\begin{equation}
 V(r)=\frac{d-1}{2}\sum_\kappa \int_Q^T\ln\big[Q_\kappa^2+m^2\big]-\frac{1}{2}\sum_\kappa\int_Q^T\ln Q_\kappa^2\,.
\end{equation}
Using ${\hat Q}_\kappa=Q_\kappa+\kappa (\hat r-r) n$ and separating the neutral and charge modes, one arrives at
\begin{eqnarray}
 V(\hat r,r) & = & \frac{m^2_0}{2g^2}\,(\hat r-r)^2\nonumber\\
& + & \frac{d-1}{2}\int_Q^T \ln\big[Q^2+m^2\big]-\frac{1}{2}\int_Q^T\ln Q^2\nonumber\\
& + & (d-2)\int_Q^T \ln\big[{\hat\omega}_+^2+q^2+m^2\big]\nonumber\\
& + & \int_Q^T \ln \left(1+\frac{m^2Q^2_+}{(Q_+^2+(\hat r-r)\omega_+)^2}\right).\label{eq:quartic}
\end{eqnarray}
If $m_0=m=0$, $V(\hat r,r)$ does not depend on $r$ and is $2\pi$-periodic in $\hat r$. In the massive case instead, $V(\hat r,r)$ is not defined for all values of $\hat r$ et $r$. The reason is the term $\ln \,\big[Q_+^2+(\hat r-r)(\omega+r)\big]$ whose argument writes
\begin{eqnarray}
 \left(\omega+\frac{\hat r+r}{2}\right)^2+q^2-\left(\frac{\hat r-r}{2}\right)^2.
\end{eqnarray}
If $0<q^2<(\hat r-r)^2/4$, the argument vanishes for real frequencies
\begin{eqnarray}
 \omega=-\frac{\hat r+r}{2}\pm\sqrt{\left(\frac{\hat r-r}{2}\right)^2-q^2}\,.
\end{eqnarray}
As $q^2$ explores the interval $[0,(\hat r-r)^2/4]$, these frequencies span the interval $[-\mbox{Max}\,(\hat r,r),-\mbox{Min}\,(\hat r,r)]$. This interval should not contain any Matsubara frequency from which we deduce that for $r\in [2\pi nT,2\pi(n+1)T]$, then $\hat r\in  [2\pi nT,2\pi(n+1)T]$.

\subsection{Numerical analysis}
The evaluation of the last Matsubara sum in Eq.~(\ref{eq:quartic}) requires solving a quartic polynomial in the frequency which is cumbersome. Instead, we perform the $q$-integral analytically and the Matsubara sum numerically. Before doing so, we need to make sure that the considered sum-integral is UV finite. To this purpose we add and subtract the leading terms contributing in the UV, which we obtain conveniently by expanding in powers of $\hat r-r$:
\begin{eqnarray}
\ln\left(1+\frac{m^2Q^2_+}{(Q_+^2+(\hat r-r)\omega_+)^2}\right) & = & \ln\left(1+\frac{m^2}{Q_+^2}\right)-\frac{2m^2\omega_+}{Q_+^2(Q_+^2+m^2)}(\hat r-r)\nonumber\\
& + & \frac{m^2\omega_+^2(m^2+3Q_+^2)}{Q_+^4(Q_+^2+m^2)^2}(\hat r-r)^2\,.
\end{eqnarray}
It is easily checked that the term in $(\hat r-r)^2$, when added to the corresponding ones from the first and third lines in Eq.~(\ref{eq:quartic}), reproduce $\chi^{-1}/(2g^2)$, as it should from (\ref{eq:susc}). Moreover the terms in $\hat r-r$ can be easily expressed in terms of the sum-integrals $\tilde J$. 
\begin{figure}[t]
  \centering
  \includegraphics[width=0.9\linewidth]{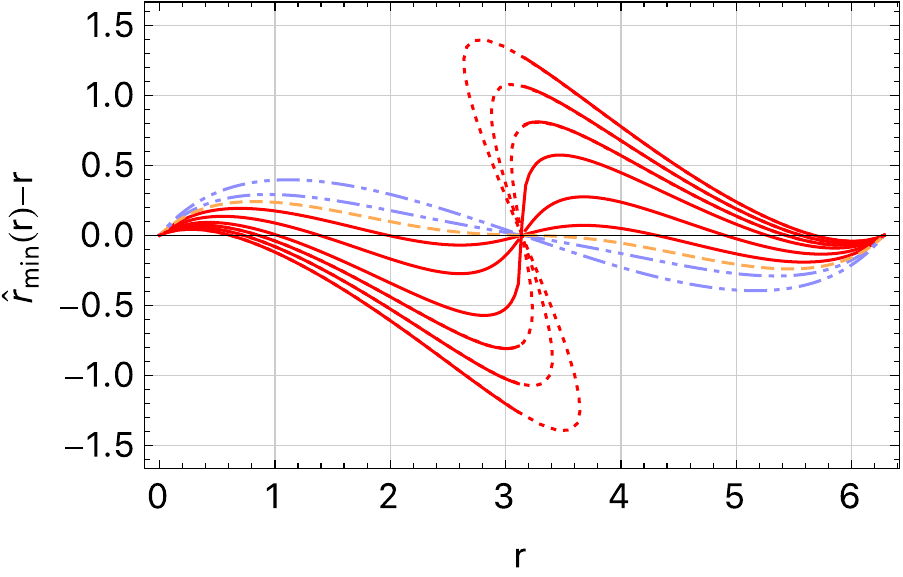}
 \caption{$\hat r_{\mbox{\scriptsize min}}(r)-r$ as a function of $r$, for increasing values of $T$. The blue (dashed-dotted), orange (dashed) and red (plaine) curves correspond respectively to temperatures below, at and above $T_c$. The dotted (red) lines correspond either to maxima or local minima as a function of $\hat r$ for fixed $r$. These additional extrema appear above $\tilde T$, the temperature at which a second order transition occurs in $V(\hat r,r=\pi T)$.}\label{fig:two}
\end{figure}
We are then left with the numerical evaluation of a convergent Matsubara sum. We have checked that only the first few terms in the sum are needed and that neglecting the sum does not change qualitatively the results, which we present in Fig.~\ref{fig:two} where identify the self-consistent backgrounds as the zeros of the function $\hat r_{\mbox{\scriptsize min}}(r)-r$. 

We find that, as long as $T<T_c$, and aside from the spurious Landau solution $r=0$, there is only one self-consistent background at $r=\pi$ corresponding to the absolute minimum of $V(r)$. In some range of temperatures $T_c<T<\tilde T$ above the transition, $r=\pi$ remains self-consistent, even though it is not anymore an absolute minimum of $V(r)$. Only above $\tilde T$ does $r=\pi$ ceases to be self-consistent. The temperature $\tilde T$ is such that $\chi^{-1}(\pi)=0$. At this temperature, a second order transition occurs in the direction of $\hat r$ at $r=\pi$ fixed and turns $\hat r=\pi$ into a maximum, explaining why $r=\pi$ ceases to be self-consistent.

\subsection{Conjecture}
It is tempting to interpret the previous tension between self-consistent backgrounds and absolute minima of $V(r)$ as originating from our approximate implementation of the gauge-fixing which does not respect the basic requirements for the generating functional $W[J,\bar A]$. As such, it should not be specific to the Curci-Ferrari approach but be also present in other practical approaches that violate these basic assumptions.\footnote{Let us mention however that this does not completely hinder the analysis of the deconfinement transition since there is a range of temperatures where the identification of self-consistent backgrounds and absolute minima of $V(r)$ holds.}

As negative as it could seem, this result can be turned into a conjecture for the actual behavior of the susceptibility in any gauge-fixing that respects the above mentioned properties, such as the lattice gauge-fixing. Indeed, in any such gauge-fixing the two temperatures $T_c$ and $\tilde T$ should coincide. One natural scenario for this to happen is that the continous phase transition in the direction $\hat r$ for $r=\pi T$ occurs exactly at $\tilde T=T_c$. In that case, not only would the curvature of $V(r)$ at $r=\pi T$ vanish, but also the susceptibility $\chi$. Our conjecture is then that, the Landau-deWitt gauge susceptibility computed on the lattice should not present a peak as in Fig.~\ref{fig_k0masses}, but rather diverge at $T_c$.

That two quantities, $V''(r_{\mbox{\scriptsize min}})$ and $\chi^{-1}(r_{\mbox{\scriptsize min}})$, vanish simultanously at $T_c$ seems highly improbable {\it a priori}. However, as we now discuss this scenario becomes viable if the basic requirements are fulfilled. The point is that, from the definition of $\hat r_{\mbox{\scriptsize min}}(r)$ and the background independence of the free energy $V(\hat r_{\mbox{\scriptsize min}}(r),r)$, follow the two equations
\begin{equation}
 0=\left.\frac{\partial V}{\partial \hat r}\right|_{\hat r_{\mbox{\tiny min}}(r),r} \quad \mbox{and} \quad  0=\left.\frac{\partial V}{\partial r}\right|_{\hat r_{\mbox{\tiny min}}(r),r}\,.\label{eq:bbb}
\end{equation}
Taking a derivative with respect to $r$, we obtain the system
\begin{eqnarray}
 0 & = & \left.\frac{\partial^2 V}{\partial\hat r^2}\right|_{\hat r_{\mbox{\tiny min}}(r),r}\hat r'_{\mbox{\tiny min}}(r)+\left.\frac{\partial^2 V}{\partial r \partial\hat r}\right|_{\hat r_{\mbox{\tiny min}}(r),r}\,,\label{eq:ccc1}\\
 0 & = & \left.\frac{\partial^2 V}{\partial r\partial\hat r}\right|_{\hat r_{\mbox{\tiny min}}(r),r}\hat r'_{\mbox{\tiny min}}(r)+\left.\frac{\partial^2 V}{\partial r^2}\right|_{\hat r_{\mbox{\tiny min}}(r),r}\,,\label{eq:ccc2}
\end{eqnarray}
which expresses that the potential is locally flat in the direction of the vector $(\hat r'_{\mbox{\tiny min}}(r),1)$, or, in other words, that the Hessian has a zero-mode in this direction. Suppose now that there exists a temperature $\hat T$ at which the second eigenvalue vanishes as well. In this case, the Hessian would be identically zero at $(\hat r,r)=(\hat r_{\mbox{\scriptsize min}}(r),r)$, and for a self consistent background $r=\hat r_{\mbox{\scriptsize min}}(r)$, we would have
\begin{equation}
 V''(r)=\left.\frac{\partial^2 V}{\partial\hat r^2}\right|_{r,r}+2\left.\frac{\partial^2 V}{\partial\hat r\partial r}\right|_{r,r}+\left.\frac{\partial^2 V}{\partial r^2}\right|_{r,r}=0\,,
\end{equation}
in addition to $\chi^{-1}(r)=0$. We mention that there is still the possibility that $V''(r)$ vanishes not simultanously to $\chi^{-1}(r)$ but to $\hat r'_{\mbox{\tiny min}}(r)-1$ (as in the Curci-Ferrari model). Indeed, using the equations above we arrive at 
\begin{equation}
 V''(r)=(1-\hat r'_{\mbox{\scriptsize min}}(r))^2\left.\frac{\partial^2 V}{\partial\hat r^2}\right|_{r,r}\,.
\end{equation}
In this scenario however $r=\pi T$ could be self-consistent while not being an absolute minimum of $V(r)$.\footnote{Unless something like a first order transition occurs in the direction of $\hat r$ at $r=\pi T$, but this is not very natural and poses certain problems of continuity of the self-consistent background, which we do not expect in the SU(2) case.}

For completeness, we mention that somewhat similar identities can be obtained in the Curci-Ferrari model, although they do not follow from the background independence of the partition function. In this case, the second equation in (\ref{eq:bbb}) is replaced by\footnote{It can also be checked that the one-loop expression for $V(\hat r,r)$ is such that $\partial V/\partial r|_{\hat r,r}=0$ for any $r$, from which it follows that $V'(r)=\partial V/\partial\hat r|_{\hat r=r,r}$ and, therefore, $V''(r)=\partial^2 V/\partial\hat r^2|_{\hat r=r,r}+\partial^2 V/\partial\hat r\partial r|_{\hat r,r}$. Combining this identity with (\ref{eq:ccc1}) which is still valid, we arrive, for a self-consistent background $r$, at $V''(r)=(1-\hat r'_{\mbox{\tiny min}}(r))\partial^2 V/\partial \hat r^2|_{\hat r=r,r}$. This formula is exact at one-loop order. It is seen to be compatible with (\ref{eq:ddd}) after expanding to leading order in $g^2$, with $1-\hat r'_{min}(r)\sim g^2$ and $g^2 \partial^2 V/\partial \hat r^2|_{\hat r=r,r}=m^2_0+{\cal O}(g^2)$.}
\begin{equation}
 \frac{m^2_0}{g^2}(r-\hat r_{\mbox{\tiny min}}(r))=\left.\frac{\partial V}{\partial r}\right|_{\hat r_{\mbox{\tiny min}}(r),r}\,,
\end{equation}
which implies
\begin{eqnarray}
 \frac{m^2_0}{g^2}(1-\hat r'_{\mbox{\tiny min}}(r)) & = & \left.\frac{\partial^2 V}{\partial r\partial\hat r}\right|_{\hat r_{\mbox{\tiny min}}(r),r}\hat r'_{\mbox{\tiny min}}(r)+\left.\frac{\partial^2 V}{\partial r^2}\right|_{\hat r_{\mbox{\tiny min}}(r),r}\,,
\end{eqnarray}
to replace (\ref{eq:ccc2}). It follows that
\begin{equation}
 V''(r)=(1-\hat r'_{\mbox{\scriptsize min}}(r))^2\left.\frac{\partial^2 V}{\partial\hat r^2}\right|_{r,r}+\frac{m^2_0}{g^2}(1-\hat r'_{\mbox{\tiny min}}(r))\,,\label{eq:ddd}
\end{equation}
which, we note, allows again for a simultanous vanishing of $V''(r)$ and $\hat r'_{\mbox{\tiny min}}(r)-1$, but not of $V''(r)$ and $\chi^{-1}(r)$ (unless there is a simultanous vanishing of three quantities).

\clearemptydoublepage
%
%
%
%

\chapter{Conclusions and Outlook}

\initial{T}ackling the most intriguing properties of Quantum Chromodynamics  and Yang-Mills theories in the continuum requires one to go beyond the standard (but incomplete) Faddeev-Popov gauge-fixing procedure which, although very efficient at high energies, is known not to be valid at low energies. The so-called Curci-Ferrari model has received a certain attention lately as a possible candidate for such an extension in the Landau gauge. Part of that attention is rooted in the impressive agreement between one-loop vaccum correlation functions computed within the model, and the most accurate determinations of the Landau gauge correlation functions on the lattice. We know now that this agreement extends and even improves at two-loop accuracy. Moreover, the Curci-Ferrari running coupling needed in these comparisons remains moderate over the whole range of scales, as opposed to the running coupling in the standard Faddeev-Popov approaches which diverges at a finite scale. These two surprising results open the exciting possibility to tackle some of the low energy properties from perturbative means.

Of course, ideally one should aim at generating the Curci-Ferrari model from first principles, starting from the QCD/YM actions and applying a {\it bona fide} gauge-fixing procedure. A clear-cut such mechanism is not known at present in the Landau gauge but various lines of investigation are being pursued. In the meantime, a more pragmatic approach is to test the perturbative predictions of the model further and confront them to existing results, in particular those from the lattice simulations. In this manuscript, we have reviewed some of these predictions with regard to the finite temperature properties of the QCD/YM system, in particular the confinement-deconfinement transition.

We have first reviewed (in chapter 2) the results for the YM correlators at finite temperature, as computed from the Curci-Ferrari model, with an emphasis on the so-called longitudinal gluon propagator. The latter was indeed foreseen for some time as providing a direct probe onto the deconfinement transition. Although the Curci-Ferrari model describes qualitatively well the lattice results for the various correlators, it fails in describing the behavior of the zero-momentum longitudinal propagator (inverse susceptibility). However, this negative result does not necessarily signal a failure of the model because the same limitations are observed within other approaches that do not rely on the Curci-Ferrari model. Rather it has been argued that the limitations originate in the use of the Landau gauge which fails in capturing the order parameter associated to the deconfinement transition. It is known that a more appropriate gauge at finite temperature is the so-called Landau-deWitt gauge, the background generalization of the Landau gauge.

We have devoted chapters 3, 4 and 5 to a thorough review of the rationale behind the use of background field gauges at finite temperature and why they are indeed the {\it good} gauges to discuss the deconfinement transition. Although the recipe for using the background field method at finite temperature is a known result, we have tried to give a self-contained and original derivation, with special emphasis and discussion of the underlying assumptions that are usually left implicit. The relevance of this discussion is that some of these assumptions are usually violated by practical implementations of the gauge-fixing (including the Curci-Ferrari modelling), with a potential impact on the interpretation of the results. Some of these assumptions can also be violated by the physical set-up, as in the presence of quarks at finite density. The corresponding impact on the background field methods has been discussed in chapter 9.

Based on this knowledge, we have extended the Curci-Ferrari model in the presence of a background, which needs to be seen as the finite temperature counterpart of the standard Curci-Ferrari model in the vacuum. We have investigated various perturbative predictions of this model with regard to the deconfinement transition. In particular, in chapter 6, we have analyzed the one-loop predictions in the YM case. We find that, already at one-loop order, the model predicts a deconfinement transition, with transition temperatures in reasonably good agreement with the results of other approaches, including lattice. Chapter 7 is devoted to testing the convergence properties of the perturbative expansion within the model. We find in particular improved values for the transition temperatures that get closer to the values determined on the lattice. Chapter 8 further deepens the connection between center symmetry and the deconfinement transition by analysing the symmetry breaking patter in the SU($4$) case. 

In chapter 10, we have extended our analysis in the presence of quarks, in the formal but interesting regime where all quarks are considered heavy. We find that most of the known qualitative and quantitative features in this case can be reproduced by using our model approach at one-loop order. Higher order corrections have also been investigated in this case and shown to improve the results.\footnote{We decided not to discuss those results in detail since they will be part of the thesis defended by my PhD student in Fall 2019.} 

In chapter 11, using the accumulated knowledge, we have revisited the question of whether the gluon susceptibility can probe the deconfinement transition. In particular, we find that, as compared to what occurred in the Landau gauge, the gluon susceptibility the Landau-deWitt gauge displays a characteristic peaked behavior at the deconfinement transition, which could be tested against lattice simulations. In fact, using the general results derived in chapter 4, we conjecture that the lattice susceptibility should diverge at the transition, the peak observed in the Curci-Ferrari model reflecting the distance from a true {\it bona fide} gauge fixing.

We have also discussed various open questions which concern not only the Curci-Ferrari model but other continuum approaches. In particular, it seems that a common mechanism for confinement in all these approaches is the ghost dominance observed at low temperatures. At first sight, a ghost dominated phase at low temperature seems worrisome for ghost degrees of freedom are unphysical and come with negative thermal distribution functions, leading potentially to inconsistent thermodynamics. We have shown, however, that the presence of a surrounding confining background leads to a transmutation of some of these negative distribution functions into positive distribution functions, with a net positive contribution to, say the entropy density at low temperatures. The same mechanism leads to a slightly negative entropy density right below the deconfinement transition at one-loop order. However, we have shown that two-loop corrections cure this inconsistent behavior. 

A problem that found no solution so far is that, in most continuum approaches in the Landau/Landau-deWitt gauges, there remain massless degrees of freedom at low temperature that contribute polynomially to the thermodynamical observables in this limit, at odds with the observations made on the lattice. This is certainly a future challenge for continuum approaches that could relate to the use of more exotic background configurations that the ones used in this work, to the way we envisage and implement gauge-fixing, or to the ability of the approach to generate the low-lying bound states, in particular glueball states in the case of YM theory. All these aspects are currently under investigation.

These problems aside, the natural question that comes next is whether the applicability of the perturbative Curci-Ferrari model (be it in the Landau gauge in the vacuum or in the Landau-deWitt gauge at finite temperature) extends to the physical QCD case, that is in the presence of light quarks, in which case the relevant symmetry is chiral symmetry and its spontanous breaking. We know already that the answer to this question is negative, the reason being that, in the presence of light quarks, the strength of the interaction between quarks and gluons in the infrared is two to three times larger than that in the YM sector of the theory, preventing a full perturbative analysis. One can however use the knowledge that the Curci-Ferrari model in the YM sector is essentially perturbative to construct a systematic expansion for the Curci-Ferrari model in the QCD case, controlled by two small parameters, the YM coupling and the inverse of the number of colors. This strategy has already been used in the vacuum where it has been shown to capture the physics of chiral symmetry breaking. It remains to be seen whether it permits to capture hadronic observables. In particular, in a current investigation we aim at the determination of the pion decay constant from an ab-initio calculation within the Curci-Ferrari model. A first investigation within this double expansion scheme has also been done at finite temperature and finite chemical potential \cite{Maelger:2019cbk}. In particular, at leading order in the double expansion, we find already a critical end-point in the phase diagram, in the same ballpark as certain low energy phenomenological approaches.\footnote{I refer to the thesis of Jan Maelger for more details.}

\clearemptydoublepage
%
%
\appendix
%
%

\chapter{BRST transformations\\ under the functional integral}
\label{chap:BRST}

\section{A simple case as a toy example}
Let us first illustrate the problem and its solution using the case of a one-dimensional numerical integral
\begin{eqnarray}
I=\int_a^b dx\,f(x)\,,
\end{eqnarray}
defined over a certain interval $[a,b]$ of the real axis.

Let us embbed the real axis in a Grassmanian algebra generated by $1$, $\theta$ et $\bar\theta$ and let us consider a change of variables of the form
\begin{eqnarray}
x=\psi(x')+\bar\theta\theta\,\varphi(x')\,,
\end{eqnarray}
where $\psi$ and $\varphi$ denote two numerical functions, with $\psi$ invertible. These functions can be extended over the Grassmanian algebra using a Taylor expansion. In particular, we find that
\begin{eqnarray}
x' & \!\!\!\!\!=\!\!\!\!\! & \psi^{-1}(x)-\bar\theta\theta\,\varphi(x')\frac{d\psi^{-1}}{dx}\nonumber\\
& \!\!\!\!\!=\!\!\!\!\! & \psi^{-1}(x)-\bar\theta\theta\,\varphi(\psi^{-1}(x))\frac{d\psi^{-1}}{dx}\,.
\end{eqnarray}
This shows that, when $\varphi\neq 0$, the new variable $x'$ cannot be numerical and it is a priori not clear how the change of variables should be applied under the integral.

We now argue that, if the following condition is satisfied
\begin{eqnarray}\label{eq:boundary}
\varphi(\psi^{-1}(b))f(b)-\varphi(\psi^{-1}(a))f(a)=0\,,
\end{eqnarray} 
one can proceed as if $x'$ were real and write
\begin{eqnarray}
I=\int_{\psi^{-1}(a)}^{\psi^{-1}(b)}dx'\,\left(\frac{d\psi}{dx'}+\bar\theta\theta\,\frac{d\varphi}{dx'}\right) f(\psi(x')+\bar\theta\theta\,\varphi(x'))\,.
\end{eqnarray}
To see this, let us start from the announced result and write
\begin{eqnarray}
& & \int_{\psi^{-1}(a)}^{\psi^{-1}(b)}dx'\,\left(\frac{d\psi}{dx'}+\bar\theta\theta\,\frac{d\varphi}{dx'}\right) f(\psi(x')+\bar\theta\theta\,\varphi(x'))\nonumber\\
& & \hspace{0.5cm}=\int_{\psi^{-1}(a)}^{\psi^{-1}(b)}dx'\,\left(\frac{d\psi}{dx'}+\bar\theta\theta\,\frac{d\varphi}{dx'}\right) \left(f(\psi(x'))+\bar\theta\theta\,\varphi(x')\left.\frac{df}{dx}\right|_{x=\psi(x')}\right)\nonumber\\
& & \hspace{0.5cm}=\int_{\psi^{-1}(a)}^{\psi^{-1}(b)}dx'\,\frac{d\psi}{dx'}\,f(\psi(x'))+\bar\theta\theta\int_{\psi^{-1}(a)}^{\psi^{-1}(b)}dx' \frac{d}{dx'}\Big(\varphi(x') f(\psi(x'))\Big)\nonumber\\
& & \hspace{0.5cm}=\int_a^b dx\,f(x)\,,
\end{eqnarray}
where we have used the condition (\ref{eq:boundary}).

\section{The case of a BRST transformations}
When a BRST transformation $\varphi=\varphi'+\bar\theta s\varphi'$ is applied under the functional integral, it is usually assumed that (`sdet' stands for the superdeterminant)
\begin{eqnarray}\label{eq:sdet}
\int{\cal D}\varphi\, {\cal I}[\varphi]=\int{\cal D}\varphi' \mbox{sdet}\left(\mathds{1}+\frac{\delta \bar\theta s\varphi'}{\delta\varphi'}\right)\, {\cal I}[\varphi'+\bar\theta s\varphi']\,,
\end{eqnarray}
where the fields that are originally numerical are assumed to remain numerical, despite the fact that, for such fields, $\varphi-\varphi'=\bar\theta s\varphi$ is non numerical. Let us now analyze under which conditions, the above identity applies. To this purpose, we shall start from the RHS of and try to go back to the LHS.

One has ($q_m=0$ for any bosonic omponent $\varphi_m$ and $q_m=1$ for any fermionic component)
\begin{eqnarray}
& & \int{\cal D}\varphi' \mbox{sdet}\left(\mathds{1}+\frac{\delta \bar\theta s\varphi'}{\delta\varphi'}\right)\, {\cal I}[\varphi'+\bar\theta s\varphi']\nonumber\\
& & \hspace{0.2cm}=\int{\cal D}\varphi' \left(1+\underbrace{\int_x(-1)^{q_m} \frac{\delta \bar\theta s\varphi'_m(x)}{\delta\varphi'_m(x)}}_{\mbox{\tiny supertrace}}\right)\left({\cal I}[\varphi']+ \int_x \bar\theta s\varphi'_m(x)\frac{\delta {\cal I}}{\delta\varphi'_m(x)} \right)\nonumber\\
& & \hspace{0.2cm}= \int{\cal D}\varphi'\,{\cal I}[\varphi']+\int{\cal D}\varphi'\left(\int_x (-1)^{q_m}\frac{\delta \bar\theta s\varphi'_m(x)}{\delta\varphi'_m(x)}\,{\cal I}[\varphi']+\int_x \bar\theta s\varphi'_m(x)\frac{\delta {\cal I}}{\delta\varphi'_m(x)}\right)\nonumber\\
& & \hspace{0.2cm}= \int{\cal D}\varphi'\,{\cal I}[\varphi']+\bar\theta\int{\cal D}\varphi' \int_x \frac{\delta}{\delta\varphi'_m(x)} \Big(s\varphi'_m(x)\,{\cal I}[\varphi']\Big)\,.\nonumber\\
\end{eqnarray}
The identity (\ref{eq:sdet}) is then valid if
\begin{eqnarray}
0=\int{\cal D}\varphi' \int_x \frac{\delta}{\delta\varphi'_m(x)} \Big(s\varphi'_m(x)\,{\cal I}[\varphi']\Big)\,,
\end{eqnarray}
which states the vanishing of the volume integral of a superdivergence (over a space made of numerical and Grassmanian fields). 

Using Stoke's theorem, we should be able to relate this to the flux of a certain current through the boundary, which would be the generalization of (\ref{eq:boundary}). To see how this works in the case where the volume contains Grassmanian directions, let us consider a simple exemple with four variables $x,y,\theta,\bar\theta$. The integral over the superspace of the superdivergence of four component field writes
\begin{eqnarray}
\int dxdyd\theta d\bar\theta \left(\frac{\partial f_x}{\partial x}+\frac{\partial f_y}{\partial y}+\frac{\partial f_\theta}{\partial\theta}+\frac{\partial f_{\bar\theta}}{\partial\bar\theta}\right)=\int dxdy\left(\frac{\partial^3 f_x}{\partial x\partial\theta\partial\bar\theta}+\frac{\partial^3 f_y}{\partial y\partial\theta\partial\bar\theta}\right),\nonumber\\
\end{eqnarray}
where we used $\int d\theta \partial/\partial\theta=0$. We obtain a standard volume integral of the divergence of a current field with components $\partial^2 f_x/\partial\theta\partial\bar\theta$ et $\partial^2 f_y/\partial\theta\partial\bar\theta$. The final result is then the flux of such vector field through the boundary of the standard volume. 

Going back to the case of Yang-Mills theory, in Chap.~\ref{chap:intro}, we applied a BRST change of variables to the integral
\begin{equation}
\int {\cal D}[A,c,\bar c,h]\,{\cal O}[A]\,\bar c\,\frac{\delta F[A]^a(x)}{\delta\lambda}e^{-S_{\mbox{\tiny FP}}[A,c,\bar c,h]}\,.
\end{equation}
In order for this change of variables to be justified, we need to show the flux of
\begin{eqnarray}
{\cal O}[A]\frac{\delta F[A]^a(x)}{\delta\lambda} \int {\cal D}[c,\bar c]\,D_\mu c\,\bar c\,e^{-S_{\mbox{\tiny FP}}[A,c,\bar c,h]}
\end{eqnarray}
vanishes through the boundary of the integration domain for the gauge-field. In fact this current is exponentially suppressed at the boundary due to the factor $e^{-S_{YM}[A]}$.\footnote{There are no flat directions due to the gauge-fixing.}
\clearemptydoublepage
%
%
%

\chapter{The su($N$) Lie algebra}
\label{app:sun}

The special unitary group SU($N$) is the group of complex $N\times N$ unitary matrices with unit determinant. The corresponding su($N$) Lie algebra is obtained by considering infinitesimal elements $U=\mathds{1}+X$ and imposing the two conditions $UU^\dagger=\mathds{1}$ and $\mbox{det}\,U=1$. One finds $X+X^\dagger=0$ and $\mbox{tr}\,X=0$, which means that su($N$) is the space of traceless anti-hermitian matrices. Its dimension is $N^2-1$ and a commonly used basis is provided by the matrices $iH_j$ ($1\leq j\leq N-1$), $iX^{jj'}$ and $iY^{jj'}$ ($1\leq j<j'\leq N$), with
\begin{equation}
H_j\equiv \frac{1}{\sqrt{2j(j+1)}}\left(
\begin{array}{ccc}
\mathds{1}_j &  & \\
& -j & \\
& & 0_{N-1-j}
\end{array}
\right)\,,
\end{equation}
and
\begin{eqnarray}
X^{jj'}_{kk'} & \!\!\!\!\!=\!\!\!\!\! & \frac{1}{2}\big(\delta_{jk}\delta_{j'k'}+\delta_{jk'}\delta_{j'k}\big)\,,\\
Y^{jj'}_{kk'} & \!\!\!\!\!=\!\!\!\!\! & -\frac{i}{2}\big(\delta_{jk}\delta_{j'k'}-\delta_{jk'}\delta_{j'k}\big)\,.
\end{eqnarray}
With this choice, the trace of the square of any basis element is $-1/2$ and the trace of the product of two distinct basis elements is $0$. In what follows, we review various important notions related to the su($N$) algebra.

\section{Defining weights of su($N$)}
The above choice of basis shows explicitly that some of the generators of the su($N$) algebra can be diagonalized simultanously. This can be rewritten formally as
\begin{equation}
H_j|\rho^{(k)}\rangle=\rho^{(k)}_j|\rho^{(k)}\rangle\,,\,\,\forall 1\leq k\leq N\,,\,\,\forall 1\leq j\leq N-1\,,
\end{equation}
where the $N$ kets $|\rho^{(k)}\rangle$ define a basis of eigenstates in the space of the defining representation of the algebra. For each eigenstate $|\rho^{(k)}\rangle$, the various eigenvalues $\rho^{(k)}_j$ are conveniently gathered into a $(N-1)$-dimensional real vector $\rho^{(k)}$ that also serves labelling the eigenstate. The $\rho^{(k)}$ are referred to as the weights of the defining representation or {\it defining weights} for short. We mention that these vectors sum up to one since 
\begin{equation}
\sum_{k=1}^N\rho^{(k)}_j=\sum_{k=1}^N \big(H_j\big)_{kk}=\mbox{tr}\,H_j=0\,.
\end{equation}
Similarly, 
\begin{equation}
\sum_{k=1}^N \rho^{(k)}_j\rho^{(k)}_{j'}=\mbox{tr}\,H_j\,H_{j'}=\frac{1}{2}\delta_{jj'}\,,
\end{equation}
from our normalization convention.

It is easy to compute the norms of the defining weights and also the angles between various such weights. To this purpose we note that
\begin{equation}
\rho^{(k)}_j=(H_j)_{kk}=\frac{1}{\sqrt{2j(j+1)}}\times\left\{
\begin{array}{l}
0\,,\,\,\mbox{for } k>j+1\\
-j\,,\,\,\mbox{for } k=j+1\\
1\,,\,\,\mbox{for } k\leq j
\end{array}\right.
\end{equation}
and write
\begin{eqnarray}
\rho^{(k)}\cdot\rho^{(k)} & \!\!\!\!\!=\!\!\!\!\! & \sum_{j=1}^{k-1} \big(H_j\big)_{kk}^2+\sum_{j=k}^{N-1} \big(H_j\big)_{kk}^2\\
& \!\!\!\!\!=\!\!\!\!\! & \frac{(k-1)^2}{2(k-1)k}+\sum_{j=k}^{N-1}\frac{1}{2j(j+1)}=\frac{1}{2}\left(1-\frac{1}{N}\right).\nonumber
\end{eqnarray}
Similarly, for $k<k'$,
\begin{eqnarray}
\rho^{(k)}\cdot\rho^{(k')} & \!\!\!\!\!=\!\!\!\!\! & \sum_{j=1}^{k-1} \big(H_{j}\big)_{kk}\big(H_{j}\big)_{k'k'}\\
& \!\!\!\!\!+\!\!\!\!\! & \sum_{j=k}^{k'-1} \big(H_{j}\big)_{kk}\big(H_{j}\big)_{k'k'}\nonumber\\
& \!\!\!\!\!+\!\!\!\!\! & \sum_{j=k'}^{N-1} \big(H_{j}\big)_{kk}\big(H_{j}\big)_{k'k'}\nonumber\\
& \!\!\!\!\!=\!\!\!\!\! & -\frac{(k'-1)}{2(k'-1)k'}+\sum_{j=k'}^{N-1}\frac{1}{2j(j+1)}=-\frac{1}{2N}\,.\nonumber
\end{eqnarray}
The defining weights of the su($N$) algebra are thus $N$ equal norm vectors of $\mathds{R}^{N-1}$ such that the angles between any two of these vectors are also all equal.

Finally, it is easily seen that, except for the full set, any subset of defining weights is linearly independent. To prove this, we choose, without loss of generality, the first $n$ weights $\rho^{(1)}$, \dots, $\rho^{(n)}$ and evaluate the Gram matrix
\begin{equation}
G_n\equiv\big(\rho^{(k)}\cdot\rho^{(l)}\big)_{(k,l)}=\frac{1}{2}\left(\mathds{1}_n-\frac{1_n}{N}\right)\,,
\end{equation}
where $1_n$ denotes the $n\times n$ matrix with all components equal to $1$. The chosen set of weights is linearly independent iff the Gram matrix is invertible. We then just need to evaluate the determinant of the Gram matrix as
\begin{eqnarray}
\ln \mbox{det}\,(2G_n) & \!\!\!\!\!=\!\!\!\!\! & \mbox{tr}\,\ln (2G_n)=-\mbox{tr}\,\sum_{k=1}^\infty \frac{1}{k}\left(\frac{1_n}{N}\right)^k\\
& \!\!\!\!\!=\!\!\!\!\! & -\sum_{k=1}^\infty \frac{1}{k}\left(\frac{n}{N}\right)^k=\ln\left(1-\frac{n}{N}\right)\,,\nonumber
\end{eqnarray}
where we have used that $1_n^k=n^{k-1}1_n$ for $k\geq 1$. Then, $\mbox{det}\,(2G_n)=1-n/N$, and, as announced, except for the full set ($n=N$), any subset of defining weights is linearly independent. In particular, any set of $N-1$ defining weights forms a basis of $\mathds{R}^{N-1}$.

\section{Roots of su($N$)}
The notion of weight is not restricted to the defining representation but applies in fact to any representation of the algebra, in particular to the adjoint representation $X\mapsto \mbox{ad}_X\equiv [X,\,\_\,]$. Since $[\mbox{ad}_{H_{j}},\mbox{ad}_{H_{k}}]=\mbox{ad}_{[H_{j},H_{k}]}=0$, it is again meaningfull to try to diagonalize simultanously all the generators $\mbox{ad}_{H_{j}}$. In fact, we know already $N-1$ eigenstates since 
\begin{equation}
{\mbox ad}_{H_{j}}H_{k}=[H_{j},H_{k}]=0\,.
\end{equation} 
The corresponding weights are all equal to the nul vector of $\mathds{R}^{N-1}$. We are thus left with determining the remaining eigenstates $E_\alpha$ such that 
\begin{equation}
[H_j,E_\alpha]=\alpha_j E_\alpha\,.
\end{equation} 
The non-zero weights $\alpha$ are referred to as the {\it roots} of the algebra and the diagonalizing basis $\{iH_j,iE_\alpha\}$ is known as a Cartan-Weyl basis.

To find the elements $E_\alpha$, consider the matrices $Z_\pm^{jj'}\equiv X^{jj'}\pm iY^{jj'}$, with $j<j'$. They are such that
\begin{eqnarray}
\big(Z_+\big)^{jj'}_{kk'}=\delta_{jk}\delta_{j'k'} \quad \mbox{and} \quad \big(Z_-\big)^{jj'}_{kk'}=\delta_{jk'}\delta_{j'k}\,.
\end{eqnarray}
Therefore,
\begin{eqnarray}
& & \big(H_{j}\big)_{\alpha\beta}\big(Z_+\big)^{kk'}_{\beta\gamma}-\big(Z_+\big)^{kk'}_{\alpha\beta}\big(H_{j}\big)_{\beta\gamma}\\
& & \hspace{2.9cm}=\,\big[\big(H_{j}\big)_{\alpha\alpha}-\big(H_{j}\big)_{\gamma\gamma}\big]\,\big(Z_+\big)^{kk'}_{\alpha\gamma}\nonumber\\
& & \hspace{2.9cm}=\,\big[\big(H_{j}\big)_{kk}-\big(H_{j}\big)_{k'k'}\big]\,\big(Z_+\big)^{kk'}_{\alpha\gamma}\,,\nonumber
\end{eqnarray}
and
\begin{eqnarray}
& & \big(H_{j}\big)_{\alpha\beta}\big(Z_-\big)^{kk'}_{\beta\gamma}-\big(Z_-\big)^{kk'}_{\alpha\beta}\big(H_{j}\big)_{\beta\gamma}\\
& & \hspace{2.9cm}=\,\big[\big(H_{j}\big)_{\alpha\alpha}-\big(H_{j}\big)_{\gamma\gamma}\big]\,\big(Z_-\big)^{kk'}_{\alpha\gamma}\nonumber\\
& & \hspace{2.9cm}=\,\big[\big(H_{j}\big)_{k'k'}-\big(H_{j}\big)_{kk}\big]\,\big(Z_-\big)^{kk'}_{\alpha\gamma}\,.\nonumber
\end{eqnarray}
This shows that the $Z_\pm^{kk'}$ (with $k<k'$) are nothing but the sought after $E_\alpha$ and the corresponding roots appear as differences of the defining weights $\alpha^{(kk')}=\rho^{(k)}-\rho^{(k')}$.\footnote{This comes as no surprise since the adjoint representation is found when decomposing the the tensor product of the defining representation and the corresponding contragredient representation, after symmetrization and elimination of the singlet trace. Since the weights of the contragredient representation are opposite to the weights of the defining representation, it is no doubt that the roots appear as differences of the defining weights.} 

We have
\begin{equation}
\alpha^{(kk')}\cdot\alpha^{(kk')}=(\rho^{(k)})^2+(\rho^{(k')})^2-2\rho^{(k)}\cdot\rho^{(k')}= 1-\frac{1}{N}+\frac{1}{N}=1\,,
\end{equation}
where we have used that $k\neq k'$. This means that the roots are all of norm unity. Scalar products between roots can be computed in a similar way, but, contrary to the defining weights, they depend on the chosen pair of roots. Finally, we have 
\begin{equation}
\sum_{k\neq k'} \alpha^{(kk')}_j\alpha^{(kk')}_{j'}=\sum_{k,k'} \Big(\rho^{(k)}_j\rho^{(k)}_{j'}-2\rho^{(k)}_j\rho^{(k')}_{j'}+\rho^{(k')}_j\rho^{(k')}_{j'}\Big)=N\delta_{jj'}\,,
\end{equation}
which is nothing but the Casimir of the adjoint representation.

\section{Relations between roots and weights}
Next, we consider the scalar product between a defining weight $\rho^{(k)}$ and a root $\alpha^{(k'k'')}$, with $k'<k''$. If $k=k'$, we have
\begin{equation}
\rho^{(k)}\cdot\alpha^{(k'k'')}=(\rho^{(k)})^2-\rho^{(k)}\cdot\rho^{(k'')}=\frac{1}{2}\left(1-\frac{1}{N}\right)+\frac{1}{2N}=\frac{1}{2}\,,
\end{equation}
while, if $k\neq k'$ and $k\neq k''$, we have
\begin{equation}
\rho^{(k)}\cdot\alpha^{(k'k'')}=\rho^{(k)}\cdot\rho^{(k')}-\rho^{(k)}\cdot\rho^{(k'')}=0\,.
\end{equation}
Consider then $N-1$ of the defining weights. Without loss of generality, we can choose $\rho^{(1)},\dots,\rho^{(N-1)}$. As we have seen above, they form a basis of the space and generate the remaining weight $\rho^{(N)}$ as a linear combination with integer coefficients. Next, consider the $N-1$ roots $\alpha^{(1)}\equiv\alpha^{1N},\dots,\alpha^{N-1}\equiv\alpha^{(N-1)N}$ which also generate the other roots as linear combinations with integer coefficients. From the above equations, it is easily seen that $4\pi\alpha^{(1)}$, \dots, $4\pi\alpha^{(N-1)}$ is a basis dual to 
$\rho^{(1)},\dots,\rho^{(N-1)}$ in the sense defined in Chapter \ref{chap:symmetries}. Similarly $4\pi\rho^{(1)},\dots,4\pi\rho^{(N-1)}$ is a basis dual to $4\pi\alpha^{(1)},\dots,4\pi\alpha^{(N-1)}$. 

These remarks are very useful because, as we have seen in Chapter \ref{chap:symmetries}, generic (resp. periodic) winding transformations are generated by vectors that generate a lattice dual to the one generated by the roots (resp. by the defining weights). Equivalently, we can now say that, up to a factor $4\pi$, generic (resp. periodic) winding transformations are generated by the defining weights (resp. by the roots). Moreover, the center element associated to a given weight $4\pi\rho^{(k)}$ is $e^{4\pi i\rho^{(k)}\cdot\rho^{(l)}}=e^{-i\frac{2\pi}{N}}$ and depends neither on $l$ nor on $k$. It can be convenient instead to find a generating set of winding transformations that covers all the center elements. This set is provided by $4\pi\rho^{(1)}$, $4\pi(\rho^{(1)}+\rho^{(2)}),\dots,4\pi(\rho^{(1)}+\cdots \rho^{(N-1)})=-4\pi\rho^{(N)}$.

\section{Complexified algebra and Killing form}
We mention that the eigenstates $iE_\alpha$ do not belong to the original (real) Lie algebra, but rather to the complexified algebra. The reason why the complexification is needed is that the operators $\mbox{ad}_{iH_j}$ are not symmetric with respect to the Killing form $(X_1;X_2)\mapsto -2\mbox{tr}\,X_1X_2$, and thus not necessarily diagonalizable over the original algebra. They become however anti-hermitian with respect to an extended version of the Killing form, and therefore diagonalizable over the complexified algebra.

To see this in more details, consider first the Killing form over the original Lie algebra. First, because
\begin{equation}
(X;X)=2\mbox{tr}\,XX^\dagger=2\sum_a |X_a|^2\,,
\end{equation}
the Killing form defines a non-degenerate, positive definite symmetric form over the original Lie algebra. Moreover, it obeys  the cyclicity property
\begin{equation}
(X_1;\mbox{ad}_{X_2}X_3)=(X_2;\mbox{ad}_{X_3}X_1)=(X_3;\mbox{ad}_{X_1}X_2)\,,
\end{equation}
which implies in particular $(X_1;\mbox{ad}_{X_2}X_3)=-(\mbox{ad}_{X_2}X_1;X_3)$. This means that the operator $\mbox{ad}_{X_2}$ is anti-symmetric, so not diagonalizable a priori over the original Lie algebra. 

To turn this antisymmetric operator into an anti-hermitian one, we now define the complexified algebra as the space of pairs $Z=(X,Y)$, which we denote also as $Z=X+IY$, equiped with the extended Lie bracket $[X_1+IY_1,X_2+IY_2]\equiv [X_1,X_2]-[Y_1,Y_2]+I([X_1,Y_2]+[Y_1,X_2])$. We also define a complex conjugation over the complexified algebra as $\overline{X+IY}\equiv X-IY$. This complex conjugation should not be mistaken with the complex conjuation of matrices. We have in fact $\bar Z=-Z^\dagger$. Finally we define 
\begin{equation}
\langle Z_1;Z_2\rangle\equiv -2\mbox{tr}\,\bar Z_1\,Z_2=\langle \bar Z_2;\bar Z_1\rangle=\langle Z_2;Z_1\rangle^*\,,
\end{equation}
which extends the Killing form over the complexified algebra (since $\bar X=X$ for elements of the original algeba). We have
\begin{equation}
\langle Z;Z\rangle=(X;X)+(Y;Y)
\end{equation}
so the extended Killing form is now hermitian positive definite. Moreover
\begin{equation}
\langle X_1;\mbox{ad}_{X_2}X_3\rangle=\langle\bar X_2;\mbox{ad}_{X_3}\bar X_1\rangle=\langle \bar X_3;\mbox{ad}_{\bar X_1}X_2\rangle\,.
\end{equation}
For an element $X_2=\bar X_2$ of the original Lie algebra, such as $iH_j$, this implies in particular $\langle X_1;\mbox{ad}_{X_2}X_3\rangle=-\langle\mbox{ad}_{X_2}X_1;X_3\rangle$, which means that the operator $\mbox{ad}_{X_2}$ is anti-hermitian and thus diagonalizable, with imaginary eigenvalues. 

This is why the $\mbox{ad}_{H_j}$ can be diagonalized simultanously and the roots are real vectors. Moreover, by conjugating $[H_j,E_\alpha]=\alpha_j E_\alpha$, we find $[H_j,\bar E_\alpha]=-\alpha_j \bar E_\alpha$ which shows why roots come by pairs and tells us that, with an appropriate choice of normalisation we have $\bar E_\alpha=-E^\dagger_\alpha=-E_{-\alpha}$ (this is preciely the choice made above for the $Z^{kk'}_\pm$). Finally, this shows, that the diagonalization basis can be chosen such that $\langle iH_j;iH_k\rangle=\delta_{jk}$, $\langle iH_j,iE_\alpha\rangle=0$ and $\langle E_\alpha;E_\beta\rangle=\delta_{\alpha\beta}$, which implies $(iH_j;iH_k)=\delta_{jk}$, $(iH_j,iE_\alpha\rangle)=0$ and $(iE_\alpha;iE_\beta)=\delta_{-\alpha,\beta}$, owing to $\bar H_j=-H_j$ and our phase convention between $\bar E_\alpha$ and $E_{-\alpha}$.\\

We mention finally that the components of an element $X$ of the original algebra in the Cartan-Weyl basis are not necessarily real. However since $\bar H_j=-H_j$ and $\bar E_\alpha=-E_{-\alpha}$, we find $X=\bar X=i(X_j^* H_j+X_\alpha^*E_{-\alpha})$ and, thus, $X_j^*=X_j$ and $X_\alpha^*=X_{-\alpha}$. This conclusion can also be reached using the hermitian conjugation which is nothing but $Z^\dagger=-\bar Z$.

\clearemptydoublepage
%
%
%

\chapter{H\"older inequality and convexity}
\label{app:convexity}

Consider two positive real numbers $p$ and $q$ such $1/p+1/q=1$ and two functions $f$ and $g$  such that $f^p$ et $g^q$ are integrable. Then, the function $fg$ is also integrable and we have the following inequality :
\begin{equation}
 \int |fg|\leq \left(\int |f|^p\right)^{1/p}\left(\int |g|^q\right)^{1/q}\,,
\end{equation}
known as {\it H\"older's inequality}. The latter can be slightly generalized as follows. Consider a positive function $h$ and assume that $hf^p$ and $hg^q$ are integrable. We can then apply the previous inequality to the functions $h^{1/p}f$ et $h^{1/q}g$. It follows that $hfg$ is integrable (this uses the assumption $1/p+1/q=1$) and
\begin{equation}
 \int h|fg|\leq \left(\int h|f|^p\right)^{1/p}\left(\int h|g|^q\right)^{1/q}\,.
\end{equation}
We assume in what follows that this result applies to functional integrals with positive measures ${\cal D}_+A$:\\
\begin{equation}\label{eq:Hoelder_func}
\int {\cal D}_+A\,\big|f[A]\,g[A]\big|\leq \left(\int {\cal D}_+A\,\big|f[A]\big|^p\right)^{1/p}\left(\int {\cal D}_+A\,\big|g[A]\big|^q\right)^{1/q}\,.
\end{equation}

Let us next consider
\begin{equation}
 e^{W[J;\bar A]}\equiv\int {\cal D}_{\mbox{\scriptsize gf}}\,[A;\bar A]\,e^{-S_{YM}[A]+J\cdot A}\,,
\end{equation}
where we assume ${\cal D}_{\mbox{\scriptsize gf}}\,[A;\bar A]$ to be positive and we have introduced the short-hand notation $J\cdot A\equiv \int d^dx\,\,(J_\mu;A_\mu)$. Given two positive real numbers $\alpha_1$ and $\alpha_2$ such that $\alpha_1+\alpha_2=1$, we can apply Eq.~(\ref{eq:Hoelder_func}) with 
\begin{equation}
 {\cal D}_+A={\cal D}_{\mbox{\scriptsize gf}}\,[A;\bar A]\,e^{-S_{YM}[A]}\,,
\end{equation}
and 
\begin{equation}
 f[A]=e^{\alpha_1J_1\cdot A}\,,\quad g[A]=e^{\alpha_2J_2\cdot A}\,,
\end{equation}
as well as $p=1/\alpha_1$ and $1=1/\alpha_2$. We find
\begin{eqnarray}
e^{W[\alpha_1J_1+\alpha_2J_2;\bar A]} & \!\!\!\!\!=\!\!\!\!\! & \int{\cal D}_{\mbox{\scriptsize gf}}\,[A;\bar A]\,e^{-S_{YM}[A]} e^{\alpha_1 J_1\cdot A}e^{\alpha_2 J_2\cdot A}\nonumber\\
& \!\!\!\!\!\leq\!\!\!\!\! & \underbrace{\left(\int {\cal D}_{\mbox{\scriptsize gf}}\,[A;\bar A]\,e^{-S_{YM}[A]} e^{J_1\cdot A}\right)^{\alpha_1}}_{\big(e^{W[J_{\tiny 1};\bar A]}\big)^{\alpha_{\tiny 1}}}\underbrace{\left(\int {\cal D}_{\mbox{\scriptsize gf}}\,[A;\bar A]\,e^{-S_{YM}[A]}e^{J_2\cdot A}\right)^{\alpha_2}}_{\big(e^{W[J_{\tiny 2};\bar A]}\big)^{\alpha_{\tiny 2}}}.
\end{eqnarray}
This implies
\begin{equation}
 W[\alpha_1 J_1+\alpha_2 J_2;\bar A]\leq \alpha_1 W[J_1;\bar A]+\alpha_2 W[J_2;\bar A]\,,
\end{equation}
which expresses the convexity of $W[J;\bar A]$ with respect to $J$.

\clearemptydoublepage
%
%
%

\chapter{Homogeneity and Isotropy modulo\\ gauge transformations}
\label{app:hom_iso}

This appendix contains unpublished work in collaboration with Marcela Pel\'aez and Nicol\'as Wschebor. The possible background configurations compatible with homogeneity and isotropy of Euclidean space-time are classified in the SU($2$) case. We find the existence of some exotic configurations satisfying these constraints, in addition to the constant temporal backgrounds used throughout the manuscript. The reason why we did not try to publish these results is that these exotic configurations do not seem to lead to a confining phase. However, it would be interesting to extend the analysis to the SU($3$) gauge group to see if there could be exotic confining configurations in this case.\\

We look for periodic Euclidean background configurations $\bar A_\mu(x)$ such that
\begin{itemize}
\item[(P1)] $\forall x,u\in[0,\beta]\times\mathds{R}^3,\,\exists U_u(x)\in {\cal G}_0, \, \bar A_\mu(x+u)=\bar A^{U_u}_\mu(x)$;
\item[(P2)] $\forall x \in[0,\beta]\times\mathds{R}^3,\,\forall R\in SO(3),\,\exists U_R(x)\in {\cal G}_0,\, \left\{\begin{array}{l}
\bar A_0(R^{-1}x)=\bar A^{U_R}_0(x)\\
R_{ij}\bar A_j(R^{-1}x)=\bar A^{U_R}_i(x)
\end{array}\right.\!,$
\end{itemize}
where we introduced the notation $R\,x\equiv (\tau,R\,\vec{x})$. We mention that the formulation of the problem is gauge-invariant in the following sense. If $\bar A_\mu(x)$ is a configuration obeying the properties (P1) and (P2), then for any periodic gauge transformation $U\in {\cal G}_0$, the transformed configuration $\bar A^U_\mu(x)$ also obeys these properties. Indeed, $\bar A_\mu^U(x)$ is periodic and
\begin{eqnarray}
\bar A_\mu^U(x+u) & \!\!\!\!\!=\!\!\!\!\! & U(x+u)\,\bar A_\mu(x+u)\,U^\dagger(x+u)-U(x+u)\,\partial_\mu U^\dagger(x+u)\nonumber\\
& \!\!\!\!\!=\!\!\!\!\! & U(x+u)U_u(x)\,\bar A_\mu(x)\,U^\dagger_u(x)U^\dagger(x+u)\nonumber\\
&  & -\,U(x+u)\,(U_u(x)\,\partial_\mu U^\dagger_u(x))\,U^\dagger(x+u)-U(x+u)\,\partial_\mu U^\dagger(x+u)\nonumber\\
& \!\!\!\!\!=\!\!\!\!\! & U(x+u)U_u(x)\,\bar A_\mu(x)\,U^\dagger_u(x)U^\dagger(x+u)\nonumber\\
&  & -\,U(x+u)U_u(x)\,\partial_\mu (U^\dagger_u(x)U^\dagger(x+u))\nonumber\\
& \!\!\!\!\!=\!\!\!\!\! & \bar A_\mu^V(x)=(\bar A_\mu^U)^{VU^{-1}}(x)\,,
\end{eqnarray}
with $V(x)\equiv U(x+u)U_u(x)$ a periodic gauge transformation. We can treat similarly the case of the spatial rotations. It follows that, in order to find configurations obeying the properties (P1) and (P2), it is enough to look for configurations in a given gauge. We shall consider a convenient choice of gauge below.

\section{Field-strength tensor}
It is convenient to note that the field-strength tensor $\bar F_{\mu\nu}(x)$ corresponding to the looked for backgroundconfigurations must be such that
\begin{itemize}
\item[(P3)] $\forall x,u\in[0,\beta]\times\mathds{R}^3,\,\exists U_u(x)\in {\cal G}_0, \,\\
\bar F_{\mu\nu}(x+u)=U_u(x)\,\bar F_{\mu\nu}(x)\,U^\dagger_u(x)$;
\item[(P4)] $\forall x\in[0,\beta]\times\mathds{R}^3,\, \forall R\in SO(3),\,\exists U_R(x)\in {\cal G}_0,\,\\ 
\vspace{0.2cm}\left\{\begin{array}{l}
R_{ij}\bar F_{j0}(R^{-1}x)=U_R(x)\,\bar F_{i0}(x)\,U^\dagger_R(x)\\
R_{ij}R_{kl}\bar F_{jl}(R^{-1}x)=U_R(x)\,\bar F_{ik}(x)\,U^\dagger_R(x)
\end{array}\right.\!\!.$
\end{itemize}
In a given basis, these properties rewrites
\begin{itemize}
\item[(P3b)] $\forall x,u\in[0,\beta]\times\mathds{R}^3,\,\exists M(u,x)\in SO(3)\,,\\
\bar F^a_{\mu\nu}(x+u)=M_{ab}(u,x)\bar F^b_{\mu\nu}(x)$;
\item[(P4b)] $\forall x\in[0,\beta]\times\mathds{R}^3,\,\forall R\in SO(3),\,\exists {\cal R}(R,x)\in SO(3),\,\\ 
\left\{\begin{array}{l}
R_{ij}\bar F^a_{j0}(R^{-1}x)={\cal R}^{-1}_{ab}(R,x)\bar F^b_{i0}(x)\\
R_{ij}R_{kl}\bar F^a_{jl}(R^{-1}x)={\cal R}^{-1}_{ab}(R,x)\bar F^b_{ik}(x)
\end{array}\right.\!\!.$
\end{itemize}
From property {(P3b)}, we have in particular that $\bar F^a_{\mu\nu}(x)=M_{ab}(x,0)\bar F^b_{\mu\nu}(0)$. So it seems that we could choose a gauge where $\bar F_{\mu\nu}(x)=\bar F_{\mu\nu}(0)$. However, this is not quite true because we are only allowed periodic gauge transformations, and we do not know whether $M(x,0)\equiv M(\tau,\vec{x})$ is periodic. What we know is that $\bar F_{\mu\nu}(\tau+\beta,\vec{x})=\bar F_{\mu\nu}(\tau,\vec{x})$ from which it follows that $M(\tau+\beta,\vec{x})\,\bar F_{\mu\nu}(0)=M(\tau,\vec{x})\,\bar F_{\mu\nu}(0)$ or, equivalently, that $M^{-1}(\tau,\vec{x})\,M(\tau+\beta,\vec{x})\bar F_{\mu\nu}(0)=\bar F_{\mu\nu}(0)$. We can then to distinguish three cases:
\begin{itemize}
\item[$(A)$] If two of the six color vectors $\bar F_{\mu\nu}(0)$ ($0\leq\mu<\nu\leq 3$) are linearly independent, we have $M(\tau+\beta,\vec{x})=M(\tau,\vec{x})$ and we can indeed choose a gauge such that $\bar F_{\mu\nu}(x)=\bar F_{\mu\nu}(0)$.
\item[$(B)$] If $\smash{\bar F^a_{\mu\nu}(0)=\bar f_{\mu\nu}n^a}$, with some of the $\bar f_{\mu\nu}\neq 0$,  we know just that this common color direction $\vec{n}$ is stabilized by $M^{-1}(\tau,\vec{x})\,M(\tau+\beta,\vec{x})$. In other words $M(\tau+\beta,\vec{x})=e^{i\theta(\tau,\vec{x})\vec{n}\cdot\vec{\sigma}}M(\tau,\vec{x})$, for a certain $\theta(\tau,\vec{x})$.
\item[$(C)$] The third case is simply $\bar F_{\mu\nu}(0)=0$. There is no constraint on $M$ but, as in case (A), we have obviously $\bar F_{\mu\nu}(x)=\bar F_{\mu\nu}(0)(=0)$.
\end{itemize}
We refer to these cases respectively as non-degenerate, degenerate and pure gauge. Below, we discuss them separately. But before we do so, let us exploit the property (P4b). Combining two rotations $R$ and $S$, we find
\begin{equation}
S_{ij}R_{jk}\bar F_{k0}^a(R^{-1}S^{-1}x)={\cal R}^{-1}_{ab}(SR,x)\bar F_{i0}^b(x)
\end{equation}
but also
\begin{eqnarray}
S_{ij}R_{jk}\bar F_{k0}^a(R^{-1}S^{-1}x) & \!\!\!\!\!=\!\!\!\!\! & {\cal R}^{-1}_{ab}(R,S^{-1}x)S_{ij}\bar F_{j0}^b(S^{-1}x)\nonumber\\
& \!\!\!\!\!=\!\!\!\!\! & {\cal R}^{-1}_{ab}(R,S^{-1}x){\cal R}^{-1}_{bc}(S,x)\bar F_{i0}^c(x)\,.
\end{eqnarray}
Applying the same argument to the magnetic sector, we arrive at
\begin{equation}
{\cal R}^{-1}_{ab}(SR,x)\bar F_{\mu\nu}^b(x)={\cal R}^{-1}_{ab}(R,S^{-1}x){\cal R}^{-1}_{bc}(S,x)\bar F_{\mu\nu}^c(x)\,.
\end{equation}
In particular, for $x=0$, we find
\begin{equation}
{\cal R}^{-1}_{ab}(SR,0)\bar F_{\mu\nu}^b(0)={\cal R}^{-1}_{ab}(R,0){\cal R}^{-1}_{bc}(S,0)\bar F_{\mu\nu}^c(0)\,.
\end{equation}
We are then lead to distinguish the three same cases as above:
\begin{itemize}
\item[$(A)$] If two of the six color vectors $\bar F_{\mu\nu}(0)$ ($0\leq\mu<\nu\leq 3$) are linearly independent, we have ${\cal R}(SR,0)={\cal R}(S,0){\cal R}(R,0)$, that is ${\cal R}$ is a real, $N^2-1$ dimensional representation of $SO(3)$.
\item[$(B)$] If $\smash{\bar F^a_{\mu\nu}(0)=\bar f_{\mu\nu}n^a}$, with some of the $\smash{\bar f_{\mu\nu}\neq 0}$, we know just that this color direction $\vec{n}$ is stabilized by ${\cal R}(SR,0)\,{\cal R}^{-1}(R,0){\cal R}^{-1}(S,0)$. In other words ${\cal R}(SR,0)=e^{i\theta(R,S)\vec{n}\cdot\vec{\sigma}}{\cal R}(S,0){\cal R}(R,0)$, for a certain $\theta(R,S)$.
\item[$(C)$] The third case is simply $\bar F_{\mu\nu}(0)=0$.
\end{itemize}
Cases $(B)$ and $(C)$ are easily handled, so we consider them first

\section{Degenerate case}
Suppose that $F_{\mu\nu}^a(0)=\bar f_{\mu\nu}n^a$ with some of the $\bar f_{\mu\nu}\neq 0$. We have, for any $R$,
\begin{equation}
R_{ij}\bar  f_{j0}n^a={\cal R}^{-1}_{ab}(R,0)\bar  f_{i0}n^b \quad \mbox{and} \quad R_{ij}R_{kl}\bar  f_{jl}n^a={\cal R}^{-1}_{ab}(R,0)f_{ik}n^b\,.
\end{equation}
This implies that, for any $R$\,,
\begin{equation}
R_{ij}\bar f_{j0}=\bar f_{i0}\,, \quad R_{ij}R_{kl}\bar f_{jl}=\bar f_{ik} \quad \mbox{and} \quad {\cal R}_{ab}(R,0)n^b=n^a\,,
\end{equation}
which is only possible if $\bar f_{\mu\nu}=0$, that is case (C) to be treated now.

\section{Pure gauge case}
If $\bar F_{\mu\nu}(x)=0$, there exists a $U\in {\cal G}$, not necessarily periodic, such that
\begin{equation}
\bar A_\mu(x)=-U(x)\,\partial_\mu U^\dagger(x)\,,
\end{equation}
with $\bar A_\mu$ periodic. Let us mention that, up to an important restriction to be discussed below, any such configuration obeys the properties (P1) and (P2). For instance
\begin{eqnarray}
\bar A_\mu(x+u) & \!\!\!\!\!=\!\!\!\!\! & -U(x+u)\,\partial_\mu U^\dagger(x+u)\nonumber\\
& \!\!\!\!\!=\!\!\!\!\! & -U(x+u)\,U^\dagger(x)\,U(x)\,\partial_\mu (U^\dagger(x)\,U(x)\,U^\dagger(x+u))\nonumber\\
& \!\!\!\!\!=\!\!\!\!\! & -U(x+u)\,U^\dagger(x)\,[U(x)\,\partial_\mu U^\dagger(x)]\,U(x)\,U^\dagger(x+u)\nonumber\\
&  & -\,U(x+u)\,U^\dagger(x)\,\partial_\mu (U(x)\,U^\dagger(x+u))\nonumber\\
& \!\!\!\!\!=\!\!\!\!\! & A^{V_u}(x)\,,
\end{eqnarray}
with $V_u(x)\equiv U(x+u)\,U^\dagger(x)$. One can treat rotations in the same way, with $V_R(x)\equiv U(R^{-1}x)\,U^\dagger(x)$. The important restriction mentioned above is that $V_u(x)$ and $V_R(x)$ should be periodic in time. This implies that
\begin{eqnarray}
& & \forall u,\, U(\tau+\beta+u_0,\vec{x}+\vec{u})\,U^\dagger(\tau+\beta,\vec{x})=U(\tau+u_0,\vec{x}+\vec{u})\,U^\dagger(\tau,\vec{x})\,,\\
& & \forall R,\, \quad U(\tau+\beta,R^{-1}\vec{x})\,U^\dagger(\tau+\beta,\vec{x})=U(\tau,R^{-1}\vec{x})\,U^\dagger(\tau,\vec{x})\,,
\end{eqnarray}
or, equivalently,
\begin{eqnarray}
& & \forall u,\, U^\dagger(\tau+\beta,\vec{x})\,U(\tau,\vec{x})=U^\dagger(\tau+\beta+u_0,\vec{x}+\vec{u})\,U(\tau+u_0,\vec{x}+\vec{u})\,,\\
& & \forall R,\, U^\dagger(\tau+\beta,\vec{x})\,U(\tau,\vec{x})=U^\dagger(\tau+\beta,R^{-1}\vec{x})\,U(\tau,R^{-1}\vec{x})\,.
\end{eqnarray}
It follows that 
\begin{equation}\label{eq:29}
U^\dagger(\tau+\beta,\vec{x})\,U(\tau,\vec{x})=V,
\end{equation} 
for a certain $V\in SU(N)$. Let us now introduce
\begin{equation}
W(\tau,\vec{x})=U(\tau,\vec{x})\,Me^{-i\frac{\tau}{\beta}r_jH_j}
\end{equation}
with $M$ and $r_j$ such that (it is always possible to find such an $M$ and $r_j$)
\begin{equation}
M^\dagger V^\dagger M=e^{i r_jH_j}\,.
\end{equation}
Using Eq.~(\ref{eq:29}), we find
\begin{eqnarray}
W(\tau+\beta,\vec{x}) & \!\!\!\!\!=\!\!\!\!\! & U(\tau+\beta,\vec{x})\,Me^{-ir_jH_j}e^{-i\frac{\tau}{\beta}r_jH_j}\\
& \!\!\!\!\!=\!\!\!\!\! & U(\tau,\vec{x})V^\dagger Me^{-ir_jH_j}e^{-i\frac{\tau}{\beta}r_jHj}\nonumber\\
& \!\!\!\!\!=\!\!\!\!\! & U(\tau,\vec{x})\,MM^\dagger V^\dagger Me^{-ir_jH_j}e^{-i\frac{\tau}{\beta}r_jH_j}\nonumber\\
& \!\!\!\!\!=\!\!\!\!\! & U(\tau,\vec{x})\,Me^{-i\frac{\tau}{\beta}r_jH_j}=W(\tau,\vec{x})\,.\nonumber
\end{eqnarray}
Moreover
\begin{equation}
We^{i\frac{\tau}{\beta}r_jH_j}\partial_\mu(We^{i\frac{\tau}{\beta}r_jH_j})^\dagger=UM\partial_\mu (UM)^\dagger=U\partial_\mu U^\dagger=-\bar A_\mu\,.
\end{equation}
In other words
\begin{eqnarray}
\beta \bar A_\mu^{W^\dagger}=-e^{i\frac{\tau}{\beta}r_jH_j}\beta\,\partial_\mu e^{-i\frac{\tau}{\beta}r_jH_j}=ir_jH_j\delta_{\mu0}\,.
\end{eqnarray}
We have thus shown that the most general periodic pure gauge configurations compatible with the properties (P1) and (P2) belong to the same ${\cal G}_0$-orbits as constant temporal backgrounds.

\section{Non-degenerate case}
This case depends on the considered group. Here we restrict to the SU($2$) gauge group and leave the SU($3$) case for a future study. In this case $\bar F_{\mu\nu}(x)=\bar F_{\mu\nu}(0)$ and ${\cal R}(R,0)$ is a real, three dimensional representation of $SO(3)$. 

If ${\cal R}(R,0)=\mathds{1}$, then
\begin{equation}
R_{ij}\bar F^a_{j0}(x)=\bar F^a_{i0}(x) \quad \mbox{and} \quad R_{ij}R_{kl}\bar F^a_{jl}(x)=\bar F^a_{ik}(x)\,.
\end{equation}
It follows that $\bar F_{i0}^a(x)=0$ and $\bar F_{ij}^a(x)=\gamma\,\delta_{ij}$.  But $\gamma$ should be zero since $\bar F_{ij}^a(x)$ is antisymmetric. This leads to the case (C) that we have already discussed. 

The other possible three dimensional representation is equivalent to the fundamental one. We can choose the color basis such that ${\cal R}(R,0)=R$. We have then
\begin{equation}
R_{ij}R_{ab}\bar F^b_{j0}(x)=\bar F^a_{i0}(x) \quad \mbox{and} \quad R_{ij}R_{kl}R_{ab}\bar F^b_{jl}(x)=\bar F^a_{ik}(x)
\end{equation}
from which we deduce that
\begin{equation}
\bar F_{i0}^a(x)=\alpha\delta_i^a \quad \mbox{and} \quad \bar F_{ij}^a(x)=\beta\varepsilon_{ija}\,.
\end{equation}

We next impose the Bianchi identity $D_\mu \bar F_{\nu\rho}+D_\nu \bar F_{\rho\mu}+D_\rho \bar F_{\mu\nu}=0$. In terms of components, we find two independent identities
\begin{eqnarray}
0 & \!\!\!\!\!=\!\!\!\!\! & \varepsilon_{abc}(\beta \bar A_0^b\varepsilon_{ijc}+\alpha \bar A_i^b \delta_{cj}-\alpha \bar A_j^b\delta_{ci})\,,\\
0 & \!\!\!\!\!=\!\!\!\!\! & \beta\varepsilon_{abc}(\bar A_i^b \varepsilon_{jkc}+\bar A_j^b \varepsilon_{kic}+\bar A_k^b \varepsilon_{ijc})\,.
\end{eqnarray}
The first identity reads
\begin{equation}
0=\beta \bar A_0^b(\delta_{ai}\delta_{bj}-\delta_{aj}\delta_{bi})+\alpha \bar A_i^b \varepsilon_{abj}-\alpha \bar A_j^b\varepsilon_{abi}\,.
\end{equation}
If $i=j$, there is no information. If $i\neq j$, we choose $a=i$ and $b=j$ and find
\begin{equation}
0=\beta \bar A_0^b\,.
\end{equation}
If we choose $a=i$ and $b\neq j$, we find
\begin{equation}
\forall i\neq j,\, 0=\alpha \bar A_i^j\,.
\end{equation}
Finally, if we choose $b=i$ and $a\neq j$, we find
\begin{equation}
\forall i,\, 0=\alpha \bar A_i^i\,.
\end{equation}
The second identity reads
\begin{equation}
0=\beta\,((\bar A_i^k-\bar A_k^i)\delta_{aj}+(\bar A_j^i-\bar A_i^j)\delta_{ak}+(\bar A_k^j-\bar A_j^k)\delta_{ai})\,,
\end{equation}
that is
\begin{equation}
\forall i\neq j,\,0=\beta\,(\bar A_i^j-\bar A_j^i)\,.
\end{equation}
The case $\alpha=\beta=0$ corresponds to the case $(C)$ already described above. If $\alpha\neq 0$ (and $\beta=0$), we have $\bar A_i^a=0$. It follows that $\bar F_{i0}^a=\partial_i \bar A_0^a=\alpha\,\delta_i^a$ and then that $\bar A_0^a(x)=\alpha x^a+\gamma^a(\tau)$. Because $\alpha\neq 0$, these configurations cannot be in the same orbits as the constant temporal backgrounds. Moreover, they obey the property (P2). However, it is easy to convince oneself that they do not obey the property (P1). To see this, let us consider an infinitesimal spatial translation. The property (P1) would read
\begin{eqnarray}
\alpha u^a & \!\!\!\!\!=\!\!\!\!\! & \partial_0\theta^a+\varepsilon^{abc}\theta^b(\alpha x^c+\gamma^c(\tau))\,,\\
0 & \!\!\!\!\!=\!\!\!\!\! &  \partial_i\theta^a\,.
\end{eqnarray}
Taking a derivative with respect to $x_i$ in the first equation, and using the second, we find
\begin{equation}
0=\alpha\varepsilon^{abi}\theta^b\,.
\end{equation}
which implies in any case $\alpha=0$ (since $\theta=0$ also implies $\alpha=0$). Finally, if $\beta\neq 0$, we find $\bar A^a_0=0$ and $\bar A_i^a=\bar A_a^i$. This case is not that interesting because it cannot lead to a confined phase at small temperatures. An example of such configurations is given by $A_0^a=0$ and $A_i^a=\sqrt{\beta}\delta_i^a$, which are easily checked to be translation and rotation invariant modulo gauge transformations.

We mention also that one can use the constraints from charge conjugation (in the pure YM case) and parity invariance. In the SU($2$) case, charge conjugation does not impose any constraint because it is tantamount to a color rotation (Weyl transformation). On the other hand, it is easily checked that partity invariance (modulo color rotations) implies that $\alpha=0$.
\clearemptydoublepage
%
%
%

\chapter{The background field potential\\ in terms of Polyakov loops}
\label{app:fund}

In this appendix, we explain how Eq.~(\ref{eq:123}) can be generalized to the SU($N$) case and how Eq.~(\ref{eq:456}) is derived in the SU($3$) case.

\section{Fields in the defining representation}
Combining the logarithms in Eq.~(\ref{eq:dV}), one arrives at
\begin{eqnarray}
& &  \ln\prod_{j=1}^N\Big[1+e^{-\beta\,(\varepsilon_{q,f}\mp\mu)\pm ir\cdot\rho^{(j)}}\Big]
\\
& & \hspace{1.5cm}=\,\ln \Big[1+\sum_{\nu=1}^N e^{-\nu\beta\,(\varepsilon_{q,f}\mp\mu)}\sum_{j_1<\cdots<j_\nu}e^{\pm i\big(\rho^{(j_1)}+\cdots+\rho^{(j_\nu)}\big)}  \Big].\nonumber
\end{eqnarray}
We have seen in chapter \ref{chap:more} that the fundamental Polyakov loops play a role in the characterization of center symmetry. It is no doubt that they will enter the generalization that we are after. At leading order, they are given by
\begin{eqnarray}
\ell_{C_N^\nu}(r) & \!\!\!\!=\!\!\!\! & \frac{1}{C_N^\nu}\sum_{j_1<\cdots<j_\nu}e^{i\big(\rho^{(j_1)}+\cdots+\rho^{(j_\nu)}\big)\cdot r}\,,\\
\ell_{\overline{C}_N^\nu}(r) & \!\!\!\!=\!\!\!\! & \frac{1}{C_N^\nu}\sum_{j_1<\cdots<j_\nu}e^{-i\big(\rho^{(j_1)}+\cdots+\rho^{(j_\nu)}\big)\cdot r}\,.
\end{eqnarray}
We then arrive at
\begin{eqnarray}
& &  \delta V_{\mbox{\scriptsize 1loop}}^{\mbox{\scriptsize SU(N)}}(\{\ell_{C_N^\nu},\ell_{\overline{C}_N^\nu}\};T,\mu)\\
& & \hspace{0.7cm}=\,-\frac{T}{\pi^2}\!\int_0^\infty \!\!dq\,q^2\Bigg\{\ln\Big[1+\sum_{\nu=1}^N C_N^\nu\ell_{C_N^\nu}e^{-\nu\beta\,(\varepsilon_{q,f}-\mu)}\Big]+\nonumber\\
& & \hspace{4.0cm}+\ln\Big[1+\sum_{\nu=1}^N C_N^\nu\ell_{\overline{C}_N^\nu}e^{-\nu\beta\,(\varepsilon_{q,f}+\mu)}\Big]\Bigg\}.\nonumber
\end{eqnarray}
At vanishing chemical potential, one can restrict to $\ell_{\overline{C}_N^\nu}=\ell_{C_N^\nu}$.

\section{Fields in the adjoint representation}
Combining the logarithms in Eq.~(\ref{eq:V222}), we obtain
\begin{eqnarray}
& &  \ln\prod_{j=1}^{8}\Big[1-e^{-\beta\varepsilon_q+ ir\cdot\kappa^{(j)}}\Big]\\
& & \hspace{1.5cm}=\ln\,(1-e^{-\beta\varepsilon_q})^{2}\prod_{j=1}^{3}\Big[1-e^{-\beta\varepsilon_q+ ir\cdot\alpha^{(j)}}\Big]\prod_{j=1}^{3}\Big[1-e^{-\beta\varepsilon_q- ir\cdot\alpha^{(j)}}\Big],\nonumber
\end{eqnarray}
with $\alpha^{(1)}=\rho^{(2)}-\rho^{(3)}$, $\alpha^{(2)}=\rho^{(3)}-\rho^{(1)}$, $\alpha^{(3)}=\rho^{(1)}-\rho^{(2)}$, and then $\alpha^{(1)}+\alpha^{(2)}+\alpha^{(3)}=0$. We have
\begin{eqnarray}
& & \prod_{j=1}^{3}\Big[1-e^{-\beta\varepsilon_q+ ir\cdot\alpha^{(j)}}\Big]=1-e^{-3\beta\varepsilon_q}-e^{-\beta\varepsilon_q}\sum_{j=1}^3 e^{ir\cdot\alpha^{(j)}}+e^{-2\beta\varepsilon_q}\sum_{j=1}^3e^{-ir\cdot\alpha^{(j)}}\,.\nonumber\\
\end{eqnarray}
Then
\begin{eqnarray}
& & \prod_{j=1}^{3}\Big[1-e^{-\beta\varepsilon_q+ ir\cdot\alpha^{(j)}}\Big]\prod_{j=1}^{3}\Big[1-e^{-\beta\varepsilon_q- ir\cdot\alpha^{(j)}}\Big]\nonumber\\
& & \hspace{1.5cm}=\,(1-e^{-3\beta\varepsilon_q})^2-e^{-\beta\varepsilon_q}(1-e^{-\beta\varepsilon_q})(1-e^{-3\beta\varepsilon_q})(8\ell_8(r)-2)\nonumber\\
& & \hspace{1.9cm}+\,3e^{-2\beta\varepsilon_q}(1+e^{-2\beta\varepsilon_q})(1+\ell_3(3r)+\ell_{\bar 3}(3r))\nonumber\\
& & \hspace{1.9cm}-\,2e^{-3\beta\varepsilon_q}(4\ell_8(2r)+8\ell_8(r)-3)\,,
\end{eqnarray}
where we have used
\begin{eqnarray}
3\ell_3(r) & \!\!\!\!=\!\!\!\! & \sum_{j=1}^3 e^{ir\cdot\rho^{(j)}}\,,\label{eq:l3}\\
8\ell_8(r) & \!\!\!\!=\!\!\!\! & 2+\sum_{j=1}^3 e^{ir\cdot\alpha^{(j)}}+\sum_{j=1}^3 e^{-ir\cdot\alpha^{(j)}}\,.
\end{eqnarray}
Now, using
\begin{equation}
8\ell_8(r)=9\ell_3(r)\ell_{\bar 3}(r)-1\,,
\end{equation}
together with\footnote{These identities are conveniently obtained by expanding $9\ell_3(r)^2$ and $27\ell_3(r)^3$ respectively, using Eq.~(\ref{eq:l3}) and then identifying the various terms of the expansions in terms of $\ell_3(r)$.}
\begin{eqnarray}
\ell_3(2r) & \!\!\!\!=\!\!\!\! & 3\ell_3(r)^2-2\ell_{\bar 3}(r)\,,\\
\ell_3(3r) & \!\!\!\!=\!\!\!\! & 1+9\ell_3(r)^3-9\ell_3(r)\ell_{\bar 3}(r)\,,
\end{eqnarray}
we finally arrive at
\begin{eqnarray}
& &  \ln\prod_{j=1}^{8}\Big[1-e^{-\beta\varepsilon_q+ ir\cdot\kappa^{(j)}}\Big]\\
& & \hspace{1.5cm}=\ln\Big[1+e^{-8\beta\varepsilon_q}-\big(9\ell_3\ell_{\bar 3}-1\big)\big(e^{-\beta\varepsilon_q}+e^{-7\beta\varepsilon_q}\big)\nonumber\\
& & \hspace{2.5cm} -\,\big(81\ell_3^2\ell_{\bar 3}^2-27\ell_3\ell_{\bar 3}+2\big)\big(e^{-3\beta\varepsilon_q}+e^{-5\beta\varepsilon_q}\big)\nonumber\\
& & \hspace{2.5cm} +\,\big(27\ell_3^3+27\ell_{\bar 3}^3-27\ell_3\ell_{\bar 3}+1\big)\big(e^{-2\beta\varepsilon_q}+e^{-6\beta\varepsilon_q}\big)\nonumber\\
& & \hspace{2.5cm} +\,\big(162\ell_3^2\ell_{\bar 3}^2-54\ell_3^3-54\ell_{\bar 3}^3+18\ell_3\ell_{\bar 3}-2\big)e^{-4\beta\varepsilon_q}\Big].
\end{eqnarray}
As in the previous section, it should be possible to generalize this formula to the SU($N$) case, in terms of the fundamental Polyakov loops.

\clearemptydoublepage
%

%
%

\end{document}